\documentclass[11pt,reqno]{amsart}
\usepackage{comment}
\usepackage[left=1in, right=1in, top=1in, bottom=1in]{geometry}
\usepackage{dsfont}
\usepackage{amsthm}
\usepackage{amsfonts}
\usepackage[foot]{amsaddr}
\usepackage{thmtools}
\usepackage{thm-restate}
\usepackage[pdftex]{hyperref}
\usepackage{physics}
\usepackage{amssymb}
\usepackage{amsmath}
\usepackage{cleveref}
\usepackage{xcolor}
\usepackage{graphicx}
\usepackage{xparse}
\usepackage{mathtools}
\usepackage{subcaption}
\PassOptionsToPackage{dvipdfmx}{graphicx}
\usepackage{sid_macros}
\usepackage{enumitem}

\usepackage[toc,page]{appendix}
\usepackage{etoolbox}
\usepackage{tcolorbox}

\usepackage{tikz}
\usetikzlibrary{arrows,backgrounds}
\newcommand{\tikzmath}[2][]
{\vcenter{\hbox{\begin{tikzpicture}[#1]#2\end{tikzpicture}}}
}

\usetikzlibrary{decorations.pathmorphing}

\tikzset{snake it/.style={decorate, decoration=snake}}

\definecolor{azure}{HTML}{007fff}
\definecolor{OliveGreen}{HTML}{6D712E}
\definecolor{boysenberry}{HTML}{873260}
\definecolor{violet}{RGB}{148,0,211}
\definecolor{salmon}{HTML}{ff8c69}
\definecolor{DarkGreen}{RGB}{0,150,0}

\hypersetup{
    colorlinks=true,                            
    linkcolor=azure,                          
    citecolor=OliveGreen,                          
    filecolor=red,                           
    urlcolor=azure                             
}

\newcommand{\DHR}{\mathsf{DHR}}

\newcommand{\nocontentsline}[3]{}
\newcommand{\tocless}[2]{\bgroup\let\addcontentsline=\nocontentsline#1{#2}\egroup}

\theoremstyle{plain}
\newtheorem{thm}{Theorem}[section]

\newtheorem{cor}[thm]{Corollary}
\newtheorem{lem}[thm]{Lemma}
\newtheorem{prop}[thm]{Proposition}

\theoremstyle{definition}
\newtheorem{defn}[thm]{Definition}
\newtheorem{nota}[thm]{Notation}

\newtheorem{facts}[thm]{Facts}
\newtheorem{rem}[thm]{Remark}

\newtheorem{asmp}{Assumption}

\newtcolorbox{thm_border}[1][]{
    colback=white, 
    colframe=cyan, 
    fonttitle=\bfseries, 
    title=Theorem, 
    sharp corners, 
    boxrule=1pt, 
    #1, 
}

\newtcolorbox{setting}[1][]{
    colback=white, 
    colframe=cyan, 
    fonttitle=\bfseries, 
    title=Setting, 
    sharp corners, 
    boxrule=1pt, 
    #1, 
}

\title{An Operator Algebraic Approach to Symmetry
Defects and Fractionalization}
\author{Kyle Kawagoe$^{1}$, Siddharth Vadnerkar$^{2}$, Daniel Wallick$^{3}$}
\address{$^1$Center for Quantum Information Science and Engineering, The Ohio State University, Columbus OH, 43210}
\address{$^{2}$Department of Physics, University of California, Davis, Davis CA, 95616}
\address{$^3$Department of Mathematics, The Ohio State University, Columbus OH, 43210}
\email{kawagoe@umd.edu, vadnerkar.sid@gmail.com, daniel.wallick@itp.uni-hannover.de}
\date{\today}

\begin{document}
\begin{abstract}
We provide a superselection theory of symmetry defects in 2+1D symmetry enriched topological (SET) order in the infinite volume setting. For a finite symmetry group $G$ with a unitary on-site action, our formalism produces a $G$-crossed braided $\rmW^*$-tensor category $\GSec$. This superselection theory is a direct generalization of the usual superselection theory of anyons, and thus is consistent with this standard analysis in the trivially graded component $\GSec_1$. This framework also gives us a completely rigorous understanding of symmetry fractionalization. To demonstrate the utility of our formalism, we compute $\GSec$ explicitly in both short-range and long-range entangled spin systems with symmetry and recover the relevant skeletal data.
\end{abstract}
\maketitle

\tableofcontents

\section{Introduction}
Long range entangled topological orders in 2+1D are characterized by Unitary Modular Tensor Categories (UMTC) which arise from the superselection theory of their emergent anyons. In many cases, this physical proposition has been rigorously verified by using DHR theory from algebraic quantum field theory on infinite lattice models \cite{MR2804555, MR3426207, 2306.13762, 2310.19661}.
Interestingly, this story changes in the presence of a finite on-site symmetry group $G$. The landmark work \cite{PhysRevB.100.115147} gave a physical justification for why this classification is given by $G$-crossed braided categories for $G$-symmetry enriched topological (SET) order. SET models and symmetry fractionalization have been studied extensively in the physics literature \cite{CHEN20173,PhysRevB.65.165113,PhysRevB.74.174423, PhysRevB.94.235136}.
Despite the impact of this work, there is currently no rigorous understanding of how these categories arise from a microscopic bulk analysis.
In particular, these SET models have not been studied before in the infinite volume setting. In this manuscript, we provide a complete formalism detailing how $G$-crossed braided fusion categories arise from a DHR style analysis of the symmetry defects of SET order. We also demonstrate our formalism in concrete examples.

The original DHR formalism comes from \cite{MR297259, MR334742}, building on \cite{MR165864}.
It was constructed to describe relativistic quantum field theory and uses finite regions of spacetime as its local regions.  
This work was later built on in \cite{MR660538} to describe states that are localized in spacelike cones instead of finite regions.  
This latter approach was then adopted to study topologically ordered quantum spin systems, starting with the Toric Code \cite{MR2804555, MR2956822, MR3135456}. 
These methods have been shown to be stable under perturbations \cite{MR4050095, MR4426734, MR4362722} and are thus an important step in understanding topological order in a model-independent way.
More recently, the DHR approach has been used to study anyons in the presence of a $U(1)$ symmetry \cite{2410.04736}. Our paper shares some aspects of their analysis, particularly in the construction of defects. However, many of their techniques and results are specific to $U(1)$ and thus not applicable to our setting since we focus on finite groups.
The DHR approach can also be generalized in the style of \cite{MR1231644}, as shown in \cite{2410.21454}.
Another DHR-inspired approach to topological order has been used in \cite{2304.00068, 2307.12552, MR4814524}.

Following these analyses, we consider a ground state $\omega_0$ of an SET and construct its GNS representation $\pi_0\colon\fA\rightarrow B(\cH)$ for the quasi-local algebra $\fA$. For each $g\in G$, we have a support preserving automorphism $\beta_g\in \Aut(\fA)$ which represents the symmetry action. We take our ground state to be symmetric in the sense that $\omega_0\circ\beta_g=\omega_0$. With such a ground state, we take inspiration from \cite{MR2183964} to define $G$-defect representations. 
These are (loosely speaking) representations housing a half-infinite-line-like symmetry defect, which are in turn truncations of symmetry domain walls analogous to the setup of \cite{else2014classifying} (cf.~discussion in Section \ref{sec:DefectAutomorphismConstructionHeuristic}). Physically, the analysis in \cite{MR2183964} should correspond to the boundary CFT at infinity surrounding the bulk SET which we study. We now state our main results. The main theoretical result is stated precisely in Theorem \ref{thm:main_result} and Corollary \ref{cor:anyon_equiv}.

\begin{thm_border}
The category of $G$-defect representations with respect to $\pi_0$ is a $G$-crossed braided $\rmW^*$-tensor category whose trivially graded component is the braided $\rmW^*$-tensor category of anyon sectors.
\end{thm_border}

This mirrors the prediction of \cite{PhysRevB.100.115147} in the operator algebraic setting. 
As this theorem suggests, their higher cohomological obstructions do not appear since we are considering strictly on-site symmetry in 2+1 dimensions.

We comment that the trivially graded component of this category being the braided tensor category of anyon sectors makes rigorous the well-known physical result that anyons can be thought of as excitations corresponding to trivial symmetry defects.
Thus the concept of a symmetry defect is strictly more general than the concept of an anyon.

We then demonstrate the utility of our formalism by computing this category in a variety of examples exhibiting both short-ranged and long-ranged order.

First, our result applies to a broad class of SPTs and gives us the following result (precisely stated in Theorem \ref{thm:SPT theorem}).

\begin{thm_border}
The category of $G$-defect representations of a $G$-SPT (Definition \ref{def:SPT_phase}) is $G$-crossed braided $\rmW^*$-tensor equivalent to $\Vect{(G, \nu)}$ where $\nu$ is a 3-cocycle. 
\end{thm_border}

We briefly discuss the context of this work and its relationship to the prior literature on SPTs. Our result is a categorical formulation of the well-known conjecture that non-anomalous SPTs with a discrete symmetry group $G$ in $2+1$D are completely classified by a cohomology class in $H^3(G,U(1))$ \cite{PhysRevB.87.155114, else2014classifying}. A cohomological index for $2+1$D SPTs was mathematically obtained in \cite{MR4354127, MR4332955} under much weaker assumptions than ours. 
The complete classification by an $H^3(G,U(1))$-valued index was recently shown to hold in the case of SPTs that satisfy the ``symmetric entangler'' property \cite{bols2026complete}, which is the case that we consider in this paper. 
These authors used boundary algebras \cite{freedman2020classification, haah2023invertible} and the methods developed in \cite{else2014classifying}. Our approach constructs symmetry defect automorphisms and cocycles following the same recipe laid out in \cite{else2014classifying}. However, we are able to use categorical methods since we have the full categorical structure of $G$-defect representations at our disposal.

We then specialize to the case of the Levin-Gu SPT \cite{PhysRevB.86.115109}, which is an example of a non-trivial $\bbZ_2$-SPT. We compute the category of $G$-defect sectors for this model and its skeletal data in the bulk and find in particular that the corresponding $3$-cocycle $\nu$ is in the non-trivial cohomology class. This result is obtained in Proposition \ref{prop:levin gu classification result}.

In addition to \cite{MR2804555, MR2956822, MR3135456}, there are several other works providing complete superselection analyses of infinite lattice models, starting with \cite{MR3426207}, which studies the abelian Quantum Double Model.  
More recently, these methods have been applied to the doubled semion model \cite{2306.13762}, the nonabelian Quantum Double Model \cite{2310.19661, MR4998297}, and the Levin--Wen model \cite{bols2025sector,2603.01936}.
A general treatment of this approach to topological order, using much weaker assumptions than those used in this paper, can be found in \cite{MR4362722}.
We remark that we expect our analysis to work in this more general setting, which uses approximate Haag duality, but we use a stronger assumption to make the analysis easier.

One of the main contributions of our work is a complete defect superselection theory analysis of a symmetry enriched model of the Toric Code. This model is defined in Section \ref{sec:SET Toric Code}. We compute the category of $G$-defect sectors of this model and analyze the resulting skeletal data. The result below is the conclusion of our analysis and shown in Sections \ref{sec:F_symbols}, \ref{sec:frac_data}.

\begin{thm_border}
    The $G$-defect representations of the symmetry-enriched Toric Code model in Section \ref{sec:SET Toric Code} form a $G$-crossed braided fusion category with trivial associators but fractionalization data with the scalars not all equal to 1.\footnote{We specifically show that the scalars are not all equal to 1 for the particular convention that we choose, although we expect that this result is not dependent on the choice of convention.}
\end{thm_border}

After posting the preprint of this manuscript, we were made aware that another research group was nearing completion of a manuscript covering many of the same topics. We encourage the interested reader to also check out \cite{2411.01210}.

Although the examples treated in this paper do not exhibit anyon permutation, the general framework is designed to accommodate such cases. The symmetry-enriched models constructed in \cite{PhysRevB.94.235136, lyons2025protocols} give explicit examples with non-trivial anyon permutation and associated defect operators. It should be easy to verify that these models satisfy our assumptions and thus that the $G$-defect representations will form a $G$-crossed braided category. 
The main difficulty lies in constructing and verifying representative $G$-defect representations and proving the complete classification of these representations, as the defects can no longer be created by finite depth circuits.

This problem is similarly faced when trying to classify anyon sectors of non-abelian Quantum Double or Levin--Wen models \cite{2310.19661, bols2025sector}. We leave the detailed analysis of these models and the explicit construction of this category to future work.

This manuscript is organized as follows. We first propose a selection criterion for representations having symmetry defects, called $g$-defect representations (Definition \ref{def:defect_sector}). In Section \ref{sec:symmetry_defects}, we then build a category of these defect representations and show that it is a $G$-crossed braided $\rmW^*$-tensor category. We then show in Section \ref{sec:coherance_data} that the coherence data of this category matches that of the algebraic theory of symmetry defects already discussed in the literature. We also show in Section \ref{sec:connection_to_anyon_sectors} that when the symmetry is trivial, this selection criterion reduces to that of anyon sectors and is thus a strict generalization.

In Section \ref{sec:general SPTs}, we treat the case of a general SPT built from a finite depth quantum circuit (FDQC), and show that there always exists a commuting projector Hamiltonian whose ground state houses a symmetry defect. As a capstone to this treatment, we verify that the category of $G$-defect sectors for any such SPT is $G$-crossed braided tensor equivalent to $\Vect(G,\nu)$ where $G$ is the underlying symmetry of the SPT state and $\nu$ is a 3-cocycle.

We specialize the treatment of Section \ref{sec:general SPTs} to $\bbZ_2$-SPTs in Sections \ref{sec:TrivialParamagnet} and \ref{sec:Levin Gu}.
In Section \ref{sec:TrivialParamagnet}, we analyze the trivial $\bbZ_2$-paramagnet and show that the category of $G$-defect sectors is $G$-crossed braided tensor equivalent to $\Vect(\bbZ_2)$.
Similarly, in Section \ref{sec:Levin Gu} we verify for the Levin-Gu SPT \cite{PhysRevB.86.115109} that the resulting category of $G$-defect sectors is $G$-crossed braided tensor equivalent to $\Vect(\bbZ_2,\nu)$ for a non-trivial cocycle $\nu$. 
In particular, we explicitly compute the cocycle $\nu$ in the bulk using automorphisms that create a symmetry defect.

In Section \ref{sec:SET Toric Code}, we explore an SET commuting projector model that is obtained from the usual Toric Code using a FDQC. 
We find a completeness result for the $G$-defect sectors of this model and explicitly compute the $F$ and $R$-symbols and the symmetry fractionalization data in the bulk.

We briefly comment on the setting and assumptions for this paper in order to summarize the main results. If the reader is not already familiar with the operator algebra formalism, a brief introduction is provided in Appendix \ref{sec:background}.

\section{Setting and main results}

\begin{figure}[!htbp]
        \centering
    \includegraphics[width=0.2\linewidth]{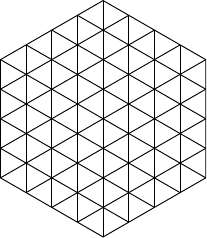}
        \caption{One commonly used lattice is the triangular lattice where a site is placed on each vertex.}
            \label{fig:example_triangular_lattice}

\end{figure}

Let $\Gamma$ be a discrete subset of points in $\mathbb{R}^2$ such that for every ball $B \subseteq \bbR^2$, the set $B \cap \Gamma$ is finite. 
We will refer to $\Gamma$ as the lattice and its elements as sites. In examples, we often associate these sites with cells in a cellulation of $\mathbb{R}^2$, such as the triangular lattice depicted in Figure \ref{fig:example_triangular_lattice}. When the sites in our examples are represented by edges or faces of a cellulation, we will take care to specify the exact locations of the sites as points in the plane.

Given a subset $\Sigma \subset \Gamma$, we denote by $\Sigma^c \subset \Gamma$ the complement of $\Sigma$ in $\Gamma$.
A (closed) \emph{geometric cone} $\Delta \subseteq \bbR^2$ is a subset of the form 
\[
\Delta
\coloneqq
\{x \in \bbR^2 : (x - a) \cdot \hat{v} \geq \|x - a\| \cos(\theta/2)\}.
\]
Here $a \in \bbR^2$ is the apex of the geometric cone, $\hat{v} \in \bbR^2$ is a unit vector specifying the axis of the cone, and $\theta \in (0, 2\pi)$ is the opening angle of the cone. We define a \emph{cone in $\Gamma$} to be a subset $\Lambda \subseteq \Gamma$ of the form $\Lambda = \Gamma \cap \Delta$, where $\Delta \subseteq \bbR^2$ is a (closed) geometric cone.\footnote{More often, open cones are used in this definition instead of closed cones (see for instance \cite{MR4362722, 2410.04736, 2511.08382}).
However, we use closed cones here, since in Remark \ref{rem:MinimalGeometricCone} we use that the arbitrary intersection of closed geometric cones is a closed geometric cone.} 
We also refer to cones in $\Gamma$ simply as \emph{cones}.

To each site $s \in \Gamma$, we associate a Hilbert space $\hilb_s = \bbC^{d_s}$, where each $d_s \in \bbN$ is greater than $1$. 
Let $\Gamma_f$ be the set of finite subsets of $\Gamma$. 
We can then define the tensor product over a finite set of sites $S \in \Gamma_f$ as $\hilb_{S} \coloneqq \bigotimes_{s \in S} \hilb_s$. Then $\cstar[S] \coloneqq B(\hilb_S)$ is a $\rmC^*$-algebra.

Now let $S,S' \in \Gamma_f$ be such that $S \subset S'$. Then we can define the canonical inclusion $\cstar[S] \hookrightarrow \cstar[S']$ by tensoring with the identity element on all $s \in S' \setminus S$. With this we can define the algebra of local observables $\cstar^{\loc}$ as $$\cstar^{\loc} \coloneqq \bigcup_{S \in \Gamma_f} \cstar[S]$$ and its norm completion, $$\cstar \coloneqq \overline{\cstar^{\loc}}^{||\cdot ||}$$ This algebra is known as the algebra of quasi-local observables, or simply, the \emph{quasi-local algebra}.

This algebra, as the name suggests, is the algebra whose elements can be approximated by strictly local observables (which are observables that act differently than the identity only on a finite subset $S \in \Gamma_f$). We say the \emph{support} of an observable $A \in \cstar$ is the smallest set $\Sigma \subset \Gamma$ such that $A \in \cstar[\Sigma]$, and we denote the support of $A$ by $\supp(A)$.

We note that we can define a quasi-local algebra $\cstar[\Sigma]$ on any (not necessarily finite) subset $\Sigma \subseteq \Gamma$ by first replacing $\Gamma$ with $\Sigma$ and then using the above procedure. We will use this fact primarily when talking about the quasi-local algebra $\cstar[\Lambda]$ of a cone $\Lambda$.

We assume that there is a symmetry action of a group $G$ on $\cstar$, i.e, a faithful homomorphism $\beta \colon G\to \Aut(\cstar)$ given by $g \mapsto \beta_g$ for all $g \in G$.
We call $\beta_g$ a \emph{symmetry automorphism}. In the cases we consider, the symmetry action is \emph{on-site}, i.e, for each $s \in \Gamma$, we assume that there is an action of $G$ on each $\hilb_s$ by unitaries $U^g_s$ acting on the site $s$. 

\begin{defn}
\label{def:GlobalSymmetryAutomorphism}
For each $A \in \cstar[S]$ with $S \in \Gamma_f$, we let $\beta_g\colon \cstar[S] \rightarrow \cstar[S]$ be the map defined by $$\beta_g (A) \coloneqq \left(\bigotimes_{s \in S} U_s^g\right) A \left(\bigotimes_{s \in S} U_s^{g}\right)^*.$$ 
We observe that this map can be uniquely extended in a norm continuous way to an automorphism $\beta_g$ acting on the whole of $\cstar$.
\end{defn}

We also sometimes consider situations where the symmetry only acts on a subset of the lattice. 

\begin{defn}
\label{def:SymmetryOnSubsets}
For any (not necessarily finite) $S \subseteq \Gamma$, we let $\beta_g^S \colon \fA \to \fA$ be the map defined by 
\[
\beta_g^S(A)
\coloneqq
\left(\bigotimes_{s \in S} U_s^g\right) A \left(\bigotimes_{s \in S} U_s^g\right)^*,
\]
More precisely, one constructs $\beta_g^S \colon \fA \to \fA$ using the method used to construct $\beta_g$. An example symmetry action is shown in Figure \ref{fig:example_symmetry_action}.

\begin{figure}[!ht]
    \centering
    \includegraphics[width=0.2\linewidth]{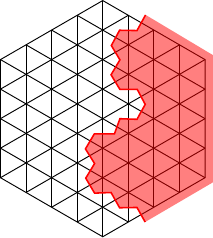}
    \caption{An example symmetry action $\beta_g^S$ on the triangular lattice, with $S$ being the region colored in red. On all sites $s$ in the red region, the symmetry acts as $U^g_s$, and $\mathds1_s$ otherwise.}
    \label{fig:example_symmetry_action}
\end{figure}
\end{defn}

\subsection{Setup and General Assumptions}
\label{sec:GCrossedAssumptions}

Fix a reference state $\omega_0$ and denote by $\pi_0\colon \cstar \rightarrow B(\hilb_0)$ its GNS representation. We now detail the assumptions that we will impose on the action by the group $G$ and on the state $\omega_0$ to ensure we obtain a $G$-crossed braided $\rmW^*$-tensor category.  

\begin{asmp}
\label{asmp:Faithfulness}
There is a fixed $n > 0$ such that for all balls $B \subseteq \mathbb{R}^2$ of radius at least $n$, the representation of $G$ given by $g \mapsto \bigotimes_{s \in B\cap\Gamma} U^g_s$ is faithful.
\end{asmp}

Note that the faithfulness assumption implies that if $\beta_g|_{\cstar[B]} = \beta_h|_{\cstar[B]}$ for any large enough finite region $B \subseteq \Gamma$, then $g = h$.
Here $\beta_g \colon \fA \to \fA$ is the symmetry automorphism from Definition \ref{def:GlobalSymmetryAutomorphism}.

We now detail our assumptions on the chosen state $\omega_0 \colon \fA \to \bbC$.

\begin{asmp}
\label{asmp:GInvariance}
The reference state $\omega_0$ is \emph{$G$-symmetric}, that is, $\omega_0 \circ \beta_g = \omega_0$ for all $g \in G$.
\end{asmp}

We observe that $G$-symmetry of $\omega_0$ implies that the map $\beta_g \colon \fA \to \fA$ is implemented by a unitary in $B(\cH_0)$, where $\cH_0$ is the usual GNS Hilbert space for $\omega_0$.  
Hence $\beta_g$ extends to a WOT-continuous automorphism of $B(\cH_0)$.

\begin{nota}
\label{nota:cone algebra}
For $\Lambda \subseteq \Gamma$, we let $\cR(\Lambda) \coloneqq \pi_0(\fA_\Lambda)''$. 
When $\Lambda$ is a cone, we call $\cR(\Lambda)$ a \emph{cone algebra}.
\end{nota}

\begin{asmp}
\label{asmp:PureState}
The reference state $\omega_0$ is a pure state.
\end{asmp}

Note that $\omega_0$ being a pure state ensures that the cone algebras $\cR(\Lambda)$ are all factors.
We actually use a stronger assumption.  

\begin{asmp}
\label{asmp:InfiniteFactor}
For every cone $\Lambda$, the algebra $\cR(\Lambda)$ is an infinite factor.  
\end{asmp}

There are various reasonable assumptions on $\omega_0$ that ensure that the cone algebras are infinite, given that $\omega_0$ is pure.
For example, this holds when $\omega_0$ is translation invariant by using a standard argument \cite{MR2281418, MR2804555}.  
This also holds when $\omega_0$ is a gapped ground state of a Hamiltonian with uniformly bounded finite range interactions \cite[Lem.~5.3]{MR4362722}.

\begin{rem}
\label{rem:MinimalGeometricCone}
    Assumption \ref{asmp:InfiniteFactor} implies that every geometric cone intersects $\Gamma$, since every cone in $\Gamma$ is non-empty due to $\cR(\Lambda)$ being an infinite factor. 
    Consequently, if $\Lambda_1,\Lambda_2$ are geometric cones such that $\Lambda_1\cap\Gamma=\Lambda_2\cap\Gamma$, then $\Lambda_1$ and $\Lambda_2$ are translates of one another.
    Otherwise, the symmetric difference of $\Lambda_1$ and $\Lambda_2$ would contain a cone, and thus an element of $\Gamma$, making $\Lambda_1\cap\Gamma\neq\Lambda_2\cap\Gamma$.
    It follows that for every cone $\Lambda\subseteq \Gamma$, there exists a unique minimal geometric cone containing $\Lambda$, namely the intersection of all such geometric cones.
    We call this geometric cone $\Lambda_{geo}\subseteq \mathbb{R}^2$.
\end{rem}

\begin{nota}\label{nota:thickening}
Let $\Omega \subseteq \bbR^2$. We denote the thickening $$\Omega^{+r} \coloneqq \{x \in \bbR^2 : \text{there exists }y \in \Omega \text{ s.t. }|x - y| \leq r\}$$ 
For cones $\Lambda \subseteq \Gamma$ specifically, we instead define $\Lambda^{+r} \coloneqq (\Lambda_{geo})^{+r} \cap \Gamma$\footnote{rather than interpret $\Lambda$ as a subset of $\bbR^2$ for the purposes of thickening.}. 
\end{nota}

We will also assume that the GNS representation $\pi_0$ satisfies a generalization of Haag duality called \emph{bounded spread Haag duality} \cite[Def.~5.2]{2410.21454}.
This definition is analogous to the definition of weak algebraic Haag duality in \cite{2304.00068}.

\begin{defn}[{\cite[Def.~5.2]{2410.21454}}]
\label{def:BSHaagDuality}
Let $\pi \colon \fA \to B(\cH)$ be a representation.  
We say that $\pi$ satisfies \emph{bounded spread Haag duality} if there exists a global constant $r \geq 0$ such that for every cone $\Lambda$,  
\[
\cR(\Lambda^c)'
\subseteq
\cR(\Lambda^{+r}).
\]
\end{defn}

\begin{asmp}
\label{asmp:BoundedSpreadHaagDuality}
The GNS representation $\pi_0$ for $\omega_0$ satisfies bounded spread Haag duality for some fixed $r \geq 0$.
\end{asmp}

\begin{rem}
\label{rem:approx_vs_bounded_spread}
    We remark that Assumption \ref{asmp:BoundedSpreadHaagDuality} can be weakened: we expect our formalism to work with the weaker notion of approximate Haag duality used in \cite{MR4362722}. We choose to use bounded spread Haag duality since the resulting analysis is much simpler.
\end{rem}

A ray $R\subseteq \mathbb{R}$ is a subset of the form
$$
R:=\{a+t\hat{v}\in\mathbb{R}^2:t\in\mathbb{R}_{\geq 0}\}.
$$
Here $a\in\mathbb{R}^2$ is the end point of $R$ and $\hat{v}\in\mathbb{R}^2$ is a unit vector.

At this stage, we fix a ray $R\subset\mathbb{R}^2$ which is disjoint from $\Gamma$ and we denote its end point by $\partial R\in\mathbb{R}^2$.
Without loss of generality, we may choose coordinates of $\mathbb{R}^2$ such that
$$
R=\{(0,y)\in\mathbb{R}^2:y\geq 0\}.
$$
The only purpose for choosing these coordinates is to simplify the following definitions.

\begin{defn}
\label{def:allowed_cone}
We say that a cone $\Lambda \subseteq \Gamma$ is \emph{allowed} if there does not exist a translate $\Delta\subseteq \mathbb{R}^2$ of $\Lambda_{geo}$ such that $R\subset \Delta$.
We take $\cL$ to be the set of allowed cones.
Likewise, we take $\cL_{geo}$ to be the set of  minimal geometric cones of allowed cones.
\end{defn}

\begin{rem}\label{rem:postitive_translate}
Definition \ref{def:allowed_cone} 
implies that for each $\Lambda_{geo}\in\cL_{geo}$, there exists some translate $\tilde{R}$ of $R$ with $\tilde{R}\cap\Lambda_{geo}=\emptyset$ and $x>0$ for each $(x,y)\in \tilde{R}$.
\end{rem}

In the following definition, we take the \textit{right ray} of a geometric cone $\Lambda_{geo}\in\cL_{geo}$ to be the first ray encountered inside $\Lambda_{geo}$ with end point as apex of $\Lambda_{geo}$ when starting from $R$ and moving clockwise.

\begin{defn}
\label{def:r(Lambda)}
For each cone $\Lambda\in\cL$, take $\tilde{R}$ to be a translate of $R$ as in Remark \ref{rem:postitive_translate}, and take $s\subset\mathbb{R}^2$ to be the line segment between, and including, the apex of $\Lambda_{geo}$ and $\partial R$.
In the following three mutually exclusive cases, we define $r(\Lambda)_{geo}\subseteq\mathbb{R}^2$ to be the unique simply connected open set obeying the specific property for each case:
\begin{enumerate}
    \item ($R\cap \Lambda_{geo}=\emptyset$): $r(\Lambda)_{geo}$ is region whose boundary is the union of the right ray of $\Lambda_{geo}$, $s$, and $R$, and satisfies $\tilde{R}\subset r(\Lambda)_{geo}$.
    \item ($\partial R \subset \Lambda_{geo}$): The complement $(R\cup\Lambda_{geo})^c\subset\mathbb{R}^2$ consists of two simply connected components. We define $r(\Lambda)_{geo}$ to be the component containing $\tilde{R}$.
    \item ($R\cap \Lambda_{geo}\neq \emptyset$ and $\partial R\notin\Lambda_{geo}$): There exists a translate $\hat{R}$ of $R$ with $\hat{R}\subset R$ and $\partial \hat{R}\in\Lambda_{geo}$. We define $r(\Lambda)_{geo}$ as in $(2)$ using $\hat{R}$ instead of $R$.
\end{enumerate}
In each case, we also define
\begin{itemize}
    \item $\ell(\Lambda)_{geo} \coloneqq (\Lambda_{geo}\cup r(\Lambda)_{geo})^c$ as a complement in $\mathbb{R}^2$,
    \item $r(\Lambda) \coloneqq r(\Lambda)_{geo}\cap\Gamma$,
    \item $\ell(\Lambda) \coloneqq \ell(\Lambda)_{geo}\cap\Gamma$.
\end{itemize}
These three cases are shown in Figure \ref{fig:defining_symmetry_action_to_the_right_of_cone}.

\begin{figure}[!ht]
    \centering
    \begin{equation*}
\tikzmath{
\fill[fill=blue!10] (1.5, -1.5) -- (0, -1.5) -- (-.25, -.25) -- (0, 0) -- (0, 1.5) -- (1.5, 1.5);
\fill[fill=red!50] (-1.5, -1.5) -- (-0.25, -0.25) -- (0, -1.5);
\draw (-1.5, -1.5) -- (-0.25, -0.25) -- (0, -1.5);
\draw[thick] (0, 0) -- (0, 1.5);
\filldraw(0, 0) circle (0.05cm);
\draw[dashed] (0, 0) -- (-0.25, -0.25);
\node at (-0.5, -1) {$\Lambda_{geo}$};
\node at (-.25, 1) {$R$};
\node at (.75, 0) {$r(\Lambda)_{geo}$};
\node at (-1, 0) {$\ell(\Lambda)_{geo}$};
}
\qquad\qquad
\tikzmath{
\fill[fill=blue!10] (1.5, -1.5) -- (0, -1.5) -- (-.25, -.25) -- (0, 0) -- (0, 1.5) -- (1.5, 1.5);
\fill[fill=red!50] (-1.5, -1.5) -- (0.25, 0.5) -- (0, -1.5);
\draw (-1.5, -1.5) -- (0.25, .5) -- (0, -1.5);
\draw[thick] (0, 0) -- (0, 1.5);
\filldraw(0, 0) circle (0.05cm);
\node at (-0.5, -1) {$\Lambda_{geo}$};
\node at (-.25, 1) {$R$};
\node at (.85, 0) {$r(\Lambda)_{geo}$};
\node at (-1, 0) {$\ell(\Lambda)_{geo}$};
}
\qquad\qquad
\tikzmath{
\fill[fill=blue!10] (1.5, -1.5) -- (-1, -1.5)-- (.25, 1) -- (0, .75) -- (0, 1.5) -- (1.5, 1.5);
\fill[fill=red!50] (-2.25, -1.5) -- (0.25, 1) -- (-1, -1.5);
\draw (-2.25, -1.5) -- (0.25, 1) -- (-1, -1.5);
\draw[thick] (0, 0) -- (0, 1.5);
\filldraw(0, 0) circle (0.05cm);
\draw[dashed] (0, 0) -- (0.25, 1);
\node at (-1.25, -1) {$\Lambda_{geo}$};
\node at (-.25, 1) {$R$};
\node at (.75, 0) {$r(\Lambda)_{geo}$};
\node at (-1.5, 0) {$\ell(\Lambda)_{geo}$};
}
    \end{equation*}
    \caption{Defining the regions $r(\Lambda)_{geo}$ required to define symmetry action $\beta_g^{r(\Lambda)}$ for different cones. We make the arbitrary choice that $R \subseteq \ell(\Lambda)_{geo}$. 
    }
    \label{fig:defining_symmetry_action_to_the_right_of_cone}
\end{figure}
\end{defn}

\begin{defn}
\label{def:defect_sector}
    Let $\pi\colon \fA\rightarrow B(\cH_0)$ be a representation. We say that $\pi$ is \emph{$g$-localized} in a cone $\Lambda \in \cL$ if 
    \begin{equation}
    \label{eq:g-localized}
    \pi|_{\cstar[\Lambda^c]}=\pi_0 \circ \mu \circ \beta_g^{r(\Lambda)}|_{\cstar[\Lambda^c]},
    \end{equation}
    where $\mu = \Ad(\bigotimes_{s \in S} U_s^{g_s})$ for some $S \in \Gamma_f$ and $g_s \in G$.
    We call such an automorphism $\mu$ a \emph{symmetry action on finitely many sites}. 
    If $\mu = \Id$, the identity automorphism, we say that $\pi$ is \emph{canonically $g$-localized}.
    We say that a $g$-localized representation $\pi$ is \emph{transportable} if for every cone $\Delta \in \cL$, there exists $\pi' \colon \fA \to B(\cH_0)$ such that $\pi' \simeq \pi$\footnote{Here $\simeq$ denotes unitary equivalence, i.e.~there exists some $U \in \cB(\cH_0)$ such that $\pi(A) = \Ad[U] \circ \pi' (A)$ for all $A \in \fA$.} and $\pi$ is $g$-localized in $\Delta$.
\end{defn}

\begin{rem}
\label{rem:glocalizedreps}
    Note that if $\Lambda_1 \subseteq \Lambda_2$ and $\pi$ is $g$-localized in $\Lambda_1$, then $\pi$ is $g$-localized in $\Lambda_2$. 
    However, if $g \neq 1$, then this does not hold if `$g$-localized' is replaced by `canonically $g$-localized'. 
    Nonetheless, if $\partial R \in (\Lambda_1)_{geo}$, then $\pi$ being canonically $g$-localized in $\Lambda_1$ implies that $\pi$ is canonically $g$-localized in $\Lambda_2 \supseteq \Lambda_1$, since in that case $r(\Lambda_1)_{geo} \subseteq r(\Lambda_2)_{geo}$.
    If $g = 1$, then the definition of canonically $g$-localized recovers the definition of localized endomorphism used in \cite{MR297259, MR660538, MR2804555}, where it is true that if $\Lambda_1 \subseteq \Lambda_2$ and $\pi$ is localized in $\Lambda_1$, then $\pi$ is localized in $\Lambda_2$. 
\end{rem}

\begin{defn}
\label{def:g-defect_rep_and_sector}
Let $\pi \colon \fA \to B(\cH_0)$ be a representation.
We say that $\pi$ is a \emph{$g$-defect representation} if it is $g$-localized and transportable.  

A \emph{$g$-defect sector} is a unitary equivalence class of $g$-defect representations. 
We say that a $g$-defect sector is \emph{irreducible} if the representations that comprise the sector are irreducible. 
\end{defn}

\begin{rem}
\label{rem:WhyNotSectorizability}
Our definitions for $g$-defect representations closely follows the definition of anyon endomorphisms in \cite{MR2804555}, or alternatively the set $\cO_0$ in \cite{MR4362722}.
It is also common to define anyon representations using the \emph{superselection criterion}, that is, as representations $\pi \colon \fA \to B(\cH)$ such that $\pi|_{\cstar[\Lambda^c]} \simeq \pi_0|_{\cstar[\Lambda^c]}$ for every cone $\Lambda$, where $\simeq$ denotes unitary equivalence. 
We adapt the former definition to obtain the $G$-grading on our category. 
Indeed, it can happen that a $g$-defect representation and an $h$-defect representation are unitarily equivalent for $g \neq h$, so if we had used a criterion analogous to the superselection criterion, our category would not be $G$-graded.\footnote{We avoid this issue with our definition of $g$-defect representation using Assumption \ref{asmp:Faithfulness}. Since there is an equality in \eqref{eq:g-localized}, Assumption \ref{asmp:Faithfulness} ensures that if $\pi$ is both a $g$-defect representation and an $h$-defect representation, then $g = h$ (see Section \ref{sec:Category_of_homogeneous_G-defect sectors}).
We then exclude intertwiners between $g$-defect representations and $h$-defect representations when $g \neq h$ to obtain a $G$-graded category.}
An explicit example exhibiting this phenomenon is discussed in Remark \ref{rem:ProductStatesAndSectorizability}.
\end{rem}

\begin{defn}
\label{def:aux algebra}
We define $\fA^a_0 \coloneqq \bigcup_{\Lambda \in \cL} \cR(\Lambda)$ and $\fA^a \coloneqq \overline{\fA^a_0}^{\| \cdot \|}$. 
Note that $\fA^a_0$ and $\fA^a$ are algebras since $\cL$ is right-directed, that is, for any $\Lambda_1, \Lambda_2 \in \cL$, there exists $\Delta \in \cL$ such that $\Lambda_1, \Lambda_2 \subseteq \Delta$. 
We term the $\rmC^*$-algebra $\fA^a$ the \emph{auxiliary algebra} (often called the \emph{allowed algebra}).
\end{defn}

In Lemma \ref{lem:GDefectsDefinedOnAuxiliaryAlgebra}, we show that these $g$-defect representations extend to endomorphisms of the auxiliary algebra $\fA^a$.

We briefly review the components of the definition of a $G$-crossed braided $\rmW^*$-tensor category; a more detailed discussion of these definitions can be found in Appendix \ref{sec:basics of cat thy}. 
We first define $\rmW^*$-tensor category. 
For us, a \emph{tensor category} is a linear monoidal category with simple unit that admits direct sums and subobjects. 
For it to be a \emph{$\rmW^*$-tensor category}, we further require that every endomorphism algebra is a von Neumann algebra (making it a \emph{$\rmW^*$-category}) and that the monoidal product functor is bi-normal \cite{MR808930, MR3687214}. 

We now explain what further structure must be added for a $\rmW^*$-tensor category to be a $G$-crossed braided $\rmW^*$-tensor category. 
The $G$-crossed braiding structure has many components. 
First, a \emph{$G$-graded tensor category} is a tensor category with a $G$-grading such that the tensor functor respects the grading (Definition \ref{def:G-graded monoidal}). 
For a $G$-graded tensor category $\cC$, we let $\cC_{\hom}$ denote the full subcategory of homogeneous objects, and for $a \in \cC_{\hom}$, we let $\partial a$ denote the grading of $a$.
A \emph{$G$-crossed $\rmW^*$-tensor category} is a $G$-graded $\rmW^*$-tensor category $\cC$ together with a tensor functor $\gamma \colon G \to \Aut_\otimes(\cC)$ (Definition \ref{def:G-crossed monoidal general v2}). 
The category $\GSec$ is actually a  \emph{strict} $G$-crossed $\rmW^*$-tensor category, meaning that each $\gamma_g \in \Aut_\otimes (\GSec)$ is a strict tensor isomorphism and $g \mapsto \gamma_g$ is a group homomorphism \cite[Def.~2.9]{MR2183964}.\footnote{In more detail, we have that $\gamma_g(\pi \otimes \sigma) = \gamma_g(\pi) \otimes \gamma_g(\sigma)$ for every $\pi, \sigma \in \GSec$ and $\gamma_{g} \circ \gamma_{h} = \gamma_{gh}$ for all $g, h \in G$.}
Finally, a \emph{braiding} on a $G$-crossed $\rmW^*$-tensor category $\cC$ is a family of unitary isomorphisms $c_{a, b} \colon a \otimes b \to \gamma_{\partial a} (b) \otimes a$ for $a \in \cC_{\hom}$ and $b \in \cC$ that satisfies coherence conditions analogous to the usual braiding conditions (Definition \ref{def:G-crossed braided general v2}).

\subsection{Main results}

We define a category $\GSec$ of $G$-defect representations whose objects are direct sums of $g$-defect representations (extended to the auxiliary algebra) for $g \in G$.
The morphisms are intertwining maps that live in $\fA^a_0$.
The category $\GSec$ will be defined precisely in Definition \ref{def:GSec}.
We also use the term ``$G$-defect sectors" to denote isomorphism classes of objects in $\GSec$. 
The $*$-operation on morphisms is the usual adjoint in $B(\cH_0)$.
For $g \in G$, the $g$-graded component of $\GSec$ is the full subcategory of $g$-defect representations. 
The $G$-action $\gamma_g$ is given on objects by $\gamma_g(\pi) \coloneqq \beta_g \circ \pi \circ \beta_g^{-1}$ and on morphisms by $\gamma_g(T) \coloneqq \beta_g(T)$.\footnote{Recall that $\beta_g$ is unitarily implemented on $\cH_0$ since $\omega_0 \circ \beta_g = \omega_0$.}

We are now ready to state the main theoretical result of this paper.

\begin{thm}[Theorem \ref{thm:GSecGCrossedBraidedWStarTensorCat}]
\label{thm:main_result}
The category $\GSec$ of $G$-defect representations with respect to $\pi_0$ is a strict $G$-crossed braided $\rmW^*$-tensor category.
\end{thm}

This result is shown in Section \ref{sec:symmetry_defects}. In Proposition \ref{prop:GCrossedMonoidal}, it is shown that this $G$-grading and $G$-action endow $\GSec$ with the structure of a $G$-crossed monoidal category. 
In Proposition \ref{prop:GCrossedBraiding}, it is shown that $\GSec$ admits a $G$-crossed braiding, which is defined analogously to the braiding of anyon representations \cite{MR4362722}. 
The fact that $\GSec$ is a $\rmW^*$-category that admits direct sums and subobjects is shown in Section \ref{sec:GSecCauchyComplete}, and the remaining details needed to show that $\GSec$ is a $G$-crossed braided $\rmW^*$-tensor category are proven in Theorem \ref{thm:GSecGCrossedBraidedWStarTensorCat}.

We now turn our attention to examples. Our techniques can be used to construct symmetry defects for SPT phases made using finite depth quantum circuits. Let $G$ be the symmetry group and $\beta_g$ the symmetry automorphism from Definition \ref{def:GlobalSymmetryAutomorphism}.

\begin{defn}
\label{def:FDQC}
    Let $N \in \bbN$ be fixed, and let $\{\cU^d\}_{d = 1}^D$ be a family of sets $\cU^d$ of unitaries $U$ in $\cstar$ such that the unitaries in $\cU^d$ have mutually disjoint supports and $\supp(U)$ is contained in a ball of diameter $N$ for each $U \in \cU^d$. 
    An automorphism $\alpha$ is a \emph{finite depth (unitary) quantum circuit (FDQC)}\footnote{We use FDQC in the spirit of \cite{MR4544190}. 
    Some authors also consider non-unitary circuit elements, namely isometries and projections. 
    See \cite{2405.17379} for an instance where both definitions are discussed.} of the family $\{\cU^d\}_{d = 1}^D$ if for all $A \in \cstar^{\loc}$ we have $$\alpha(A) = \alpha_D \circ \cdots \circ \alpha_1 (A), \qquad \qquad  \alpha_d(A) \coloneqq \Ad \! \left(\prod_{U \in \cU^d} U \right)\! (A).$$ We observe that $\alpha$ can be extended in a norm continuous way to all of $\cstar$.  
    We say $\cU^d$ is the set of \emph{entangling unitaries} of layer $d$ in the circuit. An example circuit in 1d with $N = 2$ and $D=3$ is shown in Figure \ref{fig:FDQC_example}.
    \begin{figure}[!ht]
        \centering
        \includegraphics[width=0.4\linewidth]{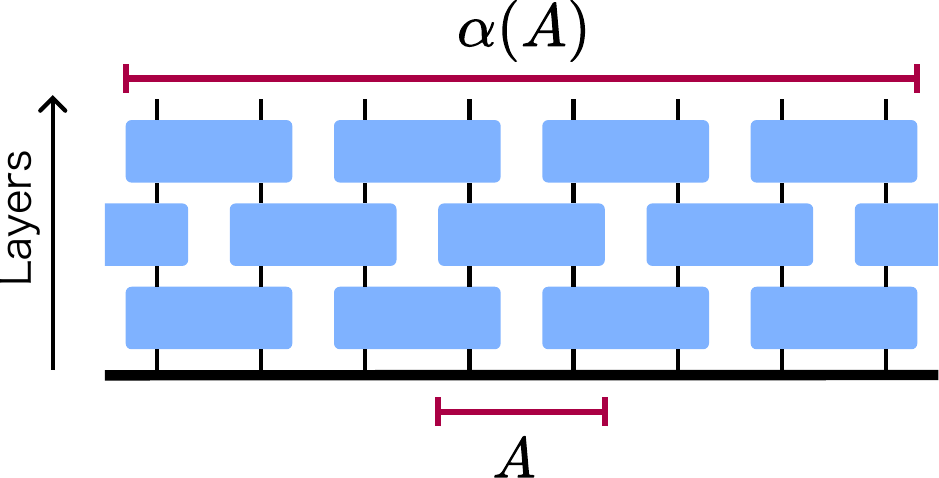}
        \caption{An example of a finite depth quantum circuit in 1 (spatial) dimension. Each block is an entangling unitary $U$ with support of 2 adjacent sites, so $N = 2$. The depth of this circuit is $D = 3$. We have $|\supp(A)| = 2$ and after the circuit, $|\supp(\alpha(A))| = 8$.}
        \label{fig:FDQC_example}
    \end{figure}
\end{defn}

\begin{defn}
\label{def:SPT_phase}
    We define a state $\omega$ to be a $G$-SPT state if there exists a pure product state $\omega_0$ satisfying $ \omega = \omega_0 \circ \alpha$, the states $\omega_0$ and $\omega$ are both invariant under the action of $\beta_g$ for all $g \in G$, and $\alpha$ is an FDQC that additionally satisfies $\alpha \circ \beta_g = \beta_g \circ \alpha$ for all $g \in G$. When $\alpha$ satisfies $\alpha \circ \beta_g = \beta_g \circ \alpha$, we call $\alpha$ a \emph{symmetric entangler}.
\end{defn}
\begin{rem}
\label{rem:ClassOfSPTsWithSymmetricEntanglers}
    We note that this definition of a $G$-SPT is more strict than others appearing in the literature \cite{MR4354127}. In particular, it discounts locally generated automorphisms (LGA) that are not FDQCs, as well as crystalline SPTs. We adopt the former restriction so that our SPTs to satisfy bounded-spread Haag duality (cf.~ Remark \ref{rem:approx_vs_bounded_spread}). As remarked earlier, we expect the analysis to work in the more general setting. The latter is because we require the $G$-action to be on-site. 
    This assumption will allow us to explicitly construct defect representations by restricting the symmetry action (cf.~ Section \ref{sec:DefectAutomorphismConstructionHeuristic}). A priori, there is no reason to expect the symmetric entangler property to hold for a generic FDQC $\alpha$. However, we include it as part of the definition of a $G$-SPT since it holds for a very general class of models \cite{PhysRevB.87.155114,PhysRevB.108.115144} which are believed to be fixed point representatives of a $G$-SPT phase.
    The class of SPTs that we consider are exactly those that were classified in \cite{bols2026complete}.
\end{rem}

\begin{thm}[{Prop.~\ref{prop:GSec is equiv to HilbGnu}}]
\label{thm:SPT theorem}
    The category of $G$-defect representations of a $G$-SPT is $G$-crossed braided $W^*$-tensor equivalent to $\Vect(G, \nu)$, where $\nu: G \times G \times G \rightarrow U(1)$ is a $3$-cocycle. 
\end{thm}

\subsection{Constructing defect automorphisms}
\label{sec:DefectAutomorphismConstructionHeuristic}

We now provide a heuristic description for how we generated the $G$-defect representations in our examples. The purpose in providing such a description is three-fold:
\begin{enumerate}
    \item to elucidate the constructions in our examples,
    \item to give some intuition to others who wish to apply our formalism to novel examples,
    \item to provide the physical intuition which motivates our formal definition of symmetry defects.
\end{enumerate}
Importantly, we do not attempt to give a rigorous algorithm for producing $G$-defect representations in an arbitrary model. We instead present a loose set of steps that we hope will be of utility to future researchers.

Recalling our earlier notation, we denote $\beta_g$ to be the on-site symmetry action and $\omega_0$ to be a pure ground state. We also assume that interactions are uniformly bounded and invariant under the symmetry action $\beta_g$.
We begin by picking a path $L \subset \bbR^2$ that is sufficiently ``nice."
A precise definition of the paths we consider is given in Section \ref{sec:paths and dual paths} for the models we consider, but for now, we just assume that $L$ is a curve that divides $\bbR^2$ into two simply connected components.\footnote{Our notation here slightly differs from the notation in Section \ref{sec:paths and dual paths} as in the models we have access to a nice lattice structure that we use when making the definitions.} 
We arbitrarily denote $r(L), \ell(L)$ as the two disjoint halves of the lattice obtained from $L$.
Since $\beta_g$ is on-site, we have $$\beta_g =  \beta^{\ell(L)}_g \otimes \beta^{r(L)}_g.$$

\begin{figure}[!ht]
\begin{subfigure}[t]{0.4\linewidth}
    \includegraphics[width=\linewidth]{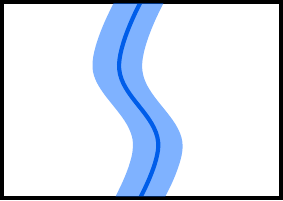}
    \caption{A dividing dual path $L$ shown in blue dividing the lattice into two halves. The strip $S_L$ is shown in light blue and centered at $L$.}
    \label{fig:strip_S}
\end{subfigure}
\hspace{5mm}
\begin{subfigure}[t]{0.4\linewidth}
    \includegraphics[width=\linewidth]{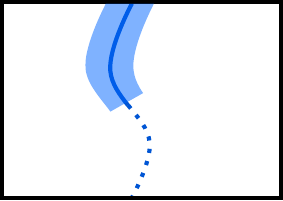}
    \caption{Half-strip $U_L$ shown in light blue. Using an automorphism $\alpha^{D_L}$ supported on half-strip $D_L = S_L \setminus U_L$, we erase the action of $\beta_g^{r(L)}$ such that the terms in the Hamiltonian supported outside $U_L$ are invariant under $\alpha^{D_L} \circ \beta_g^{r(L)}$.}
    \label{fig:partial_erasure}
\end{subfigure}
\caption{}
\end{figure}

We assume that $\omega_0$ is a ground state for some nice enough Hamiltonian (see Appendix \ref{sec:HamiltonianDynamics}) whose interaction terms are invariant under $\beta_g$.
We investigate the action of the restricted symmetry on these interaction terms. If a term in the Hamiltonian has support disjoint from $L$, this term will remain invariant under the action of the restricted symmetry. We define a strip $S_L$ as the union of the supports of all interaction terms that are not invariant under $\beta_g^{r(L)}$(See Figure \ref{fig:strip_S}).

Now our goal is (A) to find an automorphism $\alpha^L$ localized in the strip $S_L$ that can `correct' the action of this restricted symmetry action on all the Hamiltonian terms, and then (B) to split $\alpha^L$ into two disjoint halves composed with some inner automorphism implemented by a local unitary where they meet. 
We expect that this step of the algorithm breaks down for more complicated models than the ones we consider, in particular models exhibiting anyon permutation \cite{2411.01210}. 
Assuming that this can be done, we then cut $\alpha^L$ into $\Xi \circ (\alpha^{U_L} \otimes \alpha^{D_L})$ where $\Xi$ is an inner automorphism implemented by a local unitary and $\alpha^{U_L}, \alpha^{D_L}$ are disjointly supported automorphisms both supported on $S_L$ and `erase' the restricted symmetry action on the terms in the Hamiltonian along their support, in the sense that $\alpha^{D_L} \circ \beta_g^{r(L)}$ leaves the terms in the Hamiltonian supported outside $U_L$ invariant (see Figure \ref{fig:partial_erasure}). 

\begin{figure}[!ht]
\begin{subfigure}[t]{0.4\linewidth}
    \includegraphics[width=\linewidth]{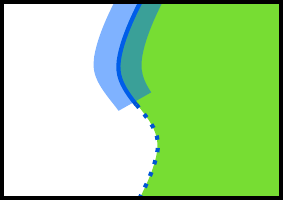}
    \caption{The symmetry defect is now given by $\kappa^{U_L} = \alpha^{D_L} \circ \beta_g^{r(L)}$ and acts trivially on Hamiltonian terms outside $U_L$ shown in light-blue.}
    \label{fig:symmetry_defect}
\end{subfigure}
\hspace{5mm}
\begin{subfigure}[t]{0.4\linewidth}
    \includegraphics[width=\linewidth]{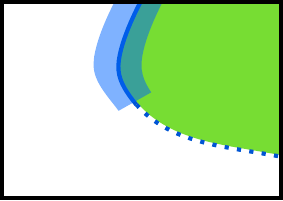}
    \caption{The dotted blue line can be freely transported while keeping the endpoint fixed, just like a string operator.}
    \label{fig:freely_transportable}
\end{subfigure}
\caption{Heuristic of a symmetry defect and its interpretation as being implemented by a string-operator.}
\end{figure}

A symmetry defect is then given by (see Figure \ref{fig:symmetry_defect}) $$\kappa^{U_L} = \alpha^{D_L}\circ \beta_g^{r(L)}.$$
We note that $\kappa^{U_L}$ depends on the entire path $L$ and not just $U_L$, but we will always omit the dependence from the notation to prevent the notation from being too cluttered.

A symmetry defect can be interpreted as being obtained by adding an automorphism to partially `erase' the action of the restricted symmetry. We contrast this with the traditional paradigm of cutting a restricted symmetry action to obtain a symmetry defect. 

We stress the word `partially' because even though the interaction terms supported away from the cut will not see the action of the restricted symmetry, there may still be observables in $\cstar$ that are transformed non-trivially under $\kappa^{U_L}$ along the erased symmetry action.

Written in this way, the symmetry defect can be interpreted as being implemented by something that behaves similarly to a string operator. More specifically, the ground state remains invariant under the action of $\kappa^{U_L}$ outside of some cone containing $U_L$. 
So the erased part of the symmetry defect can be freely transported outside of this cone (see Figure \ref{fig:freely_transportable}). A key difference between symmetry defects and anyons generated by string-operators is the presence of the $g$-action to the right of this string, so it is possible to detect the exact location of the defect line with local operators supported outside of the line, but not by the evaluation of these local operators on the ground state.

This also motivates our definition of a $G$-defect representation as a generalization of the anyon representation, in the sense that the `erased' part of the symmetry defect can always be moved into any allowed cone, as is typically done in anyon sector analysis. The key idea again is to account for the presence of the $g$-action to the right of this string.

To conclude this discussion, we summarize a simple algorithm to create symmetry defects, which we believe to be applicable to a wide variety of lattice models.

\vspace{5mm}
\centerline{%
\fbox{%
  \begin{minipage}{\dimexpr\linewidth-2\fboxsep-2\fboxrule\relax}
  \textbf{Creating a symmetry defect}
    \begin{enumerate}
    \item Observe the action of a restricted symmetry along a half-plane on the terms in the Hamiltonian.
    \item Devise an automorphism that `erases' this action on all such terms and is supported on some strip localised along the boundary of the restriction.
    \item If possible, cut this automorphism into 2 disjoint halves, possibly composed with some inner automorphism implemented by a local unitary.
    \item The symmetry defect is then given by composing the restricted symmetry action with one half of the split automorphism.
    \end{enumerate}
  \end{minipage}%
}
}

\section{\texorpdfstring{$G$}{G}-Defect Representations}
\label{sec:symmetry_defects}
Let $\omega_0 \colon \fA \to \bbC$ be a state.  
While in our examples, $\omega_0$ will usually be the unique frustration free ground state of a nice Hamiltonian (see Section \ref{sec:HamiltonianDynamics} for definitions and the assumptions on our Hamiltonian), we do not assume that in this section.
We let $\pi_0 \colon \fA \to B(\cH_0)$ be the GNS representation for $\omega_0$.  
Note that $\pi_0$ is faithful since $\fA$ is UHF algebra and hence simple. 

\subsection{Category of \texorpdfstring{$G$}{G}-defect representations}
\label{sec:GDefectRepCat}
We recall the assumptions on the symmetry action of the group $G$ and on the state $\omega_0$ that we will impose to ensure we obtain a $G$-crossed braided $\rmW^*$-tensor category (Section \ref{sec:GCrossedAssumptions}). Now we construct the category of $G$-defect representations.

Recall the definition of an allowed cone (Definition \ref{def:allowed_cone}). As before, we call $\cL$ the set of allowed cones with respect to the fixed ray $R$ with endpoint $\partial R$. 
We now recall the definition of a $g$-defect representation (Definition \ref{def:g-defect_rep_and_sector}). 
In this section, we will set $\pi_0$ as the reference representation unless stated otherwise.

\begin{facts}
\label{facts:GDefectBasics}
Here are some useful facts about $g$-defect representations (compare with \cite[Facts 2.27]{2410.21454}).
\begin{enumerate}[label=(GSec\arabic*)]

\item 
\label{GSec:MappingsOnConeAlgebras}
If $\pi$ is $g$-localized in a cone $\Lambda \in \cL$, then $\pi(\cstar[\Lambda]) \subseteq \cR(\Lambda^{+r})$ (compare with \cite[Lem.~2.12]{MR2183964}).

\begin{proof}
This result follows from bounded spread Haag duality (Assumption \ref{asmp:BoundedSpreadHaagDuality}).
Since $\pi$ is $g$-localized in $\Lambda$, we have that $\pi|_{\cstar[\Lambda^c]} = \pi_0 \circ \mu \circ \beta_g^{r(\Lambda)}|_{\cstar[\Lambda^c]}$, where $\mu = \Ad(\bigotimes_{s \in S} U^{g_s}_s)$ for some $S \in \Gamma_f$.
Let $x \in \cstar[\Lambda]$ and $y \in \cstar[\Lambda^c]$. 
Then $y = \mu \circ \beta_g^{r(\Lambda)}(z)$ for some $z \in \cstar[\Lambda^c]$, so we have that
\begin{align*}
\pi(x)\pi_0(y)
&=
\pi(x)\pi_0(\mu \circ \beta_g^{r(\Lambda)}(z))
=
\pi(x)\pi(z)
=
\pi(xz)
=
\pi(zx)
\\&=
\pi(z)\pi(x)
=
\pi_0(\mu \circ \beta_g^{r(\Lambda)}(z))\pi(x)
=
\pi_0(y)\pi(x).
\end{align*}
Thus $\pi(x) \in \pi_0(\cstar[\Lambda^c])' \subseteq \cR(\Lambda^{+r})$ by Assummption \ref{asmp:BoundedSpreadHaagDuality}.
\end{proof}

\item 
\label{GSec:GLocalizedLargerRegion}
If $\pi$ is $g$-localized in $\Lambda \in \cL$, then there is a cone $\widehat{\Lambda} \in \cL$ such that $\Lambda \subseteq \widehat{\Lambda}$ and such that $\pi$ is canonically $g$-localized in $\widetilde{\Lambda}$ for every cone $\widetilde{\Lambda} \supseteq \widehat{\Lambda}$. 

\begin{proof}
Note that $\pi|_{\cstar[\Lambda^c]} = \pi_0 \circ \Ad(U) \circ \beta_g^{r(\Lambda)}|_{\cstar[\Lambda^c]}$ for some $U \in \cstar^{\loc}$.
Let $\widehat{\Lambda} \in \cL$ such that $\Lambda \cup \supp(U) \subseteq \widehat{\Lambda}$ and such that $\partial R \in \widehat{\Lambda}_{geo}$. 
By enlarging $\widehat{\Lambda}$ if necessary (but still calling the enlarged cone $\widehat \Lambda$), we can ensure that $\beta_g^{r(\Lambda)}|_{\cstar[\widehat{\Lambda}^c]} = \beta_g^{r(\widehat{\Lambda})}|_{\cstar[\widehat{\Lambda}^c]}$.
Therefore, since $\Lambda \cup \supp(U) \subseteq \widehat{\Lambda}$, we have that
\[
\pi|_{\cstar[\widehat{\Lambda}^c]} = \pi_0 \circ \Ad(U) \circ \beta_g^{r(\Lambda)}|_{\cstar[\widehat{\Lambda}^c]}
=
\pi_0 \circ \beta_g^{r(\widehat{\Lambda})}|_{\cstar[\widehat{\Lambda}^c]}.
\]
Furthermore, since $\partial R \in \widehat{\Lambda}_{geo}$, it follows that $\pi$ is canonically $g$-localized in $\widetilde{\Lambda}$ for all $\widetilde{\Lambda} \supseteq \widehat{\Lambda}$ since $r(\widehat{\Lambda}) \subseteq r(\widetilde{\Lambda})$. 
\end{proof}

\item 
\label{GSec:IntertinersInConeAlgebras}
If $\pi, \sigma$ are canonically $g$-localized in $\Lambda \in \cL$ and $T$ is an intertwiner from $\pi$ to $\sigma$, i.e., $T\pi(-) = \sigma(-)T$, then $T \in \cR(\Lambda^{+r})$ (compare with \cite[Lem.~2.13]{MR2183964}). 
If $\pi, \sigma$ are (not canonically) $g$-localized in $\Lambda$, then $T \in \cR(\Lambda^{+t})$ for some $t \geq r$. 

\begin{proof}
The first part of this result follows from bounded spread Haag duality (Assumption \ref{asmp:BoundedSpreadHaagDuality}). 
Since $\pi, \sigma$ are canonically $g$-localized in $\Lambda$, we have that $\pi|_{\cstar[\Lambda^c]} = \pi_0 \circ \beta_g^{r(\Lambda)}|_{\cstar[\Lambda^c]} = \sigma|_{\cstar[\Lambda^c]}$. 
Let $x \in \cstar[\Lambda^c]$. 
Then $x = \beta_g^{r(\Lambda)}(y)$ for some $y \in \cstar[\Lambda^c]$, so we have that 
\[
T\pi_0(x)
=
T\pi_0(\beta_g^{r(\Lambda)}(y))
=
T\pi(y)
=
\sigma(y)T
=
\pi_0(\beta_g^{r(\Lambda)}(y))T
=
\pi_0(x)T.
\]
Thus $T \in \pi_0(\cstar[\Lambda^c])' \subseteq \cR(\Lambda^{+r})''$ by Assumption \ref{asmp:BoundedSpreadHaagDuality}.

If $\pi, \sigma$ are (not canonically) $g$-localized in $\Lambda$, then $\pi = \Ad(U_1) \circ \tilde \pi$ and $\sigma = \Ad(U_2) \circ \tilde \sigma$, where $\tilde \pi, \tilde \sigma$ are canonically $g$-localized in $\Lambda$ and $U_1, U_2 \in \cstar^{\loc}$ are unitaries.
In that case, $U_2 T U_1^{-1}$ is an intertwiner from $\tilde \pi$ to $\tilde \sigma$ and thus $U_2 T U_1^{-1} \in \cR(\Lambda^{+r})$. 
Since $U_1, U_2 \in \cstar^{\loc}$, the result follows.
\end{proof}
\end{enumerate}
\end{facts}

\begin{rem}
\label{rem:IntertwinersForThingsThatLookTheSameOutsideOfCone}
Note that the proof of \ref{GSec:IntertinersInConeAlgebras} shows that if $\pi, \sigma$ are $g$-localized in $\Lambda \in \cL$, $T$ is an intertwiner from $\pi$ to $\sigma$, and $\pi|_{\cstar[\Lambda^c]} = \sigma|_{\cstar[\Lambda^c]}$, then $T \in \cR(\Lambda^{+r})$.
\end{rem}

We now show that we can extend every $g$-localized representation to an endomorphism of an auxiliary algebra, defined as in \cite{MR660538, MR2804555}.
Recall (Notation \ref{nota:cone algebra}) that $\cR(\Lambda) = \pi_0(\cstar[\Lambda])'' \subseteq B(\cH_0)$ for $\Lambda \subseteq \Gamma$. 
Further recall (Definition \ref{def:aux algebra}) that $\fA^a_0 = \bigcup_{\Lambda\in \cL} \cR(\Lambda)$ and that the auxiliary algebra is defined to be $\fA^a = \overline{\fA^a_0}^{\| \cdot \|}$.

\begin{lem}
    \label{lem:GDefectsDefinedOnAuxiliaryAlgebra}
    Let $\pi\colon\fA\rightarrow B(\cH_0)$ be a $g$-defect representation. 
    Then there is a unique extension $\pi^a$ of $\pi$ to $\fA^a$ such that $\pi^a|_{\cR(\Lambda)}$ is WOT-continuous for all $\Lambda \in \cL$.  
    Furthermore, $\pi^a(\fA^a) \subseteq \fA^a$, that is, $\pi^a \colon \fA^a \to \fA^a$ is an endomorphism.
\end{lem}
\begin{proof}
    We proceed as in the proofs of \cite[Lem.~4.1]{MR660538} and \cite[Prop.~4.2]{MR2804555}\footnote{This proposition is numbered Proposition 4.6 in the version available from the arXiv.}.
    Let $\Lambda\in\cL$. Then there exists some other $\Delta\in\cL$ such that $\Delta \subseteq r(\Lambda)$.
    Since $\pi$ is $g$-transportable, there exists some $U\in B(\cH_0)$ such that for all $A\in \cstar[\Delta^c]$, 
    $$
    U\pi(A)U^*=\pi_0 \circ \beta_g^{r(\Delta)}(A).
    $$
    Since $\Delta \subseteq r(\Lambda)$, we have $\Lambda \subseteq \Delta^c$ and $\Lambda$ intersects $r(\Delta)$ in finitely many sites.  
    Therefore, there exists a unitary $V\in B(\cH_0)$ such that $V \pi_0(A) V^* = \pi_0 \circ \beta_g^{r(\Delta)}(A)$  for $A \in \cstar[\Lambda]$, so for all $A\in \cstar[\Lambda]$, we have that
    \[
    VU\pi(A)U^*V^*=\pi_0(A).
    \]
    Observe that we obtain a WOT-continuous formula for $\pi|_{\cstar[\Lambda]}$, namely $\pi|_{\cstar[\Lambda]} = \Ad(U^*V^*) \circ \pi_0|_{\cstar[\Lambda]}$, so $\pi|_{\cstar[\Lambda]}$ has a unique WOT-continuous extension to $\cR(\Lambda)$, which we denote $\pi_\Lambda$
    (note that we are implicitly identifying $\fA$ with $\pi_0(\fA)$, which we can do since $\pi_0 \colon \fA \to B(\cH_0)$ is faithful).
    Since $\Lambda \in \cL$ was arbitrary, we obtain a unique, well-defined extension of $\pi$ to $\fA^a_0 = \bigcup_{\Lambda \in \cL} \cR(\Lambda)$ that is WOT-continuous on each $\cR(\Lambda)$.
    Indeed, $\pi|_{\fA_\Lambda}$ has a unique WOT-continuous extension to $\cR(\Lambda)$ for every $\Lambda \in \cL$, so there is at most one possible extension of $\pi$ to $\fA^a_0$ that is WOT-continuous on each $\cR(\Lambda)$. 
    It remains to check that this extension is well-defined. 
    For any $\Lambda_1, \Lambda_2 \in \cL$,  there exists $\Delta \in \cL$ such that $\Lambda_1, \Lambda_2 \subseteq \Delta$. 
    We therefore have that $\cR(\Lambda_1) \cup \cR(\Lambda_2) \subseteq \cR(\Delta)$, and since $\pi|_{\fA_\Lambda}$ has a unique WOT-continuous extension $\pi_\Delta$ to $\cR(\Delta)$, we have that $\pi_\Delta|_{\cR(\Lambda_i)} = \pi_{\Lambda_i}$ for $i = 1, 2$. 
    In particular, for any $x \in \cR(\Lambda_1) \cap \cR(\Lambda_2)$, we have that $\pi_{\Lambda_1}(x) = \pi_\Delta(x) = \pi_{\Lambda_2}(x)$. 
    Hence the extension to $\fA^a_0$ is well-defined.
    It is also norm-continuous, so we obtain a unique extension $\pi^a$ of $\pi$ to $\fA^a$ with the desired properties.  

    It remains to show that $\pi^a(\fA^a) \subseteq \fA^a$.  
    By continuity, it suffices to show that $\pi^a(\pi_0(\cstar[\Lambda])) \subseteq \fA^a$ for all $\Lambda \in \cL$.
    Let $\Lambda \in \cL$.
    Since $\pi$ is a $g$-defect representation, $\pi$ is $g$-localized in some $\widehat{\Lambda} \in \cL$.  
    Then there exists $\Delta \in \cL$ such that $\Lambda, \widehat{\Lambda} \subseteq \Delta$.
    In particular, we have that $\pi$ is $g$-localized in $\Delta$ and $\cstar[\Lambda] \subseteq \cstar[\Delta]$.
    By \ref{GSec:MappingsOnConeAlgebras}, we have that $\pi^a(\pi_0(\cstar[\Delta])) = \pi(\cstar[\Delta]) \subseteq \cR(\Delta^{+r})$.  
    The result follows.
\end{proof}

In the remainder of the paper, we will abuse notation and identify the extension of $\pi$ to $\fA^a$ with $\pi$ for notational simplicity. The context should clarify any ambiguities.

\subsubsection{Category of homogeneous \texorpdfstring{$G$}{G}-defect representations}
\label{sec:Category_of_homogeneous_G-defect sectors}
We build a category $\GSec_{\hom}$ of homogeneous $G$-defect representations, analogously to the category in \cite{MR2183964}.\footnote{We thank David Penneys and Corey Jones for the very helpful suggestion to apply the approach of \cite{MR2183964} to this problem.}  
The objects of $\GSec_{\hom}$ are all endomorphisms of $\fA^a$ coming from $g$-defect representations for any $g \in G$. 
The morphisms between $\pi, \sigma \in \GSec_{\hom}$ are intertwiners $T$ from $\pi$ to $\sigma$ such that $T \in \fA^a_0$. 
We let $\GSec_g$ be the full subcategory of $\GSec_{\hom}$, consisting of $g$-defect representations for a fixed $g \in G$.

We make a few observations. 
First, if $\pi$ is $g$-localized in $\Lambda \in \cL$ and $h$-localized in $\Lambda$, then $g = h$ (c.f.~\cite[Rem.~2.7(3)]{MR2183964}).  
Indeed, since $\pi$ is $g$-localized in $\Lambda$ we have that $\pi|_{\cstar[\Lambda^c]} = \pi_0 \circ \mu_1\circ \beta^{r(\Lambda)}_g$, where $\mu_1$ is a symmetry action on finitely many sites (Definition \ref{def:defect_sector}).  
Similarly, $\pi|_{\cstar[\Lambda^c]} = \pi_0 \circ \mu_2\circ \beta^{r(\Lambda)}_h$, where $\mu_2$ is a symmetry action on finitely many sites.  
But $\beta_g^{r(\Lambda)}$ and $\beta_h^{r(\Lambda)}$ differ on regions of the form $B \cap \Gamma$ where $B \subseteq \bbR^2$ is a ball of arbitrarily large radius.  
By Assumption \ref{asmp:Faithfulness}, the on-site symmetry is faithful when restricted to a region of the form $B \cap \Gamma$ where $B \subseteq \bbR^2$ is a ball of sufficiently large radius. 
Therefore $g = h$. 

Next, by \ref{GSec:IntertinersInConeAlgebras}, if $\pi, \sigma \in \GSec_g$, then for any intertwiner $T$ from $\pi$ to $\sigma$, we have that $T \in \fA^a_0$. 
Indeed, $\pi$ must be $g$-localized in some $\Lambda_1 \in \cL$ and $\sigma$ must be $g$-localized in some $\Lambda_2 \in \cL$. 
Therefore, taking $\Delta \in \cL$ such that $\Lambda_1, \Lambda_2 \subseteq \Delta$, we have that $\pi, \sigma$ are both $g$-localized in $\Delta$. 
Hence by \ref{GSec:IntertinersInConeAlgebras}, we have that $T \in \cR(\Delta^{+t})$ for some $t \geq r$. 

Finally, if $\pi$ is $g$-localized in $\Lambda \in \cL$ and $\sigma$ is $h$-localized in $\Lambda$, it may be the case that there exists a nonzero intertwiner $T$ from $ \pi$ to $\sigma$ even if $g \neq h$.  
However, by the lemma below, this intertwiner cannot be in the cone algebra for any allowed cone and is therefore not a morphism in $\GSec_{\hom}$.

\begin{lem}
\label{lem:no morphisms between different gradings}
Suppose $\pi$ is $g$-localized in $\Lambda \in \cL$ and $\sigma$ is $h$-localized in $\Lambda$, and suppose that $T \colon \pi \to \sigma$ satisfies that $T \in \cR(\Delta)$ for some $\Delta \in \cL$ and $T \neq 0$.  
Then $g = h$.
\end{lem}

\begin{proof}
Note that it suffices to consider the case where $\pi$ is canonically $g$-localized in $\Lambda$ and $\sigma$ is canonically $h$-localized in $\Lambda$, since a $g$-localized representation is unitarily equivalent to a canonically $g$-localized representation using a unitary in $\fA^{\loc}$
(this assumption does not materially affect the argument, but it makes the notation easier).
Since $\Lambda, \Delta \in \cL$, there exists a cone $\widetilde{\Lambda} \in \cL$ such that $\Lambda, \Delta \subseteq \widetilde{\Lambda}$.
Now, since $T \in \cR(\Delta)$, we have that for all $A \in\cstar[\Delta^c]$, $$T\pi_0(\beta_g^{r(\Lambda)}(A)) = \pi_0(\beta_g^{r(\Lambda)}(A))T$$

Similarly, since $\pi$ is $g$-localized in $\Lambda$, $\sigma$ is $h$-localized in $\Lambda$, and $T \colon \pi \to \sigma$, we have that for all $A \in\cstar[\Lambda^c]$, 
\[
T\pi_0(\beta_g^{r(\Lambda)}(A))
=
T\pi(A)
=
\sigma(A)T
=
\pi_0(\beta_h^{r(\Lambda)}(A))T.
\]
Using $\cstar[\Delta^c], \cstar[\Lambda^c] \supseteq \cstar[\widetilde{\Lambda}^c]$ and combining these equations, we get for all $A \in\cstar[\widetilde{\Lambda}^c]$, 
\[
\pi_0\big(\beta_g^{r(\Lambda)}(A) - \beta_h^{r(\Lambda)}(A)\big)T
=
0.
\]
Now we have $T \in \cR(\widetilde{\Lambda})$ and also for $A \in\cstar[\widetilde{\Lambda}^c]$ that
\[
\pi_0\big(\beta_g^{r(\Lambda)}(A) - \beta_h^{r(\Lambda)}(A)\big)
\in\pi_0(\cstar[\widetilde{\Lambda}^c])
\subseteq
\cR(\widetilde{\Lambda})'.
\]
By \cite[Thm.~5.5.4]{MR1468229}, since $\cR(\widetilde{\Lambda})$ is a factor\footnote{We thank David Penneys for pointing out that this implies the desired result.
The key property of factors that we use here is that the only nonzero central projection is $\mathds{1}$.} and $T \neq 0$, we obtain that for all $A \in\cstar[\widetilde{\Lambda}^c]$,
\[
\pi_0\big(\beta_g^{r(\Lambda)}(A) - \beta_h^{r(\Lambda)}(A)\big) = 0.
\]
Now there exists a ball $B \subseteq (\widetilde{\Lambda}_{geo})^c$ of arbitrarily large radius such that $\beta_g^{r(\Lambda)}|_{B \cap \Gamma} = \beta_g|_{B \cap \Gamma}$ and $\beta_h^{r(\Lambda)}|_{B \cap \Gamma} = \beta_h|_{B \cap \Gamma}$.
Since $\pi_0$ is faithful, we have that $\beta_g(A) = \beta_h(A)$ for all $A \in\cstar[B]$, so by Assumption \ref{asmp:Faithfulness}, $g = h$.
\end{proof}

\subsubsection{Direct sums and subobjects of \texorpdfstring{$G$}{G}-defect representations}
\label{sec:GSecCauchyComplete}
Recall that we have assumed that the cone algebras are infinite factors.
Therefore, there exist isometries $V_1, \dots, V_n \in \cR(\Lambda)$ for all $\Lambda \in \cL$ such that $\sum_{i = 1}^n V_i V_i^* = \mathds1$ \cite[Halving Lemma 6.3.3]{MR1468230}.
We observe that the above conditions imply that $V_i^* V_j = \delta_{ij} \mathds1$.
For $\pi_1, \dots, \pi_n \in \GSec_{\hom}$, the map $\bigoplus_{i = 1}^n \pi_i \colon \fA^a \to \fA^a$ defined by $$\bigoplus_{i = 1}^n \pi_i(-) \coloneqq \sum_{i = 1}^n V_i \pi_i(-)V_i^*$$ satisfies the universal property of the direct sum.

Note that if $\Lambda \in \cL$ and $\pi_1, \dots, \pi_n \in \GSec_g$ are $g$-localized in $\Lambda$ and we choose $V_1, \dots, V_n \in \cR(\Lambda)$, then $\pi \coloneqq \bigoplus_{i=1}^n \pi_i$ is also $g$-localized in $\Lambda$  and  transportable, and hence $\pi \in \GSec_g$  
(this can be seen by adapting a standard argument; see for instance \cite[Lem.~6.1]{MR2804555}).
Following \cite[Def.~2.8]{MR2183964}, we say that $\pi \in \GSec_{\hom}$ is \emph{$G$-localized} in $\Lambda \in \cL$ if 
\[
\pi(-)
=
\sum_{i = 1}^n V_i \pi_i(-)V_i^*,
\]
where each $\pi_i \in \GSec_{\hom}$ is $g_i$-localized in $\Lambda$ for some $g_i \in G$ and $V_1, \dots, V_n \in \cR(\Lambda)$.  
Additionally, if each $\pi_i$ is canonically $g_i$-localized, we say that $\pi$ is \emph{canonically $G$-localized}.  
As before, if $\pi$ is $G$-localized in $\Lambda$ and $\Lambda \subseteq \Delta$, then $\pi$ is $G$-localized in $\Delta$, but the statement does not hold if one replaces $G$-localized with canonically $G$-localized. 

\begin{defn}
\label{def:GSec}

We define the category $\GSec$ to be the category whose objects are $\bigoplus_{i = 1}^n \pi_i$ for $\pi_1, \dots, \pi_n \in \GSec_{\hom}$ and whose morphisms are intertwiners that live in $\fA^a_0$. 
We use the term \emph{$G$-defect representations} to refer to the objects of $\GSec$, and we use the term \emph{$G$-defect sectors} to refer to the isomorphism classes of representations in $\GSec$. 
As before, we say that a $G$-defect sector is \emph{irreducible} if any representation (hence all representations) in the sector is irreducible. 
Note that every irreducible $G$-defect sector is an isomorphism class of irreducible $g$-defect representations for some $g \in G$, i.e., an irreducible $g$-defect sector.
\end{defn}

\begin{rem}
\label{rem:GSec notation reference rep}
    We note that implicit in the definition of $\GSec$ (as well as $\GSec_g$ and $\GSec_{\hom}$) is the GNS representation $\pi_0$ of the reference state $\omega_0$. This $\pi_0$ of course builds the algebra $\fA^a$ and also is used in the definition of $G$-defect sectors.

    In examples we will need to use a few different reference representations. So for an arbitrary reference representation $\pi$ we will instead notate the category as $\GSec_{\pi}$ (similarly $(\GSec_{\pi})_g$ and $(\GSec_{\hom})_g$) in order to clearly emphasize that $\pi$ is the reference representation being used.
\end{rem}

\begin{rem}
\label{rem:GLocalizedRepresentations}

Note that if $\pi \in \GSec$, then $\pi$ is $G$-localized in some cone $\Lambda$. 
Indeed, there exist $\pi_1, \dots, \pi_n \in \GSec_{\hom}$ such that $\pi(-) = \sum_{i = 1}^n V_i \pi_i(-) V_i^*$, where $V_1, \dots, V_n \in \cR(\Delta)$ are isometries satisfying $\sum_{i = 1}^n V_i V_i^* = \mathds1$ and $\Delta \in \cL$.
In particular, each $\pi_i$ is $g_i$-localized in $\Lambda_i$ for some $g_i \in G$ and $\Lambda_i \in \cL$. 
There exists $\Lambda \in \cL$ such that each $\Lambda_i \subseteq \Lambda$ and $\Delta \subseteq \Lambda$, and thus $\pi$ is $G$-localized in $\Lambda$. 
Furthermore, by \ref{GSec:GLocalizedLargerRegion}, we may take $\Lambda$ such that each $\pi_i$ is canonically $g_i$-localized in $\Lambda$, and in that case $\pi$ is canonically $G$-localized in $\Lambda$. 
\end{rem}

\begin{lem}
\label{lem:IntertwinersBetweenGLocReps}

Let $\Lambda \in \cL$ and $\pi, \sigma \in \GSec$ be canonically $G$-localized in $\Lambda$. 
If $T \colon \pi \to \sigma$, then $T \in \cR(\Lambda^{+r})$.
\end{lem}

\begin{proof}
Since $\pi$ is canonically $G$-localized in $\Lambda$, there exist isometries $V_1, \dots, V_n \in \cR(\Lambda)$ such that $\sum_{j = 1}^n V_j V_j^* = \mathds1$ and $\pi(-) = \sum_{j = 1}^n V_j \pi_j(-) V_j^*$, where each $\pi_j$ is canonically $g_j$-localized in $\Lambda$.
Similarly, since $\sigma$ is canonically $G$-localized in $\Lambda$, there exist isometries $W_1, \dots, W_m \in \cR(\Lambda)$ such that $\sum_{i = 1}^m W_i W_i^* = \mathds1$ and $\sigma(-) = \sum_{i = 1}^n W_i \sigma_i(-) W_i^*$, where each $\sigma_i$ is canonically $h_i$-localized in $\Lambda$.
Therefore, we have that $T = \sum_{i, j} W_i T_{ij} V_j^*$, where $T_{ij} \coloneqq W_i^* T V_j \colon \pi_j \to \sigma_i$. 
If $g_j = h_i$, then $T_{ij} \in \cR(\Lambda^{+r})$ by \ref{GSec:IntertinersInConeAlgebras}. 
Otherwise, $T_{ij} = 0$ by Lemma \ref{lem:no morphisms between different gradings}.
Hence $T \in \cR(\Lambda^{+r})$. 
\end{proof}

Recall the definition of a $\rmW^*$-category (Section \ref{sec:basics of monoidal cats}).
\begin{prop}
\label{prop:GSecWstarCat}
The category $\GSec$ is a $\rmW^*$-category. 
\end{prop}

\begin{proof}
First, observe that $\GSec$ is a linear dagger category, since if $T \colon \pi \to \sigma$ in $\GSec$, then $T^* \colon \sigma \to \pi$. 
Therefore, since $\GSec$ admits finite direct sums, by \cite[Lem.~2.6]{MR808930}, it suffices to show that $\End(\pi)$ is a von Neumann algebra for every $\pi \in \GSec$.\footnote{The result as stated in \cite[Lem.~2.6]{MR808930} has as an assumption that the category is a $\rmC^*$-category. However, $\GSec$ is a $\rmC^*$-category since $\GSec$ is a linear dagger category that admits finite direct sums such that $\End(\pi)$ is a C*-algebra for every $\pi \in \GSec$. 
See Remark \ref{rem:JustCheckEndsWhenDirectSums}.}
By Remark \ref{rem:GLocalizedRepresentations}, $\pi$ is canonically $G$-localized in some $\Lambda \in \cL$, so by Lemma \ref{lem:IntertwinersBetweenGLocReps}, $\End(\pi) \subseteq \cR(\Lambda^{+r})$. 
Therefore, it suffices to show that $\End(\pi)$ is closed in WOT. 
But this can easily be seen to be true from the definition of $\End(\pi)$. 
Indeed, suppose $T_i \in \End(\pi)$ and $T_i \to T \in \cR(\Lambda^{+r})$ in WOT. 
Then $T_i\pi(x) = \pi(x)T_i$ for all $x \in \fA^a$, and since multiplication is separately continuous in WOT, we get that $T\pi(x) = \pi(x)T$ for all $x \in \fA^a$. 
Thus $T \in \End(\pi)$ as desired. 
\end{proof}

We now show that our category admits subobjects using an adaptation of \cite[Lem.~5.8]{MR4362722}.  

\begin{lem}
\label{lem:subobjects}
Let $\pi \in \GSec_g$ be $g$-localized in $\Lambda$ for some $g \in G$, and $p \colon \pi \to \pi$ be a projection.  
Then there exists an isometry $v \in \cR(\Lambda^{+r})$ such that $vv^* = p$.  
It follows that the map $\widehat{\pi} \colon \fA^a \to \fA^a$ given by $\widehat{\pi}(-) = v^* \pi(-)v$ is a $g$-defect representation localized in $\Lambda^{+r}$ and that $v \colon \widehat{\pi} \to \pi$.
\end{lem}

\begin{proof}
This proof is a simplified version of the proof of \cite[Lem.~5.8]{MR4362722}.
Let $\widetilde{\Lambda}, \Delta \in \cL$ be disjoint cones such that $\widetilde{\Lambda}, \Delta \subseteq \Lambda$.
Let $\widetilde{\pi} \in \GSec_g$ be unitarily equivalent to $\pi$ and $g$-localized in $\widetilde{\Lambda}$ satisfying that $\widetilde \pi|_{\cstar[\Lambda^c]} = \pi|_{\cstar[\Lambda^c]}$. 
Such a $\widetilde \pi$ exists since $\widetilde \Lambda \subseteq \Lambda$ and $\pi$ is $g$-localized in $\Lambda$. 
Finally, let $U \colon \pi \to \widetilde{\pi}$ be a unitary implementing the unitary equivalence. 
Then $UpU^* \colon \widetilde{\pi} \to \widetilde{\pi}$ is an intertwiner. 
Since $\widetilde{\pi}$ is $g$-localized in $\widetilde{\Lambda}$, which is disjoint from $\Delta$, we have that $UpU^* \in \cR(\widetilde{\Lambda}^c)' \subseteq \cR(\Delta)'$.
Furthermore, by bounded spread Haag duality, we have that $UpU^* \in \cR(\Lambda^c)' \subseteq \cR(\Lambda^{+r})$, and additionally $\cR(\Delta) \subseteq \cR(\Lambda^{+r})$.
Thus, by \cite[Lem.~5.10]{MR4362722}, $UpU^*$ is Murray-von Neumann equivalent to $\mathds1$ in $\cR(\Lambda^{+r})$, as $\cR(\Delta), \cR(\Lambda^{+r})$ are infinite factors acting on a separable Hilbert space.  
Since $U \colon \pi \to \widetilde{\pi}$ and $\tilde \pi|_{\cstar[\Lambda^c]} = \pi|_{\cstar[\Lambda^c]}$, we have that $U \in \cR(\Lambda^{+r})$ by Remark \ref{rem:IntertwinersForThingsThatLookTheSameOutsideOfCone}, so $p$ is Murray-von Neumann equivalent to $\mathds1$ in $\cR(\Lambda^{+r})$.  
Hence there exists an isometry $v \in \cR(\Lambda^{+r})$ such that $vv^* = p$.
One verifies that the map $\widehat{\pi} \colon \fA^a \to \fA^a$ given by $\widehat{\pi}(-) \coloneqq v^* \pi(-)v$ is $g$-localized in $\Lambda^{+r}$ and transportable and that $v \colon \widehat{\pi} \to \pi$.
\end{proof}

\begin{rem}
Note that using the notation of Lemma \ref{lem:subobjects}, we can (canonically) $g$-localize $\widehat{\pi}$ in $\Lambda$. 
Let $\check{\pi}$ be (canonically) $g$-localized in $\Lambda$ with $u \colon \check{\pi} \to \widehat{\pi}$ a unitary intertwiner. 
Then $\check{\pi}(-) = (vu)^* \pi(-) vu$, and $vu$ is an isometry such that $(vu)(vu)^* = vuu^*v^* = p$. 
In particular, we still have subobjects when restricting our attention to representations (canonically) $g$-localized in $\Lambda$. 
\end{rem}

\subsection{\texorpdfstring{$G$}{G}-crossed \texorpdfstring{$\rmW^*$}{W*}-tensor and braiding structure}
In this section, we show that $\GSec$ has the structure of a $G$-crossed braided $\rmW^*$-tensor category. 
To show this, we first construct the $G$-crossed monoidal structure and then construct the braiding.

\subsubsection{\texorpdfstring{$G$}{G}-crossed monoidal structure}
We henceforth identify $\cstar$ with $\pi_0(\cstar)$, since $\pi_0$ is a faithful representation.
We show that $\GSec$ has the structure of a strict $G$-crossed monoidal category.  
For $\pi, \sigma \in \GSec$, we define $\pi \otimes \sigma \coloneqq \pi \circ \sigma$ and for $T \colon \pi \to \pi'$ and $S \colon \sigma \to \sigma'$, we define $T \otimes S \coloneqq T\pi(S) = \pi'(S)T$.\footnote{Recall that $\pi$ and $\sigma$ are endomorphisms of $\fA^a$, so this monoidal product is well-defined.}
Note that for $\pi, \sigma \in \GSec$, we have that $\pi \otimes \sigma \in \GSec$ by the following lemma.  
It follows that $\GSec$ is a strict monoidal category.  
\footnote{This monoidal structure is the one inherited from viewing $\GSec$ as a subcategory of $\End(\sB\fA^a)$, where $\sB\fA^a$ is the one-object category whose morphisms are elements of $\fA^a$.
Indeed, if one unpacks the definition of $\End(\sB\fA^a)$, one will obtain the category whose objects are endomorphisms of $\fA^a$ and whose morphisms are intertwiners in $\fA^a$. 
The category $\End(\sB\fA^a)$ naturally has a monoidal structure, which is exactly the one we define here.} 

\begin{lem}
\label{lem:TensorProductPreservesGGrading}
For $\pi \in \GSec_g$ and $\sigma \in \GSec_h$, we have that $\pi \otimes \sigma \in \GSec_{gh}$.
\end{lem}

\begin{proof}
Since $\pi \in \GSec_g$, there exists $\Lambda_1 \in \cL$ such that $\pi$ is $g$-localized in $\Lambda_1$.  
Similarly, since $\sigma \in \GSec_h$, there exists $\Lambda_2 \in \cL$ such that $\sigma$ is $h$-localized in $\Lambda_2$.  
Now, there exists $\Lambda \in \cL$ such that $\Lambda_1 \cup \Lambda_2 \subseteq \Lambda$, so $\pi$ is $g$-localized in $\Lambda$ and $\sigma$ is $h$-localized in $\Lambda$.  
We have that $\pi|_{\cstar[\Lambda^c]} = \mu_1\circ \beta^{r(\Lambda)}_g$ and $\sigma|_{\cstar[\Lambda^c]} = \mu_2\circ \beta^{r(\Lambda)}_h$, where $\mu_1$ and $\mu_2$ are symmetry actions on finitely many sites. 
We then have that $\mu_1 \circ \beta^{r(\Lambda)}_g\circ \mu_2\circ \beta^{r(\Lambda)}_h$ differs from $\beta_g^{r(\Lambda)} \circ \beta_h^{r(\Lambda)} = \beta_{gh}^{r(\Lambda)}$ at finitely many sites, and 
\[
(\pi \otimes \sigma)|_{\cstar[\Lambda^c]}
=
(\pi \circ \sigma)|_{\cstar[\Lambda^c]}
=
\mu_1 \circ \beta^{r(\Lambda)}_g\circ \mu_2\circ \beta^{r(\Lambda)}_h|_{\cstar[\Lambda^c]}
=
\mu \circ \beta_{gh}^{r(\Lambda)}|_{\cstar[\Lambda^c]},
\]
where $\mu$ is a symmetry action on finitely many sites.
Thus $\pi \otimes \sigma$ is $gh$-localized at $\Lambda$.
Now we show that $\pi \otimes \sigma$ is transportable. Choose $\Delta \in \cL$. 
Indeed, since $\pi, \sigma$ are transportable, so we have $\widehat{\pi} \simeq \pi$ and $\widehat{\sigma} \simeq \sigma$ where $\widehat{\pi}, \widehat{\sigma}$ are $g, h$-localized in $\Delta$ respectively. Let $U \colon \pi \to \widehat{\pi}$ and $V \colon \sigma \to \widehat{\sigma}$ be the unitaries implementing the unitary equivalence. Then $U \otimes V \colon \pi \otimes \sigma \to \widehat{\pi} \otimes \widehat{\sigma}$ is a unitary, and $\widehat{\pi} \otimes \widehat{\sigma}$ is $gh$-localized in $\Delta$.
\end{proof}

It remains to show that $\GSec$ is $G$-crossed monoidal.  
We define the \emph{grading} $\partial \colon \GSec_{\hom} \to G$ by $\partial\pi \coloneqq g$ for $\pi \in \GSec_g$.  
Additionally, for $g \in G$, we define the \emph{$G$-action} $\gamma_g \colon \GSec \to \GSec$ as follows. 
For $\pi \in \GSec$, we define $\gamma_g(\pi) \coloneqq \beta_g \circ \pi \circ \beta_g^{-1}$, and for $T \colon \pi \to \sigma$, we define $\gamma_g(T) = \beta_g(T)$.  
Observe that $\gamma_g(\pi)$ and $\gamma_g(T)$ are well-defined since $\pi$ and $\beta_g$ are endomorphisms of $\fA^a$.\footnote{Recall that $\beta_g$ is implemented by a unitary in $B(\cH_0)$ since $\omega_0 \circ \beta_g = \omega_0$.} 
Additionally, with the above definitions, $\gamma_g(T) \colon \gamma_g(\pi) \to \gamma_g(\sigma)$, so $\gamma_g$ is a functor.

\begin{rem}
    According to the physics literature \cite{PhysRevB.100.115147}, if there is a state housing a symmetry defect of type $h$, then under the action of the symmetry group element $g\in G$, we have a $ghg^{-1}$ defect. The physical significance of the functor $\gamma_g$ is thus the action of the symmetry $g$ on a symmetry defect. 
\end{rem}

\begin{prop}
\label{prop:GCrossedMonoidal}
The maps $\partial \colon \GSec_{\hom} \to G$ and $\gamma_g(\pi) = \beta_g \circ \pi \circ \beta_g^{-1}$ defined above equip $\GSec$ with the structure of a strict $G$-crossed monoidal category (Definition \ref{def:G-crossed monoidal general v2} and Remark \ref{rem:StrictGCrossed}).  
\end{prop}

\begin{proof}
The proof proceeds analogously to that of \cite[Prop.~2.10]{MR2183964}.
Note that by construction, every object in $\GSec$ is a direct sum of objects in $\GSec_{\hom}$.  
Therefore, in order to show that $\GSec$ is strict $G$-crossed monoidal, we have to verify the following properties of $\partial$ and $\gamma$: 
\begin{enumerate}
    \item 
$\partial$ is constant on isomorphism classes, 
\item 
$\gamma_g \colon \GSec \to \GSec$ is a strict monoidal isomorphism, 
\item 
the map $g \mapsto \gamma_g$ is a group homomorphism, 
\item 
$\partial(\pi \otimes \sigma) = \partial \pi \partial \sigma$ for all $\pi, \sigma \in \GSec_{\hom}$
\item
$\gamma_g(\GSec_h) \subseteq \GSec_{ghg^{-1}}$
\end{enumerate}

\item[(1):] 
This follows from Lemma \ref{lem:no morphisms between different gradings}.

\item[(2):]
Note that 
\(\gamma_g \circ \gamma_{g^{-1}} = \Id_{\GSec} = \gamma_{g^{-1}} \circ \gamma_g,
\)
so $\gamma_g \colon \GSec \to \GSec$ is an isomorphism.  
It remains to show strict monoidality.  
Let $\pi, \sigma \in \GSec$.  
Then we have that 
\[
\gamma_g(\pi \otimes \sigma)
=
\beta_g \circ \pi \circ \sigma \circ \beta_g^{-1}
=
\beta_g \circ \pi \circ \beta_g^{-1} \circ \beta_g \circ \sigma \circ \beta_g^{-1}
=
\gamma_g(\pi) \otimes \gamma_g(\sigma).
\]
Similarly, for $T \colon \pi \to \pi'$ and $S \colon \sigma \to \sigma'$, we have that
\begin{align*}
\gamma_g(T \otimes S)
&=
\gamma_g(T\pi(S))
=
\beta_g(T)\beta_g(\pi(S))
=
\beta_g(T)\beta_g \circ \pi \circ \beta_g^{-1} (\beta_g(S))
\\&=
\gamma_g(T) \gamma_g(\pi)(\gamma_g(S))
=
\gamma_g(T) \otimes \gamma_g(S).
\end{align*}

\item[(3):]
Let $\pi \in \GSec$. We have, 
\[
\gamma_g \circ \gamma_h (\pi) = \beta_g \circ (\beta_h \circ \pi \circ \beta_h^{-1}) \circ \beta_g^{-1} = \beta_{gh} \circ \pi \circ \beta_{gh}^{-1}
=
\gamma_{gh}(\pi),
\]
and for $T \colon \pi \to \sigma$ in $\GSec$, we have that 
\[
\gamma_g \circ \gamma_h (T) = \beta_g \circ \beta_h (T) = \beta_{gh}(T)
=
\gamma_{gh}(T).
\]

\item[(4):]
This follows from Lemma \ref{lem:TensorProductPreservesGGrading}.

\item[(5):]
Let $\pi \in \GSec_h$. 
Then there exists $\Lambda \in \cL$ such that $\pi|_{\cstar[\Lambda^c]} = \mu\circ \beta^{r(\Lambda)}_h$, where $\mu$ is a symmetry action on finitely many sites  
(we've again identified $\cstar$ with $\pi_0(\cstar)$, since $\pi_0$ is a faithful representation).
We then have that for $A \in\cstar[\Lambda^c]$, 
\[
\gamma_g(\pi)(A)
=
\beta_g \circ \pi\circ \beta_g^{-1}(A)
=
\beta_g \circ (\mu\circ \beta^{r(\Lambda)}_h) \circ \beta_g^{-1}(A), 
\]
and $\beta_g \circ \mu \circ \beta^{r(\Lambda)}_h\circ \beta_g^{-1}$ differs from $\beta_{ghg^{-1}}^{r(\Lambda)}$ by a symmetry action on finitely many sites. 
Therefore, $\gamma_g(\pi)$ is $ghg^{-1}$-localized at $\Lambda$.
Now choose $\Delta \in \cL$ and $\widehat{\pi} \simeq \pi$, with $\widehat{\pi}$ being $h$-localized in $\Delta$, and $U \colon \pi \to \widehat{\pi}$ is a unitary. Then $\gamma_g(U) \colon \gamma_g(\pi) \to \gamma_g(\widehat{\pi})$ is a unitary, and a similar argument shows that $\gamma_g(\widehat{\pi})$ is $ghg^{-1}$-localized in $\Delta$. This shows $\gamma_g(\pi)$ is transportable.
\end{proof}

\begin{rem}
\label{rem:CanonicallyLocalizedGAction}
Note that if $\pi \in \GSec_h$ is canonically $h$-localized in a cone $\Lambda \in \cL$, then it follows by the above argument that $\gamma_g(\pi)$ is canonically $ghg^{-1}$-localized in $\Lambda$.  
Thus, if $\pi \in \GSec$ is canonically $G$-localized in $\Lambda \in \cL$, then so is $\gamma_g(\pi)$.
\end{rem}

\begin{rem}
Since $\gamma_g \circ \gamma_h = \gamma_{gh}$ (strict monoidality), it may appear that we do not have any symmetry fractionalization data \cite{PhysRevB.100.115147}.
However in the case where our category is finitely semisimple, this data can be recovered from our construction in a manner similar to the one used in \cite{2306.13762} to recover $F$- and $R$-symbols.
We describe how to do this in Section \ref{sec:SymmetryFractionalization}.
\end{rem}

\subsubsection{\texorpdfstring{$G$}{G}-crossed braided structure}
\label{sec:GCrossedBraiding}
We now define the braiding on $\GSec$. 

\begin{defn}
\label{def:sufficiently_to_the_left}
Let $\Lambda, \Delta \in \cL$.
We say that $\Delta$ is \emph{sufficiently to the left} of $\Lambda$ if $\Delta^{+r} \subseteq \ell(\Lambda)$ (recall $\ell(\Lambda)$ from Definition \ref{def:r(Lambda)}). The required geometry of cones is shown in figure \ref{fig:leftward_g_localised}. 

\begin{figure}[!ht]
    \centering
    \includegraphics[width=0.5\linewidth]{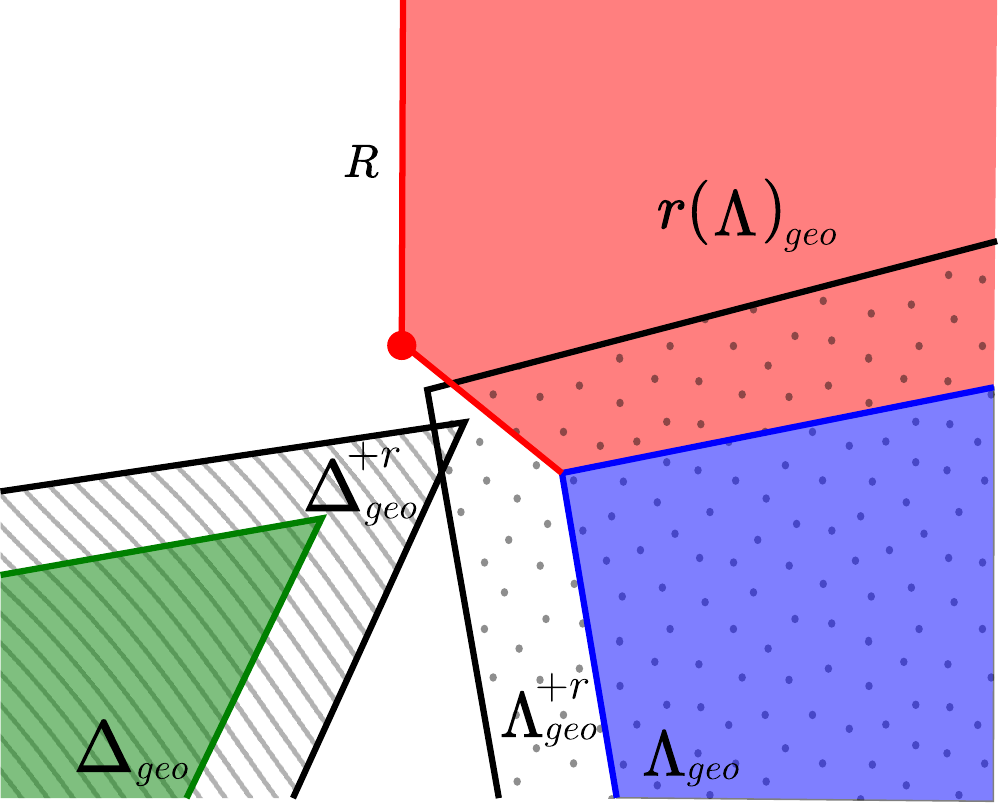}
    \caption{
    Example geometry of geometric cones $\Lambda_{geo}, \Delta_{geo}$ needed for $\Delta$ to be sufficiently to the left of 
$\Lambda$ (see Def \ref{def:sufficiently_to_the_left}). The geometric cone $\Lambda_{geo}$ is shown in blue, $\Delta_{geo}$ is shown in green, $r(\Lambda)_{geo}$ is shown in red, $\ell(\Lambda)_{geo}$ is given by $\Lambda_{geo}^c \setminus r(\Lambda)_{geo}$ (i.e.~the region not colored red or blue), and $(\Delta^{+r})_{geo}$ ($(\Lambda^{+r})_{geo}$) is shown with the striped (dotted) pattern. 
For $\Delta$ to be sufficiently to the left of $\Lambda$, it is required that $\Delta^{+r} = \Delta_{geo} \cap \Gamma$ is contained in $\ell(\Lambda) = \ell(\Lambda)_{geo} \cap \Gamma$.
It follows that $\Lambda^{+r}$ does not intersect with $\Delta$ (Remark \ref{rem:SufficientlyToTheLeftGeometry}).
}
    \label{fig:leftward_g_localised}
\end{figure}
\end{defn}

\begin{defn}
\label{def:LeftwardLocalized}
Let $\Lambda \in \cL$, and let $\Delta \in \cL$ be sufficiently to the left of $\Lambda$.  
We say that $\pi \in \GSec_g$ is \emph{leftward $g$-localized} in $\Delta$ if $\pi$ is $g$-localized in $\Delta$ and $\pi|_{\cstar[\Lambda^{+r} \cup r(\Lambda)]} = \beta_g|_{\cstar[\Lambda^{+r} \cup r(\Lambda)]}$. 
\end{defn}

\begin{rem}
\label{rem:SufficientlyToTheLeftGeometry}
We now verify that the above definition of leftward $g$-localized representations is well-defined.
To do so, we must show that the condition $\pi|_{\fA_{\Lambda^{+r} \cup r(\Lambda)}} = \beta_g|_{\fA_{\Lambda^{+r} \cup r(\Lambda)}}$ is consistent with the condition of $\pi$ being $g$-localized in $\Delta$. 
That is, we must show that it is possible for both of the following conditions to hold: $\pi|_{\fA_{\Lambda^{+r} \cup r(\Lambda)}} = \beta_g|_{\fA_{\Lambda^{+r} \cup r(\Lambda)}}$ and $\pi|_{\fA_{\Delta^c}} = \mu \circ \beta_g^{r(\Delta)}|_{\fA_{\Delta^c}}$ for some symmetry action $\mu$ on finitely many sites.
These two conditions can both hold provided that $\Lambda^{+r} \cup r(\Lambda)$ intersects $\ell(\Delta) = r(\Delta)^c \cap \Delta^c$ in a finite number of sites.
This last statement follows since $\Delta \in \cL$ is sufficiently to the left of $\Lambda \in \cL$, so $\Delta^{+r} \subseteq \ell(\Lambda)$. 

Furthermore, observe that for $\Lambda \in \cL$ and $\Delta \in \cL$ sufficiently to the left of $\Lambda$, we have that $\Lambda^{+r} \cup r(\Lambda) \subseteq \Delta^c$. 
Indeed, since $\Delta^{+r} \subseteq \ell(\Lambda)$, we have that $\ell(\Lambda) \subseteq (\Delta^{+r})^c$.
Therefore, since $(\Lambda \cup r(\Lambda))^c = \ell(\Lambda)$, we have that 
\[
\Lambda^{+r} \cup r(\Lambda)
\subseteq
((\Delta^{+r})^c)^{+r}_\Gamma
=
\Delta^c,
\]
where $((\Delta^{+r})^c)^{+r}_\Gamma \coloneqq ((\Delta^{+r})^c)^{+r} \cap \Gamma$.
Hence the definition of leftward $g$-localized, similar to the definition of $g$-localized, is a statement of how $\pi$ acts on a subalgebra of $\fA_{\Delta^c}$.
\end{rem}

\begin{facts}
Here are some facts about leftward $g$-localized representations that mirror Facts \ref{facts:GDefectBasics}.
\begin{enumerate}[label=(LgL\arabic*)]
\item
\label{LgL:LocalizedInLarger}
Let $\Lambda \in \cL$, and suppose $\Delta, \widehat{\Delta} \in \cL$ are both sufficiently to the left of $\Lambda$ and $\Delta \subseteq \widehat{\Delta}$. 
If $\pi$ is leftward $g$-localized in $\Delta$, then $\pi$ is leftward $g$-localized in $\widehat{\Delta}$.

\begin{proof}
Since $\pi$ is leftward $g$-localized in $\Delta$ and $\Delta \subseteq \widehat{\Delta}$, we have that $\pi$ is $g$-localized in $\widehat{\Delta}$, and 
\[
\pi|_{\cstar[\Lambda^{+r} \cup r(\Lambda)]} = \beta_g|_{\cstar[\Lambda^{+r} \cup r(\Lambda)]}.
\qedhere
\]
\end{proof}

\item
\label{LgL:IntertwinerLocalized}
Let $\Lambda \in \cL$, let $\Delta \in \cL$ be sufficiently to the left of $\Lambda$, and suppose $\pi, \sigma \in \GSec_g$ be leftward $g$-localized in $\Delta$. 
If $T \colon \pi \to \sigma$, then $T \in \cR(\ell(\Lambda))$.  

\begin{proof}
If $\pi|_{\cstar[\Delta^c]} = \sigma|_{\cstar[\Delta^c]}$, then $T \in \cR(\Delta^{+r}) \subseteq \cR(\ell(\Lambda))$ by Remark \ref{rem:IntertwinersForThingsThatLookTheSameOutsideOfCone}.  
Since $\pi, \sigma \in \GSec_g$ are leftward $g$-localized in $\Delta$, we have that $\pi|_{\cstar[\Delta^c]} = \Ad(U) \circ \sigma|_{\cstar[\Delta^c]}$ for some unitary $U \in \pi_0(\cstar[(\Lambda^{+r} \cup r(\Lambda))^c])$ that is of the form $U = \bigotimes_{s \in S} U^g_s$ for some $S \in \Gamma_f$. 
Thus $(\Ad(U^*) \circ \pi)|_{\cstar[\Delta^c]} = \sigma|_{\cstar[\Delta^c]}$ and $UT \colon \Ad(U^*) \circ \pi \to \sigma$. 
Therefore, by Remark \ref{rem:IntertwinersForThingsThatLookTheSameOutsideOfCone}, $UT \in \cR(\ell(\Lambda))$. 
Since $(\Lambda^{+r} \cup r(\Lambda))^c \subseteq (\Lambda \cup r(\Lambda))^c = \ell(\Lambda)$, we have that $U \in \pi_0(\cstar[(\Lambda^{+r} \cup r(\Lambda))^c]) \subseteq \cR(\ell(\Lambda))$, so $T \in \cR(\ell(\Lambda))$.
\end{proof}
\end{enumerate}
\end{facts}

The following lemma should be compared with \cite[Lem.~2.14]{MR2183964}.

\begin{lem}
\label{lem:GCrossedCommutation}
Let $\sigma$ be canonically $G$-localized in $\Lambda$, and let $\pi \in \GSec_g$ be leftward $g$-localized in $\Delta$ for some $\Delta$ sufficiently to the left of $\Lambda$.  
Then $\pi \otimes \sigma = \gamma_g(\sigma) \otimes \pi$.
\end{lem}

\begin{proof}
We adapt the proof of \cite[Lem.~2.14]{MR2183964}. 
By the definition of canonically $G$-localized, $\sigma(-) = \sum_{i = 1}^n V_i \sigma_i(-)V_i^*$ where each $\sigma_i$ is canonically $h_i$-localized in $\Lambda$ for some $h_i \in G$ and $V_1, \dots, V_n \in \cR(\Lambda)$. 
Since $\Delta$ is sufficiently to the left of $\Lambda$ and $\pi$ is leftward $g$-localized in $\Delta$, we have that $\pi(V_i) = \beta_g(V_i)$.
Therefore, it suffices to show that $\pi \otimes \sigma = \gamma_g(\sigma) \otimes \pi$ for $\sigma$ canonically $h$-localized in $\Lambda$.  
We proceed by showing that $\pi \otimes \sigma(A) = \gamma_g(\sigma) \otimes \pi(A)$ for the following cases: 
\begin{enumerate}
    \item 
$A \in\cstar[r(\Lambda)]$,
\item 
$A \in\cstar[\Lambda]$, 
\item 
$A \in\cstar[\Delta]$, and
\item 
$A \in\cstar[\ell(\Lambda) \cap \Delta^c]$.
\end{enumerate}
\item[(1):]
In this case, $\sigma(A) = \beta_h(A)$, and since $\pi$ is leftward $g$-localized in $\Delta$, we have that $\pi(A) = \beta_g(A)$.  
Therefore, we have that
\[
\pi \otimes \sigma(A)
=
\beta_g(\beta_h(A))
=
(\beta_g \circ \beta_h \circ \beta_g^{-1} \circ \beta_g)(A)
=
(\beta_g \circ \sigma \circ \beta_g^{-1})(\beta_g(A))
=
\gamma_g(\sigma) \otimes \pi(A).
\]

\item[(2):]
In this case, $\sigma(A) \in \cR(\Lambda^{+r})$ by \ref{GSec:MappingsOnConeAlgebras}.  
Since $\pi$ is leftward $g$-localized in $\Delta$, we have that 
\[
\pi \otimes \sigma(A)
=
\beta_g(\sigma(A))
=
(\beta_g \circ \sigma \circ \beta_g^{-1} \circ \beta_g(A))
=
\gamma_g(\sigma) \otimes \pi(A).
\]

\item[(3):]
Since $\Delta$ is sufficiently to the left of $\Lambda$, we have that $\Delta \subseteq \Delta^{+r} \subseteq \ell(\Lambda)$. 
Furthermore, since $A \in \cstar[\Delta]$, we have $\pi(A) \in \cR(\Delta^{+r})$ by \ref{GSec:MappingsOnConeAlgebras}. 
Now, $\sigma$ is canonically $h$-localized in $\Lambda$, and thus $\gamma_g(\sigma)$ is canonically $ghg^{-1}$-localized in $\Lambda$ by Remark \ref{rem:CanonicallyLocalizedGAction}. 
Therefore, we get that $\sigma(A) = A$ and $(\gamma_g(\sigma) \circ \pi)(A) = \pi(A)$.  
Thus, we have that 
\[
(\pi \otimes \sigma)(A)
=
\pi(\sigma(A))
=
\pi(A)
=
(\gamma_g(\sigma) \circ \pi)(A)
=
(\gamma_g(\sigma) \otimes \pi)(A).
\]

\item[(4):]
Since $A \in\cstar[\ell(\Lambda)]$, we have $\sigma(A) = A$.  
In addition, since $A \in\cstar[\Delta^c]$ and $\pi$ is $g$-localized in $\Delta$, we observe that $\pi(A)$ has the same support as $A$.  
In particular, $\pi(A) \in \cstar[\ell(\Lambda)]$.  
Therefore, we have that 
\[
\pi \otimes \sigma(A)
=
\pi(A)
=
\gamma_g(\sigma) \otimes \pi(A).
\qedhere
\]
\end{proof}

We now construct $c_{\pi, \sigma} \colon \pi \otimes \sigma \to \gamma_g(\sigma) \otimes \pi$ for $\pi \in \GSec_g$ and $\sigma \in \GSec$, using the approach of \cite[Prop.~2.17]{MR2183964}.

\begin{defn}
\label{def:GCrossedBraiding}

Let $\pi \in \GSec_g$ and $\sigma \in \GSec$. 
Note that by \ref{GSec:GLocalizedLargerRegion} and Remark \ref{rem:GLocalizedRepresentations}, there exists $\Lambda \in \cL$ such that $\pi$ is canonically $g$-localized in $\Lambda$ and $\sigma$ is canonically $G$-localized in $\Lambda$.
We also assume that $\partial R \in \Lambda_{geo}$ by enlarging $\Lambda$ if necessary
(Recall that $\partial R$ is the endpoint of $R$).
Choose $\Delta$ sufficiently to the left of $\Lambda$ and $\widetilde{\pi} \simeq \pi$ leftward $g$-localized in $\Delta$.  
Let $U \colon \pi \to \widetilde{\pi}$ be a unitary intertwiner.  
We then define the braiding isomorphism as 
\begin{equation*}
c_{\pi, \sigma}
\coloneqq
(\Id_{\gamma_g(\sigma)} \otimes U^*)(U \otimes \Id_\sigma)
=
\gamma_g(\sigma)(U^*)U 
\end{equation*}
\end{defn}

\begin{lem}
\label{lem:GCrossedBraidingIsIntertwiner}
Let $\pi \in \GSec_g$ and $\sigma \in \GSec$. 
The unitary $c_{\pi, \sigma}$ in Definition \ref{def:GCrossedBraiding} is an intertwiner from $\pi \otimes \sigma$ to $\gamma_g(\sigma) \otimes \pi$, i.e., $$c_{\pi,\sigma} (\pi \otimes \sigma)(-) = (\gamma_g(\sigma) \otimes \pi)(-) c_{\pi, \sigma}.$$ 
\end{lem}

\begin{proof}
Let $\Lambda$, $\Delta$, $\widetilde{\pi}$, and $U \colon \pi \to \widetilde{\pi}$ be as in Definition \ref{def:GCrossedBraiding}. 
We follow the schematic of the proof illustrated in Figure \ref{fig:depiction_of_braiding_isomorphism}. 
First, since $U \colon \pi \to \widetilde{\pi}$, we have that $U = U \otimes \Id_\sigma \colon \pi \otimes \sigma \to \widetilde{\pi} \otimes \sigma$. 
Now, by assumption, $\sigma$ is canonically $G$-localized in $\Lambda$ and $\widetilde \pi$ is leftward $g$-localized in $\Delta$, which is sufficiently to the left of $\Lambda$. 
Therefore, by Lemma \ref{lem:GCrossedCommutation}, $\widetilde \pi \otimes \sigma = \gamma_g(\sigma) \otimes \widetilde{\pi}$. 
Finally, $\gamma_g(\sigma)(U^*) = (\Id_{\gamma_g(\sigma)} \otimes U^*) \colon \gamma_g(\sigma) \otimes \widetilde{\pi} \to \gamma_g(\sigma) \otimes \pi$.
Putting this sequence of operations together, we get that 
\[
c_{\pi, \sigma}
=
\gamma_g(\sigma)(U^*)U
\colon 
\pi \otimes \sigma \to \gamma_g(\sigma) \otimes {\pi}.
\qedhere
\]

\begin{figure}[!ht]
\begin{subfigure}[t]{0.4\linewidth}
    \includegraphics[width=\linewidth]{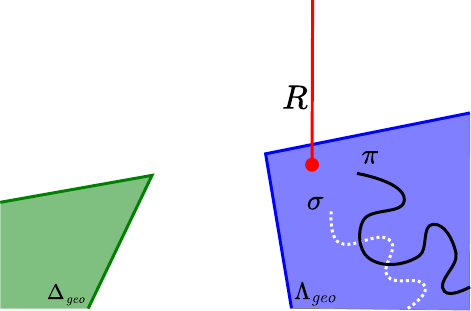}
    \caption{A cartoon of $\pi \otimes \sigma$, where the right component of the tensor is depicted by dotted lines.}
    \label{fig:defect_braid_1}
\end{subfigure}
\hspace{5mm}
\begin{subfigure}[t]{0.4\linewidth}
    \includegraphics[width=\linewidth]{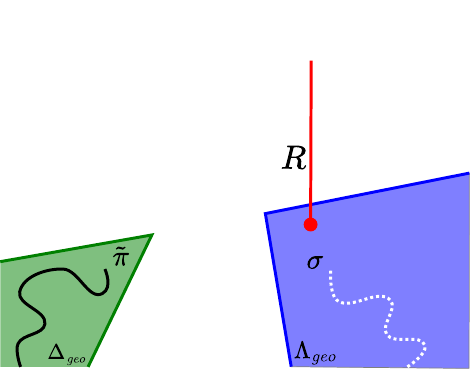}
    \caption{We conjugate by the unitary $U\colon \pi \rightarrow \tilde \pi$ to get $\tilde \pi \otimes \sigma = \Ad(U \otimes \Id_\sigma) (\pi \otimes \sigma)$ with $\tilde \pi$ leftward $g$-localised in $\Delta$.}
    \label{fig:defect_braid_2}
\end{subfigure}
\begin{subfigure}[t]{0.4\linewidth}
    \includegraphics[width=\linewidth]{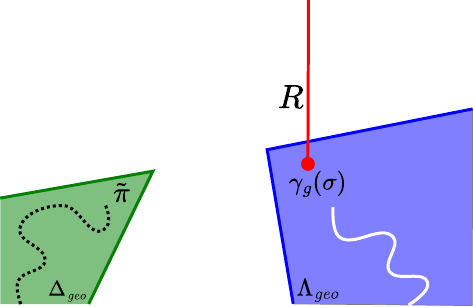}
    \caption{$\gamma_g(\sigma) \otimes \tilde \pi = \tilde \pi \otimes \sigma$ using Lemma \ref{lem:GCrossedCommutation} as $\tilde \pi$ is leftward $g$-localised in $\Delta$.}
    \label{fig:defect_braid_3}
\end{subfigure}
\hspace{5mm}
\begin{subfigure}[t]{0.4\linewidth}
    \includegraphics[width=\linewidth]{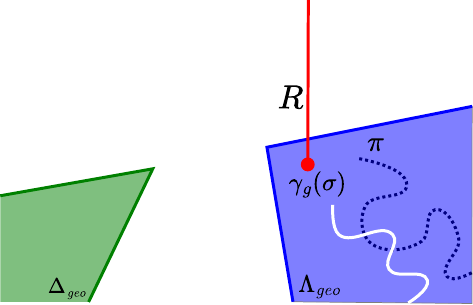}
    \caption{We can again use $U^*$ to get $\gamma_g(\sigma) \otimes \pi = \Ad(\Id_{\gamma_g(\sigma)}\otimes U^*) (\gamma_g(\sigma) \otimes \tilde \pi)$.}
    \label{fig:defect_braid_4}
\end{subfigure}
\caption{The procedure carried out by the \emph{braiding isomorphism} $c_{\pi, \sigma}$. Note that we require $\Delta, \Lambda$ to be such that $\Delta$ is sufficiently to the left of $\Lambda$ (Definition \ref{def:sufficiently_to_the_left}). We use dotted lines to represent the right component of the tensor.
}
\label{fig:depiction_of_braiding_isomorphism}
\end{figure}
\end{proof}

\begin{lem}
Let $\pi \in \GSec_g$ and $\sigma \in \GSec$.
The map $c_{\pi, \sigma}$ from Definition \ref{def:GCrossedBraiding} does not depend on the choices of $U$, $\widetilde{\pi}$, $\Delta$, and $\Lambda$.
\end{lem}

\begin{proof}
We adapt the proof in \cite[Prop.~2.17]{MR2183964}.
We first show independence of $\widetilde{\pi}$ and $U$.
Let $\widehat{\pi} \simeq \pi$ be another $g$-defect representation leftward $g$-localized in $\Delta$, and let $V \colon \pi \to \widehat{\pi}$ be a unitary. 
We wish to show that 
\[
\gamma_g(\sigma)(U^*)U
=
\gamma_g(\sigma)(V^*)V.
\]
This is equivalent to showing that 
\[
\gamma_g(\sigma)(VU^*)
=
VU^*.
\]
Now, $VU^* \colon \widetilde{\pi} \to \widehat{\pi}$. 
Therefore, by \ref{LgL:IntertwinerLocalized}, since $\Delta$ is sufficiently to the left of $\Lambda$ and $\widetilde{\pi}, \widehat{\pi}$ are both leftward $g$-localized in $\Delta$, we have that $VU^* \in \cR(\ell(\Lambda))$.  
The desired result thus holds since $\sigma$ (and hence $\gamma_g(\sigma)$) is canonically $G$-localized in $\Lambda$. 

Next, we show that $c_{\pi, \sigma}$ does not depend on $\Delta$.  
Suppose $\Delta, \widehat{\Delta} \in \cL$ are both sufficiently to the left of $\Lambda$. The existence of a single cone $\widetilde \Delta$ satisfying both $\widetilde \Delta \subset \Delta, \widehat \Delta$ as well as $\widetilde \Delta \subset \ell(\Lambda)$ is not necessary guaranteed.  
However, we are able to `zig-zag' between the two cones without leaving $\ell(\Lambda)$.\footnote{The paper \cite{2410.21454} uses zig-zags to show well-definedness of the braiding.
The idea to apply zig-zags here came from work on that paper, and more specifically from discussions with David Penneys.
Our use of zig-zags is analogous to the use of ``interpolating" or ``alternating" sequences of spacelike cones in \cite{MR660538}.}  
More precisely, a \emph{zig-zag from $\Delta$ to $\widehat{\Delta}$} is a sequence of cones $(\Delta_1, \widetilde{\Delta}_1, \dots, \Delta_n, \widetilde{\Delta}_n, \Delta_{n + 1})$ such that $\Delta_1 = \Delta$, $\Delta_{n + 1} = \widehat{\Delta}$, and for each $i = 1, \dots, n$, $\Delta_i, \Delta_{i + 1} \subseteq \widetilde{\Delta}_i$ \cite[\S 1.1]{2410.21454}. An example zig-zag with $n=2$ is shown in figure \ref{fig:zigzag}. 

Now, observe that given $\Delta, \widehat{\Delta} \in \cL$ sufficiently to the left of $\Lambda$, there exists a zig-zag from $\Delta$ to $\widehat{\Delta}$ where each cone in the zig-zag is sufficiently to the left of $\Lambda$.  
It therefore suffices to show that given $\Delta_i, \Delta_{i + 1} \subseteq \widetilde{\Delta}_i$, $g$-localizing in $\Delta_i$ and $\Delta_i$ give the same $c_{\pi, \sigma}$.  
But this follows since $g$-defect representations leftward $g$-localized in $\Delta_i$/$\Delta_{i + 1}$ are leftward $g$-localized in $\widetilde{\Delta}_i$, and we already showed independence of $\widetilde{\pi}$ leftward $g$-localized in $\widetilde{\Delta}_i$.

\begin{figure}[!ht]
    \centering
    \includegraphics[width=0.5\linewidth]{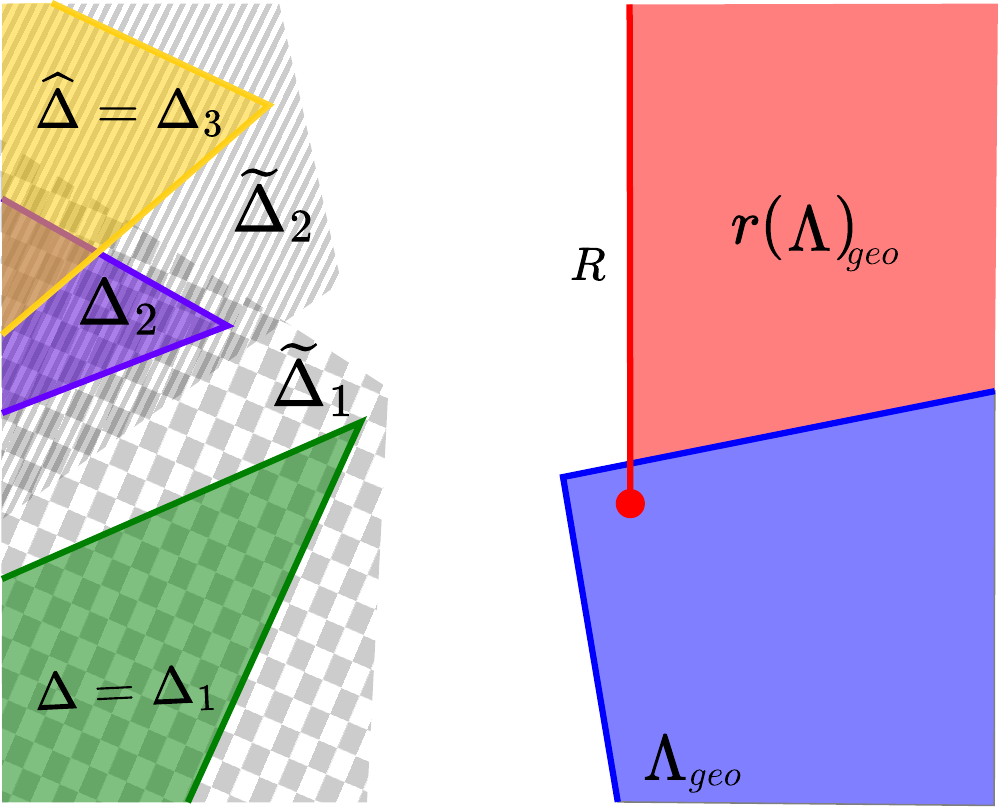}
    \caption{
    A zig-zag with $n=2$ between cones $\Delta$, $\widehat \Delta$. The subscript $geo$ has been suppressed for the cones except $\Lambda$ for clarity. Notice that given the geometry of cones $\Delta, \widehat \Delta$, it is not possible to find a single cone $\widetilde \Delta \subset \ell(\Lambda)$ that contains both $\Delta, \widehat \Delta$ and is sufficiently to the left of $\Lambda$. However we can zig-zag between them by choosing the zig-zag sequence ($\Delta_1 = \Delta, \widetilde \Delta_1, \Delta_2, \widetilde \Delta_2, \Delta_3 = \widehat \Delta)$ such that all cones in this sequence lie in $\ell(\Lambda)$ and are sufficiently to the left of $\Lambda$.}
    \label{fig:zigzag}
\end{figure}

Finally, we show that $c_{\pi, \sigma}$ does not depend on $\Lambda$. 
Let $\Lambda_1, \Lambda_2 \in \cL$ be two cones such that $\pi$ is canonically $g$-localized in $\Lambda_i$, $\sigma$ is canonically $G$-localized in $\Lambda_i$, and $\partial R \in (\Lambda_1)_{geo} \cap (\Lambda_2)_{geo}$. 
Let $\widehat{\Lambda} \in \cL$ such that $\Lambda_1, \Lambda_2 \subseteq \widehat{\Lambda}$. 
Then $\partial R \in \widehat{\Lambda}_{geo}$. 
Furthermore, by Remark \ref{rem:glocalizedreps}, $\pi$ is canonically $g$-localized in $\widehat{\Lambda}$, and by Remark \ref{rem:GLocalizedRepresentations}, $\sigma$ is canonically $G$-localized in $\widehat{\Lambda}$.
In addition, since $\Lambda_1, \Lambda_2 \subseteq \widehat{\Lambda}$ and $\partial R \in (\Lambda_1)_{geo} \cap  (\Lambda_2)_{geo} \cap \widehat{\Lambda}_{geo}$, we have that $\ell(\Lambda_1), \ell(\Lambda_2) \supseteq \ell(\widehat{\Lambda})$.
Thus, every cone $\Delta$ that is sufficiently to the left of $\widehat{\Lambda}$ is also sufficiently to the left of each $\Lambda_i$. 
Since $c_{\pi, \sigma}$ does not depend on $\Delta$, it therefore also does not depend on $\Lambda$.
\end{proof}

\begin{prop}
\label{prop:GCrossedBraiding}
The category $\GSec$ is $G$-crossed braided using the braid isomorphism from Definition \ref{def:GCrossedBraiding}.
\end{prop}

\begin{proof}
We proceed as in the proof in \cite[Prop.~2.17]{MR2183964}.
First, note that the isomorphism in Definition \ref{def:GCrossedBraiding} is an intertwiner by Lemma \ref{lem:GCrossedBraidingIsIntertwiner}.
To show that Definition \ref{def:GCrossedBraiding} gives a $G$-crossed braiding on $\GSec$, we must show that the following conditions of being a $G$-crossed braiding are satisfied:

\begin{enumerate}
    \item Naturality
    \item Monoidality
    \item Braiding is preserved by $\gamma_g$
\end{enumerate}

\item[{(1):}]
There are two naturality equations that must be verified; we verify each in turn.  
First, suppose $\pi \in \GSec_g$, and $T \colon \sigma_1 \to \sigma_2$ is a morphism in $\GSec$.  
We must show that 
\[
(\gamma_g(T) \otimes \Id_\pi)c_{\pi, \sigma_1}
=
c_{\pi, \sigma_2} (\Id_\pi \otimes T).
\]
Let $\Lambda \in \cL$ be such that $\pi$ is canonically $g$-localized in $\Lambda$, $\sigma_1$ and $\sigma_2$ are canonically $G$-localized in $\Lambda$, and $\partial R \in \Lambda_{geo}$. 
Let $\Delta$ be sufficiently to the left of $\Lambda$, $\widetilde{\pi} \simeq \pi$ be leftward $g$-localized in $\Delta$, and $U \colon \pi \to \widetilde{\pi}$ be a unitary. 
The equation to verify then becomes
\[
\gamma_g(T)\gamma_g(\sigma_1)(U^*)U
=
\gamma_g(\sigma_2)(U^*)U\pi(T).
\]
We proceed starting with the right-hand side.  
We first observe that 
\[
\gamma_g(\sigma_2)(U^*)U\pi(T)=\gamma_g(\sigma_2)(U^*)\widetilde{\pi}(T)U=\gamma_g(\sigma_2)(U^*)\beta_g(T)U.
\]
The last equality follows since $\widetilde{\pi}$ is leftward $g$-localized in $\Delta$ and $T \in \cR(\Lambda^{+r})$ by Lemma \ref{lem:IntertwinersBetweenGLocReps}. Now,
\[
\gamma_g(\sigma_2)(U^*)\beta_g(T)U
=
\beta_g(\sigma_2(\beta_g^{-1}(U^*))T)U=\beta_g(T\sigma_1(\beta_g^{-1}(U^*)))U=\gamma_g(T)\gamma_g(\sigma_1(U^*))U.
\]

Now, suppose $\sigma \in \GSec$ and $T \colon \pi_1 \to \pi_2$ is a morphism in $\GSec_g$.  
We must show that 
\[
(\Id_{\gamma_g(\sigma)} \otimes T)c_{\pi_1, \sigma}
=
c_{\pi_2, \sigma}(T \otimes \Id_\sigma).
\]
Let $\Lambda \in \cL$ be such that $\pi_1$ and $\pi_2$ are canonically $g$-localized in $\Lambda$, $\sigma$ is canonically $G$-localized in $\Lambda$, and $\partial R \in \Lambda_{geo}$. 
Let $\Delta$ be sufficiently to the left of $\Lambda$. 
For $i = 1, 2$, we let $\widetilde{\pi}_i \simeq \pi_i$ be leftward $g$-localized in $\Delta$, and $U_i \colon \pi_i \to \widetilde{\pi}_i$ be a unitary. 
The equation to verify then becomes 
\[
\gamma_g(\sigma)(T)\gamma_g(\sigma)(U_1^*)U_1
=
\gamma_g(\sigma)(U_2^*)U_2T.
\]
Note that the above equation is equivalent to 
\[
\gamma_g(\sigma)(U_2TU_1^*)
=
U_2TU_1^*.
\]
But since $U_2TU_1^* \colon \widetilde{\pi}_1 \to \widetilde{\pi}_2$ and $\widetilde{\pi}_1,\widetilde{\pi}_2$ are both leftward $g$-localized in $\Delta$, we have that $U_2TU_1^* \in \cR(\ell(\Lambda))$ by \ref{LgL:IntertwinerLocalized}. 
Therefore, since $\sigma$ (and hence $\gamma_g(\sigma)$) is canonically $G$-localized in $\Lambda$, the desired equation holds.

\item[(2):]
Again, there are two equations that we must verify.  
First, let $\pi \in \GSec_g$ and $\sigma, \tau \in \GSec$.  We must show that
\[
c_{\pi, \sigma \otimes \tau}
=
(\Id_{\gamma(\sigma)} \otimes c_{\pi, \tau})(c_{\pi, \sigma} \otimes \Id_\tau).
\]
Let $\Lambda \in \cL$ be such that $\pi$ is canonically $g$-localized in $\Lambda$, $\sigma, \tau$ are canonically $G$-localized in $\Lambda$, and $\partial R \in \Lambda_{geo}$. 
Let $\Delta$ be sufficiently to the left of $\Lambda$, $\widetilde{\pi} \simeq \pi$ be leftward $g$-localized in $\Delta$, and $U \colon \pi \to \widetilde{\pi}$ be a unitary. 
The equation to verify then becomes
\[
\gamma_g(\sigma \otimes \tau)(U^*)U
=
\gamma_g(\sigma)(\gamma_g(\tau)(U^*)U)\gamma_g(\sigma)(U^*)U.
\]
But this is easily seen to hold.  
Indeed, working from the right-hand side, we have that 
\[
\gamma_g(\sigma)(\gamma_g(\tau)(U^*)U)\gamma_g(\sigma)(U^*)U
=
(\gamma_g(\sigma) \circ \gamma_g(\tau))(U^*) \gamma_g(\sigma)(U)\gamma_g(\sigma)(U^*)U
=
\gamma_g(\sigma \otimes \tau)(U^*)U.
\]

Now, let $\pi \in \GSec_g$, $\sigma \in \GSec_h$, and $\tau \in \GSec$.
We must show that 
\[
c_{\pi \otimes \sigma, \tau}
=
(c_{\pi, \gamma_h(\tau)} \otimes \Id_\sigma)(\Id_\pi \otimes c_{\sigma, \tau}).
\]
Let $\Lambda \in \cL$ be such that $\pi$ is canonically $g$-localized in $\Lambda$, $\sigma$ is canonically $h$-localized in $\Lambda$, $\tau$ is canonically $G$-localized in $\Lambda$, and $\partial R \in \Lambda_{geo}$. 
Let $\Delta$ be sufficiently to the left of $\Lambda$, $\widetilde{\pi} \simeq \pi$ be leftward $g$-localized in $\Delta$, $\widetilde{\sigma} \simeq \sigma$ be leftward $h$-localized in $\Delta$, and $U \colon \pi \to \widetilde{\pi}$ and $V \colon \sigma \to \widetilde{\sigma}$ be unitaries. 
Note that $\widetilde{\pi} \otimes \widetilde{\sigma}$ is leftward $gh$-localized in $\Delta$, and $U \otimes V = U\pi(V) = \widetilde{\pi}(V)U \colon \pi \otimes \sigma \to \widetilde{\pi} \otimes \widetilde{\sigma}$ is a unitary.
The desired equation therefore becomes
\[
\gamma_{gh}(\tau)(U^*\widetilde{\pi}(V^*))U\pi(V)
=
\gamma_g(\gamma_h(\tau))(U^*)U\pi(\gamma_h(\tau)(V^*)V)
=
\gamma_{gh}(\tau)(U^*)U\pi(\gamma_h(\tau)(V^*))\pi(V).
\]
Note that the above equation is equivalent to 
\[
\gamma_{gh}(\tau)(\widetilde{\pi}(V^*))U
=
U\pi(\gamma_h(\tau)(V^*))
=
\widetilde{\pi}(\gamma_h(\tau)(V^*))U,
\]
so it suffices to show that $\gamma_{gh}(\tau)(\widetilde{\pi}(V^*)) = \widetilde{\pi}(\gamma_h(\tau)(V^*))$.
But this holds by Lemma \ref{lem:GCrossedCommutation} since $\gamma_h(\tau)$ is canonically $G$-localized in $\Lambda$ and $\widetilde{\pi}$ is leftward $g$-localized in $\Delta$.  

\item[(3):]
We must show that for $\pi \in \GSec_h$ and $\sigma \in \GSec$, 
\[
\gamma_g(c_{\pi, \sigma})
=
c_{\gamma_g(\pi), \gamma_g(\sigma)}.
\]
Let $\Lambda \in \cL$ be such that $\pi$ is canonically $h$-localized in $\Lambda$, $\sigma$ is canonically $G$-localized in $\Lambda$, and $\partial R \in \Lambda_{geo}$. 
Let $\Delta$ be sufficiently to the left of $\Lambda$, $\widetilde{\pi} \simeq \pi$ be leftward $h$-localized in $\Delta$, and $U \colon \pi \to \widetilde{\pi}$ be a unitary. 
Note that $\gamma_g(\widetilde{\pi})$ is leftward $ghg^{-1}$-localized in $\Delta$, and $\gamma_g(U) \colon \gamma_g(\pi) \to \gamma_g(\widetilde{\pi})$ is a unitary. 
We therefore have that 
\begin{align*}
c_{\gamma_g(\pi), \gamma_g(\sigma)}
&=
\gamma_{ghg^{-1}}(\gamma_g(\sigma))(\gamma_g(U^*))\gamma_g(U)
=
\beta_{g}(\gamma_h(\sigma)(U^*))\beta_g(U)
=
\gamma_g(c_{\pi, \sigma}).
\qedhere
\end{align*}
\end{proof}

Recall the definition of a $G$-crossed braided $\rmW^*$-tensor category (Definition \ref{def:G-crossed braided general v2}).
\begin{thm}
\label{thm:GSecGCrossedBraidedWStarTensorCat}
The category $\GSec$ is a strict $G$-crossed braided $\rmW^*$-tensor category.
\end{thm}

\begin{proof}
By Proposition \ref{prop:GSecWstarCat}, $\GSec$ is a $\rmW^*$-category, and by Proposition \ref{prop:GCrossedBraiding}, $\GSec$ is a strict $G$-crossed braided monoidal category. 
Furthermore, $\GSec$ is a tensor category since it admits direct sums and subobjects and has simple unit.\footnote{The fact that $\GSec$ has simple unit follows since the reference state $\omega_0$ is pure.}
To check that it is a $G$-crossed braided $\rmW^*$-tensor category, we must verify that the following hold: 
\begin{enumerate}
\item
The functor $- \otimes - \colon \GSec \times \GSec \to \GSec$ is bi-normal, 
\item 
For every $g \in G$, the functor $\gamma_g \colon \GSec \to \GSec$ is normal, and 
\item 
The braiding $c_{\pi, \sigma}$ for $\pi \in \GSec_{\hom}$ and $\sigma \in \GSec$ is a unitary. 
\end{enumerate}
All of these properties are easily seen to hold. 
Indeed, for $T \colon \pi \to \pi'$ and $S \colon \sigma \to \sigma'$, we have that $T \otimes S = T\pi(S)$, which is bi-normal since $\pi$ is normal on allowed cone algebras.\footnote{Note that $g$-defect representations act on the allowed cone algebras as conjugation by a unitary by the proof of Lemma \ref{lem:GDefectsDefinedOnAuxiliaryAlgebra}. 
A general object of $\GSec$ is the direct sum of $g$-defect representations and is therefore also normal on allowed cone algebras.}
In addition, for $T \colon \pi \to \sigma$ in $\GSec$, we have that $\gamma_g(T) = \beta_g(T)$. 
Thus $\gamma_g$ is normal since $\beta_g$ is implemented by a unitary on $B(\cH_0)$ by Assumption \ref{asmp:GInvariance}.
Finally, $c_{\pi, \sigma}$ is clearly unitary by construction. 
\end{proof}

\subsection{Connection to anyon representations}
\label{sec:connection_to_anyon_sectors}

We now show that $\GSec_1$ is precisely the braided $\rmC^*$-tensor category\footnote{in fact, $\GSec_1$ is a braided $\rmW^*$-tensor category.} of anyon representations with respect to $\pi_0$ \cite{MR4362722}.

\begin{defn}
\label{def:anyon representation and sector}
Let $\pi_1 \colon \fA \to B(\cH_1)$ and $\pi_2 \colon \fA \to B(\cH_2)$ be representations of $\fA$.  
We say that $\pi_1$ is an \emph{anyon representation} with respect to $\pi_2$ if it satisfies the \emph{superselection criterion} with respect to $\pi_2$, that is, if for every cone $\Lambda$ (including $\Lambda \notin \cL$), 
\[
\pi_1|_{\cstar[\Lambda^c]}
\simeq
\pi_2|_{\cstar[\Lambda^c]}.
\]
An \emph{anyon sector} with respect to $\pi_2$ is a unitary equivalence class of anyon representations with respect to $\pi_2$.
As before, we say that an anyon sector is \emph{irreducible} if the representations that comprise the sector are irreducible. 

\end{defn}

\begin{rem}
\label{rem:IrreducibilityAsAnAssumption}
We remark that we do not require anyon representations to be irreducible. 
However, the physical anyons correspond to the irreducible anyon representations, and the irreducible anyon sectors correspond to the distinct anyon types. 

\end{rem}

\begin{lem}
\label{lem:GDefectsRelationToAnyons}
The following statements are true:
\begin{enumerate}
    \item Let $\pi, \sigma \in \GSec_g$.  
Then $\pi$ satisfies the superselection criterion with respect to $\sigma$
\item Let $\sigma \in \GSec_g$, and suppose $\pi \colon \fA \to B(\cH)$ satisfies the superselection criterion with respect to $\sigma$. 
Then there exists $\widehat{\pi} \in \GSec_g$ such that $\pi \simeq \widehat{\pi}$
\end{enumerate}
\end{lem}

\begin{proof}
\item[(1):]  We must show that for every cone $\Lambda$ (including $\Lambda \notin \cL$), 
\[
\pi|_{\cstar[\Lambda^c]}
\simeq
\sigma|_{\cstar[\Lambda^c]}.
\]
Now we observe that for every cone $\Lambda$ (including $\Lambda \notin \cL$) there exists some cone $\Delta \subset \Lambda$ such that $\Delta \in \cL$.
Now since $\pi, \sigma \in \GSec_g$, we have that 
\[
\pi|_{\cstar[\Delta^c]}
\simeq
\pi_0 \circ \beta_g^{r(\Delta)}|_{\cstar[\Delta^c]}
\simeq
\sigma|_{\cstar[\Delta^c]},
\]

Noting $\cstar[\Lambda^c] \subset \cstar[\Delta^c]$, we have $\pi|_{\cstar[\Lambda^c]}
\simeq \sigma|_{\cstar[\Lambda^c]}$ as desired.

\item[(2):] 
Since $\sigma \in \GSec_g$, we have for some $\Lambda \in \cL$ that $\sigma|_{\cstar[\Lambda^c]} = \pi_0 \circ \mu\circ\beta^{r(\Lambda)}_g|_{\cstar[\Lambda^c]}$.
Noting that $\pi$ satisfies the superselection criterion with respect to $\sigma$, we have that 
\[
\pi|_{\cstar[\Lambda^c]}
\simeq
\sigma|_{\cstar[\Lambda^c]}
=
\pi_0 \circ \mu\circ \beta^{r(\Lambda)}_g|_{\cstar[\Lambda^c]}.
\]
Define $\widehat{\pi} \coloneqq \Ad(U) \circ \pi$, where $U$ is a unitary implementing the unitary equivalence 
$\pi|_{\cstar[\Lambda^c]}
\simeq
\sigma|_{\cstar[\Lambda^c]}$.
Note that $\widehat{\pi}$ is $g$-localized in $\Lambda$ by definition.
Furthermore, for all $\Delta \in \cL$ there exists $\widetilde{\sigma} \simeq \sigma$ $g$-localized in $\Delta$.  
Since $\pi$ satisfies the superselection criterion with respect to $\sigma$, we observe that $\pi$ also satisfies the superselection criterion with respect to $\widetilde{\sigma}$.
Therefore, by the same argument that we used to find $\widehat{\pi}$, we can find $\widetilde{\pi} \simeq \pi \simeq \widehat{\pi}$ $g$-localized in $\Delta$, so $\widehat{\pi} \in \GSec_g$.
\end{proof}

\begin{cor}\label{cor:anyon_equiv}
The braided $\rmW^*$-tensor category $\GSec_1$ is braided $\rmW^*$-tensor equivalent to the braided $\rmW^*$- tensor category of anyon representations with respect to $\pi_0$.

\end{cor}

\begin{proof}
The category $\GSec_1$ is equivalent to the category of anyon representations with respect to $\pi_0$ by Lemma \ref{lem:GDefectsRelationToAnyons} and the fact that $\pi_0 \in \GSec_1$. 
Furthermore when $g=1$, the tensor product reduces to the one defined in \cite{MR4362722}.
The braiding reduces to the \emph{reverse} braiding of the one defined in \cite{MR4362722}.
The result follows.
\end{proof}

\subsection{Coherence data}
\label{sec:coherance_data}

In this section we discuss how to obtain the symmetry fractionalization and other coherence data described in \cite{PhysRevB.100.115147}.
We proceed similarly to how \cite{2306.13762} obtain the $F$- and $R$-symbols in the case of anyon representations.
For this analysis, we restrict our attention to dualizable objects. 

\begin{defn}
\label{def:Dualizable}
    An object $\pi \in \GSec$ is \emph{dualizable} if there exists $\overline{\pi} \in \GSec$ and intertwiners $R \colon \pi_0 \to \overline{\pi} \otimes \pi$ and $\overline{R} \colon \pi_0 \to \pi \otimes \overline{\pi}$ such that $(\Id_{\pi} \otimes R^*)(\overline{R} \otimes \Id_{\pi}) = \Id_{\pi}$ and $(R^* \otimes \Id_{\overline{\pi}})(\Id_{\overline{\pi}} \otimes \overline{R}) = \Id_{\overline{\pi}}$.
    We define $\GSec^f$ to be the full subcategory of {dualizable objects} in $\GSec$. 
\end{defn}

\begin{lem}
\label{lem:GSecf is semisimple}
    $\GSec^f$ is a semisimple $G$-crossed braided $\rmW^*$-tensor subcategory of $\GSec$. 
\end{lem}
\begin{proof}
    We note that by \cite[Thm.~2.4]{MR1444286}, $\GSec^f$ is closed under subobjects, direct sums, and tensor products. Adding that $\GSec^f$ is a full subcategory of $\GSec$, it is automatically a $\rmW^*$-tensor category. 

    Next we check that the $G$-crossed structure restricts to $\GSec^f$. 
    First we define the $G$-grading on $\GSec^f$ as $\GSec^f_g \coloneqq \GSec^f \cap \GSec_g$. Since $\GSec$ is $G$-graded, every object $\pi \in \GSec$ decomposes as a direct sum
    $\pi \simeq \bigoplus_{g \in G} \pi_g$ with $\pi_g \in \GSec_g$. If $\pi \in \GSec^f$, then each $\pi_g$ is a subobject of $\pi$ and hence $\pi_g \in \GSec^f$. Therefore $\pi_g \in \GSec^f_g$, so $\GSec^f = \bigoplus_{g \in G} \GSec^f_g$. Moreover, if $\pi \in \GSec^f_g$ and $\sigma \in \GSec^f_h$, then $\pi \otimes \sigma \in \GSec_{gh}$ because $\GSec$ is $G$-graded, and hence $\pi \otimes \sigma \in \GSec^f_{gh}$ since $\GSec^f$ is closed under tensor products. Thus the $G$-grading on $\GSec$ restricts to a $G$-grading on $\GSec^f$.
    
    If $\pi \in \GSec^f$, then $\gamma_g(\pi)$ is again dualizable. 
    Indeed, $\gamma_g(\bar \pi)$ is the dual object, and the new intertwiners are $\gamma_g(R), \gamma_g(\bar R)$. Hence $\gamma_g$ preserves $\GSec^f$. Similarly, for $\pi \in \GSec^f_g$ and $\sigma \in \GSec^f$, we have $\pi \otimes \sigma, \gamma_g(\sigma) \otimes \pi \in \GSec_g^f$ so fullness gives that $c_{\pi,\sigma}$ is contained in $\GSec^f$. Therefore the $G$-grading, $G$-action, and $G$-crossed braiding all restrict to $\GSec^f$.

    By \cite[Lem.~3.2]{MR1444286}, $\End(\pi)$ is finite dimensional for all $\pi \in \GSec^f$, and it follows that $\GSec^f$ is a semisimple category.
\end{proof}

We let $\mathcal{K}_0(\GSec^f)$ be the fusion ring
of $\GSec^f$ and let $I$ denote the basis of $\cK_0(\GSec^f)$.\footnote{Technically, $\cK_0(\GSec^f)$ is only a unital based ring in the sense of \cite[Def.~3.1.3]{MR3242743} since $I$ need not be finite. 
However, we still use the term ``fusion ring" since this is more widely used term, and in our examples $\cK_0(\GSec^f)$ is a fusion ring since $\GSec^f$ in our example is finitely semisimple.} 
For each $i\in I$, we label the corresponding object in the category by $\pi_i$.
Note that $\pi_0$ is irreducible since $\omega_0$ is a pure state.
Since $\GSec^f$ is semisimple, every object in $\GSec^f$ is isomorphic to a finite direct sum of $\pi_i$'s.

\subsubsection{Symmetry fractionalization}
\label{sec:SymmetryFractionalization}
For every $g \in G$ and $i \in I$, we have that $\gamma_g(\pi_i)$ is irreducible, so we have that $\gamma_g(\pi_i) \simeq \pi_{i'}$ for a unique $i' \in I$. We define $g(i) := i'$ for notational clarity. We let $V_g^i : \gamma_g(\pi_i) \to \pi_{g(i)}$ be a unitary. Now, for $g, h \in G$ and $i \in I$, we have that $V_g^{h(i)} \gamma_g(V_h^i) : \gamma_g(\gamma_h(\pi_i)) \to \pi_{g(h(i))}$, since $\gamma_g(V_h^i): \gamma_g(\gamma_h(\pi_i)) \to \gamma_g(\pi_{h(i)})$. Now, since $\gamma_g(\gamma_h(\pi_i)) = \gamma_{gh}(\pi_i)$, we also have that $V_{gh}^i: \gamma_{g}(\gamma_h(\pi_i)) \to \pi_{gh(i)}$. This implies that $\pi_{gh(i)} \simeq \pi_{g(h(i))}$, so $gh(i) = g(h(i))$. Furthermore, we have that
$$V_g^{h(i)}\gamma_g(V_h^i) (V_{gh}^i)^* : \pi_{gh(i)} \to \pi_{gh(i)}$$
Therefore, since $\pi_{gh(i)}$ is irreducible, we have by Schur's lemma that 
$$V_g^{h(i)}\gamma_g(V_h^i) = \eta(g, h)_i V_{gh}^i$$
for some $\eta(g, h)_i \in U(1)$.

\begin{defn}
    \label{def:sym frac}
    For $\GSec^f$, we define the symmetry fractionalization data to be $\eta$, defined as the family of isomorphisms $\{\eta(g,h)_i\}_{g,h \in G, i \in I}$ as in the above discussion. 
\end{defn}

In fact, as shown by \cite[Lem.~1.9]{2411.01210}, the symmetry fractionalization data $\eta$ is not just a family of isomorphisms, but can be reorganized as a $2$-cocycle in group cohomology.
We reproduce the equivalent of \cite[Lem.~1.9]{2411.01210} in our Lemma \ref{lem:SymmetryFractionalizationData} in order to assist readers who are translating between these works.
We also include this to help make contact with \cite{PhysRevB.100.115147}.

Consider $M \coloneqq \operatorname{Fun}(I, U(1))$ as the abelian group of functions (with pointwise multiplication) from $I$ to $U(1)$. There exists a left $G$-action on $M$: for any $f \in M$, we define $(g \rhd f) (i) := f(g^{-1}(i))$.

With this (left) $G$-module structure on $M$, one may form the groups of $M$-valued $n$-cochains $C^n(G,M) \coloneqq \operatorname{Fun}(G^n,M)$ together with the corresponding cocycles, coboundaries, and cohomology groups. We denote the resulting cohomology by $H^n(G,M)$. We also denote the group of $n$-cocycles/coboundaries as $Z^n(G,M)$ and $B^n(G,M)$ respectively.

Then symmetry fractionalization data can be reinterpreted as an $M$-valued 2-cochain $\hat\eta \colon G \times G \to M$ with the redefinition $\hat\eta(g,h)(i) \coloneqq \eta(g,h)_{(gh)^{-1}(i)}$. The following lemma shows that in fact, $\hat\eta \in Z^2(G,M)$, and moreover, helps us connect with \cite{PhysRevB.100.115147}.

\begin{lem}
\label{lem:SymmetryFractionalizationData}
Let $i \in I$ and $g, h, k \in G$.
We have that 
$$\eta(g, h)_{k(i)} \eta(gh, k)_{i} = \eta(h, k)_i \eta(g, hk)_i.$$
Equivalently, we have $\hat\eta \in Z^2(G,M)$.
Furthermore, the cohomology class $[\hat \eta] \in H^2(G, M)$ is independent of the choices for $V_g^i$.
\end{lem}

\begin{proof}
We first observe that $$V_g^{hk(i)} \gamma_g(V_h^{k(i)}) \gamma_{gh} (V_k^i): \gamma_{ghk}(\pi_i) \to \pi_{ghk(i)},$$ so we can relate $V_g^{hk(i)} \gamma_g(V_h^{k(i)}) \gamma_{gh} (V_k^i)$ to $V_{ghk}^i$ in two different ways. Indeed, observe that 
\begin{align*}
    V_g^{hk(i)} \gamma_g(V_h^{k(i)}) \gamma_{gh} (V_k^i) &= \eta(g, h)_{k(i)} V_{gh}^{k(i)} \gamma_{gh} (V_k^i) = \eta(g, h)_{k(i)} \eta(gh, k)_{i} V_{ghk}^i, \\
\intertext{and we also have,}
V_g^{hk(i)} \gamma_g(V_h^{k(i)}) \gamma_{gh} (V_k^i) &= V_g^{hk(i)} \gamma_g((V_h^{k(i)}) \gamma_{h} (V_k^i)) = V_g^{hk(i)} \gamma_g(\eta(h, k)_i V_{hk}^i) = \eta(h, k)_i \eta(g, hk)_i V_{ghk}^i.
\end{align*}

Therefore, since \(V_{ghk}^i\) is unitary, we obtain for all $i \in I$, $$\eta(g,h)_{k(i)}\eta(gh,k)_i=\eta(h,k)_i\eta(g,hk)_i.$$ Setting \(j=ghk(i)\), and rewriting the preceding identity for $\hat \eta$ we get $$\hat{\eta}(g,h)(j)\hat{\eta}(gh,k)(j)=\hat{\eta}(h,k)(g^{-1}(j))\hat{\eta}(g,hk)(j).$$ Equivalently, $$ (g\rhd \hat{\eta}(h,k))\,\hat{\eta}(g,hk)=\hat{\eta}(g,h)\,\hat{\eta}(gh,k). $$ Which is the $G$-module $2$-cocycle equation, and thus $\hat{\eta}\in Z^2(G,M)$ as claimed.

We now show that different choices for $V^i_g$ yield equivalent cohomology classes in $H^2(G, M)$.
Let us choose another family of unitaries $\{\widetilde V_g^i: \gamma_g(\pi_i) \to \pi_{g(i)}\}_{g \in G, i \in I}$ that yield the scalars $\widetilde \eta(g, h)_i$. 
We note that $\widetilde V_g^i = \lambda(g)_i V_g^i$ for some $\lambda(g)_i \in U(1)$ by Schur's lemma. 
We then get $$\widetilde \eta(g,h)_i = \left( \lambda_{h(i)}(g) \lambda_i(h) \lambda_i(gh)^{-1} \right) \eta(g,h)_i.$$ 
We can define a 1-cochain $\hat \lambda$ as $\hat \lambda\colon G \to M$ via $\hat \lambda(g) (i) \coloneqq \lambda(g)_{g^{-1}(i)}$. 
Then the coboundary $\delta \hat \lambda \in B^2(G,M)$ is defined by $$(\delta \hat \lambda)(g,h)(i) = (g \rhd \hat \lambda(h))(i)\hat \lambda(gh)(i)^{-1}\hat \lambda(g)(i).$$ 
Thus letting $\check{\eta}(g, h)(i) \coloneqq \widetilde \eta(g, h)_{(gh)^{-1}(i)}$, we have $\check{\eta} = (\delta \hat \lambda)\hat \eta$, so $\check{ \eta}, \hat \eta$ define the same class in $H^2(G,M)$.

\end{proof}

\begin{rem}

We remark that the equation shown above, $$\hat{\eta}(g,h)(j)\hat{\eta}(gh,k)(j)=\hat{\eta}(h,k)(g^{-1}(j))\hat{\eta}(g,hk)(j),$$ is precisely \cite[Eq.~279]{PhysRevB.100.115147}. This serves as an important check that our $2$-cocycle $\hat \eta$ is indeed the same object as \cite{PhysRevB.100.115147}.

\end{rem}

\subsubsection{Other coherence data}
\label{sec:OtherCoherenceData}
We now demonstrate how to obtain the rest of the coherence data discussed in \cite{PhysRevB.100.115147}.
For computational simplicity, for the remainder of the discussion we assume the category is \emph{pointed}, meaning that $\pi_i \otimes \pi_j$ is irreducible for every $i, j \in I$.
The analysis can be done in more generality, but in that case more care must be taken.
Note that we are not constraining our general analysis with this assumption but are using it simply for demonstration purposes.

We first compute the $F$-symbols.
This proceeds exactly as done in \cite[\S2.3.1]{2306.13762}, but we repeat the discussion for convenience.  
For every $i, j \in I$, we have that $\pi_i \otimes \pi_j$ is irreducible, so $\pi_i \otimes \pi_j \simeq \pi_{ij}$ for some $ij \in I$.
Following \cite{2306.13762}, we let the tensorator $\Omega_{i, j} \colon \pi_i \otimes \pi_j \to \pi_{ij}$ be a unitary.
Now, for $i, j, k \in I$, we have that
\begin{gather*}
\Omega_{i, jk}(\Id_{\pi_i} \otimes \Omega_{j,k})
=
\Omega_{i, jk}\pi_i(\Omega_{j,k})
\colon 
\pi_i \otimes \pi_j \otimes \pi_k \to \pi_{i(jk)},
\\
\Omega_{ij,k}(\Omega_{i, j} \otimes \Id_{\pi_k})
=
\Omega_{ij,k}\Omega_{i, j}
\colon
\pi_i \otimes \pi_j \otimes \pi_k \to \pi_{(ij)k}.
\end{gather*}
Therefore, we have that $i(jk) = (ij)k \eqqcolon ijk$, and by Schur's lemma, we obtain that 
\[
\Omega_{ij,k}\Omega_{i, j}
=
F(i, j, k)
\Omega_{i, jk}\pi_i(\Omega_{j,k})
\]
for some $F(i, j, k) \in U(1)$.  
This is the $F$-symbol as defined in \cite{2306.13762}.\footnote{This convention is the adjoint of the one often used in the literature, see for instance \cite{PhysRevB.100.115147}. 
Because of this difference, the adjoint later appears in Corollary \ref{cor:identifying the cocycle} when we apply this discussion to SPTs.}
We remark that the coherence condition for the $F$-symbols holds, omitting the proof as it is shown in \cite{2306.13762}.

\begin{lem}[{\cite[Prop.~2.11]{2306.13762}}]
\label{lem:F-symbols are cocycles}
For all $i, j, k, \ell \in \{0, 1, \dots, n\}$, 
\[
F(i, j, k) F(i, jk, \ell) F(j, k, \ell)
=
F(ij, k, \ell)F(i, j, k\ell).
\]
\end{lem}

We now compute the coherence data related to the tensorator of $\gamma_g$ in the skeletalization of the category that we are now working with; this corresponds to the data defined in \cite[Eq.~269]{PhysRevB.100.115147}.  
We let $i, j \in I$ and $g \in G$.  
We first observe that the following map is a unitary intertwiner: 
\[
V^{ij}_g \gamma_g(\Omega_{i, j})
\colon
\gamma_g(\pi_i \otimes \pi_j)
\to
\pi_{g(ij)}.
\]
In addition, since $\gamma_g(\pi_i \otimes \pi_j) = \gamma_g(\pi_i) \otimes \gamma_g(\pi_j)$, we also have that the following map is a unitary intertwiner: 
\[
\Omega_{g(i), g(j)}(V^i_g \otimes V^j_g)
=
\Omega_{g(i), g(j)}V^i_g\gamma_g(\pi_i)(V^j_g)
\colon
\gamma_g(\pi_i \otimes \pi_j)
\to
\pi_{g(i)g(j)}.
\]
We therefore have that $g(ij) = g(i)g(j)$, and by Schur's lemma, we obtain that 
\[
V^{ij}_g \gamma_g(\Omega_{i, j})
=
\mu_g(i, j) \Omega_{g(i), g(j)}V^i_g\gamma_g(\pi_i)(V^j_g)
\]
for some $\mu_g(i, j) \in U(1)$.
The following lemma follows from a straightforward computation.

\begin{lem}
Let $i, j, k \in I$ and $g \in G$.
We have that
\[
F(i, j, k) \mu_g(i, jk) \mu_g(j, k)
=
\mu_g(ij, k) \mu_g(i, j) F(g(i), g(j), g(k)).
\]
\end{lem}

Finally, we compute the $R$-symbols, analogously to \cite[\S2.3.2]{2306.13762}.

Let $i, j \in I$.
Since $\pi_i \in \GSec^f$ is irreducible, $\pi_i \in \GSec^f_g$ for some $g \in G$.  
We then have that $c_{i, j} \colon \pi_i \otimes \pi_j \to \gamma_g(\pi_j) \otimes \pi_i$ is a unitary intertwiner, so we have that 
\[
\Omega_{g(j), i} (V^j_g \otimes \Id_{\pi_i})c_{\pi_i, \pi_j}
=
\Omega_{g(j), i} V^j_g c_{\pi_i, \pi_j}
\colon
\pi_i \otimes \pi_j \to \pi_{g(j)i}.
\]
On the other hand, we have that $\Omega_{i, j} \colon \pi_i \otimes \pi_j \to \pi_{ij}$ is also a unitary intertwiner.  
Therefore, $ij = g(j)i$, and by Schur's lemma, 
\[
\Omega_{g(j), i} V^j_g c_{\pi_i, \pi_j}
=
R(i, j)\Omega_{i, j}
\]
for some $R(i, j) \in U(1)$.  
This defines the $R$-symbols for our category. The $R$-symbols satisfy several coherence relations. 
We single out the heptagon equations \cite[Eq.~286 \& 287]{PhysRevB.100.115147}.

\begin{lem}
Let $i, j, k \in \{0, 1, \dots, n\}$.
Suppose that $\pi_i \in \GSec^f_g$ and $\pi_j \in \GSec^f_h$. 
We then have that 
\begin{gather*}
R(i, k) F(g(j), i, k)^* R(i, j)
=
F(g(j), g(k), i)^* \mu_g(j, k)^*R(i, jk)F(i, j, k)^*,
\\
R(i, h(k))F(i, h(k), j) R(j, k)
=
F(gh(k), i, j) \eta(g, h)_k R(ij, k) F(i, j, k).
\end{gather*}
\end{lem}

\begin{proof}
We verify the second equation, which corresponds to \cite[Eq.~287]{PhysRevB.100.115147}. The computation can be graphically represented as in Figure \ref{fig:heptagon equation}. The other equation can be verified analogously.  
We consider the unitary 
\begin{equation}
\label{eq:UnitaryForHeptagonEquation}
\Omega_{gh(k)i, j} \Omega_{gh(k), i} V_g^{h(k)} c_{\pi_i, \pi_{h(k)}} \pi_i(V_h^k c_{\pi_j, \pi_k})
\colon
\pi_i \otimes \pi_j \otimes \pi_k
\to
\pi_{gh(k)ij}.
\end{equation}
We simplify this unitary in two different ways.
To make the simplification easier to follow, we color the terms that change at each step red or blue.
Specifically, the terms colored blue are the ones that were changed in the prior step, and the terms colored red are the ones that will change in the next step.  
If a term is colored purple, that means it is involved in the changes made at consecutive steps.  
For the first simplification, we have that 
\begin{align*}
\Omega_{gh(k)i, j} \textcolor{red}{\Omega_{gh(k), i} V_g^{h(k)} c_{\pi_i, \pi_{h(k)}}} \pi_i(V_h^k c_{\pi_j, \pi_k})
&=
\textcolor{blue}{R(i, hk)} \textcolor{red}{\Omega_{gh(k)i, j}} \textcolor{blue}{\Omega_{i, h(k)}} \pi_i(V_h^k c_{\pi_j, \pi_k})
\\&=
R(i, hk) \textcolor{violet}{\Omega_{ih(k), j}} \textcolor{red}{\Omega_{i, h(k)}} \pi_i(V_h^k c_{\pi_j, \pi_k})
\\&=
R(i, hk) \textcolor{blue}{F(i, h(k), j)}\textcolor{blue}{\Omega_{i, h(k)j}} \textcolor{violet}{\pi_i(\Omega_{h(k), j})} \textcolor{red}{\pi_i(V_h^k c_{\pi_j, \pi_k})}
\\&=
R(i, hk) F(i, h(k), j)\Omega_{i, h(k)j} \textcolor{blue}{\pi_i(}\textcolor{violet}{\Omega_{h(k), j} \pi_i(V_h^k c_{\pi_j, \pi_k})}\textcolor{blue}{)}
\\&=
R(i, hk) F(i, h(k), j)\textcolor{blue}{R(j, k)}\textcolor{red}{\Omega_{i, h(k)j}} \pi_i(\textcolor{blue}{\Omega_{j, k}})
\\&=
R(i, hk) F(i, h(k), j)R(j, k)\textcolor{blue}{\Omega_{i, jk}} \pi_i(\Omega_{j, k}).
\end{align*}

\begin{figure}[!ht]
    \centering
    \includegraphics[width=0.7\linewidth]{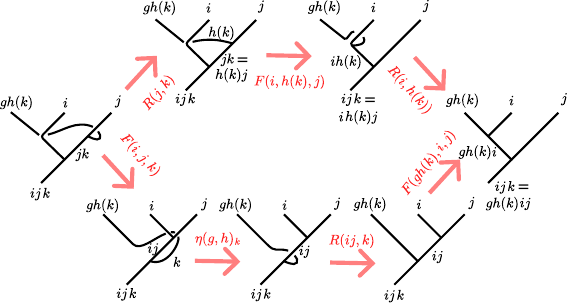}
    \caption{The graphical representation of the second heptagon equation of \cite[Eq. 287]{PhysRevB.100.115147}. We note that our definition of $R$ corresponds to $R^{-1}$ in their work. 
    For presentation purposes, some of the lines turn downwards, but one should interpret all the lines as moving upward (i.e., we are not using any evaluation/coevaluation maps).}
    \label{fig:heptagon equation}
\end{figure}

For the other simplification, we recall the following naturality and monoidality equations for the braiding (proven in Proposition \ref{prop:GCrossedBraiding}).
\begin{facts}
\label{facts:braiding}
\leavevmode
\begin{itemize}
\item 
If $\pi \in \GSec^f_g$, and $T \colon \sigma_1 \to \sigma_2$ is a morphism in $\GSec^f$, then 
\[
\gamma_g(T)c_{\pi, \sigma_1}
=
c_{\pi, \sigma_2}\pi(T).
\]
\item 
If $\sigma \in \GSec^f$ and $T \colon \pi_1 \to \pi_2$ is a morphism in $\GSec^f_g$, then
\[
\gamma_g(\sigma)(T)c_{\pi_1, \sigma}
=
c_{\pi_2, \sigma}T.
\]
\item 
If $\pi \in \GSec^f_g$, $\sigma \in \GSec^f_h$, and $\tau \in \GSec^f$, then
\[
c_{\pi \otimes \sigma, \tau}
=
c_{\pi, \gamma_h(\tau)}\pi(c_{\sigma, \tau}).
\]
\end{itemize}
\end{facts}
We now simplify the unitary in \eqref{eq:UnitaryForHeptagonEquation} using these facts, as well as the fact that $\pi_i \otimes \pi_j, \pi_{ij} \in \GSec^f_{gh}$.
In particular, we have that 
\allowdisplaybreaks
\begin{align*}
\Omega_{gh(k)i, j} \Omega_{gh(k), i} V_g^{h(k)} c_{\pi_i, \pi_{h(k)}} \textcolor{red}{\pi_i(V_h^k c_{\pi_j, \pi_k})}
&=
\Omega_{gh(k)i, j} \Omega_{gh(k), i}V_g^{h(k)} \textcolor{red}{c_{\pi_i, \pi_{h(k)}}} \textcolor{violet}{\pi_i(V_h^k)}\textcolor{blue}{\pi_i(c_{\pi_j, \pi_k})}
\\&=
\Omega_{gh(k)i, j} \Omega_{gh(k), i} V_g^{h(k)}\textcolor{blue}{\gamma_g(V_h^k)}\textcolor{violet}{c_{\pi_i, \gamma_h(\pi_k)}} \textcolor{red}{\pi_i(c_{\pi_j, \pi_k})}
\\&=
\textcolor{red}{\Omega_{gh(k)i, j} \Omega_{gh(k), i}} V_g^{h(k)} \gamma_g(V_h^k)\textcolor{blue}{c_{\pi_i \otimes \pi_j, \pi_k}}
\\&=
\textcolor{blue}{F(gh(k), i, j) \Omega_{gh(k), ij} \pi_{gh(k)}(\Omega_{i, j})} \textcolor{red}{V_g^{h(k)} \gamma_g(V_h^k)}c_{\pi_i \otimes \pi_j, \pi_k}
\\&=
F(gh(k), i, j)\textcolor{blue}{\mu(g, h)_k}\Omega_{gh(k), ij} \textcolor{red}{\pi_{gh(k)}(\Omega_{i, j})}\textcolor{violet}{V_{gh}^k}c_{\pi_i \otimes \pi_j, \pi_k}
\\&=
F(gh(k), i, j)\mu(g, h)_k\Omega_{gh(k), ij} \textcolor{blue}{V_{gh}^k}\textcolor{violet}{\gamma_{gh}(\pi_k)(\Omega_{i, j})}\textcolor{red}{c_{\pi_i \otimes \pi_j, \pi_k}}
\\&=
F(gh(k), i, j)\mu(g, h)_k\textcolor{red}{\Omega_{gh(k), ij} V_{gh}^k}\textcolor{violet}{c_{\pi_{ij}, \pi_k}}\textcolor{blue}{\Omega(i, j)}
\\&=
F(gh(k), i, j)\mu(g, h)_k\textcolor{blue}{R(ij, k)}\textcolor{violet}{\Omega(ij, k)}\textcolor{red}{\Omega(i, j)}
\\&=
F(gh(k), i, j)\mu(g, h)_kR(ij, k)\textcolor{blue}{F(i, j, k)\Omega_{i, jk} \pi_i(\Omega_{j, k})}.
\end{align*}
Comparing the two simplifications of \eqref{eq:UnitaryForHeptagonEquation}, we obtain that 
\[
R(i, h(k))F(i, h(k), j) R(j, k)
=
F(gh(k), i, j) \mu(g, h)_k R(ij, k) F(i, j, k).
\qedhere
\]
\end{proof}

\section{General SPTs}
\label{sec:general SPTs}
In this section, given an SPT we will obtain states housing defects using defect automorphisms. We will then classify all possible irreducible $g$-sectors for this SPT.

\begin{figure}[!ht]
    \centering
    \includegraphics[width=0.3\linewidth]{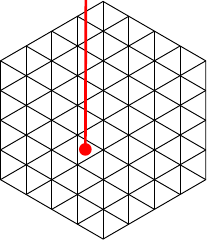}
    \caption{The ray $R$ on the triangular lattice.}
    \label{fig:chosen_ray}
\end{figure}

Take the elements of $\Gamma$ to be the vertices of the triangular lattice and let $\hilb_v \simeq \bbC^{d_v}$ with $d_v \geq 2$ for each $v \in \Gamma$.
We assume that there exists $D \in \bbN$ such that $d_v \leq D$ for all $v \in \Gamma$.
Also take $R$ to be a vertical ray that does not intersect any vertices, as illustrated in Figure \ref{fig:chosen_ray}.
Let $G$ be the symmetry group and for every $g \in G$, let $g\mapsto U^g_v$ be its unitary representation onto the vertex $v$. 
We assume that the representation on each site is faithful so that Assumption \ref{asmp:Faithfulness} is satisfied.
For each $A \in \cstar[S]$ with $S \in \Gamma_f$, we let $\beta_g$ be the automorphism from Definition \ref{def:GlobalSymmetryAutomorphism}.

Recall the definition of a $G$-SPT (Definition \ref{def:SPT_phase}). We have for a $G$-SPT state $\tilde \omega$, the existence of a symmetric entangler $\alpha$ (i.e., an FDQC satisfying that $\alpha \circ \beta_g = \beta_g \circ \alpha$ for every $g \in G$) such that $\omega_0 \circ \alpha = \tilde \omega$, where $\omega_0$ is some pure product state. 
For the entirety of this section, we let $s$ denote the spread of $\alpha$.

\begin{rem}
\label{rem:GInvarianceTrick}
In the definition of $G$-SPT (Definition \ref{def:SPT_phase}), we assume that both the product state $\omega_0$ and the state $\tilde \omega = \omega_0 \circ \alpha$ are invariant under $\beta_g$. 
However, since $\alpha$ is a symmetric entangler, $\omega_0$ is $\beta_g$-invariant if and only if $\tilde \omega$ is. 
Indeed, suppose $\omega_0$ is $\beta_g$-invariant.
Then we have that 
\[
\tilde \omega \circ \beta_g
=
\omega_0 \circ \alpha \circ \beta_g
=
\omega_0 \circ \beta_g \circ \alpha
=
\omega_0 \circ \alpha
=
\tilde \omega.
\]
The other direction follows by the same argument.
\end{rem}

We first recall the following well-known Lemma.

\begin{lem}{\cite[\S2.1.1]{MR2345476}}
\label{lem:can freely insert and remove P from the ground state.}
If $\omega(A) = 1$ for some $A \in \cstar$ satisfying that $A \leq \mathds1$, then we have for any $O \in \cstar$, $$\omega(O) = \omega(A O) = \omega(OA) = \omega(AOA).$$
\end{lem}

\begin{lem}
\label{lem:existence of gapped Hamiltonian}
    Let $\omega$ be a pure product state. Then there exists a unique set $\{P_v^\omega\}_{v \in \Gamma}$ of rank-1 projections $P_v^\omega \in \cstar[v]$ such that $\omega$ is the unique state satisfying $\omega(P_v^\omega) = 1$ for all $v \in \Gamma$. 
    Moreover, the Hamiltonian $H^\omega$ given by $H_S^\omega \coloneqq \sum_{v \in S} \mathds1 - P_v^\omega$ for $S \in \Gamma_f$ with derivation $\delta^\omega$ has $\omega$ as its unique ground state.
\end{lem}
\begin{proof}
    Since $\omega$ is a product state, $\omega^v \coloneqq \omega|_{\cstar[v]}$ is represented by a vector $\ket{\psi_v} \in \hilb_v$ for each $v \in \Gamma$. 
    Using standard arguments, it is easy to see that the rank-1 projections $P_v^\omega \in \fA_v$ onto $\ket{\psi_v}$ satisfy the desired properties.
    
\end{proof}

\begin{defn}
    For the product state $\omega_0$ satisfying $\omega_0 \circ \alpha = \tilde \omega$ for the $G-$SPT $\tilde \omega$, we define the corresponding unique rank-1 projections $P_v \coloneqq  P_v^{\omega_0}$ for $v \in \Gamma$, Hamiltonian $H^0 \coloneqq  H^{\omega_0}$, and corresponding derivation $\delta_0\coloneqq  \delta^{\omega_0}$ from Lemma \ref{lem:existence of gapped Hamiltonian}. 
    
\end{defn}

\begin{lem}
\label{lem:Pv invariant under symmetry action}
    Let $\omega$ be a pure product state satisfying for all $g \in G$ that $\omega \circ \beta_g = \omega$. Then $\beta_g(P_v^\omega) = P_v^\omega$ for all $g \in G$ and $v \in \Gamma$, where $P_v^\omega$ are the projections defined in Lemma \ref{lem:existence of gapped Hamiltonian}. In particular, $\beta_g(H_S^\omega) = H_S^\omega$ for all $V \in \Gamma_f$ and $g \in G$.
\end{lem}
\begin{proof}
    Since $\omega \circ \beta_g = \omega$, we can apply Lemma \ref{lem:can freely insert and remove P from the ground state.} with $A = P_v^\omega$ and $O = \beta_g(P_v^\omega)$ to get $\omega(\beta_g(P_v^\omega) P_v^\omega \beta_g(P_v^\omega)) = 1$. 
    The result follows since $P^\omega_v, \beta_g(P^\omega_v)$ are both rank-1 projections.
    
\end{proof}

\subsection{FDQC Hamiltonian}
We define the Hamiltonian $H$ given by $$H_V \coloneqq \sum_{v \in S} \mathds1 - Q_v, \qquad \qquad Q_v \coloneqq \alpha^{-1}(P_v)$$ for $S \in \Gamma_f$.
We define $\tilde \delta$ to be the corresponding derivation.

\begin{lem}
\label{lem:unique GS of SPT}
    The state $\tilde \omega$ is the unique ground state of derivation $\tilde \delta$. 
    In addition, $\tilde \omega$ is the unique state satisfying that $\tilde \omega(Q_v) = 1$.
    In particular, $\tilde \omega$ is pure.
\end{lem}
\begin{proof}
By \cite[Lem.~3.8]{MR3764565}, $\tilde \omega$ is a ground state for $\tilde \delta$, and by Lemma \ref{lem:QCAs preserve the ground state subspace} it is the unique ground state for this derivation.

\end{proof}

Note that the set $\{Q_v\}_{v \in \Gamma}$ is a set of commuting projections since they are the image of the projections $P_v$ under $\alpha^{-1}$. 
Therefore $H$ is a commuting projector Hamiltonian.

\begin{lem}
    \label{lem:Qv are invariant under the symmetry action}
    For all $g \in G$ and $v \in \Gamma$ we have that $\beta_g(Q_v) = Q_v$.
    In particular, this implies $\beta_g(H_S) = H_S$ for all $S \in \Gamma_f$.
\end{lem}
\begin{proof}
This follows from Lemma \ref{lem:Pv invariant under symmetry action} since $\alpha$ is a symmetric entangler.
\end{proof}

\subsection{Defects using automorphisms}
\label{sec:defects using auts}

\subsubsection{Paths and dual paths}
\label{sec:paths and dual paths}

\begin{figure}[!ht]
\centering
\begin{tikzpicture}[scale=0.6]
    \draw[black](0,0)--(0,18);
    \foreach \y in {0,...,10}{
    \draw[black](2*\y,18)--(2*\y+1,18.5);
    \draw[black](2*\y+2,0)--(2*\y+2,18);
    \draw[black](2*\y+1,.5)--(2*\y+1,18.5);
    \foreach \x in {0,...,17}{
    \draw[black](2*\y,\x)--(2*\y+1,\x+.5);
    \draw[black](2*\y,\x+1)--(2*\y+2,\x);
    \draw[black](2*\y+1,\x+.5)--(2*\y+2,\x+1);
    \draw[black](2*\y+1,\x+1.5)--(2*\y+2,\x+1);
}}
\draw[very thick,violet](2,0)--(2,1)--(4,2)--(4,5)--(2,4)--(2,8)--(3,7.5)--(4,8);
\filldraw[violet] (4,8) circle(.2cm);
\draw[very thick,violet](1,11.5)--(1,14.5)--(3,13.5)--(6,15)--(6,16)--(5,16.5)--(3,15.5)--(3,14.5)--(7,12.5)--(2,10)--(1,10.5)--(1,11.5);
\draw[very thick,blue] (12,6)--(12,7)--(13,6.5)--(13,7.5)--(14,7)--(14,8);
\draw[very thick,blue] (15,7.5)--(14,7)--(15,6.5)--(14,6)--(15,5.5)--(14,5);
\draw[very thick,blue](15,5.5)--(15,4.5)--(16,5)--(16,4);
\draw[very thick, gray,dashed](11.66,6.5)--(13.66,7.5)--(14.33,7.5)--(14.33,5.5)--(14.66,5)--(16.33,4.5);
\filldraw[blue] (11.66,6.5) circle(.2cm);
\filldraw[blue] (16.33,4.5) circle(.2cm);
\draw[very thick, blue](16,16)--(16,15)--(15,15.5)--(15,14.5)--(14,15)--(14,14)--(13,14.5)--(14,14)--(13,13.5)--(14,13)--(13,13.5)--(14,13)--(13,12.5)--(14,12)--(13,11.5)--(14,12)--(14,11)--(15,11.5)--(15,10.5)--(16,11)--(16,10)--(16,11)--(17,10.5)--(17,11.5)--(18,11)--(18,12)--(19,11.5)--(18,12)--(19,12.5)--(18,13)--(19,13.5)--(18,14)--(19,14.5)--(18,14)--(18,15)--(17,14.5)--(17,15.5)--(16,15);
\draw[very thick, orange](11,18.5)--(10,18)--(11,17.5);
\draw[very thick, orange](8,8)--(9,7.5);
\foreach \y in {1,...,7}{
\draw[very thick, orange](11,\y+10.5)--(10,\y+10)--(11,\y+9.5);
\draw[very thick, orange](9,\y+.5)--(8,\y)--(9,\y-.5);
}
\foreach \y in {1,...,10}{
\filldraw[orange](9,\y - .5) circle(.12cm);
\filldraw[orange](8,\y-1) circle(.12cm);
\filldraw[orange](10,\y-1) circle(.12cm);
\filldraw[orange](7,\y-.5) circle(.12cm);
}
\foreach \y in {1,...,10}{
\filldraw[orange](11,\y+8.5) circle(.12cm);
\filldraw[orange](10,\y+8) circle(.12cm);
\filldraw[orange](12,\y+8) circle(.12cm);
\filldraw[orange](9,\y+8.5) circle(.12cm);
}
\draw[very thick,red](10.35,18.5)--(10.35,11.5);
\filldraw[red](10.35,11.5) circle(.15);
\draw[very thick,orange](8,0)--(9,.5);
\draw[very thick,orange](11,10.5)--(10,10)--(11,9.5)--(10,10)--(10,9)--(9,9.5)--(9,8.5)--(8,9)--(9,8.5)--(8,8);
\filldraw[orange] (8, 10) circle(0.12cm);
\filldraw[orange] (11, 8.5) circle(0.12cm);

\end{tikzpicture}
\caption{Examples of the several types of paths. In violet, we show a self-avoiding half-infinite path whose initial vertex is a violet dot and a loop that is not self-avoiding.
In blue, we show a self-avoiding dual loop along with a self-avoiding dual path $\bar \gamma$ whose initial and terminal faces contain a blue dot.
Through the dual path, we also show $\bar\gamma_{dual}$ as a dashed gray curve.
We show a dividing infinite dual path $\bar L$ in orange (taking $s=2$ here) and we place an orange dot on each vertex in $\bar L^{+2}$ ($\bar L_{dual}$ not shown). 
Since we often take dividing dual paths to have infinitely many of their edges intersecting the ray $R$, we have included $R$ in red.
}
\label{fig:PathsForTriangularLattice}
\end{figure}

Let $X$ be the $1$-skeleton of the triangular lattice and let $\overline{X}$ be the $1$-skeleton of its dual lattice.
Recall that the dual lattice of the triangular lattice is the honeycomb lattice.
We now describe the various paths and dual paths we will consider in this section, as illustrated in Figure \ref{fig:PathsForTriangularLattice}.
We call a sequence of edges $\gamma_1,...,\gamma_n\in X$ a \textit{path} if it corresponds to a path in $X$ in the graph theoretic sense.
Since each edge in $\overline{X}$ each intersects precisely one edge in $X$, each graph theoretic path in $\overline{X}$ corresponds to a sequence of edges $\bar{\gamma}_1,...,\bar{\gamma}_n\in X$.
We call such a sequence a \textit{dual path}. 
We will also make use of half-infinite (dual) paths $\gamma = (\gamma_n)_{n \in \bbN}$ ($\bar \gamma = (\bar\gamma_n)_{n \in\bbN}$) and infinite (dual) paths $L = (L_n)_{n \in \bbZ}$ ($\bar L = (\bar L_n)_{n \in\bbZ}$). As a matter of convention, we will denote paths by $\gamma$ and dual paths by $\bar{\gamma}$.

Recall that $\Gamma$ is the set of vertices in the triangular lattice.
Let $F$ be the set of faces in the triangular lattice.
For a path $\gamma$, we take $\partial_0\gamma\in \Gamma$ to be the \emph{initial vertex} of $\gamma$ and $\partial_1\gamma\in \Gamma$ to be the \emph{terminal vertex} of $\gamma$.
For a dual path $\bar\gamma$, the initial and terminal vertices of $\bar\gamma$ in $\overline{X}$ correspond to faces of original triangular lattice.
As such, we take $\partial_0\bar\gamma\in F$ to be the initial face of the dual path and $\partial_1\bar\gamma\in F$ to be the terminal face of the dual path.
We use the same notation for the initial vertex (face) of a half-infinite (dual) path.
Since there is only one endpoint of a half-infinite path, we will often simply use $\partial \gamma$ ($\partial \bar \gamma$) to denote the initial vertex (face) of a half-infinite (dual) path $\gamma$ ($\bar \gamma$).

We call a path $C$ a \textit{loop} when  $\partial_0 C = \partial_1 C$.  Similarly, we define dual loops as dual paths $\bar C$ such that $\partial_0 \bar C = \partial_1 \bar C$. 

A self-avoiding (dual) path ($\bar \gamma$) $\gamma$ is one which meets any given (face) vertex at most once.
In the case of self-avoiding (dual) loops, exempt the initial (face) vertex from this requirement since (dual) loops start and end at the same location.
Given a self-avoiding  path $\bar \gamma$ (finite, half-infinite, or infinite), define $\bar \gamma_{dual}\subset\mathbb{R}^2$ to be the union of the line segments which connect the barycenters of the faces $\partial_0\bar \gamma_n\in F$ and $\partial_1\bar \gamma_n\in F$ for each edge $\bar \gamma_n$ in $\bar \gamma$.

Note that every self-avoiding loop/dual loop that is not empty divides $\bbR^2$ into 2 simply connected regions $S_1, S_2 \subset \bbR^2$, such that only one of them is bounded. 
For a dual loop $\bar \gamma$, we use $\bar \gamma_{dual}$ to make this division, \emph{not} the sequence of edges $(\bar \gamma_n)$ in $X$. 
If $S_i$ is the bounded region obtained in this way, we term the set $S_i \cap \Gamma$ the \emph{interior} of the loop/dual loop respectively.

For two paths $\gamma^1, \gamma^2$ of lengths $N,M$, respectively, with $\partial_1\gamma^1=\partial_0\gamma^2$, we define the path $\gamma^1 + \gamma^2$ by
$$
(\gamma^1+\gamma^2)_k\coloneqq\begin{cases}
\gamma^1_k &1\leq k\leq N\\
\gamma^2_{k-N} &N+1\leq k\leq N+M.
\end{cases}
$$
We also allow $\gamma^2$ to be half-infinite by allowing $M=\infty$.

For any path $\gamma$, we define $(\gamma^{+s})_{geo}\coloneqq \bigcup\limits_{n\in\mathbb{Z}}\gamma_n^{+s}\subseteq\mathbb{R}^2$ where $\gamma_n^{+s}\subseteq\mathbb{R}^2$ is the thickening of the edge $\gamma_n$ as in Notation \ref{nota:thickening}.
Similarly, for any dual path $\bar \gamma$, we define $(\bar \gamma^{+s})_{geo}\coloneqq (\bar \gamma_{dual})^{+s}\subseteq\mathbb{R}^2$, where $s$ is the spread of the entangler $\alpha$.
A \emph{dividing (dual) path} is a self-avoiding infinite (dual) path $L$ such that $\mathbb{R}^2\backslash (L^{+s})_{geo}$ consists of two unbounded simply connected components.
For ($\bar \gamma$) $\gamma$ a (dual) path, we define $\gamma^{+s} \coloneqq (\gamma^{+s})_{geo}\cap\Gamma$ and $\bar \gamma^{+s} \coloneqq (\bar \gamma^{+s})_{geo}\cap\Gamma$.
Note that we are careful not to overload Notation \ref{nota:thickening} since $\gamma$ is not a subset of $\bbR^2$.

Define an \textit{allowed} half-infinite path to be a self-avoiding half-infinite path $\gamma$ such that there exists some allowed cone $\Lambda\in\cL$ such each edge in $\gamma$ is completely contained in $\Lambda_{geo}$.
We define allowed half-infinite dual paths in a similar, but slightly different way.
We have not defined any notion of ``dual edge" so we require the edges (rather than dual edges) in $X$ intersected by an allowed $\bar \gamma$ to be completely contained in some allowed geometric cone $\Lambda_{geo}$. 
In this sense, an allowed half-infinite dual path is self-avoiding and is contained in an allowed cone.
Take $P(\Gamma)$ and $\bar P(\Gamma)$ to be the allowed half-infinite paths and dual paths, respectively.
Since we have only defined allowed half-infinite paths and dual paths, we often drop the adjective ``half-infinite" when discussing them.

We define $\bar P_R(\Gamma)$ to be the set of half-infinite self-avoiding dual paths $\bar \gamma$ which intersect $R$ on all but a finite number of edges in $\bar \gamma$.
A \emph{completion} of $\bar \gamma \in \bar P_R(\Gamma)$ is a self-avoiding infinite dual path $\bar L$ such $\bar L_k=\bar \gamma_k$ for all $k\geq 1$ and the dual path $\bar \xi_k:=\bar L_{1-k}$ for $k\geq 1$ is in $\bar P(\Gamma)$. 

For a dividing infinite dual path $\bar L$, $\mathbb{R}^2\backslash \bar L_{dual}$ has two simply connected components, one of which we call $r(\bar L)_{geo}$ and the other we call $\ell(\bar L)_{geo}$.
We define $r(\bar L):=r(\bar L)_{geo}\cap\Gamma$ and $\ell(\bar L):=\ell(\bar L)_{geo}\cap\Gamma$.
Thus far we have not set a convention for which connected component is named $r(\bar L)_{geo}$ and which is named $\ell(\bar L)_{geo}$.
However, if $\bar L$ is a completion of some $\bar\gamma\in \bar P_R(\Gamma)$, then we fix the following convention: by the definition of $\bar P_R(\Gamma)$ and completions, there exists some $\Lambda\in\cL$ such that all edges in $\bar L$ not contained in $\Lambda_{geo}$ intersect $R$.
We therefore choose $r(\bar L)_{geo}$ to be the region which intersects $r(\Lambda)$ and $\ell(\bar L)$ to be the region which intersects $\ell(\Lambda)$ (recall Definition \ref{def:r(Lambda)} for the definitions of $r(\Lambda), \ell(\Lambda)$).

\subsubsection{Construction of defect automorphisms} 
Since we consider only dual paths in this analysis, we drop $(\bar \cdot)$.

We define entangled the symmetry $\tilde \beta_g^\Sigma$ for a region $\Sigma \subset \Gamma$ as $$\tilde \beta_g^\Sigma \coloneqq  \alpha^{-1} \circ \beta_g^\Sigma \circ \alpha.$$
Recall that $\Sigma^{+s} \subset \bbR^2$ is the thickening of the points in $\Sigma$ by $s$ in the sense of Notation \ref{nota:thickening}. 
Define $\Sigma^{+s}_\Gamma \coloneqq \Sigma^{+s}\cap\Gamma$.
Note that $\Sigma^{+s}_\Gamma$ is the set of points in $\Gamma$ of distance at most $s$ from $\Sigma$.

\begin{lem}
\label{lem:entangled symmetry looks like the regular symmetry in bulk}
    Let $\Sigma \subset \Gamma$, and let $s$ denote the spread of $\alpha$ as before. Then for all $A \in \cstar[\Sigma]$ and $A \in \cstar[(\Sigma^{+2s})^c]$, we have that $$\tilde \beta_g^{\Sigma^{+s}_\Gamma}(A) = \beta_g^\Sigma(A) = \beta_g^{\Sigma^{+s}_\Gamma}(A).$$
    That is, $\alpha$ commutes with the restricted symmetry automorphism $\beta_g^{\Sigma^{+s}_\Gamma}$ when applied to $A \in \cstar[\Sigma]$ (i.e., observables in the bulk of $\Sigma^{+s}_\Gamma$) or $A \in \cstar[(\Sigma^{+2s}_\Gamma)^c]$ (i.e., observables in the bulk of $(\Sigma^{+s}_\Gamma)^c$).
\end{lem}
\begin{proof}
    First assume $A \in \cstar[\Sigma]$. 
    Then $\alpha(A) \in \cstar[\Sigma^{+s}_\Gamma]$. 
    Thus, since $\alpha$ is a symmetric entangler, we have that
    \[
    \tilde \beta_g^{\Sigma^{+s}_\Gamma}(A)
    =
    \alpha^{-1} \circ \beta_g^{\Sigma^{+s}_\Gamma} \circ \alpha(A)
    =
    \alpha^{-1} \circ \beta_g \circ \alpha(A)
    =
    \alpha^{-1} \circ \alpha \circ \beta_g(A)
    =
    \beta_g^\Sigma(A).
    \]

    Now assume $A \in \cstar[(\Sigma^{+2s}_\Gamma)^c]$.
    Then $\alpha(A) \in \cstar[(\Sigma^{+s}_\Gamma)^c]$, so $\beta_g^{\Sigma^{+s}_\Gamma}(\alpha(A)) = \alpha(A)$. Therefore, we have that 
    \[
    \tilde \beta_g^{\Sigma^{+s}_\Gamma}(A)
    =
    \alpha^{-1} \circ \beta_g^{\Sigma^{+s}_\Gamma} \circ \alpha(A)
    =
    \alpha^{-1} \circ \alpha(A)
    =
    A
    =
    \beta_g^\Sigma(A).
    \]
    Combining these two results, we have shown the full result.
\end{proof}

We define $\epsilon_g^{\partial^{s} \Sigma^{+s}_\Gamma} \coloneqq \tilde \beta_g^{\Sigma^{+s}_\Gamma} \circ (\beta_g^{\Sigma^{+s}_\Gamma})^{-1}$ to ease notation, where $\partial^{s} \Sigma^{+s}_\Gamma \coloneqq \Sigma^{+2s}_\Gamma\setminus \Sigma$. 
Intuitively, $\partial^{s} \Sigma^{+s}_\Gamma$ consists of the points in the ``thickened boundary" of $\Sigma^{+s}_\Gamma$, where the boundary is thickened by $s$. 
To motivate our notation we have the following Lemma.

\begin{lem}
\label{lem:localization along a strip}
    We have for any region $\Sigma \subset \Gamma$ that $\epsilon_g^{\partial^{s} \Sigma^{+s}_\Gamma}$ is localized in $\partial^{s} \Sigma^{+s}_\Gamma$. 
\end{lem}
\begin{proof}
    Consider $A \in \cstar[(\partial^{s} \Sigma^{+s}_\Gamma)^c]$. Then since $\beta_g$ is an on-site symmetry, we have $\beta_g^{\Sigma^{+s}_\Gamma}(A) \in \cstar[(\partial^{s} \Sigma^{+s}_\Gamma)^c]$. By Lemma \ref{lem:entangled symmetry looks like the regular symmetry in bulk} we then have,
    \[
        \epsilon_g^{\partial^{s} \Sigma^{+s}_\Gamma} (A) = \tilde \beta_g^{\Sigma^{+s}_\Gamma} \circ (\beta_g^{\Sigma^{+s}_\Gamma})^{-1}(A) = \beta_g^{\Sigma^{+s}_\Gamma} \circ (\beta_g^{\Sigma^{+s}_\Gamma})^{-1}(A) = A.
    \]

    We now show that $\epsilon_g^{\partial^{s} \Sigma^{+s}_\Gamma}(\cstar[\partial^{s} \Sigma^{+s}_\Gamma]) \subseteq \cstar[\partial^{s} \Sigma^{+s}_\Gamma]$.
    Let $A \in \cstar[\partial^{s} \Sigma^{+s}_\Gamma]$.
    Then for all $B \in \cstar[(\partial^{s} \Sigma^{+s}_\Gamma)^c]$, we have that 
    \[
    [\epsilon_g^{\partial^{s} \Sigma^{+s}_\Gamma}(A), B]
    =
    [\epsilon_g^{\partial^{s} \Sigma^{+s}_\Gamma}(A), \epsilon_g^{\partial^{s} \Sigma^{+s}_\Gamma}(B)]
    =
    \epsilon_g^{\partial^{s} \Sigma^{+s}_\Gamma}([A, B])
    =
    0.
    \]
    Hence $\epsilon_g^{\partial^{s} \Sigma^{+s}_\Gamma}(A) \in \cstar[(\partial^{s} \Sigma^{+s}_\Gamma)^c]' \cap \cstar = \cstar[\partial^{s} \Sigma^{+s}_\Gamma]$.
    Thus $\epsilon_g^{\partial^{s} \Sigma^{+s}_\Gamma}$ is localized in $\partial^{s} \Sigma^{+s}_\Gamma$.
\end{proof}

Recall that for a dividing dual path $L$, the strip $(L^{+s})_{geo}$ divides $\bbR^2 \setminus (L^{+s})_{geo}$ into two simply connected components.
On this strip, we can define a $1D$ spin chain $\fA_{L^{+s}}$ by partitioning $L^{+s}$ into disjoint finite strips $\{L^{+s}(i)\}_{i \in \bbZ}$ containing a uniformly upper bounded number of sites in $\Gamma$ such that $\bigcup_i L^{+s}(i) = L^{+s}$. We assume that this partition is chosen so that whenever $|i-j|>1$, we have
$$d(L^{+s}(i),L^{+s}(j)) \coloneqq \inf\{d(x, y): x \in L^{+s}(i), y \in L^{+s}(j)\} > 2s.$$
We now define the corresponding $1D$ spin chain $$\fA_{L^{+s}} = \overline{\bigotimes_i \fA_{L^{+s}(i)}}^{\|\cdot\|} \subset \fA$$.
Note that we are taking the nearest-neighbor lattice spacing to be 1.

By Lemma \ref{lem:localization along a strip}, the support of $\epsilon^{\partial^{s} r (L)}_g$ is contained in $L^{+s}$ since $L^{+s} = \partial^s r(L)$.\footnote{Recall that $(L^{+s})_{geo}$ divides the plane into two components, one of which is contained in $r(L)_{geo}$.
We let $\Sigma_{geo}$ denote this component. 
Then letting $\Sigma \coloneqq \Sigma_{geo} \cap \Gamma$, we have that $\Sigma^{+s}_\Gamma = r(L)$, so $\partial^sr(L) \coloneqq \partial^s \Sigma^{+s}_\Gamma$.
}
Therefore, we write $\epsilon^{\partial^{s}L}_g \coloneqq \epsilon^{\partial^{s} r (L)}_g$ for brevity.

\begin{lem}
\label{lem:epsilon is a 1D QCA}
     The automorphism $\epsilon^{\partial^s L}_g$ restricts to a $1D$-QCA of spread at most $1$ on the spin chain $\fA_{L^{+s}}$. 
\end{lem}

\begin{proof}
    Recall that $\epsilon_g^{\partial^s L} = \tilde \beta_g^{r(L)} \circ \big(\beta_g^{r(L)}\big)^{-1} = \alpha^{-1} \circ \beta_g^{r(L)} \circ \alpha \circ \big(\beta_g^{r(L)}\big)^{-1}$.
    Therefore, since $\alpha$ has spread $s$ and the symmetry action is on-site, we have that $\epsilon_g^{\partial^s L}$ is a QCA with spread $s$ on $\fA$. 
    By Lemma \ref{lem:localization along a strip}, it follows that $\epsilon^{\partial^s L}_g$ is an automorphism of $\fA_{L^{+s}}$.

    It remains to show that $\epsilon^{\partial^s L}_g(\fA_{L^{+s}(i)}) \subseteq \fA_{L^{+s}(i-1)\cup L^{+s}(i)\cup L^{+s}(i+1)}$.
    However, this follows immediately since $\epsilon^{\partial^s L}_g(\fA_{L^{+s}(i)}) \subseteq \fA_{L^{+s}(i)^{+s}}$ and $L^{+s}\cap (L^{+s}(i))^{+2s} \subset L^{+s}(i-1)\cup L^{+s}(i)\cup L^{+s}(i+1)$.
\end{proof}

The previous Lemma can actually be strengthened, as pointed out in \cite[Sec.~IIB]{zhang2023topological} (cf. \cite[Lem.~3.1]{MR4998893}) if $g \mapsto\epsilon^{\partial^{s} L}_g$ were a group homomorphism.\footnote{The stronger result was independently pointed out to us by Alex Bols and an anonymous reviewer.} 
However, since $g \mapsto\epsilon^{\partial^{s} L}_g$ is actually a twisted group homomorphism, we provide a short proof.

\begin{lem}
\label{lem:epsilon is a 1D FDQC}
    The automorphism $\epsilon^{\partial^{s} L}_g|_{\fA_{L^{+s}}}$ is a $1D$-FDQC on $\fA_{L^{+s}}$.
\end{lem}
\begin{proof}
    Since $\epsilon_g^{\partial^s L}|_{\fA_{L^{+s}}}$ is a $1D$-QCA by Lemma \ref{lem:epsilon is a 1D QCA}, $\epsilon_g^{\partial^s L}|_{\fA_{L^{+s}}}$ has a well-defined GNVW index \cite{MR2890305}. 
    We recall that the GNVW index is a group homomorphism from the space of $1D$-QCAs (with group multiplication given by composition) to positive rational numbers whose kernel is exactly the $1D$-FDQCs. Let us denote the GNVW index by $\ind{-}$. 

    Since our symmetry acts on-site, $\beta_g^{r(L)}$ clearly restricts to a $1D$-FDQC on $\fA_{L^{+s}}$. 
    Now, $\epsilon_g^{\partial^s L} = \tilde \beta_g^{r(L)} \circ (\beta_g^{r(L)})^{-1}$, where $\tilde \beta_g^{r(L)} = \alpha^{-1} \circ \beta_g^{r(L)} \circ \alpha$. 
    Therefore, since $\epsilon_g^{\partial^s L}$ and $\beta_g^{r(L)}$ restrict to $1D$-QCAs on $\fA_{L^{+s}}$, the automorphism $\tilde \beta_g^{r(L)} = \epsilon_g^{\partial^s L} \circ \beta_g^{r(L)}$ does too.
    
    Observe that $\beta_g^{r(L)}$ is obviously a group homorphism, and thus $\tilde \beta_g^{r(L)} = \alpha^{-1} \circ \beta_g^{r(L)} \circ \alpha$ is also a group homomorphism. 
    Therefore, we have that 
    \[
    \ind{\epsilon_g^{\partial^s L}|_{\fA_{L^{+s}}}}
    = 
    \ind{(\tilde \beta_g^{r(L)}|_{\fA_{L^{+s}}} \circ (\beta_g^{r(L)})^{-1}|_{\fA_{L^{+s}}})}
    =
    \ind{\tilde \beta_g^{r(L)}|_{\fA_{L^{+s}}}}\ind{\beta_g^{r(L)}|_{\fA_{L^{+s}}}}^{-1}.
    \]
    Thus, we have that 
    \begin{align*}
    \ind{\epsilon_g^{\partial^s L}|_{\fA_{L^{+s}}}}^{|G|}
    &=
    \ind{(\tilde \beta_g^{r(L)}|_{\fA_{L^{+s}}}})^{|G|}\ind{(\beta_g^{r(L)}|_{\fA_{L^{+s}}}})^{-|G|}
    \\&=
    \ind{(\tilde \beta_g^{r(L)}|_{\fA_{L^{+s}}})^{|G|}}\ind{(\beta_g^{r(L)}|_{\fA_{L^{+s}}})^{|G|}}^{-1}
    =
    1,
    \end{align*}
    so $\ind{\epsilon_g^{\partial^s L}|_{\fA_{L^{+s}}}} = 1$. 
    Since $\epsilon_g^{\partial^s L}|_{\fA_{L^{+s}}} \in \ker (\ind{-})$, the result follows. 
\end{proof}

\begin{lem}
\label{lem:strip aut is split}
    Let $L$ be a dividing dual path. 
    Let $\gamma$ be the half-infinite dual path given by $\gamma_k \coloneqq L_k$ for $k \in \bbN$, and let $\xi$ be the half-infinite dual path given by $\xi_k \coloneqq L_{1-k}$.
    Divide $(L^{+s})_{geo}$ into disjoint simply connected regions $(S^\gamma)_{geo}$ and $(S^\xi)_{geo}$ with $(S^\gamma)_{geo} \cup (S^\xi)_{geo} = (L^{+s})_{geo}$ such that $ \gamma_{dual} \subseteq (S^\gamma)_{geo} \subseteq (\gamma^{+s})_{geo}$ and $\xi_{dual} \subseteq (S^\xi)_{geo} \subseteq (\xi^{+s})_{geo}$.
    Let $S^\gamma \coloneqq (S^\gamma)_{geo} \cap \Gamma$ and $S^\xi \coloneqq (S^\xi)_{geo} \cap \Gamma$.
    We have
    $$\epsilon^{\partial^{s} L}_g = \Xi \circ (\eta^\gamma_g \otimes \eta^{\xi}_g),$$
    where $\eta^\gamma_g, \eta^{\xi}_g$ are FDQCs localized in $S^\gamma$ and $S^{\xi}$ respectively and $\Xi$ is an inner automorphism implemented by a local unitary.
\end{lem}
\begin{proof}
    By Lemma \ref{lem:epsilon is a 1D FDQC} $\epsilon^{\partial^{s} L}_g|_{\fA_{L^{+s}}}$ is an $1D$-FDQC on $\fA_{L^{+s}}$.
    Let the unitaries of the circuit be given by $\{\cB^d_g\}_{d=1}^D$, where $\cB^d_g$ is the set of unitaries that act at the depth $d$. It follows that every $U \in \bigcup_{d = 1}^D \caB^d_g$ is supported in $L^{+s}$. We can use this structure to define automorphisms implementing FDQCs localized around $\gamma$ and $\xi$ as follows.
    For $d = 1, \dots, D$, we define
    $$\eta^\gamma_{d; g}(A) \coloneqq \left(\prod_{U \in \cB^d_g, \supp(U) \subset S^\gamma} U\right) A \left(\prod_{U \in \cB^d_g, \supp(U) \subset S^\gamma} U\right)^*.$$
    Similarly, we define
    $$\eta^{\xi}_{d; g}(A) \coloneqq \left(\prod_{U \in \cB^d_g, \supp(U) \subset S^{\xi} } U\right) A \left(\prod_{U \in \cB^d_g, \supp(U) \subset S^{\xi}} U\right)^*.$$
    Now we define $$\Xi_{d; g}(A) \coloneqq \left(\prod_{U \in \cB^d_g, \supp(U) \cap S^\gamma, S^{\xi} \neq \emptyset} U\right) A \left(\prod_{U \in \cB^d_g, \supp(U) \cap S^\gamma, S^{\xi} \neq \emptyset} U\right)^*.$$
    Since the unitaries in each depth have disjoint supports and $S^\gamma \cup S^\xi = L^{+s}$, we have $$\Xi_{d;g} \circ (\eta^\gamma_{d;g} \otimes \eta^{\xi}_{d;g}) (A) = \left(\prod_{U \in \cB^d_g} U\right) A \left(\prod_{U \in \cB^d_g} U\right)^*.$$
    Additionally, $\Xi_{d;g}$ is an inner automorphism implemented by a local unitary.

    We now define $\eta^\gamma_g \coloneqq  \eta^\gamma_{1; g} \circ \cdots \circ \eta^{\gamma}_{D ; g}$ and $\eta^{\xi}_g \coloneqq  \eta^\xi_{1 ; g} \circ \cdots \circ \eta^{\xi}_{D ; g}$.
    We then get $$\epsilon^{\partial^{s}L}_g = \tilde \beta_g^{r(L)} \circ (\beta_g^{r(L)})^{-1} = \Xi_{1 ; g} \circ (\eta^\gamma_{1 ; g} \otimes \eta^{\xi}_{1 ; g})  \circ \cdots \circ \Xi_{D ; g} \circ (\eta^\gamma_{D ; g} \otimes \eta^{\xi}_{D ; g}) = \Xi \circ (\eta^\gamma_g \otimes \eta^{\xi}_g).$$
    Here we have used the fact that for any inner automorphism $\Xi$ implemented by a local unitary and FDQC $\alpha$, there exists another inner automorphism $\Xi'$ also implemented by a local unitary such that $\Xi' \circ \alpha = \alpha \circ \Xi$.
    It is clear that $\eta^\gamma_g, \eta^{\xi}_g$ are FDQCs and that they are localized in $S^\gamma$ and $S^{\xi}$ respectively. The result follows.
\end{proof}

\begin{defn}
\label{def:defect automorphisms}
    Let $L$ be a dividing dual path.
    Let $\gamma$ be the half-infinite dual path given by $\gamma_k \coloneqq L_k$, and let $\xi$ be the half-infinite dual path given by $\xi_k \coloneqq L_{1 - k}$. 
    Recall the regions $S^\gamma$ and $S^\xi$ defined in Lemma \ref{lem:strip aut is split}.
    By Lemma \ref{lem:strip aut is split}, we have $\tilde \beta_g^{r(L)} \circ (\beta_g^{r(L)})^{-1} = \Xi \circ (\eta^\gamma_g \otimes \eta_g^{\xi})$, where $\Xi$ is an inner automorphism implemented by a local unitary and $\eta_g^\gamma \in \Aut(\cstar[S^\gamma]), \eta_g^{\xi} \in \Aut(\cstar[S^{\xi}])$.
    We consider the automorphism $\alpha_\gamma^g$ given by
    \begin{equation*}
        \alpha_\gamma^g \coloneqq  \eta_g^{\xi} \circ \beta_g^{r(L)}.
    \end{equation*}
    Note that $\alpha_\gamma^g$ also depends on the path $\xi$, but we suppress this dependence for ease of notation.

    When $\gamma \in \bar P_R(\Gamma)$ (as specified in Section \ref{sec:paths and dual paths}), we will write $\alpha^g_\gamma$ in the case where $L$ is a (possibly unspecified) completion of $\gamma$. 
    In this case, we call $\alpha^g_\gamma$ a \emph{$g$-defect automorphism}.
\end{defn}

In Section \ref{sec:SPTDefectSectorReps} below, we will show that these $g$-defect automorphisms can be used to define $g$-defect representations according to Definition \ref{def:g-defect_rep_and_sector}.

\begin{lem}
\label{lem:Qv relations with defect aut}
    Let $L$ be a dividing dual path.
    Let $\gamma$ be the half-infinite dual path given by $\gamma_k \coloneqq L_k$, and let $\xi$ be the half-infinite dual path given by $\xi_k \coloneqq L_{1 - k}$. 
    Then there exists a ball $B \subset \bbR^2$ containing $\partial\gamma$ such that $\alpha_\gamma^g$ satisfies the following relations:
    \begin{equation*}
        \alpha^g_\gamma (Q_v) = \begin{cases}
            Q_v & v \in \Gamma \setminus (\gamma^{+s} \cup B) \\
            \beta_g^{r(L)}(Q_v) & v \in  \gamma^{+s} \setminus B.
        \end{cases}
    \end{equation*}
\end{lem}
\begin{proof}
    We choose $B \subset \bbR^2$ such that $\supp(Q_v) \cap \supp(U) = \emptyset$ for any $v \in B^c \cap \Gamma$, where $U \in \fA^{\loc}$ implements the automorphism $\Xi$ in Lemma \ref{lem:strip aut is split}.
    By enlarging $B$ if necessary, we may also assume that for any $v \in B^c \cap \Gamma$, we have $\supp(Q_v) \cap L^{+s} = \supp(Q_v) \cap S^\gamma$ or $\supp(Q_v) \cap L^{+s} = \supp(Q_v) \cap S^\xi$, where $S^\gamma$ and $S^\xi$ are the regions in Lemma \ref{lem:strip aut is split}.
    Furthermore, we may assume that for any $B^c \cap S^\gamma = B^c \cap \gamma^{+s}$.

    Let $v \in B^c \cap \Gamma$.
    Recall that $L^{+s}= S^\gamma  \cup S^\xi$ and that $S^\gamma$ and $S^\xi$ are disjoint. 
    Furthermore, since $v \in B^c \cap \Gamma$, we have $\supp(Q_v) \cap L^{+s} = \supp(Q_v) \cap S^\gamma$ or $\supp(Q_v) \cap L^{+s} = \supp(Q_v) \cap S^\xi$ by assumption. 
    We consider each of these cases in turn.
    
    First, suppose $\supp(Q_v) \cap L^{+s} = \supp(Q_v) \cap S^\xi$. 
    In this case, since $S^\gamma, S^\xi \subseteq L^{+s}$ are disjoint, we have that $\supp(Q_v) \cap S^\gamma = \emptyset$. 
    Thus, since $\supp(Q_v) \cap \gamma^{+s} = \supp(Q_v) \cap S^\gamma$ by assumption, we must have that $v \notin \gamma^{+s}$. 
    Therefore $v \in \Gamma \setminus (\gamma^{+s} \cup B)$.
    Now, we have also assumed that $\supp(Q_v) \cap \supp(U) = \emptyset$, where $U$ implements $\Xi$ in Lemma \ref{lem:strip aut is split}.
    Since $\supp(\beta_g^{r(L)}(Q_v)) = \supp(Q_v)$, we have that $\supp(\beta_g^{r(L)}(Q_v)) \cap S^\gamma = \emptyset$ and $\supp(\beta_g^{r(L)}(Q_v)) \cap U = \emptyset$. 
    Therefore, since $\epsilon_g^{\partial^sL} = \Xi \circ (\eta_g^\gamma \otimes \eta_g^\xi)$ by Lemma \ref{lem:strip aut is split} and $\eta_g^\gamma$ is localized in $S^\gamma$, we have that 
    \[
    \epsilon_g^{\partial^sL} \circ \beta_g^{r(L)}(Q_v)
    =
    (\Xi \circ (\eta_g^\gamma \otimes \eta_g^\xi))(\beta_g^{r(L)}(Q_v))
    =
    \eta_g^\xi(\beta_g^{r(L)}(Q_v)).
    \]
    Thus, we have that 
    \begin{align*}
        \alpha_\gamma^g( Q_v) &= \eta_g^{\xi} \circ \beta_g^{r(L)}( Q_v) = \epsilon_g^{\partial^sL} \circ \beta_g^{r(L)}( Q_v)  = \tilde \beta_g^{r(L)} \circ (\beta_g^{r(L)})^{-1}\circ \beta_g^{r(L)}( Q_v)\\
        &= \alpha^{-1} \circ \beta_g^{r(L)} \circ \alpha (Q_v)
        = \alpha^{-1} \circ \beta_g^{r(L)} (P_v) = \alpha^{-1}(P_v) = Q_v.
    \end{align*}

    Now, suppose $\supp(Q_v) \cap L^{+s} = \supp(Q_v) \cap S^\gamma$. 
    In that case, $\supp(Q_v) \cap S^\xi = \emptyset$ since $S^\gamma, S^\xi \subseteq L^{+s}$ are disjoint.
    Therefore, since $\eta_g^\xi$ is localized in $S^\xi$, we have that 
    \[
        \alpha_\gamma^g( Q_v) = \eta_g^{\xi} \circ \beta_g^{r(L)}( Q_v) = \beta_g^{r(L)}( Q_v).
    \]
    It remains to show that if $v \notin \gamma^{+s}$, then $\beta_g^{r(L)}(Q_v) = Q_v$. 
    Since $\beta_g(Q_v) = Q_v$, it suffices to show that if $v \notin \gamma^{+s}$, then $\supp(Q_v) \subseteq r(L)$ or $\supp(Q_v) \subseteq \ell(L)$. 
    We prove the contrapositive. 
    Suppose $\supp(Q_v) \not\subseteq r(L)$ and $\supp(Q_v) \not\subseteq \ell(L)$.
    Then since $r(L) \cup \ell(L) = \Gamma$, we have $\supp(Q_v) \cap r(L) \neq \emptyset$ and $\supp(Q_v) \cap \ell(L) \neq \emptyset$.
    Now $\supp(Q_v) \subseteq \{v\}^{+s}_\Gamma$ since $Q_v = \alpha^{-1}(P_v)$ where $P_v \in \cstar[v]$ and $\alpha$ is an FDQC of spread $s$.
    Therefore $\{v\}^{+s}_\Gamma \cap r(L) \neq \emptyset$ and $\{v\}^{+s}_\Gamma \cap \ell(L) \neq \emptyset$, and thus $\{v\}^{+s} \cap r(L)_{geo} \neq \emptyset$ and $\{v\}^{+s} \cap \ell(L)_{geo} \neq \emptyset$.
    Now, $\{v\}^{+s} \subset \bbR^2$ is connected, so we must have that $\{v\}^{+s} \cap L_{dual} \neq \emptyset$ since $L_{dual}$ is the boundary of the regions $r(L)_{geo}$ and $\ell(L)_{geo}$. 
    Therefore, we have that $v \in (L_{dual})^{+s} = (L^{+s})_{geo}$, and thus $v \in L^{+s}$ since $v \in \Gamma$.
    Now, since $v \in \supp(Q_v)$ and $\supp(Q_v) \cap S^\xi = \emptyset$, we must have that $v \in S^\gamma$. 
    Therefore, since $v \in B^c$ and we have assumed that $B^c \cap S^\gamma = B^c \cap \gamma^{+s}$, we have that $v \in \gamma^{+s}$ as desired.
\end{proof}

\subsection{Defect Hamiltonians}
\label{sec:defect Hamiltonians SPT}
Symmetry defects in topological order have been well explored in the literature (see for instance \cite{PhysRevLett.105.030403}, \cite{PhysRevB.100.115147}). Here we expand on the approach of \cite[\S 5]{PhysRevB.100.115147} in order to explicitly construct a \emph{defect Hamiltonian}, whose ground state has a symmetry defect. 
In particular, we give the general procedure to construct a commuting projector Hamiltonian from $H$ that houses a symmetry defect at the end-points of $\gamma$ and a domain wall along $\gamma$. 
Let $L$ be a dividing dual path, and let $\gamma$ be the half-infinite dual path given by $\gamma_k \coloneqq L_k$.
We use the results of Lemma \ref{lem:Qv relations with defect aut} in the following definition.

\begin{defn}
\label{def:SPTDefectHamiltonian}
    We define the \emph{defect Hamiltonian} $H^{(g, \gamma)}$ via $$H_S^{(g,\gamma)} \coloneqq  \sum_{v \in S} \mathds1 - \widehat Q_v^g \qquad \qquad \widehat Q_v^g \coloneqq  (\alpha_\gamma^g)^{-1}(Q_v),$$ where $S \in \Gamma_f$ and $v \in \Gamma$. 
    We let $\delta^{(g,\gamma)}$ be its corresponding derivation.
\end{defn}

It is easy to see that $H^{(g, \gamma)}$ is a commuting projector Hamiltonian with unique ground state $\tilde \omega_\gamma^g \coloneqq  \tilde \omega \circ \alpha_{\gamma}^g$. 
This ground state has the interpretation of housing a $g$-defect at the endpoints of $\gamma$.

\begin{rem}
The original construction of the defect Hamiltonian in \cite{PhysRevB.100.115147} does not necessarily yield a commuting projector Hamiltonian due to the ambiguity in how to modify the Hamiltonian at the end-points of the defect line. We use defect automorphisms to construct the defect Hamiltonian for our SPTs, so it is a commuting projector Hamiltonian. Moreover, our procedure leaves no ambiguity on how to modify the Hamiltonian around the endpoints of the defect.

\end{rem}

\subsection{Defect representations}
\label{sec:SPTDefectSectorReps}
Recall the definition of a $g$-defect representation (Definition \ref{def:g-defect_rep_and_sector}). Our reference representation will now be $(\tilde \pi, \tilde \cH)$, the GNS representation of $\tilde \omega$, unless stated otherwise. 
We also note that $\pi_0 \circ \alpha \simeq \tilde \pi$ by uniqueness of the GNS representation, where $\pi_0$ is the GNS representation for the product state $\omega_0$.

\begin{lem}
\label{lem:trivial-para-Haag-duality}
    Let $\omega$ be a pure product state and $\pi$ its GNS representation. The representation $\pi$ satisfies strict Haag duality, i.e, we have for all cones $\Lambda$ that $$\pi(\cstar[\Lambda^c])' = \pi(\cstar[\Lambda])''.$$
\end{lem}
\begin{proof}
    Since $\omega$ is a product state we can apply \cite[Lem.~4.3]{MR4426734} to get that $\pi$ satisfies strict Haag duality for any cone $\Lambda$.
\end{proof}

We now show that the assumptions in Section \ref{sec:GCrossedAssumptions} are satisfied.  
Assumptions \ref{asmp:Faithfulness} and \ref{asmp:GInvariance} are satisfied by assumption.
By Lemma \ref{lem:trivial-para-Haag-duality} we have that $\pi_0$ satisfies strict Haag duality. Using the fact that $\tilde \pi = \pi_0 \circ \alpha$ and Lemma \ref{lem:QCABSHaagDuality} we conclude that $\tilde \pi$ satisfies bounded spread Haag duality (Assumption \ref{asmp:BoundedSpreadHaagDuality}) with spread at most $2s$.
Since $\omega_0$ is pure and $\tilde \omega = \omega_0 \circ \alpha$, we have that $\tilde \omega$ is also pure (equivalently, $\tilde \pi$ is irreducible) since $\alpha$ is an automorphism of $\fA$. 
Thus Assumption \ref{asmp:PureState} is satisfied.  
Assumption \ref{asmp:InfiniteFactor} is satisfied by \cite[Lem.~5.3]{MR4362722} because $\tilde\omega$ is a gapped ground state of a Hamiltonian with uniformly bounded finite range interactions.

We say a reference representation $\pi$ has \emph{trivial superselection theory} if every irreducible anyon representation w.r.t $\pi$ (Definition \ref{def:anyon representation and sector}) is in fact unitarily equivalent to $\pi$. The following lemma shows that $\pi_0$ has trivial superselection theory since $\omega_0$ is a product state. 

\begin{lem}
\label{lem:trivial para has trivial superselection theory}
    Let $\omega$ be a pure product state. The corresponding GNS representation $\pi\colon \cstar \rightarrow \caB(\hilb)$ has trivial superselection theory. 
\end{lem}
\begin{proof}
    Since $\omega$ is a product state, for any chosen cone $\Lambda$ we have $\omega = \omega^{\Lambda} \otimes \omega^{\Lambda^c}$. Now let $\pi^\Lambda$ be the GNS representation of $\omega^{\Lambda}$ and $\pi^{\Lambda^c}$ the GNS representation of $\omega^{\Lambda^c}$. Using the uniqueness of the GNS representation, we have $\pi \simeq \pi^{\Lambda} \otimes \pi^{\Lambda^c}$. We now apply \cite[Thm.~4.5]{MR4426734} to get the required result.
\end{proof}

The following Lemma shows that $\tilde \pi$ has trivial superselection theory.

\begin{lem}
\label{lem:SPT has trivial superselection theory}
    Let $\pi$ be the GNS representation of a pure product state $\omega$, and let $\alpha$ be a quasi-factorizable automorphism. Then the representation $\widehat \pi\coloneqq \pi \circ \alpha$ has trivial superselection theory.
\end{lem}
\begin{proof}
    Let $\pi'$ be an anyon representation with respect to $\widehat \pi$. We apply \cite[Theorem ~4.7]{MR4426734} to get that $\pi' \circ \alpha^{-1}$ must be an anyon representation with respect to $\widehat \pi \circ \alpha^{-1} = \pi$. But from Lemma \ref{lem:trivial para has trivial superselection theory} we have that $\pi$ has trivial superselection theory, implying $\pi' \circ \alpha^{-1} \simeq \pi$. 
    We thus have $$\pi' \simeq \pi \circ \alpha \simeq \widehat \pi,$$ giving us the required result.
\end{proof}

We now show that the symmetry defect states $\tilde\omega_\gamma^g$ are finitely transportable. 
Let us first prepare an important lemma.

\begin{nota}
Let $r \geq 0$. 
For $S, W \subseteq \bbR^2$, we say that $S \Subset_r W$ if $S^{+r} \subseteq W$ 
(this notation should be compared to $\ll_r$ in \cite{2307.12552}).
\end{nota}
\begin{rem}
\label{rem:SubsetSpreadComplements}
If $S \subseteq \bbR^2$ is simply connected and $S \Subset_r S'$, then $S'^c \Subset_r S^c$.  
Indeed, we have that $((S^{+r})^c)^{+r} = S^c$, so we have that
\[
(S'^c)^{+r}
\subseteq
((S^{+r})^c)^{+r}
=
S^c.
\]
\end{rem}

\begin{lem}
\label{lem:restrictions of ground states are equal SPT generalization}
    Let $L$ be a dividing dual path, and let $\gamma$ be the half-infinite dual path given by $\gamma_k \coloneqq L_k$. 
    Let $\alpha_\gamma^g$ be the automorphism given in Definition \ref{def:defect automorphisms}, and let $\widehat Q_v^g = (\alpha^g_\gamma)^{-1}(Q_v)$ as in Definition \ref{def:SPTDefectHamiltonian}.
    Let $S, S' \subseteq \bbR^2$ be balls satisfying $S \Subset_{s^\gamma} S'$, where $s^\gamma$ is the spread of $(\alpha^g_\gamma)^{-1} \circ \alpha^{-1}$.
    Let $\omega_1, \omega_2 \in \cS(\cstar)$ be two states such that $\omega_1(\widehat Q_v^g) = \omega_2(\widehat Q_v^g) = 1$ for all $v \in S^c \cap \Gamma$. Then we have $$\omega_1 |_{\cstar[S'^c \cap \Gamma]} = \omega_2 |_{\cstar[S'^c \cap \Gamma]}.$$
\end{lem}
\begin{proof}
    Define the product state $\omega_0^{S^c} \coloneqq  \omega_0 |_{\cstar[S^c \cap \Gamma]} \in \cS(\cstar[S^c \cap \Gamma])$, where $\cS(A)$ denotes the state space of $A$ for a $\rmC^*$-algebra $A$. 
    Suppose $\omega \in \cS(\cstar)$ satisfies that $\omega(P_v) = 1$ for every $v  \in S^c \cap \Gamma$.
    We claim that $\omega|_{\cstar[S^c] \cap \Gamma} = \omega_0^{S^c}$.
    Indeed, let $A \in \cstar[S^c \cap \Gamma]^{\loc}$ be a simple tensor. 
    That is, $A = \bigotimes_{v \in W} A_v$ for some $W \in (S^c \cap \Gamma)_f$, where $A_v \in \cstar[v]$.
    We then have that 
    \[
    \omega(A)
    =
    \omega\!\left(\bigotimes_{v \in W} A_v\right)
    =
    \omega\!\left(\bigotimes_{v \in W} P_v A_v P_v \right).
    \]
    Now, $P_v A_v P_v \in \bbC P_v$ for all $v \in \Gamma$, so $\bigotimes_{v \in W} P_v A_v P_v = \lambda \bigotimes_{v \in W} P_v$ for some $\lambda \in \bbC$.
    Therefore, we have that 
    \[
    \omega(A)
    =
    \omega\!\left(\bigotimes_{v \in W} P_v A_v P_v \right)
    =
    \lambda \omega\!\left(\bigotimes_{v \in W} P_v\right)
    =
    \lambda.
    \]
    By the same argument, $\omega_0^{S^c}(A) = \lambda$.
    Since the simple tensors span a dense subset of $\cstar[S^c \cap \Gamma]$, we get that $\omega = \omega_0^{S^c}$.
    
    Now, suppose $\omega_1, \omega_2 \in \cS(\cstar)$ satisfy that $\omega_1(\widehat Q_v^g) = \omega_2(\widehat Q_v^g) = 1$ for all $v \in S^c \cap \Gamma$.
    In that case for $i = 1, 2$, we have that $\tilde \omega_i \coloneqq \omega_i  \circ (\alpha_\gamma^g)^{-1} \circ \alpha^{-1}$ satisfies that for all $v \in S^c \cap \Gamma$, 
    \[
    \tilde \omega_i(P_v)
    =
    \omega_i \circ (\alpha^g_\gamma)^{-1}(Q_v)
    =
    \omega_i(\widehat{Q}_v)
    =
    1.
    \]
    Thus by the previous paragraph, we have that $\tilde \omega_1(A) = \tilde \omega_2(A)$ for every $A \in \cstar[S^c \cap \Gamma]$.
    Now, since $S \Subset_{s^\gamma} S'$, we have that $S'^c \Subset_{s^\gamma} S^c$ by Remark \ref{rem:SubsetSpreadComplements}. 
    Thus, if $A \in \cstar[S'^c \cap \Gamma]$, we have that $ \alpha \circ \alpha^g_\gamma(A) \in \cstar[S^c \cap \Gamma]$, so we have that 
    \[
    \omega_1(A)
    =
    \tilde \omega_1( \alpha(\alpha^g_\gamma(A)))
    =\tilde \omega_2( \alpha(\alpha^g_\gamma(A)))
    =
    \omega_2(A).
    \qedhere
    \]
\end{proof}

\begin{lem}
\label{lem:defects are finitely transportable SPT case}
    Let $L^1$ and $L^2$ be dividing dual paths, and for $i = 1, 2$, let $\gamma^i$ be the half-infinite dual path given by $\gamma^i_k \coloneqq L^i_k$.
    Suppose that there exists some $N \in \bbN$ and $m \in \bbZ$ such that $\gamma^1_k=\gamma^2_{k + m}$ for all $k\geq N$ (see Figure \ref{fig:paths that only differ finitely}). 
    Then $\tilde \omega_{\gamma^1}^g \simeq \tilde \omega_{\gamma^2}^g$.
        \begin{figure}[!ht]
        \centering
        \includegraphics[width=0.2\linewidth]{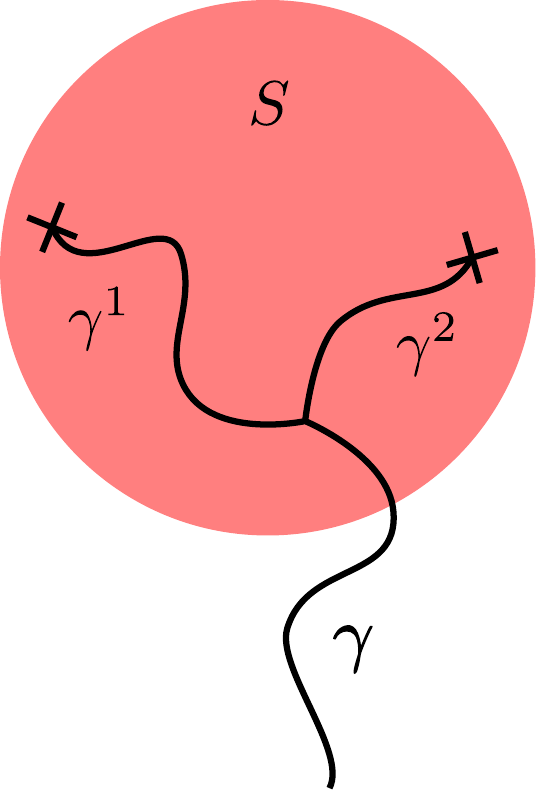}
        \caption{An example of two half-infinite dual paths $\gamma^1, \gamma^2 \in \bar P(\Gamma)$ such that $\gamma^1 \cap \gamma^2 = \gamma \in \bar P(\Gamma)$ is another half-infinite dual path, i.e, $\gamma^1, \gamma^2$ differ only in a finite region $V$. The region $V$ is designed such that $V^c \cap \gamma^1 \subset \gamma$ and $V^c \cap \gamma^2 \subset \gamma$}
        \label{fig:paths that only differ finitely}
    \end{figure}
\end{lem}
\begin{proof}
    Since $\tilde \omega_{\gamma^1}^g$ and $\tilde \omega_{\gamma^2}^g$ are pure states, $\tilde \omega_{\gamma^1}^g$ and $\tilde \omega_{\gamma^2}^g$ are unitarily equivalent if and only if they are quasi-equivalent \cite[Prop.~10.3.7]{MR1468230}.
    Since there exists some $N \in \bbN$ and $m \in \bbZ$ such that $\gamma^1_k=\gamma^2_{k + m}$ for all $k\geq N$, by Lemma \ref{lem:Qv relations with defect aut}, there exists a ball $B \subseteq \bbR^2$ such that $\alpha_{\gamma^1}^g(Q_v) = \alpha_{\gamma^2}^g(Q_v)$ for every $v \in B^c \cap \Gamma$.  
    Therefore, by Lemma \ref{lem:restrictions of ground states are equal SPT generalization}, $\tilde \omega_{\gamma^1}^g|_{\cstar[(B^{+r})^c \cap \Gamma]} = \omega_{\gamma^2}^g|_{\cstar[(B^{+r})^c \cap \Gamma]}$ for some $r \geq 0$, so by \cite[Cor.~2.6.11]{MR887100}, $\tilde \omega_{\gamma^1}^g \simeq \tilde \omega_{\gamma^2}^g$.
\end{proof}

We will now show that for $\gamma \in \bar P_R(\Gamma)$ and $g \in G$, the representation $\tilde \pi_\gamma^g \coloneqq \tilde \pi \circ \alpha_\gamma^g$ is a $g$-defect representation with respect to $\tilde \pi$.
Note that $\tilde \pi_\gamma^g$ is a GNS representation of $\tilde\omega_\gamma^g = \tilde \omega \circ \alpha_\gamma^g$.

\begin{lem}
\label{lem:defect state GNS reps are sectorizable (SPT case)}
    For all $\gamma \in \bar P_R(\Gamma)$ and $g \in G$, the representation $\tilde\pi^g_\gamma$ is an irreducible $g$-defect representation with respect to $\tilde \pi$.
\end{lem}
\begin{proof}
    We let $L$ be a dividing completion of $\gamma$, and let $\xi \in \bar P(\Gamma)$ be the half-infinite path given by $\xi_k \coloneqq L_{1 - k}$.
    Since $\xi \in \bar P(\Gamma)$, there exists a cone $\Lambda_0 \in \cL$ such that $\xi_{dual} \subset(\Lambda_0)_{geo}$.
    We let $\Lambda \in \cL$ be a cone such that $((\Lambda_0)_{geo})^{+s} \subseteq \Lambda_{geo}$, so that $(\xi^{+s})_{geo} = (\xi_{dual})^{+s} \subset ((\Lambda_0)_{geo})^{+s} \subseteq \Lambda_{geo}$. 
    We show that $\tilde \pi_\gamma^g$ is $g$-localized in $\Lambda$.
    Recall that $\alpha^g_\gamma = \eta_g^{\xi} \circ \beta_g^{r(L)}$, where $\eta_g^{\xi}$ only acts nontrivially on $S^\xi \subseteq \xi^{+s}$.  
    Therefore, we have that $\eta_g^{\xi}|_{\cstar[\Lambda^c]} = \Id$, so we have that 
    \[
    \tilde \pi_\gamma^g|_{\cstar[\Lambda^c]}
    =
    \tilde \pi \circ \alpha_\gamma^g|_{\cstar[\Lambda^c]}
    =
    \tilde \pi \circ \eta_g^{\xi} \circ \beta_g^{r(L)}|_{\cstar[\Lambda^c]}
    =
    \tilde \pi \circ \beta_g^{r(L)}|_{\cstar[\Lambda^c]}.
    \]
    Since $\gamma \in \bar P_R(\Gamma)$, we have that $r(L) \cap \Lambda^c$ differs from $r(\Lambda)$ by finitely many vertices.  
    Therefore, $\tilde \pi_\gamma^g$ is $g$-localized in $\Lambda$.

    It remains to show that $\tilde \pi_\gamma$ is transportable.  
    Let $\Lambda' \in \cL$ be another cone.  
    We choose a dual path $\widehat{\gamma} \in \bar P_R(\Gamma)$ and a dividing completion $\widehat{L}$ such that the half-infinite dual path $\xi$ given by $\xi_k \coloneqq L_{1 - k}$ satisfies that $\xi^{+2s} \subset (\Lambda')_{geo}$.
    Then by the preceding argument, $\tilde \pi_{\widehat{\gamma}}^g$ is $g$-localized in $\Lambda'$.  
    Furthermore, by Lemma \ref{lem:defects are finitely transportable SPT case}, $\tilde \pi_\gamma^g \simeq \tilde \pi_{\widehat{\gamma}}^g$. 
    Thus $\tilde \pi_\gamma^g$ is transportable, and therefore $\tilde \pi_\gamma^g$ is a $g$-defect representation with respect to $\tilde \pi$.
\end{proof}

\begin{lem}
\label{lem:classification of defect sectorizable representations SPT}
    Let $g \in G$. Every irreducible $g$-defect representation with respect to $\tilde \pi$ is unitarily equivalent to the representation $\tilde \pi_\gamma^g$ for some $\gamma \in \bar P_R(\Gamma)$.
\end{lem}
\begin{proof}
    Let $\pi$ be an irreducible $g$-defect representation with respect to $\tilde \pi$. From Lemma \ref{lem:defect state GNS reps are sectorizable (SPT case)} we have that $\tilde \pi_\gamma^g$ is a $g$-defect representation with respect to $\tilde \pi$. We have from Lemma \ref{lem:GDefectsRelationToAnyons} that $\pi$ is an anyon representation with respect to $\tilde \pi_\gamma^g$. Now, $\tilde \pi_\gamma^g = \tilde \pi \circ \alpha_\gamma^g$, and $\alpha_\gamma^g$ is an FDQC and thus quasi-factorizable by Lemma \ref{lem:FDQCQuasi-Factorizable}. Furthermore, by Lemma \ref{lem:SPT has trivial superselection theory}, $\tilde \pi$ has trivial superselection theory. Therefore, $\tilde \pi_\gamma^g$ has trivial superselection theory by the proof of Lemma \ref{lem:SPT has trivial superselection theory}.
    Putting these results together, we have $\tilde \pi^g_\gamma \simeq \pi$ as desired.
\end{proof}

We have shown the following classification result.

\begin{prop}
    \label{prop:general SPT defect sector classification}
    Let $\tilde \omega$ be a $G$-SPT (Definition \ref{def:SPT_phase}). 
    Let $\tilde \pi$ be the GNS representation of $\tilde \omega$ and define $\tilde \pi_{{R}}^g \coloneqq  \tilde \pi \circ \alpha_{R}^g$, where $\alpha_{R}^g$ is the defect automorphism in Definition \ref{def:defect automorphisms} and we abuse notation slightly by identifying $R$ with the dual path composed of the edges intersected by $R$.

    Each representation $\tilde \pi_{{R}}^g$ corresponds to a distinct irreducible $g$-defect sector with respect to $\tilde \pi$, and these sectors comprise all of the irreducible $G$-defect sectors with respect to $\tilde \pi$.
\end{prop}

We let $\GSec_{\tilde \pi}$ (Definition \ref{def:GSec}, cf.~Remark \ref{rem:GSec notation reference rep}) to be the category of $G$-defect representations with respect to reference representation $\tilde \pi$, extended to be endomorphisms of the auxiliary algebra $\fA^a$. 
For the remainder of this section, we will use $\tilde \pi_\gamma^g$ to denote the corresponding extension to the auxiliary algebra.

\begin{lem}
\label{lem:tensor of defect reps SPT}
For every $\gamma \in \bar P_R(\Gamma)$ and $g,h \in G$, we have
\(
\tilde \pi_\gamma^g \otimes \tilde \pi_\gamma^h \simeq \tilde \pi_\gamma^{gh}.
\)

\end{lem}

\begin{proof}

Since $\fA \subseteq \fA^a$ via the identification $\fA = \tilde \pi(\fA)$, we get that $\tilde \pi_\gamma^g (\tilde \pi(A)) = \tilde \pi \circ \alpha_\gamma^g(A)$ for all $A \in \fA$. Since $\tilde \pi_\gamma^g = \tilde \pi \circ \alpha_\gamma^g $ and $ \tilde \pi_\gamma^h = \tilde \pi \circ \alpha_\gamma^h$, we have
\[
(\tilde \pi_\gamma^g \otimes \tilde \pi_\gamma^h)(\tilde \pi(A)) 
= \tilde \pi_\gamma^g (\tilde \pi \circ \alpha_\gamma^h (A))  = \tilde \pi \circ \alpha_\gamma^g \circ \alpha_\gamma^h(A).
\]
Because $\alpha_\gamma^g \circ \alpha_\gamma^h$ is an automorphism of $\cstar$, and $\tilde \pi$ is irreducible, it follows that $\tilde \pi_\gamma^g \otimes \tilde \pi_\gamma^h$ is irreducible.

Now, by Proposition \ref{prop:GCrossedMonoidal}, $\tilde \pi_\gamma^g \otimes \tilde \pi_\gamma^h$ is a $gh$-defect representation since $\tilde \pi^g_\gamma$ is a $g$-defect representation and $\tilde \pi^h_\gamma$ is an $h$-defect representation.
Therefore, since $\tilde \pi_\gamma^g \otimes \tilde \pi_\gamma^h$ is irreducible, $\tilde \pi_\gamma^g \otimes \tilde \pi_\gamma^h \simeq \tilde \pi^{gh}_\gamma$ by Lemma \ref{lem:classification of defect sectorizable representations SPT}.
\end{proof}

We recall that $\GSec^f_{\tilde \pi}$ denotes the category of dualizable $G$-defect representations and that $(\GSec^f_{\tilde \pi})_g$ denotes the category of dualizable $g$-defect representations (Definition \ref{def:Dualizable}). 

\begin{cor}
\label{lem:IrreducibleGDefectsDualizableSPT}
For every $\gamma \in \bar P_R(\Gamma)$ and $g \in G$, we have that $\tilde \pi^g_\gamma \in (\GSec^f_{\tilde \pi})_g$. 
\end{cor}

\begin{proof}
By Lemma \ref{lem:defect state GNS reps are sectorizable (SPT case)}, $\tilde \pi^g_\gamma \in \GSec_g$. 
Furthermore, by Lemma \ref{lem:tensor of defect reps SPT}, $\tilde \pi^g_\gamma \otimes \tilde \pi^{g^{-1}}_\gamma \simeq \tilde \pi^1_\gamma = \tilde \pi$. 
Thus $\tilde \pi^g_\gamma$ is invertible and in particular dualizable.
\end{proof}

We give an introduction to the category $\Vect(G,\nu)$ in Appendix \ref{sec:VecG} and refer the reader to Section \ref{sec:basics of cat thy} for the definitions since we make heavy use of basic category theoretical language and machinery in this proof. 

\begin{prop}
\label{prop:GSec is equiv to HilbGnu}
    There exists a normalized $3$-cocyle $\nu \in Z^3(G,U(1))$ such that the category $\GSec_{\tilde\pi}^f$ is $G$-crossed braided $\rmW^*$-tensor equivalent to $\Vect(G,\nu)$.
\end{prop}
\begin{proof}
    We set $\pi_g \coloneqq \tilde\pi_{R}^g$ to ease notation. 
    Recall that $\pi_g \in \GSec^f_{\tilde \pi}$ by Corollary \ref{lem:IrreducibleGDefectsDualizableSPT}.

By Lemma \ref{lem:GSecf is semisimple}, $\GSec^f_{\tilde \pi}$ is a semisimple $G$-crossed braided $\rmW^*$-tensor category. By Lemma~\ref{lem:defect state GNS reps are sectorizable (SPT case)} and Lemma~\ref{lem:classification of defect sectorizable representations SPT}, each homogeneous component $(\GSec_{\tilde\pi}^f)_g$ contains exactly one simple object up to isomorphism, namely $\pi_g$, implying that the $G$-grading is faithful. 
In particular, $\GSec_{\tilde \pi}^f$ has exactly $|G|$-many irreducible $G$-defect sectors.

These results imply that $\GSec_{\tilde \pi}^f$ is a pointed unitary fusion category. 
By Lemma \ref{lem:tensor of defect reps SPT}, the irreducible $G$-defect sectors in $\GSec_{\tilde \pi}$ form the group $G$ under tensor product. 
By \cite[Prop.~2.2]{galindo2012clifford} in the unitary setting, we have that $\GSec^f_{\tilde \pi}$ is $\rmW^*$-tensor equivalent to $\Vect(G, \nu)$ for some $\nu \in Z^3(G, U(1))$.\footnote{An explicit construction of this equivalence is given in the exposition following this proof.
It is also clear from the construction of this equivalence that it preserves the $G$-grading.}

Now, it is well known that the $G$-crossed braided structure on $\Vect(G,\nu)$ is unique up to $G$-crossed braided $\rmW^*$-tensor equivalence (see for instance \cite[Cor.~7.7]{jones2022extension}).
Therefore, since $\GSec^f_{\tilde \pi}$ is $\rmW^*$-tensor equivalent to $\Vect(G, \nu)$, we must further have that these two categories are $G$-crossed braided $\rmW^*$-tensor equivalent.
Indeed, if $F \colon \cC \to \cD$ is a ($\rmW^*$-)tensor equivalence and $\cC$ has a $G$-crossed braided monoidal structure, then $\cD$ has a $G$-crossed monoidal structure induced from that of $\cC$.
Thus, given a $\rmW^*$-tensor equivalence $\Phi \colon \GSec^f_{\tilde \pi} \to \Vect(G, \nu)$, the $G$-crossed braided structure on $\GSec^f_{\tilde \pi}$ induces the unique such structure on $\Vect(G, \nu)$, so $\Phi$ is necessarily also a $G$-crossed braided $\rmW^*$-tensor equivalence.

\end{proof}

\begin{cor}
\label{cor:identifying the cocycle}
    The $3$-cocycle $\nu$ in Proposition \ref{prop:GSec is equiv to HilbGnu} is given by the adjoint of the $F$-symbol of $\GSec_{\tilde \pi}$ (Defined in Section \ref{sec:OtherCoherenceData}).
\end{cor}
\begin{proof}
We again set $\pi_g \coloneqq \tilde \pi^g_{R}$ for notational simplicity.
Following the notation of Section \ref{sec:OtherCoherenceData}, we choose unitary intertwiners $\Omega_{g,h}\colon \pi_g\otimes \pi_h\to \pi_{gh}$ and consequently get the $F$-symbol $F(g,h,k)\in U(1)$ via 
\[
\Omega_{gh, k} \Omega_{g, h}
=
F(g, h, k)
\Omega_{g, hk} \pi_g(\Omega_{h, k}).
\]
By Lemma \ref{lem:F-symbols are cocycles}, $F\in Z^3(G,U(1))$. 
We claim that $\nu'(g,h,k) \coloneqq F(g,h,k)^*$ is the desired $3$-cocycle.

We define a functor $\Phi \colon \Vect(G,\nu')\to \GSec_{\tilde\pi}^f$ by setting  $$\Phi\left(\bigoplus_{g} (\bbC_g)^{n_g}\right) \coloneqq \bigoplus_{g} (\pi_g)^{\otimes n_g} \qquad \qquad \Phi(T) \coloneqq \bigoplus_g T_g \otimes \Id_{\pi_g}$$ for a general object $\bigoplus_{g} (\bbC_g)^{n_g} \in \Vect(G,\nu)$ and morphism $T = \bigoplus_g T_g$ with $T_g \in M_{m_g \times n_g}(\bbC)$. 
It is easily verified that $\Phi$ defines a linear dagger functor, and $\Phi$ is normal since all hom spaces are finite dimensional. 

It is easy to see that $\Phi$ is an equivalence.
Indeed, $\Phi$ is fully faithful since
\begin{align*}
    \Hom{\bbC_g\to \bbC_h} =
\begin{cases}
\bbC, & g=h,\\
0, & g\neq h,
\end{cases}
\qquad
\Hom{\pi_g\to \pi_h} =
\begin{cases}
\bbC, & g=h,\\
0, & g\neq h.
\end{cases}
\end{align*}
Furthermore, $\Phi$ is essentially surjective because every object of
$\GSec_{\tilde\pi}^f$ is a finite direct sum of the simple objects
$\{\pi_g\}_{g\in G}$.
Note that $\Phi$ preserves the $G$-grading since $\Phi(\bbC_g)=\pi_g\in(\GSec_{\tilde\pi}^f)_g$. 

We now equip $\Phi$ with a tensorator $\Phi^2_{a,b}\colon \Phi(a\otimes b)\to \Phi(a)\otimes \Phi(b)$. On simple
objects, we define $\Phi^2_{\bbC_g,\bbC_h}\coloneqq \Omega_{g,h}^{-1}$ and for arbitrary objects, $\Phi^2_{a,b}$ is defined componentwise. The unitor $\Phi^0$ is the identity. The monoidal coherence condition for $\Phi$ is precisely the equation defining the $F$-symbol, together with the identity $\nu'=F^*$. 
Thus $\Phi$ is a $\rmW^*$-tensor equivalence, and $\nu' = F^*$ is the desired 3-cocycle.
\end{proof}

\begin{cor}
\label{cor:SPT symmetry fractionalization is trivial}
    The symmetry fractionalization data for $\Vect(G, \nu) \simeq \GSec_{\tilde \pi}^f$ satisfies $\eta(g, h)_k = 1$ for all $g, h, k \in G$. 
    
\end{cor}
\begin{proof}
    
The $G$-action $\gamma \colon G \to \Aut_\otimes(\Vect(G, \nu))$ satisfies that $\gamma_g(\bbC_h) = \bbC_{ghg^{-1}}$ (see Appendix \ref{sec:VecG}).
Therefore, recalling the notation in Section \ref{sec:SymmetryFractionalization}, we can set the unitary $V^h_g \colon \gamma_g(\bbC_h) \to \bbC_{ghg^{-1}}$ to be $V^h_g = \Id_{\bbC_{ghg^{-1}}}$.
We thus have that for all $g, h, k \in G$
\[
\eta(g, h)_k \Id_{ghg^{-1}}
=
V^{hkh^{-1}}_g \gamma_g(V^k_h) (V^k_{gh})^{-1}
=
\Id_{ghg^{-1}},
\]
so $\eta(g, h)_k = 1$ for all $g, h, k \in G$.
\end{proof}

\begin{rem}
\label{rem:ProductStatesAndSectorizability}
    Let $\omega$ be a product state such that $\omega \circ \beta_g = \omega$ and let $\pi$ be its GNS representation. Then for all $V \subseteq \Gamma$ and $g \in G$, we have that $\pi \circ \beta_g^V$ is unitarily equivalent to $\pi$.
    Indeed, by Lemma \ref{lem:Pv invariant under symmetry action} we have $\beta_g(P_v^\omega) = P_v^\omega$ for all $v \in \Gamma$. Now, given $V \subseteq \Gamma$ and $v \in \Gamma$, we have $$\omega \circ \beta_g^V (P_v^\omega) = \omega (P_v^\omega) = 1.$$ By Lemma \ref{lem:existence of gapped Hamiltonian} we thus have that $\omega \circ \beta_g^V = \omega$ for all $g$. So by uniqueness of the GNS representation, we get that $\pi \simeq \pi \circ \beta_g^V$. 

    In particular, we see that for any cone $\Lambda \in \cL$, we have that $\pi \circ \beta_g^{r(\Lambda)}$ is $g$-localized in $\Lambda$ and transportable, so it is a $g$-defect representation. 
    But for any $g, h \in G$, we have that 
    \[
    \pi \circ \beta_g^{r(\Lambda)}
    \simeq
    \pi
    \simeq
    \pi \circ \beta_h^{r(\Lambda)}.
    \]
    So it is unclear whether one should classify $\pi$ as a $g$-defect sector or $h$-defect sector (or even the $1$-defect sector).
    Thus we can see clearly that a criterion analogous to the superselection criterion is insufficient to form a $G$-graded category!

\end{rem}

\section{\texorpdfstring{$\bbZ_2$}{Z2} SPTs: Trivial \texorpdfstring{$\bbZ_2$}{Z2} paramagnet}

Having contructed the $G$-defect category for general $G$-SPTs in Section \ref{sec:general SPTs} and showing that this category is $G$-crossed braided $\rmW^*$-tensor equivalent to $\Vect(G,\nu)$ for some $3$-cocycle $\nu$, we will now verify the precise cocycle for the two $\bbZ_2$-SPTs: the trivial paramagnet in this section, and the Levin-Gu SPT in Section \ref{sec:Levin Gu}.

As before, we take the sites of $\Gamma$ to be vertices of the regular triangular lattice and $R$ to be a vertical ray as in Figure \ref{fig:chosen_ray}.
On each site, we place one qubit.
Let $\sigma^x_v, \sigma^y_v, \sigma^z_v$ denote the Pauli matrices in $\cstar[v]$. 
Then $\{\mathds1, \sigma^x_v, \sigma^y_v, \sigma^z_v\}$ is a basis of $\cstar[v]$ for each $v$.

We now define a symmetry action on $\cstar$. Let $G = \bbZ_2$ and let $g\mapsto U^g_v$ be its unitary representation onto the vertex $v$, with $U^g_v = \sigma^x_v$ for the non-trivial group element $g \in \bbZ_2$ and $U^1_v = \mathds1$ for the trivial group element $1 \in \bbZ_2$. We can then define $\beta_g$ as in Definition \ref{def:GlobalSymmetryAutomorphism}.

\label{sec:TrivialParamagnet}
\subsection{Hamiltonian and ground state}
We define the Hamiltonian $H^0$ by $$H^0_S \coloneqq \sum_{v \in S} (\mathds1 - \sigma^x_v)/2, \qquad\qquad S \in \Gamma_f$$ 
It is easy to verify that $H^0$ is a commuting projector Hamiltonian, with the projections $(\mathds1 - \sigma^x_v)/2$ trivially commuting since they have disjoint supports for different $v, v'$. Let $\delta_0$ be the corresponding generator of dynamics.

We state the following lemma without proof, as it is easy to verify and well known to experts.

\begin{lem}
\label{lem:GS of trivial para is unique}
    There is a unique state $\omega_0$ satisfying that $\omega_0(\sigma_v^x) = 1$ for all $v \in \Gamma$. Moreover, $\omega_0$ is pure and a product state.
\end{lem}

Note that the Hamiltonian $H^0$ is translation invariant.
Additionally, observe that $\omega_0(H^0_S) = 0$ for all $S \in \Gamma_f$, and $\omega_0$ is translation invariant since for any translation $\tau$, we have that $\omega_0 \circ \tau(\sigma^x_v) = 1$ for all $v \in \Gamma$.  
Thus by \cite[Thm.~6.2.58]{MR1441540}, $\omega_0$ is a ground state of $\delta_0$ and hence a frustration free one. 
We again state the following lemma without proof, as it can be easily verified.

\begin{lem}
    \label{lem:GS of trivial para is unique actual lemma}
    The state $\omega_0$ is the unique ground state of $\delta_0$.
\end{lem}

Instead of starting with $H^0$ and obtaining $\omega_0$ as its unique ground state, we could instead have proceeded in the opposite direction of constructing a product state $\omega_0$, then working out a commuting projector Hamiltonian whose ground state is $\omega_0$. We now adopt the latter approach as it better allows us to connect to Section \ref{sec:general SPTs}.

Let $\omega_0$ be the product state defined in Lemma \ref{lem:GS of trivial para is unique}. By Lemma \ref{lem:existence of gapped Hamiltonian} we have the existence of unique projections $P_v \coloneqq  P_v^{\omega_0}$, Hamiltonian $H^0 \coloneqq  H^{\omega_0}$, and corresponding derivation $\delta_0 \coloneqq  \delta^{\omega_0}$. Uniqueness of $P_v$ implies that $P_v = S_v = (\mathds1+ \sigma^x_v)/2$.

For $g \in G$ and $S \subseteq \Gamma$, we recall the symmetry automorphism $\beta_g^{S}\colon \cstar \rightarrow \cstar$ from Definition \ref{def:SymmetryOnSubsets}. 
We observe that $\beta_g(\sigma^x_v) = \sigma^x_v$ for every $v \in \Gamma$, so $\omega_0 \circ \beta_g = \omega_0$ by Lemma \ref{lem:GS of trivial para is unique}.
Hence the assumptions of Lemma \ref{lem:Pv invariant under symmetry action} hold.

\subsection{Category of $G$-defect representations}
\begin{figure}[!ht]
    \centering
    \includegraphics[width=0.2\linewidth]{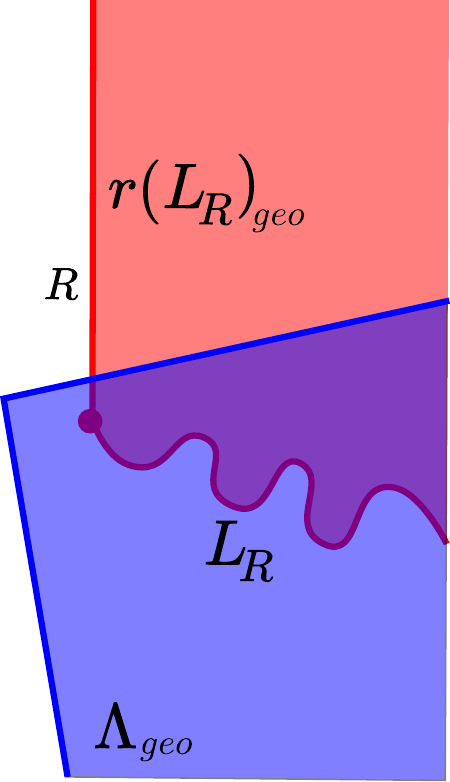}
    \caption{
    An example of $L_{R}$. We require the infinite dual path $L_{R} \in \bar P(\Gamma)$ to be the completion of the dual path composed of the edges intersected by $R$. 
    We further require that the half-infinite dual path $\xi$ defined by  $\xi_k \coloneqq (L_R)_{1 - k}$ satisfies that $\xi \subset \Lambda$ for the chosen cone $\Lambda$ depicted in blue. The region $r(L_{R})$ is shown in red.}
    \label{fig:example of completion of gamma}
\end{figure}

We define $\pi_0$ to be the GNS representation of the ground state $\omega_0$. Note that by Lemma \ref{lem:GS of trivial para is unique} we have that $\omega_0$ is pure, so $\pi_0$ is an irreducible representation. We note that the assumptions in Section \ref{sec:GCrossedAssumptions} are satisfied in this setup, as it is a special case of the discussion in Section \ref{sec:general SPTs}. 
We now compute the dualizable $G$-defect representations with respect to $\pi_0$.
In particular, we will show that the category $\GSec^f_{\pi_0}$ of such representations is $G$-crossed braided $\rmW^*$-tensor equivalent to $\Vect(\bbZ_2) \coloneqq \Vect(\bbZ_2, \nu = 1)$, where $\nu$ is the trivial $3$-cocycle.
Recall that $R$ is the vertical ray as shown in Figure \ref{fig:chosen_ray}. 
Let $L_R$ be a dividing completion of the dual path composed of the edges intersected by $R$ and take $r(L_{R}) \subseteq \Gamma$ be the set of vertices to the right of $L_R$, using the conventions in Section \ref{sec:paths and dual paths}.
An example of $L_{R}$ is shown in Figure \ref{fig:example of completion of gamma}.
By virtue of $L_R$ being a completion, there exists a cone $\Lambda\in\cL$ such that $\Lambda_{geo}$ contains all edges in $L_R$ that do not intersect $R$.
We then have that $$\pi_0\circ \beta_g^{r(L_{R})}\big|_{\cstar[\Lambda^c]} = \pi_0 \circ \beta_g^{r(\Lambda)}\big|_{\cstar[\Lambda^c]},$$ so $\pi_0\circ \beta_g^{r(L_{R})}$ is $g$-localized in $\Lambda$.  
Furthermore, by Remark \ref{rem:ProductStatesAndSectorizability}, $\pi_0\circ \beta_g^{r(L_{R})} \simeq \pi_0 \simeq \pi_0 \circ \beta_g^{S}$ for any $S \subseteq \Gamma$, so $\pi_0\circ \beta_g^{r(L_{R})}$ is transportable. 
Thus $\pi_0\circ \beta_g^{r(L_{R})}$ is a $g$-defect representation.
By Proposition \ref{prop:general SPT defect sector classification}, $\pi_0$ is the only 1-defect representation and $\pi_0\circ \beta_g^{r(L_{{R}})}$ is the only $g$-defect representation, up to unitary equivalence.

By Proposition \ref{prop:GSec is equiv to HilbGnu} we know that $\GSec^f_{\pi_0} \simeq \Vect(\bbZ_2, \nu)$ for some $3$-cocycle $\nu$. 
We now compute the $F$-symbols for this category as done in Section \ref{sec:OtherCoherenceData} to determine the cocycle $\nu$, since $\nu = F^*$ by Corollary \ref{cor:identifying the cocycle}.
We observe that 
\[
\pi_0 \otimes \left(\pi_0\circ \beta_g^{r(L_{R})}\right) = \left(\pi_0\circ \beta_g^{r(L_{R})}\right) \otimes \pi_0 = \pi_0\circ \beta_g^{r(L_{R})}.
\]
Additionally, if $g \neq 1$, then 
\[
\left(\pi_0\circ \beta_g^{r(L_{R})}\right) \otimes \left(\pi_0\circ \beta_g^{r(L_{R})}\right)
=
\pi_0.
\]
Therefore, $\Omega_{1, 1} = \Omega_{g, 1} = \Omega_{1, g} = \Omega_{g, g} = \Id$. Therefore by definition we have that all the $F$-symbols are trivial. By Corollary \ref{cor:identifying the cocycle}, we see that actually $\GSec^f_{\tilde \pi} \simeq \Vect(\bbZ_2)$, i.e., the $3$-cycle $\nu$ is trivial.

\section{\texorpdfstring{$\bbZ_2$}{Z2} SPTs: Levin--Gu SPT}
\label{sec:Levin Gu}
In this section, we verify the well-known result that the category of $G$-defect representations for the Levin--Gu SPT corresponds to the nontrivial element of $H^3(\bbZ_2, U(1))$.

To define the non-trivial $\bbZ_2$-paramagnet, we follow \cite{PhysRevB.86.115109}. We import the setup from Section \ref{sec:TrivialParamagnet}. 
Given neighboring vertices $v,q,q' \in \Gamma$, we let $<vqq'>$ denote the (elementary) face formed by them.
As before, we denote the set of faces in the triangular lattice by $F$.
Let $\triangle_v$ denote the set of all triangles that the vertex $v$ belongs to and let $V(\triangle)$ denote the set of vertices in a face $\triangle$.

\subsection{Hamiltonian and ground state} We define the Hamiltonian $H$ given by 
$$H_{S} \coloneqq \sum_{v \in S} (\mathds1 - B_v)/2 \in \cstar[S^{+1}], \qquad \qquad B_v \coloneqq - \sigma^x_v \prod_{<vqq'> \in \triangle_v} i^{\frac{1 - \sigma^z_q \sigma^z_{q'}}{2}}, \qquad\qquad S \in \Gamma_f.$$ 
Observe that $B_v, B_{v'}$ satisfy the following properties for all $v,v' \in \Gamma$:
\begin{align*}
    B_v^2 = 1, && B_v^* = B_v, && [B_v, B_{v'}] = 0.
\end{align*}
With the above properties, it is easily checked that $(\mathds1 - B_v)/2$ is a projection, so the Hamiltonian $H$ is a commuting projector Hamiltonian for all $S \in \Gamma_f$. 
Let $\tilde \delta$ be the corresponding generator of dynamics.

We recall and rigorously define a useful `entangling' unitary given in \cite[Appendix A]{PhysRevB.86.115109}.  For each face $\triangle \in F$, we define a unitary $U_{\triangle} \in \cstar[\triangle]$ given by $$U_{\triangle} \coloneqq e^{i \frac{\pi}{24} \left( 3 \prod_{v \in V(\triangle)} \sigma^z_v - \sum_{v \in V(\triangle)} \sigma^z_v \right)}.$$ 
Observe that we have $[U_{\triangle}, U_{\triangle'}] = 0$ for all faces $\triangle, \triangle' \in F$.

\begin{defn}
\label{def:LevinGuEntangler}
    For each $A \in \cstar[S']$ with $S' \in \Gamma_f$, let $S \in \Gamma_f$ be a sufficiently large supset of $S'$, i.e., for each site $v' \in S'$, we have that $V(\triangle) \subseteq S$ for all $\triangle \in \triangle_{v'}$. We define a map $\alpha \colon \cstar[S'] \rightarrow \cstar[S]$ given by 
    \begin{align*}
        \alpha (A) \coloneqq \left(\prod_{\triangle \in F(S)} U_{\triangle}\right)^* A \left(\prod_{\triangle \in F(S)} U_{ \triangle}\right),
    \end{align*} 
    where $F(S)\subseteq F$ consists of all $\triangle\in F$ with $V(\triangle)\subseteq S$.
    This map can be uniquely extended in a norm continuous way to an automorphism $\alpha$ of $\cstar$.
\end{defn}

We remark that $\alpha$ is an FDQC of depth 3 with spread $s = 1$, as shown in detail in Lemma \ref{lem:alpha_theta is a bounded spread automorphism}.
In particular, $\alpha$ is a quasi-factorizable QCA.

Let $\omega_0$ be the unique ground state of the trivial $\bbZ_2$ paramagnet as defined in Lemma \ref{lem:GS of trivial para is unique}. We define $\tilde\omega \coloneqq  \omega_0 \circ \alpha.$

\begin{lem}
\label{lem:Levin-Gu ground-state is SPT}
    The state $\tilde \omega$ is a $\bbZ_2$-SPT.
\end{lem}
\begin{proof}
    Recall that $\omega_0$ is $\beta_g$-invariant, and by Lemma \ref{lem:alpha_theta is a bounded spread automorphism}, $\alpha$ is an FDQC.
    Therefore, by Remark \ref{rem:GInvarianceTrick}, $\tilde \omega$ is a $\bbZ_2$-SPT provided that $\alpha$ is a symmetric entangler. 
    We now show this fact.
    By Lemma \ref{lem:automorphism connecting trivial para GS and Levin Gu GS}, we have that for every $v \in \Gamma$, 
    \begin{gather*}
    \alpha(\beta_g(\sigma^x_v))
    =
    \alpha(\sigma^x_v)
    =
    B_v
    =
    \beta_g(B_v)
    =
    \beta_g(\alpha(\sigma^x_v)),
    \\
    \alpha(\beta_g(\sigma^z_v))
    =
    \alpha(-\sigma^z_v)
    =
    -\sigma^z_v
    =
    \beta_g(\sigma^z_v)
    =
    \beta_g(\alpha(\sigma^z_v)).
    \end{gather*}
    Since $\{ \sigma^x_v, \sigma^z_v : v \in \Gamma\}$ generates $\cstar$, we have that $\alpha \circ \beta_g = \beta_g \circ \alpha$, as desired.
\end{proof}

\begin{lem}
\label{lem:facts about Levin-Gu ground-state}
    We have the following facts about $\tilde \omega$:
    \begin{enumerate}
        \item $\tilde \omega$ is the unique state satisfying $\tilde \omega(B_v) = 1$ for all $v \in \Gamma$
        \item $\tilde \omega \circ \beta_g = \tilde \omega$ for all $g \in G$
        \item $\tilde \omega$ is the unique (hence pure) ground state of $\tilde \delta$
        \item $\tilde \omega$ is translation invariant
    \end{enumerate}
\end{lem}
\begin{proof}
Since $\omega_0$ is a product state, from Lemma \ref{lem:existence of gapped Hamiltonian} we have corresponding projections $P_v \coloneqq P_v^{\omega_0}$ and Hamiltonians $H^0 \coloneqq H^{\omega_0}$. From Lemmas \ref{lem:existence of gapped Hamiltonian}, \ref{lem:GS of trivial para is unique} we get that $P_v = (\mathds1 + \sigma^x_v)/2$ for all $v \in \Gamma$. 

Now, by Lemma \ref{lem:unique GS of SPT} we have that $\tilde\omega$ is the unique ground state of the Hamiltonian $H'$ given by
$$H_S' = \sum_{v \in S} \mathds1 - Q_v \qquad \qquad Q_v \coloneqq  \alpha(P_v)$$ 
and additionally that $\tilde \omega$ is the unique state satisfying $\tilde \omega(Q_v) = 1$ for all $v \in \Gamma$. 
By Lemma \ref{lem:automorphism connecting trivial para GS and Levin Gu GS}, $Q_v = \alpha(P_v) = (\mathds1+B_v)/2$, so $H' = H$.
Thus $\tilde \omega$ is the unique ground state of $\tilde \delta$ and is the unique state satisfying $\tilde \omega(B_v) = 1$ for all $v \in \Gamma$. 
This proves (1) and (3) above. 
The statement in (2) is a direct consequence of Lemma \ref{lem:Levin-Gu ground-state is SPT}.
Finally, (4) follows from (1) since for any translation $\tau$, we have that $\omega_0 \circ \tau(B_v) = 1$ for any $v \in \Gamma$.
\end{proof}

\subsection{Defect Hamiltonian}
\label{sec:Defect auts Hamiltonian Levin-Gu}
We construct a defect Hamiltonian by first constructing a defect automorphism that will give us a $g$-defect representation.
Recall the discussion in Section \ref{sec:paths and dual paths} about paths and dual paths, in particular the definition of $\bar P_R(\Gamma)$ and the definition of a dividing completion $L$ of a dual path $\gamma \in \bar P_R(\Gamma)$. 

We consider the defect automorphism $\alpha_\gamma^g$ from Definition \ref{def:defect automorphisms} with $g \in \bbZ_2$ being the non-trivial element. For the Levin--Gu SPT, it is possible to explicitly compute $\alpha_\gamma^g$. We do this computation in Appendix \ref{sec:LGDefectSectorComputationAppendix}.

\begin{defn}
    Let $\gamma \in \bar P_R( \Gamma)$ be a dual path with a dividing completion $L$.
    Let $\alpha^\gamma \coloneqq  \alpha_\gamma^g$ be the defect automorphism (Definition \ref{def:defect automorphisms}) for the Levin-Gu SPT as constructed in Appendix \ref{sec:LGDefectSectorComputationAppendix}. 
    For each $v \in \Gamma$, we define the operator $\widehat B_v^\gamma$ to be 
    \[
    \widehat B_v^\gamma
    \coloneqq
    (\alpha^\gamma)^{-1}(B_v).
    \]
    In cases where the dual path $\gamma$ is clear from context, we may simply write $\widehat B_v$ instead of $\widehat B_v^\gamma$.
\end{defn}

\begin{nota}
\label{nota:VerticesInDualPath}
Recall the terminology and conventions in Section \ref{sec:paths and dual paths}.
For a (half-infinite) dual path $\gamma$, we use $V(\gamma)$ to refer to the set of vertices $v \in \Gamma$ such that $v$ is contained in an edge comprising $\gamma$.
We use $\widehat{V}(\gamma)$ to refer to the subset of $V(\gamma)$ that excludes those vertices adjacent to the face $\partial_0 \gamma$.
\end{nota}

Specializing Lemma \ref{lem:Qv relations with defect aut} to the case of the Levin-Gu SPT, we have the following.
\begin{lem}
Let $\gamma \in \bar P_R(\Gamma)$.
If $v \notin V(\gamma)$, then $\widehat{B}_v^\gamma = B_v$.
If $v \in \widehat{V}(\gamma)$, then $\widehat{B}_v^\gamma = \beta_g^{r(L_\gamma)}(B_v)$, where $g \in \bbZ_2$ is the non-identity element.
\end{lem}

\begin{rem}
Let $\gamma \in \bar P_R(\Gamma)$.  
If $v \in \widehat{V}(\gamma)$, then $\widehat{B}_v^\gamma = \beta_g^{r(L)}(B_v)$ does not depend on the choice of dividing completion $L$.  
Indeed, if $L'$ is another dividing completion of $\gamma$, then $\beta_g^{r(L)}(B_v) = \beta_g^{r(L')}(B_v)$ since $r(L)$ and $r(L')$ only differ outside the support of $B_v$.
\end{rem}

\begin{defn}
\label{def:DefectHamiltonianLG}
For $\gamma \in \bar P(\Gamma)$, we define the \emph{defect Hamiltonian} $H^\gamma$ as follows: for $S \in \Gamma_f$, 
$$H_S^\gamma \coloneqq \sum_{v \in S} (\mathds1 - \widehat B_v^\gamma)/2 \in \cstar[S^{+1}].$$ 
We denote the corresponding derivation by $\tilde \delta^\gamma$. 
\end{defn}

Note that $H^\gamma$ is a commuting projector Hamiltonian for all $\gamma \in \bar P(\Gamma)$ since it is the image of the commuting projector Hamiltonian $H$ under $(\alpha^\gamma)^{-1}$.

We note that $H^\gamma$ is not invariant under the action of $\beta_g$. However we observe the following fact. Let $\Lambda \subseteq \Gamma$ be a cone such that $\gamma \subset \Lambda$, which is guaranteed to exist since $\gamma \in \bar P_R(\Gamma)$.\footnote{Since $\gamma \in \bar P_R(\Gamma)$, the cone $\Lambda$ is \emph{not} an allowed cone.}
Then for all $S \in \Gamma_f$ satisfying $S^{+1} \cap \Lambda = \emptyset$, we have that $H_S^\gamma = H_S$ and thus $\beta_g(H_S^\gamma) = H_S^\gamma$. 
Therefore, even though $\tilde \omega_\gamma$ is not a ground state of $\tilde \delta$, the state $\tilde \omega_\gamma$ still satisfies $\tilde \omega_\gamma(H_S) = \tilde \omega(H_S) = 0$ for all $S$ as above. In other words, $\tilde \omega_\gamma$ `looks like' $\tilde \omega$ outside of $\Lambda$.

\subsection{Category of $G$-defect representations}
\label{sec:Defect sectors in the Levin-Gu SPT}
Recall the definition of a $g$-defect representation (Definition \ref{def:g-defect_rep_and_sector}). We now set $\tilde \pi$, the GNS representation of the ground state $\tilde \omega$ of $\tilde \delta$ as our reference representation. 
Note that since $\tilde \omega$ is pure, $\tilde \pi$ is an irreducible representation. 
In addition, the assumptions in Section \ref{sec:GCrossedAssumptions} are satisfied by the fact that Levin--Gu SPT is a $\bbZ_2$-SPT (Lemma \ref{lem:Levin-Gu ground-state is SPT}).

Noting that $\alpha$ is a quasi-factorizable automorphism (Lemma \ref{lem:FDQCQuasi-Factorizable}) and applying Lemma \ref{lem:SPT has trivial superselection theory} gives us that $\tilde \pi$ has trivial superselection theory. We observe that by definition, $\tilde \pi$ is a $1$-defect representation. By specializing Lemmas \ref{lem:defect state GNS reps are sectorizable (SPT case)}, \ref{lem:classification of defect sectorizable representations SPT} to the case of Levin-Gu SPT we get the following proposition.

\begin{prop}
    The representation $\tilde \pi_\gamma \coloneqq \pi_0 \circ \alpha^\gamma$ is an irreducible $g$-defect representation for all $\gamma \in \bar P_R(\Gamma)$, and it corresponds to the unique irreducible $g$-defect sector with respect to $\tilde \pi$.
\end{prop}

Let us define $\GSec_{\tilde \pi}$ to be the category of $\bbZ_2$-defect representations with respect to the reference representation $\tilde \pi$ (in this case the Levin--Gu SPT ground state), and $\GSec_{\tilde \pi}^f$ the category of dualizable $G$-defects. From Proposition \ref{prop:general SPT defect sector classification}, we get that $\GSec^f_{\tilde \pi} \simeq \Vect(\bbZ_2,\nu)$ for some $3$-cocycle $\nu\colon \bbZ_2 \times \bbZ_2 \times \bbZ_2 \to U(1)$.  

We now want to compute cocycle, relying on Corollary \ref{cor:identifying the cocycle}. To do this, we compute the $F$-symbols in Appendix \ref{sec:LGDefectSectorComputationAppendix} using the explicit form of the $g$-defect automorphisms.
In Lemma \ref{lem:F-symbols are cocycles Levin Gu}, we obtain that $F(g,g,g)=-1$ while all other $F$-symbols are trivial, hence $\nu = F^*$ is the nontrivial $3$-cocycle on $\bbZ_2$.

We have shown the following. 
\begin{prop}
\label{prop:levin gu classification result}
The category $\GSec_{\tilde \pi}^f$ is $G$-crossed braided $\rmW^*$-tensor equivalent to  $\Vect(\bbZ_2, \nu)$, where $\nu$ is the nontrivial cocycle on $\bbZ_2$. 
\end{prop}

\section{A \texorpdfstring{$\bbZ_2$}{Z2}-Symmetry Enriched Toric Code}
\label{sec:SET Toric Code}
In this section, we will apply our general formalism to give a complete analysis of an infinite lattice model with $\mathbb{Z}_2$ on-site symmetry whose underlying topological order is that of the Toric Code. Our version of this model is closely related to the construction of \cite{PhysRevB.108.115144}.

Given the level of detail required in this type of analysis, we now give a brief outline of the following subsections. In Subsection \ref{sec:TC_review}, we quickly review the traditional Toric Code model.  This is followed by a presentation in Subsection \ref{sec:SET_model} of the $\mathbb{Z}_2$ symmetric Toric Code model and general analysis which shows that it has the same underlying topological order as the traditional Toric Code. We are then in a position to define the defects of this theory in Section \ref{sec:SET_defects} and provide a proof that they obey our selection criterion. This is followed by the calculation of the $F$-symbols, the symmetry fractionalization data, and the $G$-crossed braiding in sections \ref{sec:F_symbols}, \ref{sec:frac_data}, and \ref{sec:braid_data}.

\subsection{Review of Toric Code}
\label{sec:TC_review}


\subsubsection{Toric Code}

We begin by recapitulating the construction and properties of the Toric Code model \cite{MR1951039} which is the string net condensate generated by the unitary fusion category $\Vect(\mathbb{Z}_2)$ \cite{PhysRevB.71.045110, MR3204497, 10.22331/q-2024-03-28-1301}.
A thorough operator algebraic treatment can be found in \cite{MR2804555, MR2956822, MR3135456}.

The Toric Code is usually defined on the square lattice with a qubit on each edge, i.e., the sites $e\in\Gamma$ are represented by edges of the square lattice and $\hilb_e \simeq \bbC^2$.
We can thus define the quasi-local algebra $\cstar$.
As elements of $\mathbb{R}^2$, we take the edge sites to be located at the midpoint of each edge.
To be specific, we are taking the vertices of the square lattice to be $V=\mathbb{Z}^2\subset\mathbb{R}^2$, so $\Gamma=((\mathbb{Z}+1/2)\times \mathbb{Z})\cup(\mathbb{Z}\times (\mathbb{Z}+1/2))$.
We fix the ray $R\coloneqq \{1/4\}\times \mathbb{R}_{\geq 1/4}\subset\mathbb{R}^2$.

We denote the vertices, edges, and faces of the square lattice by $V,$ $E,$ and $F$, respectively.
Since the sites in $\Gamma$ are represented by edges in $E$, for $e \in E$, we denote the Pauli operators acting on the site at the midpoint of $e$ by $\{\mathds1_e, \sigma^x_e, \sigma^z_e, \sigma^y_e\}$.
For brevity, we will henceforth assume that $v\in V$ refers to a vertex and $f\in F$ to a face  whenever it is clear from the context.
We can define the \emph{star} operator $A_v$ and \emph{plaquette} operator $B_f$ as follows: $$A_v \coloneqq  \prod_{e \ni v} \sigma^x_e \qquad \qquad B_f \coloneqq  \prod_{e \in f} \sigma^z_e$$
where the first product is over all edges $e\in E$ whose boundary contains $v\in V$ and the second product is over all edges $e\in E$ contained in the boundary of $f\in F$.
As a matter of terminology, we refer to the above as $v$ \textit{neighboring} $e$ and $f$ neighboring $e$. 
It is easily checked for all $v\in V$ and $f \in F$ that $[A_v, B_f] = 0$.

Let $S\in\Gamma_f$ be finite, let $V(S)\subset V$ be the set of vertices whose neighboring edges are all in $S$, and let $F(S)\subset F$ be the set of faces whose neighboring edges are all in $S$.
Our Hamiltonian $H^{TC}$ is then given by $$H_S^{TC} \coloneqq  \sum_{v \in S} (\mathds1 - A_v)/2 + \sum_{f \in S} (\mathds1 - B_f)/2, \qquad\qquad S \in \Gamma_f,$$ 
which is a commuting projector Hamiltonian.

\begin{lem}[\cite{MR2345476}]
    \label{lem:UniqueFrustrationFreeGSToricCode}
    The Toric Code Hamiltonian $H^{TC}$ has a unique frustration-free ground state $\omega_0^{TC}$, which is pure.
    
\end{lem}

We define $\pi_0^{TC}$ to be the GNS representation of $\omega_0^{TC}$.
Note that $\pi_0^{TC}$ is irreducible since $\omega_0^{TC}$ is pure. 

\begin{lem}[{\cite{MR2956822}}]
\label{lem:TC Haag duality}
    The representation $\pi_0$ satisfies strict Haag duality. 
\end{lem}

We define our various types of paths and dual paths as in Section \ref{sec:paths and dual paths}.
The only difference is that we are now working on the square lattice and its dual which is also a square lattice.

Some important objects in the study of the Toric Code are the \emph{string operators}. There are 2 different types of string operators, $F^\epsilon_\gamma, F^m_{\bar\gamma}$. 

\begin{defn}
    Let $\gamma$ be a path and $\bar \gamma$ be a dual path. The \emph{string operators} are defined as $$F^\epsilon_\gamma \coloneqq  \prod_{e \in \gamma} \sigma^z_e \qquad \qquad F^m_{\bar \gamma} \coloneqq  \prod_{e \in \bar \gamma} \sigma^x_e.$$
\end{defn}

We have the relations $$F^\epsilon_\gamma F^m_{\bar \gamma} = (-1)^{c(\gamma, \bar \gamma)} F^m_{\bar \gamma} F^\epsilon_\gamma$$ where $c(\gamma, \bar \gamma)$ counts the number of shared edges between $\gamma, \bar \gamma$. Using these string operators, we define the automorphisms which create the anyon representations of the Toric Code.

\begin{defn}
\label{def:ToricCodeAnyonAutomorphisms}
    Let $\gamma\in P(\Gamma)$ and let $\gamma^i$ for $i\in\mathbb{N}$ be the path consisting of the first $i$ elements of $\gamma$.
    Define $\bar{\gamma}^i$ with respect to a half-infinite dual path $\bar{\gamma}\in \bar P(\Gamma)$ similarly.
    We may then define the automorphisms $\alpha_\gamma^\epsilon, \alpha^m_{\bar \gamma}$ for all $A \in \cstar$ as  $$\alpha_\gamma^\epsilon(A) \coloneqq  \lim_{\gamma^i \uparrow \gamma} F^\epsilon_{\gamma^i} A F^\epsilon_{\gamma^i} \qquad \qquad \alpha_{\bar \gamma}^m(A) \coloneqq  \lim_{\bar \gamma^i \uparrow \bar \gamma} F^m_{\bar \gamma^i} A F^m_{\bar \gamma^i}.$$ as the charge/flux automorphisms respectively.
    Define also the following automorphism $$\alpha^\psi_{\gamma, \bar \gamma}(A) \coloneqq  \alpha_\gamma^\epsilon \circ \alpha_{\bar \gamma}^m(A).$$
\end{defn}

The following result is due to \cite{MR2804555}.

\begin{lem}[{\cite[Thm.~3.1]{MR2804555}}]
\label{lem:TC states are independent of paths}
    Let $\gamma, \gamma'\in P(\Gamma)$ and $\bar \gamma, \bar \gamma'\in \bar P(\Gamma)$. We have, $$\omega_0^{TC} \circ \alpha_\gamma^\epsilon \simeq \omega_0^{TC} \circ \alpha_{\gamma'}^\epsilon \qquad \omega_0^{TC} \circ \alpha_{\bar \gamma}^m \simeq \omega_0^{TC} \circ \alpha_{\bar \gamma'}^m \qquad \omega_0^{TC} \circ \alpha_{\gamma, \bar \gamma}^\psi \simeq \omega_0^{TC} \circ \alpha_{\gamma', \bar \gamma'}^\epsilon$$
\end{lem}

To prove the above result, \cite{MR2804555} uses the following lemma (\ref{lem:Naa31}), which is of independent interest.

\begin{lem}[{\cite[Lem.~3.1]{MR2804555}}]\label{lem:Naa31}
    Let $\gamma, \gamma'\in P(\Gamma)$ and $\bar \gamma, \bar \gamma'\in \bar P(\Gamma)$ such that $\partial_0 \gamma = \partial_0 \gamma'$ and $\partial _0 \bar\gamma' = \partial_0 \bar \gamma'$. Then we have, $$\omega_0^{TC} \circ \alpha_\gamma^\epsilon = \omega_0^{TC} \circ \alpha_{\gamma'}^\epsilon \qquad \omega_0^{TC} \circ \alpha_{\bar \gamma}^m = \omega_0^{TC} \circ \alpha_{\bar \gamma'}^m \qquad \omega_0^{TC} \circ \alpha_{\gamma, \bar \gamma}^\psi = \omega_0^{TC} \circ \alpha_{\gamma', \bar \gamma'}^\psi$$
\end{lem}

\begin{lem}
    We have that $\pi_0^{TC} \circ \alpha_\gamma^\epsilon, \pi_0^{TC} \circ \alpha_{\bar \gamma}^m, \pi_0^{TC} \circ \alpha_{\gamma, \bar \gamma}^\psi$ are all localized in some cone and transportable for any chosen half-infinite paths/dual paths $\gamma/ \bar \gamma$ in $P(\Gamma)/\bar P(\Gamma)$. 
\end{lem}
\begin{proof}
    Straightforward from Lemma \ref{lem:TC states are independent of paths} and the definitions of the automorphisms.
\end{proof}

We now fix a path $\gamma_0\in P(\Gamma)$ and dual path $\bar \gamma_0\in \bar P(\Gamma)$ and define the representations $$\pi^\epsilon \coloneqq  \pi_0^{TC} \circ \alpha^\epsilon_{\gamma_0}, \qquad  \pi^m \coloneqq  \pi_0^{TC} \circ \alpha^m_{\bar \gamma_0}, \qquad \pi^\psi \coloneqq  \pi_0^{TC} \circ \alpha^\psi_{\gamma_0, \bar \gamma_0}$$

\begin{lem}[{\cite{MR2804555, MR3135456}}]
    \label{lem:ToricCodeAnyonSectors}
    The representations $\{\pi_0^{TC}, \pi^\epsilon, \pi^m, \pi^\psi\}$ are anyon representations, and these representations comprise the distinct irreducible anyon sectors with respect to $\pi_0$.
    
\end{lem}

Recall Definition \ref{def:allowed_cone} of the set of allowed cones $\cL$, and recall that the auxiliary algebra was defined to be 
\[
\fA^a
\coloneqq
\overline{\bigcup_{\Lambda \in \cL} \cR(\Lambda)}^{\| \cdot \|}
\subset
B(\cH_0).
\]

By \cite[Prop.~4.2]{MR2804555}, the maps $\pi_0^{TC} \circ \alpha_\gamma^\epsilon, \pi_0^{TC} \circ \alpha_{\bar \gamma}^m, \pi_0^{TC} \circ \alpha_{\gamma, \bar \gamma}^\psi$ all have a unique extension to $\cstar^a$ such that on any allowed cone $\Lambda$ the extension is weakly continuous. 
Furthermore, all these extensions are endomorphisms of $\cstar^a$.

\begin{defn}
    An endomorphism $\rho$ of $\cstar^a$ is \emph{localized} in cone $\Lambda$ if for all $A \in \cR(\Lambda^c)$ we have $\rho(A) = A$. We say $\rho$ is \emph{transportable} if for any allowed cone $\Lambda'$ there exists an endomorphism $\rho'$ of $\cstar^a$ localized in $\Lambda'$ and satisfying such that $\rho \simeq \rho'$.
    We denote by $\DHR_{\pi_0^{TC}}$ the category of localized transportable endomorphisms of $\cstar^a$ that are localized in some cone $\Lambda$, where the morphisms are intertwiners.
    We use the $\pi_0^{TC}$ subscript to indicate the dependence of this category on the reference representation $\pi_0^{TC}$.
\end{defn}

\begin{rem}
    In \cite{MR2804555, MR3135456}, these localized and transportable endomorphisms are extended to the auxiliary algebra $\cstar^a$.
    One can then show that $\DHR_{\pi_0^{TC}}$ is a braided monoidal category.
\end{rem}

\begin{thm}[{\cite[Thm.~6.2]{MR2804555}}]
    The category $\DHR_{\pi_0^{TC}}$ is a braided monoidally equivalent to $\Rep(D(\bbZ_2))$, where $D(\bbZ_2)$ is the quantum double of $\bbC[\bbZ_2]$ \cite[Def.~IX.4.1]{MR1321145}.
\end{thm}

Specifically, if the simple objects in $\Rep (D(\bbZ_2))$ are denoted $1, e, m , \psi$, then we can make the following identifications: 
$$\pi_0^{TC} \mapsto 1 \qquad \pi^\epsilon \mapsto e \qquad \pi_0^m \mapsto m \qquad \pi^\psi \mapsto \psi$$

\subsection{SET Toric Code model}
\label{sec:SET_model}

In order to define the SET Toric Code mode, we include vertex spins in addition to the edge spins from the traditional Toric Code.
As such, we take the sites of $\Gamma$ to be the vertices and edges of the square lattice and we denote the vertices $V$, edges $E$, and faces $F$ of the square lattice as before.
As elements of $\mathbb{R}^2$, we take the edge sites to be located at the midpoint of each edge.
We again take $R=\{1/4\}\times\mathbb{R}_{\geq 1/4}$. We use the same notation for paths and dual paths established in Section \ref{sec:paths and dual paths},  but now we consider these objects on the square lattice.
For this model, however, we will find it useful to assign an orientation to each edge.
Vertical edges are oriented from down to up and horizontal edges are oriented from left to right.
This model is based on the models described in \cite{PhysRevB.108.115144}. For $i\in\{x,y,z\}$, the Pauli operators on the vertex spins will be denoted as $\tau^i_v$ for $v\in V$ and we will continue to denote the $i$th Pauli operator on the edge $e\in E$ by $\sigma^i_e$.
Since these operators have disjoint support, they must commute. 
On each finite region $S \in \Gamma_f$ with vertices $V(S)\subset V$ defined as before, we have the Hilbert spaces $$\hilb^V_S \coloneqq  \bigotimes_{v \in V(S)} \hilb_v  \qquad\qquad \hilb^E_S \coloneqq  \bigotimes_{e \in S} \hilb_e \qquad\qquad \hilb_S \coloneqq  \hilb^V_S \otimes \hilb^E_S.$$
We also define the following algebras for $S \in \Gamma_f$: $$\cstar[S]^V \coloneqq  B(\hilb^V_S) \qquad\qquad \cstar[S]^E \coloneqq  B(\hilb_S^E) \qquad\qquad \cstar[S] \coloneqq  B(\hilb_S).$$

We can define the following quasi-local algebras in the usual way as $$\cstar^V \coloneqq  \overline{\bigcup_{S \in \Gamma_f} \cstar[S]^V }^{||\cdot||} \qquad \cstar^E \coloneqq  \overline{\bigcup_{S \in \Gamma_f} \cstar[S]^E }^{||\cdot||} \qquad \cstar \coloneqq  \overline{\bigcup_{S \in \Gamma_f} \cstar[S] }^{||\cdot||}.$$
We note that in the previous section our quasi-local operator algebra is what we are now calling $\cstar^E$. We also note that $\cstar = \cstar^E \otimes \cstar^V$.

On this new spin lattice, we are able to define the SET Toric Code Hamiltonian by finding an FDQC and applying it to a modified version of the traditional Toric Code Hamiltonian which accounts for the new vertex spins. 
For any edge $e$, we let $\partial_0 e\in V$ represent the source vertex and $\partial_1e\in V$ represent the target vertex with respect to the orientation of the edge.
On each edge $e\in E$, define 

$$W_e \coloneqq  i^{(1 - \sigma^x_e)(\tau^z_{\partial_1 e} - \tau^z_{\partial_0 e})/4} \in \cstar[e]\otimes \cstar[\partial_0 e] \otimes \cstar[\partial_1 e],$$ which gives rise to the maps $\alpha_S \colon \fA_S \to \fA_{S^{+1}_\Gamma}$ given by $$\alpha_S(A) = \left(\bigotimes_{e \in S^{+1}_\Gamma} W_e\right) A \left(\bigotimes_{e \in S^{+1}_\Gamma} W_e^*\right), 
\qquad\qquad
A \in \cstar[S].$$
Recall that $S^{+1}_\Gamma$ is the set of sites in $\Gamma$ that are distance at most one away from the sites in $S$.
These maps can be norm-continuously extended to obtain an automorphism $\alpha$ of $\cstar$. This automorphism is an FDQC by the following lemma.
\begin{lem}
\label{lem:SETToricCodeFDQC}
    The automorphism $\alpha$ is an FDQC of depth 4 with spread 1.
\end{lem}
\begin{proof}
    Each $W_e$ acts only on an edge and its bounding vertices. The 4-coloring of the edges of the square lattice immediately tells us how to construct our FDQC.
    The fact that $\alpha$ has spread 1 follows immediately from the definition. 
\end{proof}

\subsubsection{Hamiltonian}
We can now define the SET Toric Code Hamiltonian $H$ by
$$
H_S 
\coloneqq 
\sum_{v \in V(S)} \left(\dfrac{\mathds1 - A_v}{2} + \dfrac{\mathds1 - \tilde Q_v}{2}\right) + \sum_{f \in F(S)} \frac{\mathds1 - \tilde B_f}{2},
\qquad\qquad
S \in \Gamma_f,
$$
where we have denoted the vertices and faces in $S$ by $V(S)$ and $F(S)$, respectively, and we are defining 
$$
\tilde Q_v\coloneqq \dfrac{\mathds1+A_v}{2}\alpha(\tau_v^x)\quad\quad\text{and}\quad\quad \tilde{B}_f\coloneqq \alpha(B_f).
$$
We denote the corresponding derivation by $\tilde \delta$. We will first show that $H$ is a commuting projector Hamiltonian.
Then we show that there is a unique frustration-free ground state $\tilde \omega$ of $\tilde \delta$ by using our FDQC $\alpha$ to relate $\tilde{\omega}$ to the ground state of the traditional Toric Code. Finally, we will use this relationship to prove that the underlying braided fusion category of anyons in this theory is braided monoidally equivalent to that of the Toric Code.

\begin{lem}
\label{lem:ImagesOfTCExtensionHamiltonianTermsUnderEntangler}
    For $v \in V$ and $f \in F$,
    $$\alpha(A_v) = A_v \qquad\qquad \tilde{B}_f = i^{- \sum_{e \in f}\sigma^x_e(\tau^z_{\partial_1 e} - \tau^z_{\partial_0 e})/2} B_f
    \qquad\qquad \alpha(\tau^x_v) = \tau^x_v i^{-\tau^z_v \sum_{e \ni v} f(e,v) \sigma^x_e/2} \eqqcolon Q_v,$$
    where $f(e,v) = 1$ if $v = \partial_0 e$ and $f(e,v) = -1$ if $v = \partial_1 e$. 
\end{lem}

\begin{proof}
    
    \item[\underline{($\alpha(A_v) = A_v$):}] Since $W_e$ commutes with $A_v$ for all $e$, we straightforwardly have $\alpha(A_v) = A_v$.

    \item[\underline{($\tilde B_f$ formula):}] We have for each $e \in f$, 
    \begin{align*}
    W_e B_f W_e^* 
    &= 
    i^{(1 - \sigma^x_e)(\tau^z_{\partial_1 e} - \tau^z_{\partial_0 e})/4} B_f i^{-(1 -\sigma^x_e)(\tau^z_{\partial_1 e} - \tau^z_{\partial_0 e})/4}
    \\&= i^{(1 - \sigma^x_e)(\tau^z_{\partial_1 e} - \tau^z_{\partial_0 e})/4} i^{-(1 + \sigma^x_e)(\tau^z_{\partial_1 e} - \tau^z_{\partial_0 e})/4} B_f  
    \\&= i^{- \sigma^x_e(\tau^z_{\partial_1 e} - \tau^z_{\partial_0 e})/2} B_f
    \end{align*}
    which gives us the required result after taking the product over all edges $e \in f$.

    \item[\underline{($\alpha(\tau^x_v)$ formula):}] Consider $e_1, e_2 \ni v$ such that $\partial_1 e_1 = \partial_0 e_2 = v$. Then we have,
    \begin{align*}
        W_{e_1} W_{e_2} \tau^x_v W_{e_2}^* W_{e_1}^* &= \Ad[W_{e_1}] (i^{(1 - \sigma^x_{e_2})(\tau^z_{\partial_1 e_2} - \tau^z_{v})/4} \tau_v^x i^{-(1 - \sigma^x_{e_2})(\tau^z_{\partial_1 e_2} - \tau^z_{v})/4})\\
        &= \Ad[W_{e_1}] ( \tau_v^x i^{(1 - \sigma^x_{e_2})(\tau^z_{\partial_1 e_2} + \tau^z_{v})/4} i^{-(1 - \sigma^x_{e_2})(\tau^z_{\partial_1 e_2} - \tau^z_{v})/4})\\
        &= \Ad[W_{e_1}] ( \tau_v^x i^{(1 - \sigma^x_{e_2}) \tau^z_{v}/2})\\
        &= \tau_v^x i ^{-(1 - \sigma^x_{e_1})\tau^z_v/2} i^{(1 - \sigma^x_{e_2}) \tau^z_{v}/2}\\
        &= \tau_v^x i^{- \tau_v^z(f(e_1,v)\sigma^x_{e_1}+ f(e_2)\sigma^x_{e_2,v})/2}
    \end{align*}
    Performing this conjugation on $\tau_v^x$ for both pairs of neighboring edges, we obtain the desired result.
\end{proof}

We observe that Lemma \ref{lem:ImagesOfTCExtensionHamiltonianTermsUnderEntangler} implies that $H$ is a commuting projection Hamiltonian.
In the standard physics presentation, the Hamiltonian of an SET should be symmetric. Our analysis only directly requires that the ground state be symmetric. Nevertheless, it will be useful to prove the following lemma which implies the symmetry of our Hamiltonian.

\begin{defn}
The global symmetry action $\beta_g$ for the SET Toric Code is given by the formula in Definition \ref{def:GlobalSymmetryAutomorphism} with $U_v^g = \tau^x_v$ and $U^e_g = \mathds1_v$, i.e., for $S \in \Gamma_f$ and $A \in \cstar[S]$, 
\[
\beta_g(A)
=
\left(\bigotimes_{v \in V(S)} \tau^x_v\right) A \left(\bigotimes_{v \in V(S)} \tau^x_v\right).
\]
This formula can be norm-continoulsy extended to obtain an automomorphism of $\fA$.
\end{defn}

\begin{lem}
\label{lem:SET Hamiltonian terms are symmetric}
    The operators $A_v, \tilde B_f, \tilde Q_{v}$ are invariant under the action of the symmetry $\beta_g$.
\end{lem}
\begin{proof}
    Using the definition of $A_v$, it is straightforward to verify that $\beta_g(A_v) = A_v$.

    We now show that $\tilde B_v$ is symmetric under $\beta_g$.
    \begin{align*}
        \beta_g(\tilde B_f) &= \left(\prod_{v \in f} \tau_v^x\right)\! i^{- \sum_{e \in f}\sigma^x_e(\tau^z_{\partial_1 e} - \tau^z_{\partial_0 e})/2} B_f \!\left(\prod_{v \in f} \tau_v^x\right)\\ &= i^{ \sum_{e \in f}\sigma^x_e(\tau^z_{\partial_1 e} - \tau^z_{\partial_0 e})/2} B_f\\
        &= i^{ \sum_{e \in f}\sigma^x_e(\tau^z_{\partial_1 e} - \tau^z_{\partial_0 e})} \tilde B_f.
        \intertext{We now use the fact that $(\tau^z_{\partial_1 e} - \tau^z_{\partial_0 e})$ always has eigenvalues $\pm 2,0$ and $\sigma^x_e$ has eigenvalues $\pm 1$, to observe that $i^{\sigma^x_e(\tau^z_{\partial_1 e} - \tau^z_{\partial_0 e})}$ has exactly the same spectral decomposition as $i^{(\tau^z_{\partial_1 e} - \tau^z_{\partial_0 e})}$. Therefore, using the fact that $i^{\pm\tau^z}=\pm i\tau^z$,}
        \beta_g(\tilde B_f)&= i^{\sum_{e \in f}(\tau^z_{\partial_1 e} - \tau^z_{\partial_0 e})} \tilde B_f = i^4(-i)^4\prod_{v \in f}(\tau^z_v)^2 \tilde B_f = \tilde B_f.
    \end{align*}
    
    Now we turn to $\tilde Q_v$. Consider the following calculation.
    \begin{align*}
        \beta_g(Q_v)&=\tau^x_v \tau^x_v i^{-\tau^z_v \sum_{e \ni v} f(e,v) \sigma^x_e/2} \tau^x_v = \left(i^{-\tau^z_v \sum_{e \ni v} f(e,v) \sigma^x_e}\right)\tau^x_v \left(i^{-\tau^z_v \sum_{e \ni v} f(e,v) \sigma^x_e/2}\right)\\
        &=  i^{\tau^z_v \sum_{e \ni v} f(e,v) \sigma^x_e} Q_v=  A_vQ_v
    \end{align*}
    where in the last step we have used that $\tau_v^z\sum_{e \ni v} f(e,v) \sigma^x_e \in \{\pm 4, 0\}$ on states where $A_v=1$ and $\tau^z_v\sum_{e \ni v} f(e,v) \sigma^x_e \in \{\pm 2\}$ on states where $A_v=-1$. Therefore, 
    \[
    \beta(\tilde{Q}_v)=\beta\!\left(\dfrac{\mathds1+A_v}{2}Q_v\right)=\dfrac{\mathds1+A_v}{2}A_vQ_v=\tilde{Q}_v.
    \qedhere
    \]
\end{proof}

By Lemma \ref{lem:SET Hamiltonian terms are symmetric} it follows that the Hamiltonian $H$ is invariant under the $\bbZ_2$ symmetry.

\subsubsection{Relation to Toric Code}
We now define an augmented version of the traditional Toric Code Hamiltonian $H^0$ as follows: for $S \in \Gamma_f$, we have that $$H_S^0 \coloneqq  H_S^{TC} + \sum_{v \in V(S)} \dfrac{\mathds1 - \tau_v^x}{2}, $$
where $H_S^{TC} \in \cstar^E$ is the Toric Code Hamiltonian on $S$.
We have simply added ancilla spins on the vertices and energetically enforced them to be in a product state.
Let $\delta^0$ be the derivation corresponding to this new Hamiltonian. It is easy to see that $H^0$ is still a commuting projector Hamiltonian. Let $\omega_0$ be the state on $\cstar$ defined by $$\omega_0 \coloneqq  \omega^E_{TC} \otimes \omega^V_0$$ where $\omega^E_{TC}$ (defined on $\cstar^E$) is the Toric Code frustration-free ground state and $\omega^V_0$ is defined on $\cstar^V$ as a product state given by $\omega^V_0(A) \coloneqq  \inner{\bigotimes_{v \in \Gamma}\psi_v}{A \bigotimes_{v \in \Gamma}\psi_v}$ and $\ket{\psi_v} \in \hilb_v$ satisfies $\ket{\psi_v} = \tau_v^x \ket{\psi_v}$.

Then it is easy to see that $\omega_0 $ is a frustration-free ground state of $H^0$. We now list some useful facts about $\omega_0$, which have been shown in Appendix \ref{app:TC with ancillary vertex spins}.

\begin{prop}
    
    The state $\omega_0 = \omega^E_{TC}\otimes \omega_0^V$ is the unique frustration free ground state of $\delta^0$ and is pure.
    
\end{prop}
This Proposition is proved in Lemma \ref{lem:TC extension FF GS is unique}.

We now define $\pi_0$ as the GNS representation of $\omega_0$. 

\begin{prop}
\label{prop:StackedTCBasicInfo}
    The representation $\pi_0$ satisfies the following:
    \begin{enumerate}
        \item $\pi_0$ is irreducible.
        \item $\pi_0$ satisfies Haag Duality.
        \item The representations $\{\pi_0 \circ\zeta\}_\zeta$ comprise the distinct irreducible anyon sectors with respect to $\pi_0$, where $\zeta \in \{\Id, \alpha_\gamma^\epsilon, \alpha_{\bar \gamma}^m, \alpha_{\gamma, \bar \gamma}^\psi\}$ for fixed half-infinite $\gamma \in P(\Gamma), \bar \gamma \in \bar P(\Gamma)$.
    \end{enumerate}
\end{prop}
The fact that $\pi_0$ is irreducible follows from $\omega_0$ being pure. The rest of this proposition is proved in parts in Lemmas \ref{lem:TC extension Haag duality}, \ref{lem:TC extension anyon sectors bound}, \ref{lem:TC extension anyon sectors constructed} and Corollary \ref{cor:TC extension anyon sector classification}.\footnote{The recent result due to \cite{2511.08382} provides a shorter proof of Proposition \ref{prop:StackedTCBasicInfo}.}

We now define the category $\DHR_{\pi_0}$ as the braided $C^*$ tensor category of endomorphisms of $\cstar^a$ that are localized in some cone $\Lambda \in \cL$ (with respect to $\pi_0$) and transportable. 
Note that we have $\DHR_{\pi_0} \simeq \DHR_{\pi_0^{TC}}$.

\subsubsection{Ground state}
We now investigate the frustration-free ground state of $\tilde \delta$, which we show is the state $\tilde \omega \coloneqq  \omega_0 \circ \alpha^{-1}$.
\begin{lem}
\label{lem:SET toric code unique FF GS}
    The state $\tilde \omega$ is the unique state satisfying for all $v\in V,f \in F$ 
    \begin{equation}
    \label{eq:SETTCFrustrationFreeGroundState-Condition1}
    \tilde \omega(A_v) = \tilde \omega(\tilde B_f) = \tilde \omega(Q_v) = 1.
    \end{equation}
    Additionally, $\tilde \omega$ is the unique state satisfying for all $v\in V$ and $f \in F$,
    \begin{equation}
    \label{eq:SETTCFrustrationFreeGroundState-Condition2}
    \tilde \omega(A_v) = \tilde \omega(\tilde B_f) = \tilde \omega(\tilde Q_v) = 1.
    \end{equation}
    In particular, $\tilde \omega$ is the unique frustration-free ground state of $\tilde \delta$.
    Moreover, this state is symmetric under $\beta_g$, as are the dynamics generated by $\tilde \delta$.
\end{lem}
\begin{proof}
We first show that $\tilde \omega = \omega_0 \circ \alpha^{-1}$ satisfies \eqref{eq:SETTCFrustrationFreeGroundState-Condition1}.
Indeed, by Lemma \ref{lem:ImagesOfTCExtensionHamiltonianTermsUnderEntangler}, we have for every $v\in V,f\in F$ that
\begin{gather*}
\tilde \omega(A_v)
=
\omega_0 \circ \alpha^{-1} \circ \alpha(A_v)
=
\omega_0(A_v)
=
1,
\\
\tilde \omega(\tilde B_f)
=
\omega_0 \circ \alpha^{-1} \circ \alpha(B_f)
=
\omega_0(B_f)
=
1,
\\
\tilde \omega(Q_v)
=
\omega_0 \circ \alpha^{-1} \circ \alpha(\tau^x_v)
=
\omega_0(\tau^x_v)
=
1.
\end{gather*}
Now, suppose $\omega \in \cS(\cstar)$ is another state satsifying \eqref{eq:SETTCFrustrationFreeGroundState-Condition1}.
Then by Lemma \ref{lem:ImagesOfTCExtensionHamiltonianTermsUnderEntangler}, we have that 
\[
\omega \circ \alpha(A_v)
=
\omega \circ \alpha(B_f)
=
\omega \circ \alpha(\tau^x_v)
=
1.
\]
Therefore, by Lemma \ref{lem:TC extension FF GS is unique}, we have that $\omega \circ \alpha = \omega_0$, from which it follows that $\omega = \tilde \omega$.

We now show that $\tilde \omega$ is the unique state satisfying \eqref{eq:SETTCFrustrationFreeGroundState-Condition2}.
It suffices to show that a state $\omega$ satisfies \eqref{eq:SETTCFrustrationFreeGroundState-Condition1} if and only if it satisfies \eqref{eq:SETTCFrustrationFreeGroundState-Condition2}.
First, suppose that $\omega$ satisfies \eqref{eq:SETTCFrustrationFreeGroundState-Condition2}.
Then by Lemma \ref{lem:can freely insert and remove P from the ground state.}, we have that 
\[
\omega(Q_v)
=
\omega\!\left(\frac{\mathds1 + A_v}{2}Q_v \right)
=
\omega(\tilde Q_v)
=
1,
\]
so $\omega$ satisfies \eqref{eq:SETTCFrustrationFreeGroundState-Condition1}.
Now, suppose $\omega$ satisfies \eqref{eq:SETTCFrustrationFreeGroundState-Condition1}.
We then have by Lemma \ref{lem:can freely insert and remove P from the ground state.} that 
\[
\omega(\tilde Q_v)
=
\omega\!\left(\frac{\mathds1 + A_v}{2}Q_v \right)
=
\omega(Q_v)
=
1,
\]
so $\omega$ satisfies \eqref{eq:SETTCFrustrationFreeGroundState-Condition2}.
It follows that $\tilde \omega$ is the unique frustration-free ground state of $\tilde \delta$ by \cite[Lem.~3.8]{MR3764565}.

Also note that since $A_v,\tilde B_v,\tilde Q_v$ are symmetric by Lemma \ref{lem:SET Hamiltonian terms are symmetric}, $\tilde \omega\circ\beta_g$ satisfies \eqref{eq:SETTCFrustrationFreeGroundState-Condition2}.
By uniqueness, $\tilde \omega=\tilde\omega\circ\beta_g$, so $\tilde\omega$ is symmetric.
Lemma \ref{lem:SET Hamiltonian terms are symmetric} also directly implies that the dynamics generated by $\tilde\delta$ are symmetric.
\end{proof}

\begin{rem}
    The Hamiltonian $H$ was chosen specifically to be symmetric under the action of $\beta_g$, which we elaborate on below. However, it is also natural to consider the Hamiltonian $H' \coloneqq  \alpha(H^0)$ instead. Notice that $\tilde \omega$ is the unique frustration-free ground state for both Hamiltonians, which follows from Lemma \ref{lem:SET toric code unique FF GS} and \cite[Lem.~3.8]{MR3764565}. This of course means that $\tilde \omega$ is $\beta_g$ invariant, since it is a ground state of $H$, and can be obtained using a FDQC. These are the only properties required to completely determine the $g$-defect representations with respect to $\tilde \omega$, and the choice of the dynamics is irrelevant to our story.
\end{rem}

Given that $\tilde \omega = \omega_0 \circ \alpha^{-1}$, we now let $\tilde \pi \coloneqq  \pi_0 \circ \alpha^{-1}$ be the GNS representation of $\tilde \omega$, where $\pi_0$ is the GNS representation of $\omega_0$, the frustration-free ground state of $\delta^0$.

\subsubsection{Anyon sectors}
We now begin our analysis of the anyon sectors of the SET Toric Code.

We may easily obtain the new string operators by applying the entangling automorphism $\alpha$ to the old string operators. 
Recall the definition of string operators on the original Toric Code. For a path $\gamma$ and dual path $\bar \gamma$, we have $$F^\epsilon_\gamma \coloneqq  \prod_{e \in \gamma} \sigma^z_e \qquad \qquad F^m_{\bar \gamma} \coloneqq  \prod_{e \in \bar \gamma} \sigma^x_e.$$
We now define the entangled string operators as $$\tilde F^\epsilon_\gamma \coloneqq  \alpha(F^\epsilon_\gamma) = F^\epsilon_\gamma \prod_{e \in \gamma} i^{\sigma^x_e(\tau^z_{\partial_1 e} - \tau_{\partial_0 e}^z)/2} \qquad \tilde F^m_{\bar \gamma} \coloneqq  \alpha(F^m_{\bar \gamma}) = F^m_{\bar \gamma}$$ 
The string operators still satisfy the identities of the Toric Code string operators, $$\tilde F^\epsilon_\gamma \tilde F^m_{\bar \gamma} = (-1)^{c(\gamma, \bar \gamma)} \tilde F^m_{\bar \gamma} \tilde F^\epsilon_\gamma$$ where $c(\gamma, \bar \gamma)$ counts the number of shared edges between $\gamma, \bar \gamma$.

\begin{lem}
\label{lem:action of symmetry on string ops}
    We have the following identities: $$\beta_g(\tilde F^\epsilon_\gamma) = \tau^z_{\partial_1 \gamma} \tau^z_{\partial_0 \gamma} \tilde F^\epsilon_\gamma \qquad \qquad \beta_g(\tilde F^m_{\bar \gamma}) = \tilde F^m_{\bar \gamma}$$
\end{lem}
\begin{proof}
    The second identity is trivial. We prove the first identity. 
    \begin{align*}
        \beta_g(\tilde F^\epsilon_\gamma) &= \left( \prod_{v \in \gamma} \tau_v^x \right) F^\epsilon_\gamma \prod_{e \in \gamma} i^{\sigma^x_e(\tau^z_{\partial_1 e} - \tau^z_{\partial_0 e})/2} \left( \prod_{v \in \gamma} \tau_v^x \right)\\
        &= F^\epsilon_\gamma \prod_{e \in \gamma} i^{-\sigma^x_e(\tau^z_{\partial_1 e} - \tau^z_{\partial_0 e})/2}= \tilde F^\epsilon_\gamma \prod_{e \in \gamma} i^{-\sigma^x_e(\tau^z_{\partial_1 e} - \tau^z_{\partial_0 e})}\\
        \intertext{We now use the fact that $\sigma_e^x(\tau^z_{\partial_1 e} - \tau^z_{\partial_0 e})$ always has eigenvalues $\pm 2,0$ to observe that $i^{-\sigma^x_e(\tau^z_{\partial_1 e} - \tau^z_{\partial_0 e})}$ has the same spectral decomposition as $i^{(\tau^z_{\partial_1 e} - \tau^z_{\partial_0 e})}$. We then obtain}
        \beta_g(\tilde F^\epsilon_\gamma)&= \tilde F^\epsilon_\gamma \prod_{e \in \gamma} i^{(\tau^z_{\partial_1 e} - \tau^z_{\partial_0 e})}= \tilde F^\epsilon_\gamma i^{(\tau^z_{\partial_1 \gamma} - \tau^z_{\partial_0 \gamma})} = \tilde F^\epsilon_\gamma i{\tau^z_{\partial_1 \gamma} i(- \tau^z_{\partial_0 \gamma})} = \tilde F^\epsilon_\gamma \tau^z_{\partial_1 \gamma} \tau^z_{\partial_0 \gamma}.
    \end{align*}
    In the second to last equality we have used the fact that $i^{\pm\tau^z_v}=\pm i\tau^z_v$.
\end{proof}

We obtain new automorphisms for the anyon representations as follows. 
Let $\gamma \in P(\Gamma)$ and $\bar \gamma \in \bar P(\Gamma)$.
Taking $\zeta_\gamma \in \{\alpha_\gamma^\epsilon, \alpha_{\bar \gamma}^m, \alpha_{\gamma, \bar \gamma}^\psi\}$, the new automorphism $\tilde \zeta_\gamma$ is given by $$\tilde \zeta_\gamma \coloneqq  \alpha \circ \zeta_\gamma \circ \alpha^{-1}.$$
Here we note the dependence of $\zeta$ on $\gamma$ since the paths $\gamma, \bar \gamma$ are not fixed in this instance. 
Often, we will be considering fixed paths $\gamma, \bar\gamma$, and in that case we will drop the subscript $\gamma$ on $\zeta$.

Define the representations $\tilde \pi_{\zeta_\gamma} \coloneqq \pi_{\zeta_\gamma} \circ \alpha^{-1} = \tilde \pi \circ \tilde \zeta_\gamma$, which are irreducible since $\zeta_\gamma \circ \alpha^{-1}$ is an automorphism and $\pi_0$ is irreducible.

\begin{rem}\label{rem:SETTCAutomorphismDefn}
It is easy to verify that for $A \in \cstar$,
\[
\tilde \alpha^\epsilon_\gamma(A)
=
\lim_{n \to \infty}
\Ad(\tilde F^\epsilon_{\gamma^n})(A),
\qquad
\tilde \alpha^m_{\bar\gamma}(A)
=
\lim_{n \to \infty}
\Ad(\tilde F^m_{\bar\gamma^n})(A),
\qquad
\tilde \alpha^\psi_{\gamma, \bar\gamma}(A)
=
\lim_{n \to \infty}
\Ad(\tilde F^\epsilon_{\gamma^n}\tilde F^m_{\bar\gamma^n})(A),
\]
where the superscript $n$ denotes truncation after the first $n$ edges.
\end{rem}

\begin{lem}\label{lem:TCSymmetry_permute}
    The action of the symmetry does not permute anyon types, i.e, $\gamma_g(\tilde \pi_{\zeta_\gamma}) \simeq \tilde \pi_{\zeta_\gamma}$.
\end{lem}

\begin{proof}
    We have for any observable $A \in \cstar^{\loc}$,
    \[
        \gamma_g(\tilde \pi_{\zeta_\gamma})(A) = \beta_g \circ (\tilde \pi \circ \tilde\zeta_\gamma) \circ \beta_g(A) = \tilde \pi \circ \Ad[\beta_g(\tilde F_{\gamma'})](A),
    \]
    where $\gamma'$ is a truncation of $\gamma$ such that $\supp(A) \cap (\gamma - \gamma') = \emptyset$.
    Here, $\tilde F_{\gamma'}$ is either $\tilde F^\epsilon_{\gamma'}$, $\tilde F^m_{\bar\gamma'}$, or $\tilde F^\epsilon_{\gamma'} \tilde F^m_{\bar\gamma'}$ depending on which automorphism $\zeta_\gamma$ is.
    Now, there exists $U \in \cstar^{\loc}$ such that $\Ad[\beta_g(\tilde F_{\gamma'})](A) = \Ad[U\tilde F_{\gamma'}](A)$ for any such truncation $\gamma'$ of $\gamma$. 
    By continuity, we get that $\gamma_g(\tilde \pi_{\zeta_\gamma})(A) = \Ad[\tilde \pi(U)] \circ \tilde \pi_{\zeta_\gamma}(A)$ for all $A \in \cstar$.
\end{proof}

For the remainder of this subsection, we fix paths $\gamma, \bar \gamma$, so we drop subscripts on $\zeta$. 

\begin{lem}
\label{lem:anyon sectors of GS SET}
    The representations given by $\tilde \pi_\zeta \circ \alpha^{-1}$ for $\zeta \in \{\Id, \alpha^\epsilon, \alpha^m, \alpha^\psi\}$ comprise the distinct irreducible anyon sectors with respect to $\tilde \pi$.
    
\end{lem}
\begin{proof}
    Note that $\alpha$ is an FDQC. The result follows from Corollary \ref{cor:TC extension anyon sector classification}, Lemma \ref{lem:FDQCQuasi-Factorizable}, and \cite[Thm.~4.7]{MR4426734}.
\end{proof}

In fact, we can apply a theorem from \cite{MR4362722} to obtain a stronger result. 

\begin{prop}\label{prop:equiv_to_TC}
    The category $\DHR_{\pi_0}$ is braided monoidally equivalent to $\DHR_{\tilde \pi}$. In particular, $\DHR_{\tilde \pi}$ is braided monoidally equivalent to $\DHR_{\pi_0^{TC}}$. 
    
\end{prop}
\begin{proof}
    This follows from noting that $\alpha$ is an FDQC and then applying Proposition \ref{prop:TC extension braided monoidal}, Lemma \ref{lem:FDQCQuasi-Factorizable}, and \cite[Thm.~6.1]{MR4362722}. 
\end{proof}

\subsection{Symmetry defects}
\label{sec:SET_defects}
Recall that $ A_v, \tilde B_f, \tilde Q_v$, the terms of the SET Toric Code Hamiltonian, are invariant under the action of the symmetry (Lemma \ref{lem:SET Hamiltonian terms are symmetric}).
Our defect construction strategy will be similar to sections \ref{sec:Defect auts Hamiltonian Levin-Gu} and \ref{sec:defects using auts}. 
The idea is to observe the action of $\beta_g^{\bar L}$ along some dividing dual path $\bar{L}$ on the terms of the SET Toric Code Hamiltonian. 
We then erase the action of the symmetry along some $\bar{\gamma} \subset \bar{L}$ using an explicit automorphism. 
However, we do not directly use these results concerning the symmetry action when showing that the representations we define are $G$-defect representations. 
We therefore relegate this discussion to Appendix \ref{sec:SETToricCodeDefectHamiltonian}.

Let $\bar \gamma$ be a self-avoiding dual path, and consider the curve $\bar \gamma_{dual} \subset \bbR^2$ (see Section \ref{sec:paths and dual paths}). 
We use the ordering of the edges $\bar \gamma_n$ to define an orientation of the curve $\bar \gamma_{dual}$.
We then define the operator 
$$F_{\bar \gamma}^\sigma \coloneqq  \prod_{e \in \bar \gamma} e^{-i\pi p(e) \sigma^x_e/4}$$ where $p(e) = +1$ if $\partial_1 e$ is to the right of $\bar \gamma_{dual}$ and $p(e) = -1$ otherwise. Note that `right' and `left' are considered with respect to the orientation of $\overline{\gamma}_{dual}$. 
\begin{center}
    \begin{tikzpicture}
        \draw[thick,black,->](0,0)--(0,2);
        \draw[thick,black,->](4,1)--(6,1);
        \node at (-.8,1){Left};
        \node at (.8,1){Right};
        \node at (5,1.5){Left};
        \node at (5,.5){Right};
    \end{tikzpicture}
\end{center}

Now, let $\bar \gamma$ be a half-infinite self-avoiding dual path.
Let $A \in \cstar$ and consider a sequence of dual finite paths $\{\bar \gamma^n\}_{n \in \bbN}$ such that $\bar\gamma^n \subset \bar\gamma^{n+1}$ and $\partial_0\bar \gamma^n$ is constant for all $n$. Define the automorphism $\alpha^\sigma_{\bar \gamma}$ as $$\alpha^\sigma_{\bar \gamma} \coloneqq  \lim_{\bar \gamma^n \uparrow \bar \gamma} F_{\bar \gamma^n}^\sigma A (F_{\bar \gamma^n}^\sigma)^*.$$

Let $\overline{\xi}^1, \overline{\xi}^2$ be two self-avoiding dual paths such that $\partial_i \overline{\xi}^1 = \partial_i\overline{\xi}^2$ for $i = 0, 1$.
Then the loop $\overline{\xi}^1_{dual} \cup \overline{\xi}^2_{dual} \subset \bbR^2$ encloses a finite set of vertices $S(\overline{\xi}_1, \overline{\xi}_2)\subset V$. 
We can define the unitary $$\tilde F^\sigma_{\overline{\xi}_1, \overline{\xi}_2} \coloneqq  F^\sigma_{\overline{\xi}_1} (F^\sigma_{\overline{\xi}_2})^* \left(\bigotimes_{v \in S(\overline{\xi}_1, \overline{\xi}_2)} \tau^x_v \right).$$ 
\begin{lem}\label{lem:boundary_string}
    Let $\tilde \gamma$ be a self-avoiding dual loop and take $S\subset V$ to be the set of vertices enclosed by $\tilde \gamma_{dual}$. Then
    $$
    P_S\left(\prod\limits_{v\in S} Q_v\right)=P_S F^\sigma_{\tilde{\gamma}}\left(\prod\limits_{v\in S}\tau^x_v\right),
    $$
    where $P_S$ is the projection onto the $A_v=1$ subspace for all $v\in S$.
\end{lem}
\begin{proof}
    We do this analysis in the case where $S$ is a single vertex spin $v$. The general case follows inductively by gluing smaller regions together and seeing that paths in opposite directions cancel.

    Starting with the edge directly above $v$ and going around clockwise, we label the edges neighboring $v$ as $1,2,3,4$. Then,
    $$
        \dfrac{\mathds1+A_v}{2}Q_v=\dfrac{\mathds1+A_v}{2}\tau^x_v i^{-\tau_v^z(\sigma^x_1+\sigma_2^x-\sigma^x_3-\sigma^x_4)/2}=\dfrac{\mathds1+A_v}{2} i^{\tau_v^z(\sigma^x_1+\sigma_2^x-\sigma^x_3-\sigma^x_4)/2}\tau^x_v\\
    $$
    Note that the spectrum of the exponent is always even, so we may remove $\tau^z_v$ from this expression. So we have that
    \[
    \dfrac{\mathds1+A_v}{2}Q_v=\dfrac{\mathds1+A_v}{2}e^{i\frac{\pi}{4}\sigma^x_1}e^{i\frac{\pi}{4}\sigma^x_2}e^{-i\frac{\pi}{4}\sigma^x_3}e^{-i\frac{\pi}{4}\sigma^x_4}\tau_v^x=\dfrac{\mathds1+A_v}{2}F^\sigma_{1\rightarrow 2\rightarrow 3\rightarrow 4\rightarrow 1}\tau_v^x.
    \qedhere
    \]
\end{proof}

\begin{lem}
\label{lem:the pizza operators eval to 1 SET}
    Let $\overline{\xi}^1, \overline{\xi}^2$ be two self-avoiding dual paths such that $\partial_i \overline{\xi}^1 = \partial_i\overline{\xi}^2$ for $i = 0, 1$.
    Then
    $$\tilde \omega(\tilde F^\sigma_{\overline{\xi}^1, \overline{\xi}^2}) = 1.$$ 
\end{lem}
\begin{proof}

We prove this in the case where the region enclosed by $\overline\xi^1_{dual}\cup \overline\xi^2_{dual}$ is simply connected. Call the set of vertices in that simply connected region $S\subset V$. The non-simply connected case follows inductively. 

Let $\tilde{\gamma}$ be a self-avoiding dual path such that $\tilde{\gamma}_{dual}$ encloses $S$ and $\tilde{\gamma}$ runs parallel to $\overline{\xi}^1$ and anti-parallel to $\overline{\xi}^2$. If an edge $e$ is traversed in the same direction by the dual paths $\overline{\xi}^1$ and $\overline{\xi}^2$, so that $p(e)$ has the same sign with respect to $\overline{\xi}^1$ and $\overline{\xi}^2$, then the operator $\tilde F^\sigma_{\overline{\xi}^1, \overline{\xi}^2}$ acts trivially on that edge. Therefore, we may write
$$
\tilde F^\sigma_{\overline{\xi}^1, \overline{\xi}^2}=F^\sigma_{\tilde{\gamma}}\prod\limits_{v\in V(S)}\tau_v^x.
$$
We may then use Lemma \ref{lem:boundary_string} to see that
$$
\tilde \omega(\tilde F^\sigma_{\overline{\xi}^1, \overline{\xi}^2}) =\tilde\omega\left(P_S F^\sigma_{\tilde{\gamma}}\prod\limits_{v\in V(S)}\tau_v^x\right)=\tilde\omega\left(P_S\prod\limits_{v\in V(S)}Q_v\right)=1
$$
where $P_S$ is the projection to the $A_v=1$ subspace for each $v\in S$ and Lemma \ref{lem:can freely insert and remove P from the ground state.} is used to obtain the first and last equalities.
\end{proof}

In what follows, we let $\tilde \cH$ denote the GNS Hilbert space corresponding to $\tilde \pi$ and let $\tilde \Omega$ denote a unit cyclic vector for $\tilde \pi$ representing $\tilde \omega$.

\begin{lem}
\label{lem:defects are cone transportable}
    Pick $\bar\gamma \in \bar P_R(\Gamma)$ and suppose that there exist two completions $\bar L^1, \bar L^2$ of $\bar \gamma$. 
    For $i = 1, 2$, let $\bar \eta^i \in \bar P(\Gamma)$ be the half-infinite path given by $\bar \eta^i_k \coloneqq \bar L^i_{1 - k}$.
    Then $$\tilde \pi\circ \alpha_{\bar \eta^1}^\sigma \circ \beta^{r(\bar L^1)}_g \simeq \tilde \pi \circ \alpha_{\bar \eta^2}^\sigma \circ \beta^{r(\bar L^2)}_g.$$
    
    In fact, there is a unique unitary $V \in B(\tilde \cH)$ witnessing the above unitary equivalence such that $V\tilde \Omega = \tilde \Omega$.  
    This unitary is the WOT-limit of the sequence $V_n \coloneqq \tilde \pi(\tilde F_{(\bar \eta^2)^n, \overline{\xi}^n}^\sigma)$. 
    Here, for $i = 1, 2$, $(\bar\eta^i)^n$ is the truncation of $\bar \eta^i$ after the first $n$ edges. 
    Furthermore, $\overline{\xi}^n$ is a self-avoiding dual path defined by $\overline{\xi}^n \coloneqq \bar \eta^1 + \bar \varsigma^n$, where $\bar \varsigma^n$ is a self-avoiding dual path satisfying $\partial_0 \bar \varsigma^n = \partial_1 (\bar \eta^1)^n$ and $\partial_1 \bar \varsigma^n = \partial_1 (\bar \eta^2)^n$, and $d(\bar \varsigma^n_{dual}, \partial \bar \gamma) \to \infty$ as $n \to \infty$ (Figure \ref{fig:pizza}). 
    \begin{figure}[!ht]
\begin{center}
\begin{tikzpicture}
    \draw[thick,black](0,0)--(0,1)--(1,2);
    \draw[thick,blue](2,0)--(2,1)--(1,2);
    \draw[thick,red](1,2)--(1,3);
    \draw[thick,purple,snake it](0,.5)--(2,.5);
    \node at (1,0){$\bar \varsigma^n$};
    \node at (0,-.2){$\overline{\eta}^1$};
    \node at (2,-.2){$\overline{\eta}^2$};
    \node at (1.2, 3) {$\overline{\gamma}$};
    \draw[thick,orange,dashed,<-](2,.65)--(.1,.65)--(.1,.92)--(1,1.8);
    \node at (.8,1.1){$\overline{\xi}^n$};
\end{tikzpicture}
\end{center}
\caption{The geometry of the paths in Lemma \ref{lem:defects are cone transportable}.}
\label{fig:pizza}
\end{figure}
\end{lem}
\begin{proof}
    We consider states $\omega_{i} \coloneqq  \tilde \omega \circ \alpha_{\bar \eta^i}^\sigma \circ \beta^{r(L^i)}_g$.
    Observe that for $i = 1, 2$, $\tilde \pi\circ \alpha_{\bar \eta^i}^\sigma \circ \beta^{r(\bar L^i)}_g$ is a GNS representation for $\omega_i$. 
    Therefore, by uniqueness of the GNS representation, it suffices to show that $\omega_1 = \omega_2$.
    Let $A \in \cstar^{\loc}$. Choose $n$ large enough such that $\supp(A)$ does not intersect $\bar \varsigma^n$. We have,
    \begin{align*}
        \omega_{1}(A) &= \inner{\tilde \Omega}{\tilde \pi \circ  \alpha^\sigma_{\bar \eta^1} \circ \beta_g^{r(\bar L^1)}(A)\tilde \Omega}= \inner{\tilde \Omega}{ \tilde \pi(\tilde F^\sigma_{(\bar \eta^2)^n, \overline{\xi}^n}) \tilde \pi \circ \alpha^\sigma_{\bar \eta^1} \circ \beta_g^{r(\bar L^1)}(A) \tilde \pi(\tilde F^\sigma_{(\bar \eta^2)^n, \overline{\xi}^n})^*\tilde \Omega} \\
        &= \inner{\tilde \Omega}{ \tilde \pi(\tilde F^\sigma_{(\bar \eta^2)^n, \overline{\xi}^n}( \alpha^\sigma_{\bar \eta^1} \circ \beta_g^{r(\bar L^1)}(A)) (\tilde F^\sigma_{(\bar \eta^2)^n, \overline{\xi}^n})^*)\tilde \Omega}= \inner{\tilde \Omega}{ \tilde \pi \circ \alpha^\sigma_{\bar \eta^2} \circ \beta_g^{r(\bar L^2)}(A)\tilde \Omega} =\omega_{2}(A).
    \end{align*}
    Note that the second equality above follows by Lemma \ref{lem:the pizza operators eval to 1 SET}. 
    Indeed, this lemma implies that $\tilde \omega(\tilde F^\sigma_{(\bar \eta^2)^n, {\overline \xi}^n}) = 1$ and thus $\tilde \pi(\tilde F^\sigma_{(\bar \eta^2)^n, {\overline \xi}^n}) \tilde \Omega = \tilde \Omega$.

    To show the second half of the lemma, we proceed as in \cite[Lem.~4.1]{MR2804555}.
    There exists a unitary $V \in B(\tilde\cH)$ satisfying intertwining $\tilde \pi\circ \alpha_{\bar \eta^1}^\sigma \circ \beta^{r(\bar L^1)}_g $ and $ \tilde \pi \circ \alpha_{\bar \eta^2}^\sigma \circ \beta^{r(\bar L^2)}_g$ satisfying that $V\tilde \Omega = \tilde \Omega$ by uniqueness of the GNS representation.  
    This unitary is unique by Schur's lemma, since $\tilde \pi\circ \alpha_{\bar \eta^1}^\sigma \circ \beta^{r(\bar L^1)}_g $ and $ \tilde \pi \circ \alpha_{\bar \eta^2}^\sigma \circ \beta^{r(\bar L^2)}_g$ are irreducible representations.
    We now show that the sequence $V_n = \tilde \pi(\tilde F_{(\bar \eta^2)^n, \overline{\xi}^n}^\sigma)$ converges in the WOT to $V$.
    Let $A, B \in \cstar^{\loc}$.
    Let $n$ be large enough so that $\supp(B)$ does not intersect $\bar \varsigma^n$.
    For ease of notation, we define $\tilde \pi_i \coloneqq \tilde \pi\circ \alpha_{\bar \eta^i}^\sigma \circ \beta^{r(\bar L^i)}_g$ for $i = 1,2$.
    Then by the same argument as before, we have that
    \begin{align*}
    \langle \tilde \pi_1(A) \tilde \Omega, V_n \tilde \pi_1(B) \tilde \Omega \rangle
    &=
    \langle \tilde \pi_1(A) \tilde \Omega, \tilde \pi(\tilde F_{(\bar \eta^2)^n, \overline{\xi}^n}^\sigma) \tilde \pi_1(B) \tilde \Omega \rangle
    =
    \langle \tilde \pi_1(A) \tilde \Omega, \tilde \pi_2(B) \tilde \pi(\tilde F_{(\bar \eta^2)^n, \overline{\xi}^n}^\sigma)\tilde \Omega \rangle    
    \\&=
    \langle \tilde \pi_1(A) \tilde \Omega, \tilde \pi_2(B) \tilde \Omega \rangle 
    =
    \langle \tilde \pi_1(A) \tilde \Omega, \tilde \pi_2(B) V\tilde \Omega \rangle 
    =
    \langle \tilde \pi_1(A) \tilde \Omega, V\tilde \pi_1(B) \tilde \Omega \rangle,
    \end{align*}
    where the third equality follows by Lemma \ref{lem:the pizza operators eval to 1 SET} similarly to before. 
    Now, since $\alpha_{\bar \eta^i}^\sigma \circ \beta^{r(\bar L^i)}_g$ is an automorphism of $\cstar$, we have that $\tilde \pi_i(\cstar^{\loc})$ is dense in $\cstar$.
    Therefore, since $(V_n)$ is a uniformly bounded sequence, $V_n \to V$ in the WOT.
\end{proof}

Armed with these results, we now define the defect automorphisms.
\begin{defn}
\label{def:SETToricCodeDefectAutomorphism}
    Let $\bar \gamma \in \bar P_R(\Gamma)$, $\bar L$ a completion of $\bar \gamma$, and $\bar \eta \in \bar P(\Gamma)$ be the half-infinite path given by $\bar \eta_k \coloneqq \bar L_{1 - k}$. Define the \emph{defect automorphism} $\tilde \alpha_{\bar \gamma}^\sigma$ to be $\tilde \alpha_{\bar \gamma}^\sigma \coloneqq \alpha_{\bar \eta}^\sigma \circ \beta^{r(\bar L)}_g.$
    Observe that $\tilde \alpha^\sigma_{\bar \gamma}$ depends on the completion $\bar L$ of $\bar \gamma$, but we suppress this dependence for notational convenience.
\end{defn}

\begin{rem}
    Note that by Lemma \ref{lem:defects are cone transportable}, we get that $\tilde\pi \circ \tilde \alpha^{\sigma}_{\bar \gamma}$ are all unitarily equivalent for different completions of $\bar \gamma$. 
    
\end{rem}

\begin{rem}
    Physically, the defect automorphism $\tilde \alpha_{\bar \gamma}^\sigma$ creates a defect whose endpoint lives near $\partial_0 \bar \gamma$ and has a domain wall along the dual path $\bar \gamma$.
\end{rem}

\begin{rem}\label{rem:sigma2ism}

We have $$\tilde \alpha^\sigma_{\bar \gamma} \circ  \tilde \alpha^\sigma_{\bar \gamma} = \alpha_{\bar \gamma}^m,$$ which can be seen immediately by noting that $\beta_g^{r(\bar L)} \circ \alpha_{\bar \eta}^\sigma = \alpha_{\bar \eta}^\sigma \circ \beta_g^{r(\bar L)}$ and $\alpha^\sigma_{\bar \eta} \circ \alpha^\sigma_{\bar \eta} = \alpha_{\bar \eta}^m$.
\end{rem}

Now we define the following defect states as $$\tilde \omega_{\bar \gamma}^\sigma \coloneqq  \tilde \omega \circ \tilde \alpha_{\bar \gamma}^\sigma, \qquad \qquad \tilde \omega^{\sigma, \zeta_\gamma}_{\bar \gamma} \coloneqq  \tilde \omega^\sigma_{\bar \gamma} \circ \zeta_\gamma.$$
Similarly, we define $\tilde \pi^\sigma_{\bar \gamma} \coloneqq \tilde \pi \circ \tilde \alpha^\sigma_{\bar \gamma}$ and $\tilde \pi^{\sigma, \zeta_\gamma}_{\bar \gamma} \coloneqq \tilde \pi^\sigma_{\bar \gamma} \circ \zeta_\gamma$.

\begin{lem}
\label{lem:SET toric code defects are finitely transportable}
    For $i = 1, 2$, let $\bar \gamma^i\in \bar P_R(\Gamma)$ to be half-infinite dual paths for $i = 1,2$.
    Then $\tilde \pi^\sigma_{\bar \gamma^1} \simeq \tilde \pi^\sigma_{\bar \gamma^2}.$
\end{lem}

Before we prove this lemma, we remark that Lemma \ref{lem:defects are cone transportable} only shows the equivalence in the case where $\bar \gamma^1 = \bar \gamma^2$. 
We now prove the more general case. 

\begin{proof}[Proof of Lemma \ref{lem:SET toric code defects are finitely transportable}]
    We consider the pure states $\tilde \omega^\sigma_{\bar \gamma^i}$ for $i = 1, 2$. 
    We wish to show that $\tilde \omega^\sigma_{\bar \gamma^1} \simeq \tilde \omega^\sigma_{\bar \gamma^2}$.
    Note that by the proof of Lemma \ref{lem:defects are cone transportable}, the state $\tilde \omega^\sigma_{\bar \gamma^i}$ is independent of the completion $\bar L^i$ of $\bar \gamma^i$ used to construct the automorphism $\tilde \alpha^\sigma_{\bar \gamma^i}$.
    We therefore take completions $\bar L^1$ of $\bar \gamma^1$ and $\bar L^2$ of $\bar \gamma^2$ such that $\bar L^1_{dual}$ and $\bar L^2_{dual}$ only differ in a finite region. 
    As before, for $i = 1, 2$, we let $\bar \eta^i \in \bar P(\Gamma)$ be the half-infinite dual path given by $\bar \eta^i_k \coloneqq \bar L^i_{1 - k}$.
    Since $\bar L^1_{dual}$ and $\bar L^2_{dual}$ only differ in a finite region, there exists $S \in \Gamma_f$ such that for all observables $A \in \cstar[S^c]$,
    $$\tilde \alpha_{\bar \gamma^1}^\sigma(A) = \alpha_{\bar \eta^1} \circ \beta_g^{r(\bar L^1)}(A) = \alpha_{\bar \eta^2} \circ \beta_g^{r(\bar L^2)}(A) = \tilde \alpha_{\bar \gamma^2}^\sigma(A).$$
    
    Note that since $\tilde \omega_{\bar \gamma^1}^\sigma$ and $\tilde \omega_{\bar \gamma^2}^\sigma$ are pure states, $\tilde \omega_{\bar \gamma^1}^\sigma$ and $\tilde \omega_{\bar \gamma^2}^\sigma$ are unitarily equivalent if and only if they are quasi-equivalent \cite[Prop.~10.3.7]{MR1468230}.
    We can therefore apply \cite[Cor.~2.6.11]{MR887100}.  Observe, for all $A \in \cstar[S^c]$ we have $$\tilde \omega_{\bar \gamma^1}^\sigma(A) = \tilde \omega \circ \tilde \alpha_{\bar \gamma^1}^\sigma(A) = \tilde \omega \circ \tilde \alpha_{\bar \gamma^2}^\sigma(A) =  \tilde \omega_{\bar \gamma^2}^\sigma(A),$$ so we have that $\tilde \omega^\sigma_{\bar \gamma^1} \simeq \tilde \omega^\sigma_{\bar \gamma^2}$.
    
\end{proof}

\subsection{\texorpdfstring{$G$}{G}-defect representations}
We set $\tilde\pi$, the GNS representation of state $\tilde \omega$, as the reference representation. 
We first verify that all the assumptions given in Section \ref{sec:GCrossedAssumptions} hold.
Since the representation of $G$ given by $g \mapsto U^g_v$ is faithful for every vertex $v \in \Gamma$, we have that Assumption \ref{asmp:Faithfulness} holds.
By Lemma \ref{lem:SET toric code unique FF GS}, Assumption \ref{asmp:GInvariance} holds.
By Lemmas \ref{lem:TC extension Haag duality}, \ref{lem:SETToricCodeFDQC}, and \ref{lem:QCABSHaagDuality}, Assumption \ref{asmp:BoundedSpreadHaagDuality} is satisfied using the fact that $\tilde \pi \simeq \pi_0 \circ \alpha^{-1}$.
By Lemma \ref{lem:SET toric code unique FF GS}, $\tilde \omega$ is pure, so Assumption \ref{asmp:PureState} holds.
Note that this implies that $\tilde \pi$ is irreducible.
In addition, $\tilde \omega$ is translation-invariant, so Assumption \ref{asmp:InfiniteFactor} holds by a standard argument \cite{MR2804555, MR2281418}.

\begin{lem}
\label{lem:defect sector TC case}
    For all paths $\bar \gamma^0 \in \bar P_R(\Gamma)$, the representation $\tilde \pi^{\sigma}_{\bar \gamma^0}$ is a $g$-defect representation with respect to $\tilde \pi$.
\end{lem}
\begin{proof}
    Let $\bar L^0$ be a completion of $\bar \gamma^0$ and let $\bar \eta^0 \in \bar P(\Gamma)$ be the half-infinite dual path defined by $\bar \eta_k^0 \coloneqq \bar L_{1 - k}^0$. Choose a cone $\Lambda \in \cL$ such that $\bar \eta^0$ is contained in $\Lambda$ and let $A \in \cstar[\Lambda^c]$. We then have
    \begin{align*}
        \tilde \pi^\sigma(A) &= \tilde \pi \circ \tilde \alpha_{\bar \gamma^0}^{\sigma}(A) = \tilde \pi \circ \alpha_{\bar \eta^0}^{\sigma} \circ \beta_g^{r(\bar L^0)}(A) = \tilde \pi \circ \beta_g^{r(\bar L^0)}(A).
    \end{align*}
    In the last equality we used that $\beta_g(A) \in \cstar[\Lambda^c]$ since $\beta_g$ is an on-site symmetry. 
    We also used that for any observable $A' \in \cstar[\Lambda^c]$, we have $\alpha^\sigma_{\bar \eta^0}(A') = A'$ since $\alpha^\sigma_{\bar \eta^0}$ is localized inside $\Lambda$.
    Since $\bar \gamma^0 \in \bar P_R(\Gamma)$, we have that $r(\bar L^0) \cap \Lambda^c$ differs from $r(\Lambda)$ by finitely many vertices.
    This shows $\tilde \pi^\sigma_{\bar \gamma^0}$ is localized in $\Lambda$. 
    
    It remains to be shown that $\tilde \pi_{\bar \gamma^0}^\sigma$ is transportable. Choose another cone $\Lambda'$. We now choose another path $\bar \gamma^1 \in \bar P_R(\Gamma)$ and completion $\bar L^1$ such that the path $\bar \eta^1 \in \bar P(\Gamma)$ defined by $\bar \eta'_k \coloneqq \bar L^1_{1 - k}$ lies entirely in $\Lambda'$. By the above argument, we have that $\tilde \pi^\sigma_{\bar \gamma^1}$ is $g$-localized in $\Lambda'$. Using Lemma \ref{lem:SET toric code defects are finitely transportable} we get that $\tilde \pi^\sigma_{\bar \gamma^0} \simeq \tilde \pi^\sigma_{\bar \gamma^1}$ giving us that $\tilde \pi^\sigma_{\bar \gamma^0}$ is transportable.
\end{proof}

We define some more representations given by $\tilde \pi^{\sigma, \zeta}_{\bar \gamma^0} \coloneqq  \tilde \pi_{\bar \gamma}^\sigma \circ \zeta$ for $\bar \gamma^0 \in \bar P_R(\Gamma)$.
We omit subscripts on $\zeta$ since we are fixing the paths $\gamma \in P(\Gamma)$ and $ \bar \gamma \in \bar P(\Gamma)$ defining the automorphisms in this instance.

\begin{lem}
\label{lem:anyon sectors of defect SET}
    Let $\bar \gamma^0 \in \bar P_R(\Gamma)$.
    The representations given by $\{\tilde \pi^{\sigma, \zeta}_{\bar \gamma^0}\}_{\zeta}$ comprise the distinct irreducible anyon sectors with respect to $\tilde \pi^\sigma_{\bar \gamma^0}$.
    
\end{lem}
\begin{proof}
    Since $\tilde \alpha^\sigma_{\bar \gamma^0}$ is an FDQC, the result follows from Lemmas \ref{lem:anyon sectors of GS SET}, \ref{lem:FDQCQuasi-Factorizable}, and \cite[Thm.~4.7]{MR4426734}.
\end{proof}

\begin{lem}
    Let $\bar \gamma^0 \in \bar P_R(\Gamma)$. 
    For $\zeta \in \{\Id, \alpha_\gamma^\epsilon, \alpha_{\bar \gamma}^m, \alpha_{\gamma, \bar \gamma}^\psi\}$, the representations $\tilde \pi^{\sigma, \zeta}_{\bar \gamma^0}$ are $g$-defect representations with respect to $\tilde \pi$ and the representations $\tilde \pi^{\zeta}$ are $1$-defect representations with respect to $\tilde \pi$.
\end{lem}
\begin{proof}
    
    From Lemma \ref{lem:defect sector TC case} we know that $\tilde \pi_{\bar \gamma^0}^{\sigma}$ is a $g$-defect representation, and $\tilde \pi$ is obviously a $1$-defect representation. 
    Since $\tilde \pi^\zeta$ are anyon representations with respect to $\tilde \pi$ (Lemma \ref{lem:anyon sectors of GS SET}) it follows that $\pi^\zeta$ are unitarily equivalent to $1$-defect representations by Lemma \ref{lem:GDefectsRelationToAnyons}.
    Similarly, since $\tilde \pi_{\bar \gamma^0}^{\sigma, \zeta}$ are anyon representations with respect to $\tilde \pi_{\bar \gamma^0}^{\sigma}$, it follows that $\tilde \pi_{\bar \gamma^0}^{\sigma, \zeta}$ are unitarily equivalent to $g$-defect representations.

    Since $\gamma, \bar \gamma \subset \Lambda$, we have that $\zeta$ is localized in $\Lambda$. 
    Thus for all observables $A \in \cstar[\Lambda^c]$ we have $\tilde \pi^{\zeta}(A)= \tilde \pi (A)$, so $\tilde \pi^{\zeta}$ is localized in $\Lambda$. Therefore, $\tilde \pi^{\zeta}$ is a $1$-defect representation. 
    Similarly, if we choose a completion $\bar L^0$ of $\bar \gamma^0$ and let $\bar \eta^0 \in \bar P(\Gamma)$ be given by $\bar \eta^0_k \coloneqq \bar L^0_{1 - k}$, then we have that $\tilde \pi_{\bar \gamma_0}^{\sigma, \zeta}$ is $g$-localized in some $\Lambda' \in \cL$ containing $\gamma$, $\bar \gamma$, and $\bar \eta^0$. 
    Therefore, $\tilde \pi_{\bar \gamma^0}^{\sigma, \zeta}$ is a $g$-defect representation.
\end{proof}

\begin{prop}
\label{lem:classification of defect sectorizable representations TC}
    Let $\bar \gamma^0 \in \bar P_R(\Gamma)$ and $\zeta \in \{\Id, \alpha_\gamma^\epsilon, \alpha_{\bar \gamma}^m, \alpha_{\gamma, \bar \gamma}^\psi\}$. The representations $\{\tilde \pi^\zeta_{\bar \gamma^0}, \tilde \pi^{\sigma,\zeta}_{\bar \gamma^0}\}_{\zeta}$ comprise the distinct irreducible $G$-defect sectors with respect to $\tilde \pi$.
    
\end{prop}
\begin{proof}
    Note that $\{\tilde \pi^\zeta_{\bar \gamma^0}, \tilde \pi^{\sigma,\zeta}_{\bar \gamma^0}\}_{\zeta}$ forms a collection of mutually disjoint irreducible $G$-defect sectors by Lemmas \ref{lem:anyon sectors of GS SET} and \ref{lem:anyon sectors of defect SET}
    (recall that $G$-defect sectors denote isomorphism classes of objects in $\GSec$).
    Now, let $\pi$ be a $g$-defect representation for $g$ being the non-trivial group element. 
    From Lemma \ref{lem:defect state GNS reps are sectorizable (SPT case)} we have that $\tilde \pi_{\bar \gamma^0}^{\sigma}$ is a $g$-defect representation. 
    We have from Lemma \ref{lem:GDefectsRelationToAnyons} that $\pi$ is an anyon representation with respect to $\tilde \pi^\sigma_{\bar \gamma^0}$. But by Lemma \ref{lem:anyon sectors of defect SET} $\pi \simeq \tilde \pi_{\bar \gamma^0}^{\sigma, \zeta}$ for some $\zeta$ in the set. Repeating the same analysis for $g=1$ and using Lemmas \ref{lem:GDefectsRelationToAnyons}, \ref{lem:anyon sectors of GS SET} gives us the other case and hence the result.
\end{proof}

\begin{rem}
    For $\bar \gamma^0 \in \bar P_R(\Gamma)$, can be shown that the category $\DHR_{\tilde \pi^\sigma_{\bar \gamma^0}}$ is braided monoidally equivalent to $\DHR_{\tilde \pi}$. Since $\tilde \alpha_{\bar \gamma^0}^\sigma$ is an FDQC, the result immediately follows from \cite[Thm. ~6.1]{MR4362722}.

    Even though the above result is stronger than Proposition \ref{lem:classification of defect sectorizable representations TC}, we note that $\DHR_{\tilde \pi^\sigma_{\bar \gamma^0}}$ is the category of anyon representations with respect to $\tilde \pi^\sigma_{\bar \gamma^0}$ as the reference state. While the objects in this category are the objects we are ultimately interested in, we note that we do not want to inherit the fusion and braiding structure from this category as it disregards the presence of the defect. Below we construct the $G$-crossed braided monoidal category that we are interested in.
\end{rem}

\subsection{Defect tensor category}
\label{sec:F_symbols}

Recall that $\GSec_{\tilde \pi}$ is the category of $G$-defect representations with respect to $\tilde \pi$, extended to endomorphisms of the auxiliary algebra $\fA^a \coloneqq  \overline{\bigcup_{\Lambda \in \cL} \cR(\Lambda)}^{\|\cdot \|}$.
Further recall that $\GSec_{\tilde \pi}^f$ denotes the full subcategory of dualizable $G$-defect representations (Definition \ref{def:Dualizable}).
In this subsection, we describe $\GSec_{\tilde \pi}^f$ as a $\mathbb{Z}_2$-graded tensor category. 
In the subsections which follow, we will give the rest of the $G$-crossed braiding data.

We now fix some geometric notation that will be used for the remainder of this section.

\begin{nota}\label{notation:defects}

    We use Figure \ref{fig:strings} to fix our geometric notation.
    First, we remark that the origin vertex $0$ corresponds to the large yellow dot in the figure. 
    As always, the orientation of each vertical edge is upward and the orientation of each horizontal edge is to the right.
    We take $\overline{L}$ to be the infinite dual path whose $n$th edge site $\overline{L}_n \in \Gamma$ corresponds to the midpoint of $[0, 1] \times \{n\}$.\footnote{Recall that when we view edges of the square lattice as elements of $\Gamma \subset \bbR^2$, we take each edge site to be located at the midpoint of the edge.} 
    The path $\overline{L}_{dual} \subset \bbR^2$ is exactly the vertical gray line in the figure. 
    Note that $\beta^{r(\overline{L})}$ is the symmetry action on $\Gamma \cap \bbR_{> 1/2} \times \bbR$, which is the region to the right of the vertical gray line. 
    The red ray is $R = \{1/4\} \times \bbR_{\geq 1/4}$. 
    We fix the path $\bar \gamma^0 \in \bar P_R(\Gamma)$ to be the dual path whose edges are exactly those intersected by $R$. 
    We fix $\gamma \in P(\Gamma)$ to be the path whose $n$th edge site $\gamma_n \in \Gamma$ corresponds to the midpoint of $\{0\} \times [-n + 1, -n]$. 
    Note that $\gamma$ corresponds in the figure to the wiggling ray extending downward from $0$. 
    From this, we have $\partial\gamma=0$. 
    We also fix $\bar \gamma \in \bar P(\Gamma)$ to be the dual path whose $n$th edge site $\bar \gamma_n \in \Gamma$ corresponds to the midpoint of $[0, 1] \times \{-n + 1\}$. 
    Note that $\bar \gamma$ correspond in the figure to the orange dual path.
    Observe that $\overline{L}$ is a completion of $\bar \gamma^0$ and that $\bar \gamma_k = \overline{L}_{1 - k}$.
    We fix $\Lambda \in \cL$ to be the cone given by the intersection of $\bbR_{\geq -5/4} \times \bbR_{\leq 7/4}$ (the blue region in the figure) with $\Gamma$.

    \begin{figure}[!ht]
\begin{center}
\begin{tikzpicture}[scale=0.6]
    \filldraw[blue!30,thick,fill=blue!10](2.75,0)--(2.75,7.75)--(9,7.75)--(9,0);

    \foreach \x in {0,...,9}{
    \draw[thick,black](\x,0)--(\x,10);
    }
    \foreach \y in {0,...,10}{
    \draw[thick,black](0,\y)--(9,\y);
    }
    \draw[thick,black!50](4.5,10)--(4.5,0);
    
     \draw[very thick, violet,snake it](4,0)--(4,6);
    \foreach \x in {0,...,6}{
    \draw[very thick,orange](4,\x)--(5,\x);
    }
   \filldraw[violet,thick,fill=yellow] (4,6) circle(.2cm);
    \draw[thick,red](4.25,6.25)--(4.25,10);
    \filldraw[thick,red,fill=red](4.25,6.25) circle(.09cm);
\end{tikzpicture}
\end{center}
    \caption{The ray $R$ and $\epsilon, m, \psi$ strings. 
    }
    \label{fig:strings}
\end{figure}
\end{nota}

Note that if $\pi \in \GSec_{\tilde \pi}$, then by Definition \ref{def:GSec}, $\pi \cong \pi_1 \oplus \pi_g$, where $\pi_1 \in (\GSec_{\tilde \pi})_1$ and $\pi_g \in (\GSec_{\tilde \pi})_g$. 
By Lemma \ref{lem:GDefectsRelationToAnyons}, $\pi_1$ is an anyon representation with respect to $\tilde \pi$ and $\pi_g$ is an anyon representation with respect to $\tilde \pi^\sigma_{\bar \gamma^0}$.

We denote the irreducible anyon representations with respect to $\tilde \pi$ by $\{1, \epsilon, m , \psi\},$ where, for example, $\epsilon$ corresponds to extension of the automorphism $\alpha_{\gamma}^\epsilon$ to $\cstar^a$ (Lemma \ref{lem:GDefectsDefinedOnAuxiliaryAlgebra}). 
Likewise, the irreducible $g$-defect representations with respect to $\tilde \pi$ are denoted by $\{1^\sigma, \epsilon^\sigma, m^\sigma , \psi^\sigma\}$.
By Lemma \ref{lem:anyon sectors of defect SET}, we may define these distinct $G$-defect representations for each $a\in\{1, \epsilon, m , \psi\}$ by the unique extension of  $\tilde{\pi}^\sigma_{\bar \gamma^0}\circ a|_{\cstar}$ to $\cstar^a$. 
In other words, we are defining
$a^\sigma \coloneqq 1^\sigma\otimes a$ for each $a\in\{1, \epsilon, m , \psi\}$ where $1^\sigma$ is the extension of $\tilde{\pi}^\sigma_{\bar \gamma^0}$ to $\cstar^a$.

\begin{nota}
To ease notation, from this point forward we omit the (dual) path subscript on automorphisms when the (dual) path is the fixed (dual) path ($\bar \gamma$) $\gamma$.
In more detail, we adopt the following: 
\[
\alpha^\epsilon 
\coloneqq 
\alpha^\epsilon_\gamma,
\qquad
\alpha^m
\coloneqq
\alpha^m_{\bar \gamma},
\qquad
\alpha^\psi
\coloneqq
\alpha^\psi_{\bar \gamma},
\qquad
\alpha^\sigma
\coloneqq
\alpha^\sigma_{\bar\gamma}.
\]
Note that since $\overline{L}$ is a completion of $\bar \gamma^0 \in \bar P_R(\Gamma)$ and $\bar \gamma \in \bar P(\Gamma)$ satisfies that $\bar \gamma_k = \overline{L}_{1 - k}$, we have that $\tilde \alpha^\sigma_{\bar \gamma^0} = \alpha^\sigma \circ \beta_g^{r(L)}$. 
We realize this notation is confusing, but we will only have reason to refer to $\alpha^\sigma$ (and not $\tilde \alpha^\sigma_{\bar \gamma^0}$) in the remainder of this section.
    
\end{nota}

Note that the path $\gamma$ was defined in Notation \ref{notation:defects} so that $\alpha^\epsilon$ is localized outside of $r(\overline{L})$.
Therefore, we have that $\alpha^\epsilon\circ \beta^{r(\overline{L})}_g =\beta^{r(\overline{L})}_g \circ\alpha^\epsilon$.

\begin{lem}\label{lem:TC_Fusion}
The fusion rules for the irreducible objects of $\GSec_{\tilde \pi}$ are determined by the usual Toric Code fusion rules in $\DHR_{\tilde{\pi}}$, as well as the equalities $a^{\sigma}\otimes b= (ab)^\sigma$ for $a,b\in\{1,\epsilon,m,\psi\}$ and $1^\sigma\otimes 1^{\sigma}= m$. 
In particular, every irreducible object in $\GSec_{\tilde \pi}$ is dualizable. 
\end{lem}
\begin{proof}
    We already proved in Proposition \ref{prop:equiv_to_TC} that $\DHR_{\tilde{\pi}}$ obeys the usual fusion rules for the Toric Code.
    The second statement follows directly from the fact that $a^{\sigma}= 1^\sigma\otimes a$.
    Finally, we may notice that $\beta_g^{r(\overline{L})}\circ \alpha^\sigma=\alpha^\sigma\circ \beta_g^{r(\overline{L})}$ and by Remark \ref{rem:sigma2ism} $\alpha^\sigma\circ \alpha^\sigma=\alpha^m$. Therefore, 
    $$
    1^{\sigma}\circ 1^{\sigma}|_{\fA}=\alpha^\sigma\circ \beta_g^{r(\overline{L})}\circ \alpha^\sigma\circ \beta_g^{r(\overline{L})}=\alpha^\sigma\circ  \alpha^\sigma\circ \beta_g^{r(\overline{L})}\circ \beta_g^{r(\overline{L})}=\alpha^m.
    $$
    As endomorphisms of the auxiliary algebra, $1^\sigma \otimes 1^\sigma=m$.

    Note that the remaining fusion rules all follow from the existence of a $G$-crossed braiding and the fact that the symmetry acts trivially on the anyons, as was shown in Lemma \ref{lem:TCSymmetry_permute}.
\end{proof}

For the rest of this section, we restrict our attention to $\GSec_{\tilde \pi}^f$, the $G$-crossed braided $\rmW^*$-tensor category of dualizable objects in $\GSec_{\tilde \pi}$ (Definition \ref{def:Dualizable}).
Note that we have already implicitly chosen the basis of $\mathcal{K}_0(\GSec_{\tilde \pi}^f)$, the fusion ring for $\GSec_{\tilde \pi}^f$, to be $I\coloneqq \{1, \epsilon, m , \psi,1^\sigma, \epsilon^\sigma, m^\sigma , \psi^\sigma\}$.
We will use this basis for the remainder of this section. 

\begin{prop}\label{prop:triv_assoc}
    The tensorators $\Omega_{i,j}$ are all the identity.
    Therefore the skeletalization of the tensor category $\GSec_{\tilde \pi}^f$ is strict.
\end{prop}
\begin{proof}
    To prove this, we simply need to show that we may pick representative endomorphisms for each isomorphism class so that the composition of any two is also a representative. After considering Lemma \ref{lem:TC_Fusion}, all that remains is to show that $a\otimes 1^{\sigma}=a^{\sigma}$ for $a\in\{1,\epsilon,m,\psi\}$. 
    The remaining equalities all easily follow from this fact.

    For $a\in \{1,\epsilon,m,\psi \}$, we have that 
    $$
    \alpha^a\circ\alpha^\sigma=\alpha^\sigma\circ\alpha^a\quad\quad \text{and}\quad\quad\alpha^a\circ\beta_g^{r(\overline{L})}=\beta_g^{r(\overline{L})}\circ\alpha^a.
    $$
    Therefore, we have that
    $$
    a\otimes 1^\sigma|_{\fA}=\alpha^a\circ\alpha^\sigma\circ\beta_g^{r(\overline{L})}=\alpha^\sigma\circ\beta_g^{r(\overline{L})}\circ \alpha^a=a^\sigma|_{\fA}.
    $$
    Extending this endomorphism to the auxiliary algebra $\cstar^a$ gives that $a\otimes 1^\sigma=a^\sigma$.
    Therefore, all of the tensorators are the identity.
\end{proof}

\subsection{Symmetry fractionalization}
\label{sec:frac_data}

We now compute the symmetry fractionalization data for $\GSec_{\tilde \pi}^f$ following our prescription in Section \ref{sec:SymmetryFractionalization}.
Recall that we have chosen our basis of $\cK_0(\GSec^f_{\tilde \pi})$ to be $I = \{1, \epsilon, m , \psi,1^\sigma, \epsilon^\sigma, m^\sigma , \psi^\sigma\}$.
For each $h\in \mathbb{Z}_2$ and $i\in I$, we have the unitary intertwiner $V_h^i\colon\gamma_h(\pi_i)\rightarrow \pi_{h(i)}$.
We reserve $g\in \mathbb{Z}_2$ to be the non-trivial element.
When $h\neq g,$ $V^i_h=1$.

\begin{lem}\label{lem:symmetry_on_e}
    
    For any $x\in\cstar^a$, we have that
    $$
    \tau^z_{0} \gamma_g(\epsilon)(x)(\tau^z_{0})^*=\epsilon(x),
    \qquad\qquad
    \tau^z_{0} \gamma_g(\psi)(x)(\tau^z_{0})^* = \psi(x).
    $$
\end{lem}
\begin{proof}
    By continuity, it is sufficient to prove this statement in the case where $x$ is a local operator.
    In that case, $\epsilon(x) = \alpha^\epsilon(x)$ and $\psi(x) = \alpha^\psi(x)$. 
    We begin by proving the first equation. 
    Using the definition of $\gamma_g$, we have
    $$
    \tau^z_{0} \gamma_g(\epsilon)(x)(\tau^z_{0})^*=\tau^z_{0} \beta_g \epsilon(\beta_g^{-1}(x))(\tau^z_{0})^*.
    $$
    Since $x$ is local and $\beta_g^{-1}$ preserves the support of local operators, $\beta^{-1}_g(x)$ is also local.
    Take $\gamma'$ to be a finite subpath of the path $\gamma$ which defines $\alpha^\epsilon$ with $\partial_0\gamma'=\partial\gamma$.
    Using Remark \ref{rem:SETTCAutomorphismDefn}, since $\alpha^\epsilon$ is an FDQC, we may choose $\gamma'$ to be long enough such that
    $$
    \epsilon(\beta_g^{-1}(x))
    =
    \alpha^{\epsilon}(\beta_g^{-1}(x))=\alpha^{\epsilon}_{\gamma'}(\beta_g^{-1}(x))
    $$
    and such that the support of $\alpha^\epsilon(x)$ is disjoint from $\partial_1\gamma'$.
    Finally, this reasoning along with Lemma \ref{lem:action of symmetry on string ops} implies that
    \begin{align*}
    \tau^z_{0} \gamma_g(\epsilon)(x)(\tau^z_{0})^*&=\tau^z_{0} \beta_g(\alpha^\epsilon_{\gamma'}(\beta^{-1}_g(x)))(\tau^z_{0})^*\\
    &=\tau^z_{0} \beta_g(\tilde{F}^\epsilon_{\gamma'})x\beta_g(\tilde{F}^\epsilon_{\gamma'})^*(\tau^z_{0})^*\\
    &=\tau^z_{\partial_1 \gamma'}\tilde{F}^\epsilon_{\gamma'}x(\tilde{F}^\epsilon_{\gamma'})^*(\tau^z_{\partial_1 \gamma'})^*\\
    &=\tau^z_{\partial_1 \gamma'}\alpha_{\gamma'}^\epsilon(x)(\tau^z_{\partial_1 \gamma'})^*\\
     &=\tau^z_{\partial_1 \gamma'}\alpha^\epsilon(x)(\tau^z_{\partial_1 \gamma'})^*\\
    &=\alpha^\epsilon(x)\\
    &= \epsilon(x).
    \end{align*}

    To see the second equation, recall that by Lemma \ref{lem:action of symmetry on string ops}, $\beta_g(\tilde F^m_{\bar \gamma'}) = \tilde F^m_{\bar \gamma'}$ for any finite truncation $\bar \gamma'$ of $\bar \gamma$. 
    Similar to before, we choose $\bar \gamma'$ long enough so that 
    \[
    \alpha^m(\beta_g^{-1}(x)) = \alpha^m_{\bar \gamma'}(\beta_g^{-1}(x)),
    \qquad\qquad
    \alpha^m(x) = \alpha^m_{\bar \gamma'}(x).
    \]
    We then have that 
    \begin{align*}
    \tau^z_{0} \gamma_g(\psi)(x)(\tau^z_{0})^*
    &=
    \tau^z_{0} \gamma_g(\epsilon)(\gamma_g(m)(x))(\tau^z_{0})^*
    =
    \epsilon(\gamma_g(m(x)))
    =
    \epsilon(\beta_g(\alpha^m_{\bar \gamma'}(\beta_g^{-1}(x))))
    \\&=
    \epsilon(\beta_g(\tilde F^m_{\bar \gamma'}) x \beta_g(\tilde F^m_{\bar \gamma})^*)
    =
    \epsilon(\tilde F^m_{\bar \gamma'} x (\tilde F^m_{\bar \gamma'})^*)
    =
    \epsilon(\alpha^m_{\bar \gamma'}(x))
    =
    \epsilon(m(x) 
    =
    \psi(x).
    \qedhere
    \end{align*}
\end{proof}

This lemma shows that we may take $V^\epsilon_g=\tau^z_{0}$ where $g$ is the non-trivial element of $\mathbb{Z}_2$.
Similar computations reveal that $$V^1_g=V^m_g=V^{1^\sigma}_g=V^{m^\sigma}_g=\mathds1$$ and $$V^\psi_g=V^{\epsilon^\sigma}_g=V^{\psi^\sigma}_g=\tau^z_{\partial \gamma}.$$
Noting that $1^\sigma$ is invariant under the symmetry, we have that the only non-identity values of $\eta$ from Section \ref{sec:SymmetryFractionalization} are 
$$
\eta(g,g)_\epsilon=\eta(g,g)_\psi=\eta(g,g)_{\epsilon^\sigma}=\eta(g,g)_{\psi^\sigma}=-1.
$$

Since $V_1=\mathds1$, we have that $\mu_1(a,b)=1$ for all basis elements $a,b\in \mathcal{K}_0(\GSec)$.

Now we compute $\mu_g(a,b)$ where $g$ is the non-trivial element of $\mathbb{Z}_2$. Since Proposition \ref{prop:triv_assoc} tells us that all the tensorators ($\Omega_{i,j}$) are the identity, we have that
$$
V_g^{ij}=\mu_g(i,j)V_g^i\gamma_g(\pi_i)(V_g^j).
$$
In this model, it can be easily checked that $V^{ij}_g=V^i_gV^j_g$.
Therefore,
$$\mu_g(i,j)\mathds1=\gamma_g(\pi_i)(V^j_g)^*V^j_g=V^i_g\pi_i(V^j_g)^*(V^i_g)^* V^j_g.$$
However, using the fact that $\pi_i(\tau^z_{0})=\tau^z_{0}$ and $[V_g^i,V_g^j]=0$ for all basis elements $i,j\in I$, we have that $\mu_g(i,j)=1$.

\subsection{\texorpdfstring{$G$}{G}-crossed braiding}
\label{sec:braid_data}

We are now in a position to compute the $G$-crossed braiding data. 

In Notation \ref{notation:braiding}, we will define the operators $U^\pi_N$ which limit to the operator $U^\pi$ which transports $\pi$ from $\Lambda$ to $\Delta$ for certain simple $\pi$ in $\GSec_{\tilde \pi}^f$.

\begin{nota}\label{notation:braiding}
    We will consider the two diagrams (Figures \ref{fig:sigma} and \ref{fig:U}) below to define $U_N^\pi$ for $\pi \in I$, our chosen basis of $\cK_0(\GSec^f_{\tilde \pi})$. 
    The figures depict the case where $N=4$. 
    We start with the first diagram (Figure \ref{fig:sigma}). 
    Refer to Notation \ref{notation:defects} for the definitions of $\overline{L}$, $\bar \gamma^0$, $\bar \gamma$, $\gamma$, and $\Lambda$. 
    Recall that we take the vertex $0$ to be the large yellow dot. 
    As before, $\overline{L}_{dual}$ is the vertical gray line just to the right of the vertex $0$ in the figure so that $\beta^{r(\overline{L})}_g$ is the symmetry action on the vertices to the right of this line. 
    Also just as before, the half-infinite path $\gamma$ is given by the purple wiggling ray and $\bar \gamma$ is the half-infinite dual path just to the right of $\gamma$ beginning at the edge neighboring the origin.
    To be precise, the midpoint of the edge $\bar\gamma_n$ is $(1/2,-n+1)$ for each $n\in\mathbb{N}$.
    The cone $\Lambda$ is given blue shading whereas another cone $\Delta \in \cL$ is shaded in red.
    In particular, $\Delta = (\bbR_{\leq -3.75} \times \bbR_{\leq 1.75}) \cap \Gamma$. 
    Since $R = \{1/4\} \times \bbR_{\geq 1/4}$ (the red ray in the figure), we have that $\partial R = (1/4, 1/4) \in \Lambda_{geo}$.
    Further, $\Delta$ is sufficiently to the left of $\Lambda$ (Definition \ref{def:sufficiently_to_the_left}). 
    Indeed,  $\Delta^{+2} \subseteq \ell(\Lambda)$ since $\Delta^{+2} \subseteq \bbR_{\leq -1.75} \times \bbR_{\leq 2.75} \subseteq \ell(\Lambda)_{geo}$, and by Lemmas \ref{lem:TC extension Haag duality}, \ref{lem:SETToricCodeFDQC}, and \ref{lem:QCABSHaagDuality}, $\tilde \pi$ satisfies bounded spread Haag duality with spread at most $2$.
    
    Take the path $\bar{\xi}^N$ to be the black wiggling dual path going clockwise around the large black dots. The edges in orange are the edges traversed by $\bar{\xi}^N$. This dual path is parameterized by $N$ so as to intersect $\gamma$ at the $N$th edge below the origin, that is, at the midpoint of $\{0\} \times [N - 1, N]$. Take the region $A_N \coloneqq [-5, 0] \times [-N + 1, 3] \cap \Gamma^V$, that is, $(N+3)\times 6$ rectangular grid of vertices positioned such that the origin is the vertex in the third row and fifth column of this rectangle. 
    The vertices in $A_N$, other than the origin, are shown in the diagram as large black dots.
    We are now able to define
    $$
    U_N^\sigma
    \coloneqq 
    F^\sigma_{\bar{\xi}^N}\prod\limits_{v\in A_N}\tau_v^x.
    $$
    We also define
    $$
    U_N^m\coloneqq F^m_{\bar{\xi}^N}.
    $$
    \begin{figure}[!ht]
    \begin{center}
\begin{tikzpicture}[scale=0.8]
    \filldraw[blue!30,thick,fill=blue!10](7.75,0)--(7.75,7.75)--(16,7.75)--(16,0);
    \filldraw[red!30,thick,fill=red!10](0,7.75)--(5.25,7.75)--(5.25,0)--(0,0);
    \foreach \x in {0,...,16}{
    \draw[thick,black](\x,0)--(\x,11);
    }
    \foreach \y in {0,...,11}{
    \draw[thick,black](0,\y)--(16,\y);
    }
    \draw[thick,black!50](9.5,11)--(9.5,0);
    \draw[thick,black,snake it](9.5,6.5)--(9.5,2.5)--(3.5,2.5)--(3.5,6.5);
     \draw[very thick, violet,snake it](9,0)--(9,6);
    \foreach \x in {0,...,3}{
    \draw[very thick,orange](3,3+\x)--(4,3+\x);
    \draw[very thick,orange](9,3+\x)--(10,3+\x);
    \draw[very thick,orange](\x+4,2)--(\x+4,3);
    }
    \draw[very thick,orange](8,2)--(8,3);
    \draw[very thick, orange] (9, 2) -- (9, 3);
    \foreach \x in {0,...,5}{
    \foreach \y in {0,...,6}{
    \filldraw[thick,black,fill=black](4+\x,3+\y)circle(.15cm);
    }
    }
   \filldraw[violet,thick,fill=yellow] (9,6) circle(.2cm);
   \draw[thick,red](9.25,6.25)--(9.25,11);
    \filldraw[thick,red,fill=red](9.25,6.25) circle(.09cm);
    
\end{tikzpicture}
\end{center}
    \caption{Geometry of $1^\sigma$ strings and $U_N^m$ and $U_N^\sigma$ operators in the case where $N=4$.}
     \label{fig:sigma}
\end{figure}

    In the second diagram (Figure \ref{fig:U}), we have drawn the path $\zeta^N$ in purple in the case where $N=4$. It extends $N$ edges down from $0$, travels along $5$ edges toward $\Delta$, and then extends upward $N$ edges. Note that $\zeta^N$ and $\overline{\xi}^N$ share the $N$th edge $e_N$ below the origin, that is, the site that corresponds to the midpoint of $\{0\} \times [-N + 1, -N]$.  From $\zeta^N$, we define
    $$
    U_N^\epsilon\coloneqq \tilde{F}^\epsilon_{\zeta^N}.
    $$
    We also define
    $$
    U_N^\psi\coloneqq U_N^\epsilon U_N^m.
    $$

    \begin{figure}[!ht]
\begin{center}
\begin{tikzpicture}[scale=0.8]
    \filldraw[blue!30,thick,fill=blue!10](7.75,0)--(7.75,7.75)--(16,7.75)--(16,0);
    \filldraw[red!30,thick,fill=red!10](0,7.75)--(5.25,7.75)--(5.25,0)--(0,0);
    \foreach \x in {0,...,16}{
    \draw[thick,black](\x,0)--(\x,11);
    }
    \foreach \y in {0,...,11}{
    \draw[thick,black](0,\y)--(16,\y);
    }
    \draw[thick,black!50](9.5,11)--(9.5,0);
    \draw[thick,black,snake it](9.5,6.5)--(9.5,2.5)--(3.5,2.5)--(3.5,6.5);

    \foreach \x in {0,...,5}{
    \foreach \y in {0,...,6}{
    \filldraw[thick,black,fill=black](4+\x,3+\y)circle(.15cm);
    }
    }
   \draw[very thick, violet](9,6)--(9,2)--(3,2)--(3,6);
   \filldraw[violet,thick,fill=yellow] (9,6) circle(.3cm);
       
\end{tikzpicture}
\end{center}
    \caption{Geometry of $U_N^\epsilon$ and $U_N^\psi$ operators in the case where $N=4$.}
     \label{fig:U}
\end{figure}

\end{nota}

\begin{nota}
\label{notation:DeltaSectors}

Observe that for that for $a \in \{1, \epsilon, m, \psi\}$, we have that $a \in (\GSec^f_{\tilde \pi})_1$ is canonically $1$-localized in $\Lambda$ and $a^\sigma \in (\GSec^f_{\tilde \pi})_g$ is canonically $g$-localized in $\Delta$.
We now define $a_\Delta \in (\GSec^f_{\tilde \pi})_1$ that are leftward $1$-localized in $\Delta$ and $a^\sigma_\Delta \in (\GSec^f_{\tilde \pi})_g$ that are leftward $g$-localized in $\Delta$ (Definition \ref{def:LeftwardLocalized}).
Recall the notation from Notation \ref{notation:defects} and Notation \ref{notation:braiding}
We let $\gamma' \in P(\Gamma)$ be the purple path in the figure. 
In particular, for $k = 1, \dots, 6$, we have that the edge site $\gamma'_k$ corresponds to the midpoint of $[-k, -k + 1] \times \{0\}$, and for $k \geq 7$, we have that the edge site $\gamma'_k$ corresponds to the midpoint of $\{-6\} \times [7-k, 6-k]$. 
We let $\bar \upsilon^1$ denote the cyan wiggling dual path in the figure. 
More precisely, for $k = 1, \dots, 6$, we have that the edge site $\bar \upsilon^1_k$ corresponds to the midpoint of $\{-k + 1\} \times [0, 1]$. 
Next, we let $\bar \eta \in \bar P(\Gamma)$ denote the black wiggling path in the figure. 
In particular, for $k \in \bbN$, we have that the edge site $\bar \eta_k$ corresponds to the midpoint of $[-6, -5] \times \{-k + 1\}$. 
We then let $\bar \gamma' \coloneqq \bar \upsilon^1 + \bar \eta$. 
For $a \in \{\epsilon, m, \psi\}$, we define $a_\Delta$ to be the extension to $\fA^a$ of $\alpha^\epsilon_{\gamma'}$, $\alpha^m_{\bar \gamma'}$, or $\alpha^\psi_{\gamma', \bar \gamma'}$ as appropriate.
Note that $a_\Delta$ is leftward $1$-localized in $\Delta$.

We will now define $1^\sigma_\Delta \in (\GSec^f_{\tilde \pi})_g$ that is leftward $g$-localized in $\Delta$. 
We let $\bar \upsilon^2$ be the green wiggling dual path in the figure. 
In particular, for $k = 1, 2, 3$, the edge site $\bar \upsilon^2_k$ corresponds to the midpoint of $[-6, -5] \times \{k\}$, and for $k = 4, \dots, 9$, the edge site $\bar \upsilon^2_k$ corresponds to the midpoint of $\{k - 9\} \times [3, 4]$.
Recall from Notation \ref{notation:defects} that $\bar \gamma^0$ is the dual path consisting of the edges intersected by $R = \{1/4\} \times \bbR_{\geq 1/4}$ (the red ray in the figure).
We then define $\bar \gamma^1 \coloneqq \bar \upsilon^2 + \bar \gamma^0 \in \bar P_R(\Gamma)$,  and we define $\bar{L}'$ to be the dividing dual path given by 
\[
\bar L'_k
\coloneqq
\begin{cases}
\bar \gamma^1_{k} & k \geq 1,
\\
\bar \eta_{1 - k} & k \leq 0.
\end{cases}
\]
We define $1^\sigma_\Delta$ to be the extension to $\fA^a$ of $\alpha^\sigma_{\bar \eta} \circ \beta_g^{r(\bar L')}$.
Furthermore, for $a \in \{\epsilon, m, \psi\}$, we define $a^\sigma_\Delta \coloneqq 1^\sigma_\Delta \otimes a_\Delta$. 
Note that each $a^\sigma_\Delta$ is leftward $g$-localized in $\Delta$ since $\Lambda^{+2} \cup r(\Lambda) \subseteq r(\bar L')$ and $\alpha^\sigma_{\bar \eta}$ is localized in $\Delta$.
    \begin{figure}[!ht]
    \begin{center}
\begin{tikzpicture}[scale=0.8]
    \filldraw[blue!30,thick,fill=blue!10](7.75,0)--(7.75,7.75)--(16,7.75)--(16,0);
    \filldraw[red!30,thick,fill=red!10](0,7.75)--(5.25,7.75)--(5.25,0)--(0,0);
    \foreach \x in {0,...,16}{
    \draw[thick,black](\x,0)--(\x,11);
    }
    \foreach \y in {0,...,11}{
    \draw[thick,black](0,\y)--(16,\y);
    }
    \draw[thick,black!50](9.5,11)--(9.5,0);
    \draw[thick,black,snake it](3.5,6.5)--(3.5,0);
    \draw[thick, cyan,snake it] (9.5, 6.5) -- (3.5, 6.5);
    \draw[thick, DarkGreen, snake it] (9.5, 9.5) -- (3.5, 9.5) -- (3.5, 6.5);
    \draw[very thick, violet] (9, 6) -- (3, 6) -- (3, 0);
   \filldraw[violet,thick,fill=yellow] (9,6) circle(.2cm);
   \draw[thick,red](9.25,6.25)--(9.25,11);
    \filldraw[thick,red,fill=red](9.25,6.25) circle(.09cm);
    
\end{tikzpicture}
\end{center}
    \caption{Geometry of $a_\Delta$ and $a^\sigma_\Delta$ for $a \in \{1, \epsilon, m, \psi\}$.}
     \label{fig:delta_loc_reps}
\end{figure}
\end{nota}

\begin{prop}\label{prop:well-defined_transport}

    For $a \in \{\epsilon, m, \psi\}$, the sequence $(U_N^a)$ converge in WOT to a unitary $U^a \colon a \to a_\Delta$. 
    Similarly, the sequence $(U_N^\sigma)$ converges in WOT to a unitary $U^\sigma \colon 1^\sigma \to 1^\sigma_\Delta$.
    
\end{prop}
\begin{proof}
The statement for $U^\epsilon, U^m, U^\psi$ follows by appropriately modifying the proof of \cite[Lem.~4.1]{MR2804555} to take into account that the string operators for our model are not exactly the usual Toric Code string operators.
We now prove this statement for $U^\sigma$. 
Recall the definitions of $\bar \upsilon^1, \bar \upsilon^2, \bar \eta, \bar \eta^1, \bar L'$ from Notation \ref{notation:DeltaSectors}.
We define $\bar L''$ to be the dividing dual path given by 
\[
\bar L''_k
\coloneqq
\begin{cases}
\bar \gamma^0_{k} & k \geq 1,
\\
\bar \gamma_{1 - k} & k \leq 0,
\end{cases}
\]
Observe that the unitary $U \coloneqq F^\sigma_{\bar{\upsilon}^1} \prod\limits_{v \in S} \tau^x_v$ intertwines $1^\sigma_\Delta$ and $\alpha^\sigma_{\bar \gamma} \circ \beta_g^{r(\bar L'')}$ (extended to $\fA^a$).
Note that Lemma \ref{lem:defects are cone transportable} provides the construction of a sequence of unitaries $\tilde U^\sigma_N$ that converge to a unitary intertwiner $\tilde U^\sigma \colon \alpha^\sigma_{\bar \gamma}$ from $1^\sigma$ to $\alpha^\sigma_{\bar \gamma} \circ \beta_g^{r(\bar L'')}$. 
In particular, $U^\sigma_N = \tilde U^\sigma_N U^*$, from which the result follows. 

\end{proof}

In principle, we can define more unitary intertwiner to transport the other symmetry defects from $\Lambda$ to $\Delta$.
We omit the definition of these operators because they will not be used in the computation of the $G$-crossed braiding data.

In what follows, we will use the definition of the anyon automorphisms in terms of conjugating by string operators, as per Remark \ref{rem:SETTCAutomorphismDefn}.

We remark that the braiding we use on the category of anyon representations is the reverse of that used by \cite{MR2804555}.

\begin{prop}\label{prop:TCbraiding}
    The only non-identity braiding isomorphisms among $\{1, \epsilon, m, \psi\}$ are given by
    $$
    c_{m,\epsilon}=c_{m,\psi}=c_{\psi,\epsilon}=c_{\psi,\psi}=-1.
    $$
\end{prop}
\begin{proof}
    Based on the geometry of our set up, we have that whenever $a\in\{1,\epsilon\}$ or $b\in \{1,m\}$,
    $$
    c_{a,b}=b((U^a)^*)U^a=1.
    $$
    Let $e_N$ denote the $N$th edge below $0$, that is, the site corresponding to the midpoint of $\{0\} \times [-N + 1, -N]$. 
    For ease of notation, for $i \in \{X, Y, Z\}$, we write $\sigma^i_N$ instead of $\sigma^i_{e_N}$.
    Using the definition $\alpha^m_{\bar \gamma}$ and $U_N^\epsilon$, we have that 
    $$
    m((U^\epsilon_N)^*)U^\epsilon_N=\sigma^x_{N} (U^\epsilon_N)^* \sigma^x_{N} U^\epsilon_N=\sigma^x_{N}\sigma^z_{N}\sigma^x_{N}\sigma^z_{N}=-1.
    $$
    Therefore, $c_{m,\epsilon}=-1$.

    By continuity, we have
    $$
    c_{m,\psi}=\psi((U^m)^*)U^m=\epsilon(m((U^m)^*))U^m=\epsilon((U^m)^*)U^m=c_{m,\epsilon}=-1.
    $$
    Using Facts \ref{facts:braiding}, we have that
    $$
    c_{\psi,b}=c_{m,b}c_{\epsilon,b}=c_{m,b}.
    $$
    Therefore, we also have $c_{\psi,\epsilon}=c_{\psi,\psi}=-1$.
\end{proof}

\begin{prop}\label{prop:sigma_epsilon}
We have the braiding $c_{1^\sigma,\epsilon}=\tau^z_{0}$. 
\end{prop}
\begin{proof}
    For this proof, take the $\gamma^N$ to be the path made up of the first $N$ edges of $\gamma$. Define the self-adjoint operator $X_N\coloneqq \prod\limits_{v\in A_N}\tau^x_v$, where $A_N$ is the region defined in Notation \ref{notation:braiding}.\footnote{In particular, the set $A_N$ corresponds to the grid of large black/yellow vertices in Figure \ref{fig:U}.}
    Then
    $$
    U_N^\sigma=X_N F^\sigma_{\bar{\xi}^N}=F^\sigma_{\bar{\xi}^N}X_N.
    $$

    As before, let $e_k$ denote the $k$th edge below $0$, that is, the site corresponding to the midpoint of $\{0\} \times [-k + 1, -k]$. 
    We label the $N+1$ vertices going from the top of edge $e_1 = \gamma_1^N$ to the bottom of edge $e_N = \gamma_N^N$ as $v_0,v_1, \dots, v_N$.
    In particular, $v_k = (0, k)$, and we have that   $v_0=\partial\gamma=\partial_0\gamma^N$ and $v_N=\partial_1\gamma^N$. 
    As before, for ease of notation, we write $\sigma^i_k$ instead of $\sigma^i_{e_k}$ and $\tau^i_{k}$ instead of $\tau^i_{v_k}$ for $i \in \{X, Y, Z\}$.
    Using this notation, we define
    $$
    S_N\coloneqq\prod\limits_{k=1}^N \sigma_{k}^z,
    \qquad\qquad
    T_n\coloneqq\prod\limits_{k=1}^{n}i^{\sigma^z_{k}(\tau^z_{k-1}-\tau^z_k)/2},
    $$
    so that $\tilde{F}^\epsilon_{\gamma^N}=S_N T_N$.
    A simple calculation reveals that $T_n^2=\tau_0^z\tau_n^z$.
    Also note that
    $$
    X_NT_N^*=X_NT_{N-1}^*i^{-\sigma^x_{N}(\tau_{N-1}^z-\tau_N^z)/2}=T_{N-1}X_N i^{-\sigma^x_{N}(\tau_{N-1}^z-\tau_N^z)/2}.
    $$
    We use these facts and notation to compute the following:
    \begin{align*}
    \epsilon((U^\sigma_{N})^*)&=\tilde{F}_{\gamma_N}^\epsilon (U^\sigma_{\gamma^N_N})^* (\tilde{F}_{\gamma^N}^\epsilon)^*\\
    &=\left[\tilde{F}_{\gamma^N}^\epsilon (F^\sigma_{\bar{\xi}^N})^* (\tilde{F}_{\gamma^N}^\epsilon)^* \right]\left[\tilde{F}_{\gamma^N}^\epsilon X_N (\tilde{F}_{\gamma^N}^\epsilon)^*\right]\\
    &=\left[\sigma^z_{N} (F^\sigma_{\bar{\xi}^N})^* \sigma^z_{N} \right]\left[\tilde{F}_{\gamma^N}^\epsilon X_N (\tilde{F}_{\gamma^N}^\epsilon)^*\right]\\
    &=\left[e^{-i\frac{\pi}{2}\sigma_{N}^x}(F^\sigma_{\bar{\xi}^N})^*\right]\left[S_NT_N X_N T_N^* S_N\right]\\
    &=\left[e^{-i\frac{\pi}{2}\sigma_{N}^x}(F^\sigma_{\bar{\xi}^N})^*\right]\left[S_NT_{N-1}^2 i^{\sigma^x_{N}(\tau_{N-1}^z-\tau_N^z)/2} X_N  i^{-\sigma^x_{N}(\tau_{N-1}^z-\tau_N^z)/2}S_N\right]\\
    &=\left[-i\sigma_{N}^x(F^\sigma_{\bar{\xi}^N})^*\right]\left[S_NT_{N-1}^2 i^{\sigma^x_{N}\tau_{N-1}^z}  X_N S_N\right]\\
    &=\left[-i\sigma_{N}^x(F^\sigma_{\bar{\xi}^N})^*\right]\left[S_N\tau^z_0\tau^z_{N-1} (i\sigma^x_{N}\tau_{N-1}^z)  X_N S_N\right]\\
    &=\sigma_{\gamma^N_N}^x\tau_0^z (F^\sigma_{\bar{\xi}^N})^* S_N\sigma^x_{\gamma^N_N} S_N X_N\\
    &=-\tau^z_0(F^\sigma_{\bar{\xi}^N})^* X_N\\
    &=-\tau_0^z (U_N^\sigma)^*.
    \end{align*}
    
    Using Lemma \ref{lem:symmetry_on_e}, we then have
    $$
    \gamma_g(\epsilon)((U_N^\sigma)^*)=\tau_0^z(-\tau^z_0(U_N^\sigma)^*)\tau_0^z=\tau_0^z(U_N^\sigma)^*.
    $$
    Taking the appropriate limits and using continuity, we obtain
    \[
    c_{1^\sigma,\epsilon}=\gamma_g(\epsilon)((U^\sigma)^*) U^\sigma=\tau_0^z.
    \qedhere
    \]    
\end{proof}

\begin{lem}\label{lem:sigma_triv_braids}
    In $\GSec_{\tilde \pi}^f$, we have $c_{1^\sigma,1}=c_{1^\sigma,m}=c_{1^\sigma,1^\sigma}=1$.
\end{lem}
\begin{proof}
    Take $a\in\{1,m,1^\sigma\}$. Then $\gamma_g(a)=a$ and $a((U^\sigma_N)^*)=(U^\sigma_N)^*$. Therefore,
    \[
    c_{1^\sigma,a}=\gamma_g(a)((U^\sigma)^*)U^\sigma=1.
    \qedhere
    \]
\end{proof}

\begin{lem}\label{lem:sigma_psi}
    We have the braiding $c_{1^\sigma,\psi}=\tau_{0}^z$.
\end{lem}
\begin{proof}
    By Lemma \ref{lem:symmetry_on_e}, we have that $\gamma_g(\psi)(-)=\tau_0^z\psi(-)\tau_0^z$.
    We may then perform the following computation:
    \begin{align*}
        c_{1^\sigma,\psi}&=\gamma_g(\psi)((U^\sigma)^*)U^\sigma =\tau^z_{0}\epsilon(m((U^\sigma)^*))\tau^z_{0}U^\sigma =\tau^z_{0}\epsilon(c_{1^\sigma,m}(U^\sigma)^*)\tau^z_{0}U^\sigma\\
        &=\tau^z_{0}\epsilon((U^\sigma)^*)\tau^z_{0}U^\sigma =\gamma_g(\epsilon)((U^\sigma)^*)U^\sigma =c_{1^\sigma,\epsilon} =\tau^z_{0}.
    \end{align*}
    where the fourth equality follows from the fact that $c_{1^\sigma,m}=1$ from Lemma \ref{lem:sigma_triv_braids}, the fifth equality follows from Lemma \ref{lem:symmetry_on_e}, and the last equality follows from Proposition \ref{prop:sigma_epsilon}.
\end{proof}

\begin{lem}\label{lem:finish_fifth_row}
    For $b\in \{1,\epsilon,m,\psi\}$, we have that $c_{1^\sigma,b^\sigma}=c_{1^\sigma,b}$.
\end{lem}
\begin{proof}
    Take $n=0$ if $b\in\{1,m\}$ and $n=1$ if $b\in\{\epsilon,\psi\}$. Using that $c_{1^\sigma,1^\sigma}=1$ by Lemma \ref{lem:sigma_triv_braids} and that $\gamma_g(\psi)(-)=\tau_0^z\psi(-)\tau_0^z$ by Lemma \ref{lem:symmetry_on_e}, we have that
    \begin{align*}
        c_{1^\sigma,b^\sigma}&=\gamma_g(b^\sigma)((U^\sigma)^*)U^\sigma =(\tau^z_0)^n b(1^\sigma((U^\sigma)^*))(\tau^z_0)^n U^\sigma\\
        &=(\tau^z_0)^n b(c_{1^\sigma,1^\sigma}(U^\sigma)^*)(\tau^z_0)^n U^\sigma =\gamma_g(b)(c_{1^\sigma,1^\sigma}(U^\sigma)^*) U^\sigma =c_{1^\sigma,b}.
        \qedhere
    \end{align*}
\end{proof}

\begin{lem}\label{lem:triv_a_sigma}
    We have $c_{a,1^\sigma}=1$ for all $a\in\{1,\epsilon,m,\psi\}$.
\end{lem}
\begin{proof}

    We first show that $1^\sigma((U^a_N)^*)=\beta_g^{r(\bar L)}((U^a_N)^*)=(U^a_N)^*$ for any $a\in\{1,\epsilon,m,\psi\}$.
    Indeed, for any $a\in\{1,\epsilon,m,\psi\}$, $U^a_N$ is supported outside of $r(\bar L)$, and $\beta_g^{r(\bar L)}$ acts trivially on all sites outside of $r(\bar L)$. 
    Taking the appropriate limits, we then have that for $a\in\{1,\epsilon,m,\psi\}$,
    \[
    c_{a,1^\sigma}=1^\sigma((U^a)^*)U^a=1.
    \qedhere
    \]
\end{proof}

\begin{lem}\label{lem:braid_absigma}
    For $a,b\in\{1,\epsilon,m,\psi\}$, we have that $c_{a,b^\sigma}=c_{a,b}$. 
\end{lem}
\begin{proof}
    Using the fact that $c_{a,b}$ is a scalar multiple of the identity from Proposition \ref{prop:TCbraiding},
    \begin{align*}
        c_{a,b^\sigma}&=b^\sigma((U^a)^*)U^a =1^\sigma(b((U^a)^*))U^a =1^\sigma(c_{a,b}(U^a)^*)U^a\\
        &=c_{a,b}1^\sigma((U^a)^*)U^a =c_{a,b}c_{a,1^\sigma} =c_{a,b}
        \qedhere
    \end{align*}
    where we have used and the fact that $c_{a,1^\sigma}=1$ from Lemma \ref{lem:triv_a_sigma}.
\end{proof}

\begin{lem}\label{lem:bottomleft}
    For $a,b\in\{1,\epsilon,m,\psi\}$, we have that $c_{a^\sigma,b}=c_{a,b}c_{1^\sigma,b}$
\end{lem}
\begin{proof}
    This is a special case of Facts \ref{facts:braiding} by the following:
    \[
    c_{a^\sigma,b}=c_{1^\sigma,b}1^\sigma(c_{a,b})=c_{a,b}c_{1^\sigma,b}.
    \qedhere
    \]
\end{proof}

\begin{lem}\label{lem:bottomright}
    For $a,b\in\{1,\epsilon,m,\psi\}$, we have that $c_{a^\sigma,b^\sigma}=c_{a,b}c_{1^\sigma,b}$.
\end{lem}
\begin{proof}
By Facts \ref{facts:braiding},
    $$
    c_{a^\sigma,b^\sigma}=c_{1^\sigma,b^\sigma}1^\sigma(c_{a,b^\sigma})=c_{1^\sigma,b^\sigma}c_{a,b^\sigma},
    $$
    where the second equality uses the fact that $c_{a,b^\sigma}$ is a scalar multiple of the identity operator by Lemma \ref{lem:braid_absigma} and Proposition \ref{prop:TCbraiding}. We then use Lemma \ref{lem:finish_fifth_row} to see that
    \[c_{a^\sigma,b^\sigma}=c_{1^\sigma,b}c_{a,b}.\qedhere\]
\end{proof}

\begin{thm}
    The $G$-crossed braiding of $\GSec_{\tilde \pi}^f$ is determined by the data in the following table:
\begin{center}
$$
\begin{array}{|c||c|c|c|c|c|c|c|c|}\hline
    c_{\pi_1,\pi_2} & \pi_2=1 &\epsilon & m &\psi & 1^\sigma & \epsilon^\sigma & m^\sigma & \psi^\sigma \\
    \hline\hline
  \pi_1=1   & 1&1 &1 &1 &1 &1 &1 &1 \\
  \hline
  \epsilon & 1&1 & 1 &1 &1 &1 &1 &1\\
  \hline
  m & 1&-1 & 1 &-1 &1 &-1 &1 &-1\\
  \hline
  \psi & 1&-1 & 1 &-1 &1 &-1 &1 &-1\\
  \hline
  1^\sigma   & 1& \tau^z_0 & 1&\tau^z_0&1 &\tau_0^z &1 &\tau^z_0 \\
  \hline
  \epsilon^\sigma & 1&\tau^z_0 &1 & \tau^z_0&1 &\tau^z_0 & 1&\tau^z_0 \\
  \hline
  m^\sigma & 1&-\tau^z_0 &1 &-\tau^z_0 &1 &-\tau^z_0 &1 &-\tau^z_0\\
  \hline
  \psi^\sigma & 1 &-\tau^z_0  &1 &-\tau^z_0 &1 &-\tau^z_0 &1 &-\tau^z_0 \\\hline
\end{array}
$$
\end{center}

\end{thm}
\begin{proof}
    The top left quadrant of this table is given in Proposition \ref{prop:TCbraiding}. The top right quadrant is then obtained from the top left quadrant and Lemma \ref{lem:braid_absigma}. The fifth row follows from Proposition \ref{prop:sigma_epsilon}, Lemma \ref{lem:sigma_triv_braids}, Lemma \ref{lem:sigma_psi}, and Lemma \ref{lem:finish_fifth_row}. The remainder of the bottom half is given by the top half, the fifth row, and Lemmas \ref{lem:bottomleft} and \ref{lem:bottomright}.

    We now show that this data determines the braiding on $\GSec_{\tilde \pi}^f$. 
    Let $\pi \in (\GSec_{\tilde \pi}^f)_{\hom}$ and $\sigma \in \GSec_{\tilde \pi}^f$. 
    Then there exist isometries $V_1, \dots, V_n, W_1, \dots, W_m \in \fA^a_0$ satisfying that $\sum_i V_i V_i^* = \mathds{1} = \sum_j W_j W_j^*$ such that $\pi(-) = \sum_i V_i \pi_i(-) V_i^*$ and $\sigma(-) = \sum_j W_j \sigma_j(-) W_j^*$, where each $\pi_i, \sigma_j$ is in our chosen basis for $\cK_0(\GSec_{\tilde \pi}^f)$. 
    By naturality of the braiding, we have that 
    \[
    c_{\pi, \sigma}
    =
    \sum_{i, j}c_{\pi, \sigma}(V_i V_i^* \otimes W_j W_j^*)
    =
    \sum_{i, j}(\gamma_{\partial \pi}(W_j) \otimes V_i)c_{\pi_i, \sigma_j}(V_i^* \otimes W_j^*),
    \]
    so $c_{\pi, \sigma}$ is determined by the values in the table. 
\end{proof}

\section{Discussion}

In this manuscript, we rigorously proved the expectation from \cite{PhysRevB.100.115147} that the symmetry defects of a 2+1D SET form a $G$-crossed braided tensor category. To do this, we defined symmetry defects in accord with the DHR paradigm. We demonstrated the utility of this definition by computing the defect category associated with SPTs and a lattice model of the $\mathbb{Z}_2$ symmetric Toric Code.

One potential direction for future work is to understand the role of antiunitary symmetries such as time reversal symmetry. A discussion of such SETs can be found in \cite{Barkeshli2020}, \cite{PhysRevLett.119.136801}, and \cite{PhysRevB.98.115129}. However, a detailed microscopic understanding of such bulk defects is missing, especially in the context of DHR theory.

We expect that there are many other lattice models which are amenable to our analysis. In particular, the models of SETs presented in \cite{PhysRevB.108.115144} give an extremely general class of models which are obtained by sequentially gauging abelian quotient groups of a global symmetry.
In addition to providing a large class of models to study, this research also suggests that it may be fruitful to understand the superselection theory in terms of gauging.

Finally, \cite{PhysRevB.94.235136} presents a model of the $\mathbb{Z}_2$-symmetric Toric Code where the symmetry swaps the anyons $\epsilon$ and $m$. 
In that example, the $\mathbb{Z}_2$-symmetry defects have non-integer quantum dimension, which provides an interesting challenge in terms of a DHR-style analysis.
This manuscript also presents a wide variety of other exactly solvable SETs which are related to string-nets by gauging the global symmetry.

\section*{Acknowledgments}

We would like to thank 
Juan Felipe Ariza Mej\'ia,  
Sven Bachmann, 
Alex Bols,
Tristen Brisky,
Chian Yeong Chuah, 
Martin Fraas,
Brett Hungar, 
Corey Jones, 
Michael Levin, 
Pieter Naaijkens, 
Bruno Nachtergaele, 
David Penneys, 
Sean Sanford, 
Wilbur Shirley, 
Daniel Spiegel, 
Dominic Williamson
for helpful conversations. 
We would also like to thank the referees for their suggestions and comments on the preprint. 
SV was funded by the NSF grant number DMS-2108390. 
SV thanks the VILLUM FONDEN for its support with a Villum Young Investigator (Plus) Grant (Grant No. 25452 and Grant No. 60842) as well as via the QMATH Centre of Excellence (Grant No. 10059).
DW was funded by the NSF grant number DMS-2154389.
KK was funded by NSF DMS 2231533, NSF DMS 1654159, and the Center for Emergent Materials at The Ohio State University, an NSF-funded MRSEC, under Grant No. DMR-2011876.

\begin{appendix}

\addtocontents{toc}{\protect\setcounter{tocdepth}{1}}
\section{Introduction to operators algebras and category theory}
\label{sec:background}

\subsection{Operator algebras}
\label{sec:operator algebra basics}
In this section we provide a brief introduction to the operator algebraic approach to quantum spin systems on infinite lattices.  
For more detail, we refer the reader to \cite{MR3617688, MR887100, MR1441540}.

\subsubsection{States and representations} Let $\omega$ be a state on $\cstar$, meaning a positive linear functional of norm $1$. We denote by $\cS(\cstar)$ the space of all states on $\cstar$. 

Using a construction by Gelfand, Naimark, Segal (the GNS construction for short) one can associate to $(\omega, \cstar)$ a GNS triple $(\pi, \hilb, \ket{\Omega})$ where $\hilb$ is a Hilbert space, $\pi\colon \cstar \rightarrow B(\hilb)$ is a $^*$-representation on $\hilb$ and $\ket{\Omega} \in \hilb$ is a cyclic vector, such that for all $A \in \cstar$ we have $\omega(A) = \inner{\Omega}{\pi(A) \Omega}$. The GNS triple for any state $\omega$ is unique up to unitary equivalence. 

We say that a state is \emph{pure} if for every $\phi \colon \fA \to \bbC$ satisfying that $0 \leq \phi \leq \omega$, we have that $\phi = \phi(\mathds1)\omega$. If $\omega$ is a pure state, then its GNS representation $\pi$ is irreducible.

If two representations $(\pi_1, \hilb_1)$ and $(\pi_2, \hilb_2)$ are unitarily equivalent, then we write $\pi_1 \simeq \pi_2$. Two states $\omega_1, \omega_2$ of $\cstar$ are unitarily equivalent (denoted again by $\omega_1 \simeq \omega_2$) if their GNS representations are unitarily equivalent.

\subsubsection{Dynamics}
\label{sec:HamiltonianDynamics}
For us, a Hamiltonian $H$ is a collection of self-adjoint local Hamiltonians $H_S \in \cstar[S]$ for $S \in \Gamma_f$. 
In our examples, the Hamiltonians $H$ will be of the form $H_S = \sum_{Z \subseteq S} \Phi(Z)$ for $S \in \Gamma_f$.  
Here $\Phi \colon \Gamma_f \to \fA^{\loc}$ is a map that satisfies the following conditions:
\begin{itemize}
\item 
$\Phi(Z) \in \cstar[Z]$ for $Z \in \Gamma_f$, and 
\item 
$\Phi(Z) \geq 0$ for all $Z \in \Gamma_f$.
\end{itemize}
We call the $\Phi(Z)$ \emph{interactions}.
We call the interactions \emph{finite range} if there exists $n > 0$ such that $\Phi(Z) = 0$ if $Z$ is not contained in a ball of radius $n$. We say that interactions are \emph{uniformly bounded} if there is some $N > 0$ such that for every $Z \in \Gamma_f$, we have that $\|\Phi(Z)\| \leq N$.

In the infinite volume limit, the net of operators $(H_S)$ does not converge in norm. However, under the finite range assumption, for any observable $A \in \cstar^{\loc}$, the limit $\delta(A) \coloneqq \lim_{S \rightarrow \Gamma} i [H_S, A]$ exists and extends to a densely defined unbounded $^*$-derivation $\delta$ on $\cstar$.
Note that in this case, we have that for $A \in \fA^{\loc}$ with $\supp(A) = S$, 
\[
\delta(A)
=
i\left[\sum_{Z \cap S \neq \emptyset} \Phi(Z), A\right],
\]
and the sum is finite since the interactions are finite range. In our examples, the interactions will be uniformly bounded and finite range.

A state $\omega_0$ is called a \emph{ground state} of $\delta$ if for all $A \in \cstar^{\loc}$ we have $$-i \omega_0(A^* \delta(A)) \geq 0,$$ and it is \emph{gapped} if there is some $g > 0$ such that for all $A \in \cstar^{\loc}$ satisfying $\omega_0(A)=0$, we have $$-i \omega_0(A^* \delta(A)) \geq g \omega_0 (A^*A).$$

Recall that we are working in the setting where our interactions are positive operators. 
We then have that the local Hamiltonians $H_S$ are also positive. 
In this context, a ground state $\omega_0$ is called \emph{frustration free} if for all $S \in \Gamma_f$ we have $\omega_0(H_S) = 0$.\footnote{This definition is applicable in the more general context of frustration free interactions.} 
Note that by \cite[Lem.~3.8]{MR3764565}, a state $\omega_0 \colon \fA \to \bbC$ is a frustration free ground state if and only if $\omega_0(H_S) = 0$ for all $S \in \Gamma_f$. In our examples, we will have that there is a unique frustration free ground state $\omega_0$ for the derivation under consideration. 
By a standard argument (see for instance \cite[Cor.~2.24]{2307.12552}), $\omega_0$ must be a pure state.  
Indeed, suppose $\phi \colon \fA \to \bbC$ satisfies that $0 \leq \phi \leq \omega_0$.
Then for all $S \in \Gamma_f$ we have that
\[
0 
\leq
\phi(H_S)
\leq
\omega_0(H_S)
=
0,
\]
so $\phi(H_S) = 0$ for all $S \in \Gamma_f$.
Therefore, the map $\omega \colon \fA \to \bbC$ given by $\omega(A) = \frac{1}{\phi(\mathds1)}\phi(A)$ for $A \in \fA$ is a state satisfying that $\omega(H_S) = 0$ for all $S \in \Gamma_f$, so $\omega$ is a frustration free ground state.  
Thus, $\omega = \omega_0$ and hence $\phi = \phi(\mathds1)\omega_0$.

We let $(\pi_0, \hilb_0)$ be the GNS representation of $\omega_0$.

\subsubsection{von Neumann algebras} Let $(\pi_0, \hilb_0)$ be the GNS representation of the state $\omega_0 \colon \cstar \to \bbC$. For each set $S \subseteq \Gamma$ we can denote $\caR(S) \coloneqq \pi_0 (\cstar[S])'' \subseteq B(\hilb_0)$ where ($'$) denotes the commutant in $B(\hilb_0)$.
Equivalently, $\cR(S)$ is the closure of $\pi_0(\cstar[S])$ in the WOT-topology.  
In more detail, if $(A_i)$ is a net in $B(\cH_0)$, then $A_i \to A$ if for all $\overline{\xi}, \eta \in \cH_0$, we have that $\langle \eta, A_i \overline{\xi} \rangle \to \langle \eta, A\overline{\xi}\rangle$.
In the case that the state $\omega_0$ is pure, the algebras $\cR(S)$ are factors, meaning that they have trivial center.  

There is a useful notion of two projections in a von Neumann algebra $M$ being equivalent.
If $p, q \in M$ are two projections, we say that $p, q$ are \emph{Murray von-Neumann equivalent}, denoted $p \sim q$, if there exists $v \in M$ such that $v^*v = p$ and $vv^*=q$.
A von Neumann algebra $M$ is said to be \emph{infinite} if there exists $p \in M$ such that $p \neq \mathds1$ but $p \sim \mathds1$ in $M$.
There is a more specific notion of a von Neumann algebra $M$ being properly infinite; however, in the case that $M$ is a factor, this is equivalent to being infinite.
We will consider regions $\Lambda \subseteq \Gamma$ (specifically cones) such that the algebras $\cR(\Lambda)$ are infinite factors.

\subsubsection{Automorphisms of the quasi-local algebra}
\label{sec:AutomorphismsOfQLAlgebra}

In this subsection, we discuss various types of automorphisms that preserve the structure of the quasi-local algebra. 
Often, we wish to consider automorphisms that preserve locality up to some spread; these are termed \emph{quantum cellular automata} \cite{quant-phys/0405174}. 

\begin{defn}
\label{def:QCA definition}
A $*$-automorphism $\alpha \colon \fA \to \fA$ is a \emph{quantum cellular automaton} (\emph{QCA} for short) if there exists $s > 0$ such that $\alpha(\cstar[S]) \subseteq \cstar[S^{+s}_\Gamma]$ and $\alpha^{-1}(\cstar[S]) \subseteq \cstar[S^{+s}_\Gamma]$, where $S^{+s}_\Gamma$ is the set of sites in $\Gamma$ that are distance at most $s$ from $S$.
The minimum such $s$ is called the \emph{spread} of the QCA $\alpha$.
\end{defn}

\begin{rem}
\label{rem:InverseQCA}
Observe that if $\alpha \colon \fA \to \fA$ is a QCA with spread $s$, then so is $\alpha^{-1}$. 
Indeed, let $y \in \fA_{(S^{+s}_\Gamma)^c}$ and $x \in \fA_S$.
We wish to show that $[\alpha^{-1}(x), y] = 0$. 
Note that this will imply the desired result, since in this case we have that 
\[
\alpha^{-1}(\fA_S)
\subseteq
\fA_{(S^{+s}_\Gamma)^c}' \cap \fA
=
\fA_{S^{+s}_\Gamma}.
\]
To see that $[\alpha^{-1}(x), y] = 0$, observe that $\alpha(y) \in \fA_{((S^{+s}_\Gamma)^c)^{+s}_\Gamma} = \fA_{S^c}$, so $[x, \alpha(y)] = 0$.
Applying $\alpha^{-1}$, we obtain the desired result.
\end{rem}

\begin{lem}
\label{lem:QCAs preserve the ground state subspace}
    Let $H_1$ be a Hamiltonian with finite range interactions $\Phi_1(Z)$ for $Z \in \Gamma_f$, and let $\delta_1$ the corresponding derivations. 
    
    Let $\alpha \colon \fA \to \fA$ be a QCA with spread $s$, and for $Z \in \Gamma_f$, define $\Phi_2(Z) \coloneqq \alpha(\Phi_1(Z))$.
    Let $H_{2}$ be the Hamiltonian corresponding to the interactions $\Phi_2(Z)$, and let $\delta_2$ the corresponding derivation. 
    If $\omega_2$ is a ground state of derivation $\delta_2$, then $\omega_1 \coloneqq \omega_2 \circ \alpha$ is a ground state of $\delta_1$.
\end{lem}
\begin{proof}
    Since $\omega_2$ is a ground state of $\delta_2$, we have for all $A \in \cstar^{\loc}$ that $$-i \omega_2(A^* \delta_2(A)) \geq 0.$$
    Now let $A \in \cstar^{\loc}$ with $\supp(A) = S$. Then we have,
    \begin{align*}
        -i \omega_1(A^* \delta_1(A)) &= -i \omega_1\left(A^*i\left[\sum_{Z \cap S \neq \emptyset} \Phi_1(Z), A\right]\right) = \omega_2 \circ \alpha \left(A^*\left[\sum_{Z \cap S \neq \emptyset} \Phi_1(Z), A\right]\right)\\
        &= \omega_2\left(\alpha(A)^*\left[\sum_{Z \cap S \neq \emptyset} \alpha(\Phi_1(Z)), \alpha(A)\right]\right) \\
        &= \omega_2\left(\alpha(A)^*\left[\sum_{Z \cap S \neq \emptyset} \Phi_2(Z), \alpha(A)\right]\right),
    \end{align*}
    where the last line follows by how $\Phi_2(Z)$ is defined. 
    On the other hand,
    \[
    \delta_2(\alpha(A))
    =
    i \left[\sum_{Z \cap S \neq \emptyset} \Phi_2(Z), \alpha(A)\right]
    \]
    Therefore, we have that 
    \[
    -i \omega_1(A^* \delta_1(A))
    =
    -i\omega_2(\alpha(A)^*\delta_2(\alpha(A))
    \geq
    0.
    \]
    Thus $\omega_1$ is indeed a ground state of $\delta_1$.
\end{proof}

By Remark \ref{rem:InverseQCA}, the set of QCAs forms a group.
A special type of QCA is a \emph{finite depth quantum circuit} (FDQC for short); see Definition \ref{def:FDQC}.

\begin{lem}
    \label{lem:FDQCBoundedSpread}
    Let $\alpha \colon \fA \to \fA$ be the FDQC built from $\{\cU^d\}_{d = 1}^D$, where each unitary in $\cU^d$ has support contained in a ball of diameter $N$. Then $\alpha$ is a QCA with spread at most $ND$.
\end{lem}

\begin{proof}
It suffices to show that $\alpha_d$ is a QCA with spread at most $N$ for all $d = 1, \dots, D$.  
Suppose that $S \subseteq \Gamma$ and $A \in \cstar[S]^{\loc}$
Then we have that 
\[
\alpha_d(A)
=
\Ad \! \left(\prod_{U \in \cU^d} U \right)\!(A).
\]
Since the support of each $U \in \cU^d$ is contained in a ball of diameter $N$, we have that $$\supp(\Ad \! \left(\prod_{U \in \cU^d} U \right)\!(A)) \subseteq \supp(A)^{+N}_\Gamma.$$  
Thus $A \in \cstar[S^{+N}_\Gamma]$, so $\alpha_d$ is a QCA of spread at most $N$.
\end{proof}

Another useful notion is the notion of a \emph{quasi-factorizable} automorphism; these have been studied in \cite{MR4426734,MR4362722} as maps that preserve the anyon data when precomposed with the ground state.

\begin{defn}
    Let $\alpha$ be an automorphism of $\cstar$ and consider an inclusion of cones
    \[
    \Gamma_1' \subset \Lambda \subset \Gamma_2'
    \]
    We say that $\alpha$ is \emph{quasi-factorizable} with respect to this inclusion if there is a unitary $u \in \cstar$ and automorphisms $\alpha_\Lambda$ and $\alpha_{\Lambda^c}$ of $\cstar[\Lambda]$ and $\cstar[\Lambda^c]$ respectively, such that
    \[
    \alpha = \operatorname{Ad}(u) \circ \widetilde{\Xi} \circ (\alpha_\Lambda \otimes \alpha_{\Lambda^c}),
    \]
    where $\widetilde{\Xi}$ is an automorphism on $\cstar[\Gamma_2' \setminus \Gamma_1']$.
\end{defn}

\begin{lem}
\label{lem:FDQCQuasi-Factorizable}
If $\alpha \colon \fA \to \fA$ is a finite depth quantum circuit, then for every cone $\Lambda$, we have that $\alpha$ is quasi-factorizable with respect to some inclusion of cones $\Gamma'_1 \subset \Lambda \subset \Gamma_2'$.
\end{lem}

\begin{proof}
We first observe that for each $d = 1, \dots, D$, we may assume that $\bigcup_{U \in \cU^d} \supp(U) = \Gamma$.  
Indeed, if this is not the case, we can always include $\mathds1_s$ for every $s \notin \bigcup_{U \in \cU^d} \supp(U)$ to $\cU^d$.
We now let $\Lambda$ be a cone.  
We define ${\cU}^1_{\mathrm{in}} \coloneqq \{U \in \cU^1 : U \in \cstar[\Lambda]\}$ and ${\cU}^1_{\mathrm{out}} \coloneqq \{U \in \cU^1 : U \in \cstar[\Lambda^c]\}$.
We also define $\Lambda_0 \coloneqq \Lambda$ and $\Lambda_0' \coloneqq \Lambda^c$.
For $d = 1, \dots, D - 1$, we inductively define 
\begin{align*}
\Lambda_d 
&\coloneqq 
\bigcup_{U \in {\cU}^d_{\mathrm{in}}} \supp(U), 
&
\Lambda_d' 
&\coloneqq 
\bigcup_{U \in {\cU}^d_{\mathrm{out}}} \supp(U),
\\
{\cU}^{d + 1}_{\mathrm{in}}
&\coloneqq
\{U \in \cU^{d + 1} : U \in \cstar[\Lambda_d]\},
&
{\cU}^{d + 1}_{\mathrm{out}}
&\coloneqq
\{U \in \cU^{d + 1} : U \in \cstar[\Lambda_d']\}.
\end{align*}
Observe that for all $d = 1, \dots, D - 1$, we have that $\Lambda_{d - 1} \subseteq \Lambda_d$ and $\Lambda_{d - 1}' \subseteq \Lambda_d'$. 
In particular, for all $d = 0, 1, \dots, D - 1$, we have that $\Lambda_d \subseteq \Lambda$ and $\Lambda_d' \subseteq \Lambda^c$.

For $d = 1, \dots, D$, we define $\alpha_d^{\mathrm{in}} \colon \cstar \to \cstar$ and $\alpha_d^{\mathrm{out}} \colon \cstar \to \cstar$ by
\[
\alpha_d^{\mathrm{in}}(A)
\coloneqq
\Ad \! \left(\prod_{U \in {\cU}^d_{\mathrm{in}}} U \right)\!(A),
\qquad\qquad
\alpha_d^{\mathrm{out}}(A)
\coloneqq
\Ad \! \left(\prod_{U \in {\cU}^d_{\mathrm{out}}} U \right)\!(A).
\]
for $A \in \cstar^{\loc}$.
Note that since $\Lambda_d \subseteq \Lambda$ and $\Lambda_d' \subseteq \Lambda^c$ for all $d = 0, 1, \dots, D - 1$, we have that $\alpha_d^{\mathrm{in}}$ is an automorphism of $\cstar[\Lambda]$ and $\alpha_d^{\mathrm{out}}$ is an automorphism of $\cstar[\Lambda^c]$ for all $d = 1, \dots, D$.
We therefore have that 
\[
\alpha_\Lambda 
\coloneqq
\alpha_D^{\mathrm{in}} \circ \dots \circ \alpha_1^{\mathrm{in}}, 
\qquad\qquad
\alpha_{\Lambda^c} 
\coloneqq
\alpha_D^{\mathrm{out}} \circ \dots \circ \alpha_1^{\mathrm{out}}
\]
are automorphisms of $\cstar[\Lambda]$ and $\cstar[\Lambda^c]$ respectively.  

We now consider the automorphism 
\[
\alpha_\Lambda \otimes \alpha_{\Lambda^c}
=
(\alpha_D^{\mathrm{in}} \otimes \alpha_D^{\mathrm{out}}) \circ \dots \circ (\alpha_1^{\mathrm{in}} \otimes \alpha_1^{\mathrm{out}}).
\]
We observe that for all $d = 1, \dots, D$, 
\[
\alpha_d^{\mathrm{in}} \otimes \alpha_d^{\mathrm{out}}(A)
=
\Ad \! \left(\prod_{U \in {\cU}^d_{\mathrm{in}} \cup {\cU}^d_{\mathrm{out}}} U \right)\!(A),
\]
for $A \in \cstar^{\loc}$.
For $d = 1, \dots, D$, we define $\widehat{\cU}^d \coloneqq \cU^d \setminus ({\cU}^d_{\mathrm{in}} \cup {\cU}^d_{\mathrm{out}})$, and we define $\Xi_d \colon \cstar \to \cstar$ by 
\[
\Xi_d(A)
\coloneqq
\Ad \! \left(\prod_{U \in \widehat{\cU}^d} U \right)\!(A)
\]
for $A \in \cstar^{\loc}$.
We similarly define $\Xi \colon \cstar\to\cstar$ by 
\(
\Xi
\coloneqq
\Xi_D \circ \dots \circ \Xi_1.
\)
Note that $\alpha_d = \Xi_d \circ (\alpha_d^{\mathrm{in}} \otimes \alpha_d^{\mathrm{out}})$.
By how $\alpha_d^{\mathrm{in}}$ and $\alpha_d^{\mathrm{out}}$ were defined, we have that $\Xi_d$ commutes with $\Xi_{d'}$ for all $d' \geq d$.  
Therefore, we have that 
\[
\Xi \circ (\alpha_\Lambda \otimes \alpha_{\Lambda^c})
=
(\Xi_D \circ (\alpha_D^{\mathrm{in}} \otimes \alpha_D^{\mathrm{out}})) \circ \dots \circ (\Xi_1 \circ (\alpha_1^{\mathrm{in}} \otimes \alpha_1^{\mathrm{out}}))
=
\alpha_D \circ \dots \circ \alpha_1
=
\alpha.
\]

It remains to show that there exists an inclusion of cones 
\(
\Gamma'_1 \subset \Lambda \subset \Gamma'_2
\)
such that $\Xi$ is an automorphism on $\cstar[\Gamma_2' \setminus \Gamma_1']$.
At this point, we use the assumption that $\bigcup_{U \in \cU^d} \supp(U) = \Gamma$ for each $d = 1, \dots, D$.
We also use the fact that every $U \in \bigcup_{d = 1}^D \cU^d$ has support at most $N$.
By these two facts, the unitaries in $\widehat{\cU}^1 = \cU^1 \setminus (\cU^1_{\mathrm{in}} \cup \cU^1_{\mathrm{out}})$ are all supported in the strip $\Delta_1 \coloneqq \Lambda^{+N} \cap (\Lambda^c)^{+N}_\Gamma$.  
Similarly, for each $d = 2, \dots, D$, we have that the unitaries in $\widehat{\cU}^d$ are supported in the strip $\Delta_d \coloneqq (\Delta_{d - 1})^{+N}_\Gamma = \Lambda^{+dN} \cap (\Lambda^c)^{+dN}_\Gamma$.
Therefore, we have that all unitaries in $\bigcup_{d = 1}^D \widehat{\cU}^d$ are supported in the strip $\Delta_D = \Lambda^{+DN} \cap (\Lambda^c)^{+DN}_\Gamma$.
In particular, $\Xi$ is an automorphism on $\cstar[\Delta_D]$.
Now, if we let $\Gamma'_1 \coloneqq \left((\Lambda^c)^{+DN}_\Gamma\right)^c$ and $\Gamma'_2 \coloneqq \Lambda^{+DN}$, then $\Gamma'_1 \subset \Lambda \subset \Gamma'_2$ and $\Delta_D = \Gamma_2' \setminus \Gamma_1'$.
The result follows.
\end{proof}

We recall the notion of bounded spread Haag duality (Definition \ref{def:BSHaagDuality}).
We then have the following result, which is a special case of \cite[Prop.~5.10]{2410.21454}. 

\begin{lem}
\label{lem:QCABSHaagDuality}
If $\pi \colon \cstar \to B(\cH)$ satisfies strict Haag duality and $\alpha \colon \cstar \to \cstar$ is a QCA with spread $s$, then $\pi \circ \alpha$ satisfies bounded spread Haag duality with spread at most $2s$.
\end{lem}

\begin{proof}
This result is analogous to \cite[Prop.~5.10]{2410.21454}, but we repeat the proof for clarity and to avoid subtle geometric issues.
Note that for every cone $\Lambda$, 
\begin{align*}
\pi \circ \alpha(\cstar[\Lambda^{+2s}])'
&\subseteq
\pi \circ \alpha(\alpha^{-1}(\cstar[\Lambda^{+s}]))'
&&
\text{($\alpha^{-1}$ has spread $s$ by Remark \ref{rem:InverseQCA})}
\\&=
\pi(\cstar[\Lambda^{+s}])'
\\&=
\pi(\fA_{(\Lambda^{+s})^c})''
&&
\text{($\pi$ satisfies strict Haag duality)}
\\&\subseteq
\pi \circ \alpha(\fA_{((\Lambda^{+s})^c)^{+s}_\Gamma})''
&&
\text{($\alpha$ has spread $s$)}
\\&=
\pi \circ \alpha(\fA_{\Lambda^c})''.
\end{align*}
Thus $\pi \circ \alpha(\fA_{\Lambda^c})' \subseteq \pi \circ \alpha(\cstar[\Lambda^{+2s}])''$, so $\pi \circ \alpha$ satisfies bounded spread Haag duality with spread at most $2s$.
\end{proof}

\subsection{Category theory}
\label{sec:basics of cat thy}
In this section we introduce the main category-theoretic notions used in this paper. 
For more details, the reader can consult \cite{MR3242743} for the algebraic setting and \cite{MR808930, MR3687214} for the $\rmC^*$-/$\rmW^*$-setting.
\subsubsection{Basics of tensor categories}
\label{sec:basics of monoidal cats}
In our examples, we will be working with a category $\cC$ that is a linear $\rmW^*$-category. A \emph{linear category} is a category $\cC$ such that $\Hom{a \to b}$ is a (complex) vector space for all $a, b \in \cC$ and composition is bilinear.  
Following \cite[Def.~1.1]{MR808930}, a linear category $\cC$ is a $\rmC^*$-category if:
\begin{itemize}
    \item $\Hom{a \to b}$ is a Banach space with a norm $||\cdot||$ for all $a, b \in \cC$.
    \item Composition is submultiplicative, i.e., $||g \circ f|| \leq ||g|| \,\,||f||$ for all $f \in \Hom{a \to b}, g \in \Hom{b \to c}$.
    \item There is an involutive anti-linear map $(-)^* \colon \Hom{a \to b} \to \Hom{b \to a}$ such that for all $f \colon a \to b$ and $g \colon b \to c$ in $\cC$, we have $(g \circ f)^* = f^* \circ g^*$.
    \item Morphisms satisfy the $\rmC^*$-identity, i.e., $||f^* \circ f|| = ||f||^2$ for all $a,b \in \cC$ and $f \in \Hom{a\to b}$.
    \item $\End(a) \coloneqq \Hom{a \to a}$ is a $\rmC^*$-algebra for all $a \in \cC$.
\end{itemize}
If only the third bullet point is satisfied, we say that $\cC$ is a \emph{dagger category}.
Furthermore, following \cite[Def.~2.1]{MR808930}, a $\rmC^*$-category is a \emph{$\rmW^*$-category} if:
\begin{itemize}
    \item $\Hom{a \to b}$ is a dual Banach space (i.e., it has a specified predual) for all $a,b \in \cC$.
    \item The involution $f \mapsto f^*$ is normal (i.e. continuous in the weak$^*$-topology).
    \item Composition of morphisms is bi-normal (i.e. normal in each variable).
\end{itemize}
If $\cC$ is a $\rmW^*$-category, it follows that $\End(a)$ into a von Neumann algebra for all $a \in \cC$ in a $\rmW^*$-category $\cC$. 

The $\rmC^* / \rmW^*$-categories we consider are \emph{orthogonal Cauchy complete} \cite{2411.01678}, meaning that they admit all finite orthogonal direct sums and subobjects. Given $a_1, \dots, a_n \in \cC$, the \emph{orthogonal direct sum} of $a_1, \dots, a_n$ is an object $\bigoplus_{i = 1}^n a_i$ along with morphisms $v_j \colon a_j \to \bigoplus_{i = 1}^n a_i$ for all $j \in \{1, \dots, n\}$ that satisfy the following properties: 
\begin{itemize}
\item $v_i^*v_j = \delta_{ij} \Id_{a_i}$ for all $i \in \{1, \dots, n\}$, and
\item $\sum_{i = 1}^n v_i v_i^* = \Id_{\bigoplus_{i = 1}^n a_i}$.
\end{itemize}
Note that the orthogonal direct sum $\bigoplus_{i = 1}^n a_i$ is unique up to a unique isomorphism. We will also often omit the word `orthogonal' for simplicity. Similarly, we say that $\cC$ \emph{admits all subobjects} if for every orthogonal projection $p \colon a \to a$ in $\cC$ (that is, a morphism satisfying that $p^* = p = p^2$), there exists an object $b \in \cC$ (called a \emph{subobject}) and an isometry $v \colon b \to a$ such that $v^*v = \Id_b$ and $vv^* = p$. The property of admitting subobjects is also called \emph{projection complete}, although we do not use this term in this paper. As with direct sums, given a projection $p \colon a \to a$ in $\cC$, any two subobjects corresponding to $p$ are isomorphic.

\begin{rem}
\label{rem:JustCheckEndsWhenDirectSums}
As noted in \cite{MR3687214}, a dagger category $\cC$ that admits orthogonal direct sums is a $\rmC^*$-category if $\End(a)$ is a $\rmC^*$-algebra for all $a \in \cC$. 
Furthermore, by \cite[Lem.~2.6]{MR808930}, a $\rmC^*$-category $\cC$ that admits orthogonal direct sums is a $\rmW^*$-category if $\End(a)$ is a $\rmW^*$-algebra for all $a \in \cC$. 
\end{rem}

A (linear) \emph{functor} $F\colon \cC\to\cD$ (for $\cC, \cD$ linear categories) consists of the following:
\begin{itemize}
    \item 
    a map $a\mapsto F(a)$ from objects $a \in \cC$ to objects $F(a) \in \cD$, and
    \item 
    (linear) maps $F \colon \Hom{a \to b} \to \Hom{F(a) \to F(b)}$ for all $a, b \in \cC$ that satisfy $F(\Id_a)=\Id_{F(a)}$ and $ F(g\circ f)=F(g)\circ F(f)$.
\end{itemize} 
A \emph{natural transformation} $\eta\colon F\Rightarrow G$ between functors $F,G\colon \cC\to\cD$ is a collection of morphisms $\eta_a\colon F(a)\to G(a)$ such that $G(f)\circ \eta_a=\eta_b\circ F(f)$ for every $f\colon a\to b$. If each $\eta_a$ is an isomorphism, then $\eta$ is called a \emph{natural isomorphism}. 

An equivalence between (linear) categories $\cC, \cD$ is a (linear) functor $F \colon \cC \to \cD$ such that there exists a functor $G \colon \cD \to \cC$ such that $F \circ G  \simeq \Id_{\cD}$ and $G \circ F \simeq \Id_{\cC}$ where $\simeq$ denotes natural isomorphism. Equivalently, $F$ is an equivalence iff it is fully faithful and essentially surjective.

For $\rmC^*$-categories and dagger categories more generally, the functors we consider are linear \emph{dagger functors}, which are functors $F \colon \cC \to \cD$ such that $F(f^*) = F(f)^*$ for any $f \colon a \to b$ in $\cC$. 
Additionally, for $\rmW^*$-categories, we consider \emph{normal} dagger functors, that is, functors for which the induced maps $$\Hom{a\to b}\to \Hom{F(a)\to F(b)}$$ are weak$^*$-continuous.

The categories we consider will also be monoidal categories. 
A \emph{monoidal category} with a strict unit is a category $\cC$ with a bilinear functor $-\otimes -\colon \cC \times \cC \rightarrow \cC$, often termed \emph{tensor}, along with a family of natural isomorphisms $\alpha_{a, b, c} \colon (a \otimes b) \otimes c \to a \otimes (b \otimes c)$, called the \emph{associator}, that satisfies the pentagon equation (as given in \cite[Eqn.~(2.2)]{MR3242743}). Moreover, \emph{unitality} is satisfied, i.e., there is a distinguished object $\mathds1 \in \cC$ called the \emph{tensor unit}, satisfying $\mathds1 \otimes a = a = a \otimes \mathds1$ for all $a \in \cC$ and the associator $\alpha_{a,b,c}$ satisfies $\alpha_{\mathds1, a ,b } = \alpha_{a, \mathds1, b} = \alpha_{a,b, \mathds 1} = \Id_{a\otimes b}$.\footnote{Technically speaking, these are strictly unital monoidal categories, and ``monoidal category" is usually taken to be a slightly more general notion. 
However, since all the monoidal categories we will consider have a strict unit, we define monoidal categories in this way.}
A category $\cC$ is a \emph{strict monoidal category} if the associator is trivial.
We define a \emph{tensor category} to be a Cauchy complete linear monoidal category such that the tensor unit is simple, that is, $\End(\mathds{1}) \simeq \bbC$. 
However, other definitions of tensor category exist in the literature (see for instance \cite{MR3242743}).

We now define $\rmC^*$- and $\rmW^*$-tensor categories. 
A tensor category $\cC$ is a \emph{$\rmC^*$-tensor category} if it is a $\rmC^*$-category, the associators $\alpha_{a, b, c}$ are unitaries, and $(f \otimes g)^* = f^* \otimes g^*$ for any $f \colon a \to c$ and $g \colon b \to d$. 
This definition almost precisely agrees with \cite[Def.~2.1.1]{MR3204665}; the only difference is that we are assuming strict unitality and we do not address size considerations explicitly in this paper.
A $\rmC^*$-tensor category $\cC$ is a \emph{$\rmW^*$-tensor category} if additionally the functor $- \otimes -$ is bi-normal~\cite{MR3687214}.

If $\cC$ and $\cD$ are monoidal categories, then we say that a functor $F \colon \cC \to \cD$ is \emph{monoidal} if there are natural \emph{tensorator} isomorphisms $F^2_{a,b} \colon F(a \otimes b) \to F(a) \otimes F(b)$ and a \emph{unitor} isomorphism $F^0 \colon F(\mathds1_\cC) \to \mathds1_\cD$ satisfying the usual coherence conditions. We will sometimes consider monoidal functors that are \emph{strict}, meaning that the tensorators and unitors are trivial. A \emph{tensor functor} between tensor categories is a linear monoidal functor. If $\cC$ and $\cD$ are $\rmC^*$-tensor categories, we will want to consider \emph{dagger tensor functors}, that is, tensor functors that are dagger functors and such that the tensorator and unitor isomorphisms are unitaries. 
Furthermore, if $\cC$ and $\cD$ are $\rmW^*$-tensor categories, we will want to consider \emph{normal} dagger tensor functors.
If $F,G$ are tensor functors, then the natural transformation $\eta\colon F \Rightarrow G$ is a \emph{monoidal natural transformation} if $(\eta_a\otimes \eta_b)\circ F^2_{a,b}=G^2_{a,b}\circ \eta_{a\otimes b}$ and $G^0\circ \eta_{\mathds1}=F^0$. A \emph{monoidal natural equivalence} is a monoidal natural transformation whose components are isomorphisms. 
A (normal dagger) tensor equivalence is a (normal dagger) tensor functor that is also an equivalence.
We will use the term \emph{$\rmW^*$-tensor equivalence} to denote a normal dagger tensor equivalence.

\subsubsection{\texorpdfstring{$G$}{G}-crossed braided structure}
We now define the notion of $G$-crossed monoidal categories and $G$-crossed braided monoidal categories. If $\cA, \cB$ are subcategories of a category $\cC$, we say that $\cA$ and $\cB$ are \emph{disjoint} if $\Hom{a \to b} = \{0\}$ for every $a \in \cA$ and $b \in \cB$. 
For a (finite) set $S$, we say that a category $\cC$ is \emph{$S$-graded} if $\cC = \bigoplus_{s \in S} \cC_s$, where $\{\cC_s\}_{s \in S}$ is a collection of mutually disjoint subcategories of $\cC$. 
That is, we require that every object $a \in \cC$ is of the form $a = \bigoplus_{s \in G} a_s$, where $a_s \in \cC_g$. We we say that $a \in \cC$ is \emph{homogeneous} if $a \in \cC_s$ for some $s \in S$, and we let $\cC_{\hom}$ denote the full subcategory of homogeneous elements of $\cC$.
If $\cC$ is also a $\rmC^*/\rmW^*$-category, then $\cC_s$ is a $\rmC^*/\rmW^*$-subcategory of $\cC$ for all $s \in S$. Following \cite{MR2183964}, we let $\partial \colon \cC_{\hom} \to S$ be the map (called the \emph{grading}) defined by $\partial a \coloneqq s$ if $a \in \cC_s$.

\begin{defn}
\label{def:G-graded monoidal}
A category $\cC$ is a \emph{$G$-graded monoidal category} if $\cC$ is a $G$-graded category that is monoidal for which $\mathds{1} \in \cC_{\hom}$ and the grading $\partial\colon \cC_{\hom}\rightarrow G$ obeys $\partial(a \otimes b) = \partial a \partial b$ for all $a, b \in \cC_{\hom}$. If $\cC$ is additionally strict as a monoidal category, then it is called a $G$-graded strict monoidal category. 
\end{defn}

\begin{rem}
Note that if $\cC$ is a $G$-graded monoidal category, then $\mathds{1} \in \cC_1$. 
Indeed, we have that $\partial \mathds{1} = \partial (\mathds{1} \otimes \mathds{1}) = \partial \mathds{1} \partial \mathds{1}$, so $\partial \mathds{1} = 1$. 
Furthermore, if $\cC$ is a tensor category, then the condition $\mathds{1} \in \cC_{\hom}$ is automatically satisfied since $\mathds{1}$ is simple. 
\end{rem}

We now define the notion of a $G$-crossed braided monoidal category. 

\begin{defn}[{\cite[\textsection 3.1]{galindo2017coherence}}]
\label{def:G-crossed monoidal general v2}

A \emph{$G$-crossed monoidal category} is a $G$-graded monoidal category $\cC$ together with a monoidal functor
\[
\gamma \colon {G} \to \Aut_\otimes(\cC), \qquad g \mapsto \gamma_g,
\]
called the \emph{$G$-action}, such that $\gamma_g(\cC_h) \subseteq \cC_{ghg^{-1}}$ for all $g,h \in G$. Here $\Aut_\otimes(\cC)$ denotes the monoidal natural equivalences of $\cC$ to itself. 
Following \cite{MR2183964, galindo2017coherence}, 
we let ${}^g a \coloneqq \gamma_g(a)$ for $g \in G$ and $a \in \cC$, and for $f \colon a \to b$ in $\cC$, we let ${}^g f \coloneqq \gamma_g(f)$.
We denote the \emph{$G$-action tensorator} by
\[
(\gamma^2_g)_{a,b}\colon \gamma_g(a \otimes b) \to \gamma_g(a)\otimes \gamma_g(b),
\]
the \emph{$G$-action unitor} by
\[
(\gamma^0_g)\colon \gamma_g(\mathds1)\to \mathds1,
\]
and the \emph{$G$-action compositor} by
\[
(\gamma_{g,h})_a\colon \gamma_{gh}(a)\to \gamma_g(\gamma_h(a)).
\]
We also have a monoidal natural isomorphism
\[
\gamma_0 \colon \gamma_1 \Rightarrow \Id_\cC,
\qquad
(\gamma_0)_a \colon {}^1 a \to a .
\]

In order for $\gamma \colon G \to \Aut_\otimes (\cC)$ to be a monoidal functor, the coherence isomorphisms must satisfy the following constraints (c.f.~\cite[Def.~2.1]{delaney2024g}):
\begin{enumerate}
\item {(Monoidality of $\gamma_g$)}
For all $a, b, c \in \cC$ and $g \in G$,
$$
\alpha_{{}^g a,{}^g b,{}^g c}\circ\big((\gamma_g^2)_{a,b}\otimes \Id_{{}^g c}\big)\circ (\gamma_g^2)_{a\otimes b,c} = (\Id_{{}^g a}\otimes (\gamma_g^2)_{b,c})\circ (\gamma_g^2)_{a,b\otimes c}\circ {}^g(\alpha_{a,b,c}).
$$

\item {(Coherence of compositors)}
For all $a \in \cC$ and $g, h, k \in G$,
\[
{}^g\big((\gamma_{h,k})_a\big)\circ (\gamma_{g,hk})_a = (\gamma_{g,h})_{{}^k a}\circ (\gamma_{gh,k})_a.
\]

\item {(Monoidality of the compositor)}
For all $a, b \in \cC$ and $g, h \in G$,
\[
\big((\gamma_g^2)_{{}^h a,{}^h b}\big)\circ {}^g\big((\gamma_h^2)_{a,b}\big)\circ (\gamma_{g,h})_{a\otimes b} = \big((\gamma_{g,h})_a\otimes (\gamma_{g,h})_b\big)\circ (\gamma_{gh}^2)_{a,b}.
\]

\item {(Unit coherence for $\gamma_g$)}
For all $a \in \cC$ and $g \in G$,
\[
(\gamma_g^0 \otimes \Id_{{}^g a}) \circ (\gamma_g^2)_{\mathds1,a} = \Id_{{}^g a}, \qquad (\Id_{{}^g a} \otimes \gamma_g^0) \circ (\gamma_g^2)_{a,\mathds1} = \Id_{{}^g a}.
\]

\item {(Unit coherence for the $G$-action)}
For all $a\in \cC$ and $g\in G$,
\[
{}^g\big((\gamma_0)_a\big)\circ (\gamma_{g,1})_a = \Id_{{}^g a}, \qquad (\gamma_0)_{{}^g a}\circ (\gamma_{1,g})_a = \Id_{{}^g a}.
\]
\end{enumerate}

We say that $\cC$ is a \emph{$G$-crossed tensor category} if $\cC$ is additionally a tensor category. 
We say that $\cC$ is a \emph{$G$-crossed $\rmC^*$-tensor category} if $\cC$ is additionally a $\rmC^*$-tensor category, $\gamma_g$ is a dagger monoidal functor for all $g \in G$, and all coherence isomorphisms are unitaries. 
Finally, we say that $\cC$ is a \emph{$G$-crossed $\rmW^*$-tensor category} if $\cC$ is additionally a $\rmW^*$-tensor category and $\gamma_g$ is a normal dagger functor for all $g \in G$.
\end{defn}

\begin{defn}[{\cite[Def.~8.24.1]{MR3242743}}]
\label{def:G-crossed braided general v2}
A $G$-crossed braided monoidal category is a $G$-crossed monoidal category $\cC$ together with a collection of isomorphisms
\[
c_{a,b}: a\otimes b \to {}^{\partial a}b\otimes a
\]
for $a\in \cC_{\hom}$ and $b\in \cC$ that satisfy the following coherence conditions:
\begin{itemize}
\item (Naturality) For all homogeneous $a,b\in \cC_{\hom}$, all $f_1\colon a\to b$, and all $f_2\colon c\to d$ in $\cC$,
\[
({}^{\partial a}f_2 \otimes \Id_a)\circ c_{a,c} = c_{a,d}\circ (\Id_a\otimes f_2), \qquad\qquad (\Id_{{}^{\partial a}c}\otimes f_1)\circ c_{a,c} = c_{b,c}\circ (f_1\otimes \Id_c).
\]

\item
(Equivariance of the $G$-crossed braiding) For $t\in G$, homogeneous $a\in \cC_g$, and $b\in \cC$,
\[
\Big(\big((\gamma_{tgt^{-1},t})_b\circ (\gamma_{t,g})_b^{-1}\big)\otimes \Id_{{}^t a}\Big) \circ (\gamma_t^2)_{{}^g b,a} \circ {}^t(c_{a,b}) = c_{{}^t a,{}^t b}\circ (\gamma_t^2)_{a,b}.
\]

\item (First $G$-crossed hexagon axiom) For homogeneous $x\in \cC_k$ and $a,b\in \cC$,
\[
\big((\gamma_k^2)_{a,b}^{-1}\otimes \Id_x\big)\circ \alpha^{-1}_{{}^k a,{}^k b,x}\circ (\Id_{{}^k a}\otimes c_{x,b})\circ \alpha_{{}^k a,x,b}\circ (c_{x,a}\otimes \Id_b)\circ \alpha^{-1}_{x,a,b} = c_{x,a\otimes b}.
\]

\item (Second $G$-crossed hexagon axiom)
For homogeneous $a\in \cC_g$, $b\in \cC_h$, and $c\in \cC$,
\[
\big((\gamma_{g,h})_c^{-1}\otimes \Id_{a\otimes b}\big)\circ \alpha_{{}^g({}^h c),a,b}\circ (c_{a,{}^h c}\otimes \Id_b)\circ \alpha^{-1}_{a,{}^h c,b}\circ (\Id_a\otimes c_{b,c})\circ \alpha_{a,b,c} = c_{a\otimes b,c}.
\]
\end{itemize}
We call $\cC$ a $G$-crossed braided $(\rmC^*/\rmW^*)$-tensor category if $\cC$ is additionally a $G$-crossed $(\rmC^*/\rmW^*)$-tensor category. In the case of $G$-crossed braided $\rmC^*/\rmW^*$-tensor categories, we require that the isomorphisms $c_{a, b}$ are unitaries.
\end{defn}

\begin{rem}
\label{rem:StrictGCrossed}
    We observe that if $\cC$ is strict monoidal and the $G$-action is strict (i.e.\ $\gamma_g^2$, $\gamma_g^0$, and $\gamma_{g,h}$, $\gamma_0$ are identities), then Definitions \ref{def:G-crossed monoidal general v2} and \ref{def:G-crossed braided general v2} reduce to \cite[Def.~2.9]{MR2183964} and \cite[Def.~2.16]{MR2183964}, respectively. 
\end{rem}

\begin{defn}[{\cite[Def.~2.6]{delaney2024g}}]
\label{def:G_crossed_equivalence}
Let $\cC,\cD$ be $G$-crossed braided monoidal categories with $G$-crossed braidings $c^\cC,c^\cD$ and $G$-actions $\gamma_g$ as above.
A \emph{$G$-crossed monoidal functor} from $\cC$ to $\cD$ consists of a monoidal functor $(F,F^2,F^0)\colon \cC\to \cD$, and a monoidal natural isomorphism for each $g\in G$
    \[
    \varepsilon_g\colon \gamma_g^\cD\circ F \Rightarrow F\circ \gamma_g^\cC, \qquad (\varepsilon_g)_a\colon {}^gF(a)\to F({}^ga).
    \]
such that the following hold:
\begin{itemize}
    \item (Preservation of grading) $F(\cC_g)\subseteq \cD_g$ for all $g\in G$,.
    \item (Coherence with $G$-action compositors) For all $g,h\in G$ and all $a\in \cC$,
    \[
    F\big((\gamma^\cC_{g,h})_a\big)\circ (\varepsilon_{gh})_a = (\varepsilon_g)_{{}^h a}\circ {}^g\big((\varepsilon_h)_a\big)\circ (\gamma^\cD_{g,h})_{F(a)}.
    \]
    \item (Unit coherence for $\varepsilon$) For all $a\in \cC$,
    \[
    (\varepsilon_1)_a = F\big((\gamma^\cC_0)_a\big)^{-1}\circ (\gamma^\cD_0)_{F(a)}.
    \]
\end{itemize}
The functor $F$ is said to be \emph{braided} if for all homogeneous $a\in \cC_{\hom}$ and all $b\in \cC$ we have
\[
F^2_{{}^{\partial a} b,a}\circ F\left(c^\cC_{a,b}\right) = \big((\varepsilon_{\partial a})_b\otimes \Id_{F(a)}\big)\circ c^\cD_{F(a),F(b)}\circ F^2_{a,b}.
\]
If $F$ is additionally an equivalence of categories, then it is a \emph{$G$-crossed braided monoidal equivalence}.
\end{defn}

If $\cC, \cD$ are $G$-crossed (braided) tensor categories, then a \emph{$G$-crossed (braided) tensor functor} $\cC \to \cD$ is a linear $G$-crossed (braided) monoidal functor. 
If $\cC, \cD$ are $G$-crossed (braided) $\rmC^*$-tensor categories, then a \emph{dagger $G$-crossed (braided) tensor functor} $\cC \to \cD$ is $G$-crossed (braided) tensor functor that is also a dagger tensor functor for which $(\varepsilon_g)_a$ is a unitary for all $a \in \cC$ and $g \in G$. 
Finally, if $\cC, \cD$ are $G$-crossed (braided) $\rmW^*$-tensor categories, then we will want to consider dagger $G$-crossed (braided) tensor functors $\cC \to \cD$ that are normal. 
As before, we will term a normal dagger $G$-crossed (braided) tensor functor a \emph{$G$-crossed (braided) $\rmW^*$-tensor equivalence}.

By \cite[Thm.~1.1]{galindo2017coherence}, every $G$-crossed (braided) monoidal category is $G$-crossed (braided) monoidal equivalent to a strict $G$-crossed (braided) monoidal category. 
We make the following remark related to this result.

\begin{rem}
We remark that the symmetry fractionalization data as described in \cite{PhysRevB.100.115147} is invariant under $G$-crossed braided monoidal equivalence. Thus even in the case of a strict presentation of $\cC$, the fractionalization data is not lost and can be recovered using an approach similar to the one used in \cite{2306.13762} to compute $F$- and $R$-symbols for anyon representations. 
We illustrate this computation in Section \ref{sec:SymmetryFractionalization}. 
\end{rem}

\subsubsection{G-crossed braided structure on $\Vect(G,\nu)$}
\label{sec:VecG}

Let $G$ be a finite group and let $\nu\in Z^3(G,U(1))$ be a normalized $3$-cocycle. We briefly outline the standard $G$-crossed braided $\rmW^*$-tensor structure on the pointed category $\Vect(G,\nu)$.

The category $\Vect(G,\nu)$ has as its objects the finite-dimensional $G$-graded Hilbert spaces
\[
a=\bigoplus_{g\in G} a_g.
\]
Let $a_g$ denote the $g$-graded component of $a$. A morphism $f\colon a\to b$ is a grading-preserving linear map (i.e., $f = \oplus_{g} f_g$ and $f_g\colon a_g \to b_g$ for all $g\in G$). The $(-)^*$ operation is the Hilbert space adjoint $f \mapsto f^*$ and the norm is taken to be the operator norm. The simple objects of $\Vect(G,\nu)$ are given by $\{\bbC_g\}_{g \in G}$, where $(\bbC_g)_h \simeq \delta_{g,h} \bbC$. For each $g\in G$, let $\Vect(G,\nu)_g$ be the full subcategory of all objects $a \in \Vect(G,\nu)$ such that $a_h =0$ for all $h \neq g$. Then
\[
\Vect(G,\nu)=\bigoplus_{g\in G}\Vect(G,\nu)_g,
\]
and for $a\in\Vect(G,\nu)_g$, we have $\partial a=g$.

Note that $\Vect(G, \nu)$ is orthogonally Cauchy complete. 
Also, for every $a \in \Vect(G, \nu)$, we have that $\End(a) \cong \bigoplus_{g \in G} B(a_g)$, which is a von Neumann algebra. 
Therefore, by \cite[Lem.~2.6]{MR808930}, $\Vect(G, \nu)$ is a $\rmW^*$-category.

We now elaborate on the various structures on $\Vect(G,\nu)$:

\begin{itemize}
\item (Monoidal structure) The tensor product is defined by
\[
(a\otimes b)_g \coloneqq \bigoplus_{hk=g} a_h\otimes b_k.
\]
In particular, we have $\bbC_g \otimes \bbC_h \simeq \bbC_{gh}$. The unit is $\mathds1=\bbC_1$ lying in the component $\Vect(G,\nu)_1$. We take the unitors to be trivial since $\nu$ is normalized. The associator is defined on homogeneous elements $x\in a_g$, $y\in b_h$, $z\in c_k$ by
\[
\alpha_{a,b,c}\big((x\otimes y)\otimes z\big)=\nu(g,h,k) \cdot x\otimes (y\otimes z).
\]
The pentagon equation holds precisely because $\nu$ is a $3$-cocycle. The associator is unitary since $\nu$ lies in $U(1)$. Since the tensor product is bi-normal and $\mathds{1}$ is simple, $\Vect(G,\nu)$ is a $\rmW^*$-tensor category.\footnote{In fact, $\Vect(G, \nu)$ is a unitary fusion category.}

\item ($G$-action) Following \cite[Ex.~2.4]{delaney2024g},\footnote{The example in \cite{delaney2024g} uses a right $G$-action, whereas here we use a left $G$-action to match Definition \ref{def:G-crossed monoidal general v2}, following \cite{galindo2017coherence}. } we define the functor $\gamma_t\colon \Vect(G,\nu)\to \Vect(G,\nu)$ for each $t\in G$ as follows:
\[
(\gamma_t(a))_g \coloneqq a_{t^{-1}gt}, \qquad\text{so that}\qquad \gamma_t\big(\Vect(G,\nu)_h\big)\subseteq \Vect(G,\nu)_{tht^{-1}}.
\]
Additionally $\gamma_t$ applied to a morphism gives the same underlying linear map. Thus, on simple objects,
\[
{}^t\bbC_g\coloneqq\gamma_t(\bbC_g)=\bbC_{tgt^{-1}}.
\]
Note that the functors $\gamma_t$ satisfy that $\gamma_t \circ \gamma_s = \gamma_{ts}$ and $\gamma_1 = \Id_{\Vect(G, \nu)}$.

To make this into a monoidal $G$-action compatible with the $\nu$-twisted associator, one must equip the functors $\gamma_g$ with nontrivial tensorators and compositors in general. We take the $G$-action unitors $\gamma_g^0, \gamma_0$ to be identities. On simple homogeneous objects, the $G$-action tensorators $(\gamma_t^2)_{\bbC_g,\bbC_h}$
and compositors $(\gamma_{r,s})_{\bbC_g}$ are given by
\[
(\gamma_t^2)_{\bbC_g,\bbC_h} = \frac{\nu(g,h,t^{-1})\nu(t^{-1},tgt^{-1},tht^{-1})} {\nu(g,t^{-1},tht^{-1})}  \Id_{\bbC_{tgh t^{-1}}},
\]
and
\[
(\gamma_{r,s})_{\bbC_g} = \frac{\nu(s^{-1},sgs^{-1},r^{-1})} {\nu(s^{-1},r^{-1},rsgs^{-1}r^{-1})\nu(g,s^{-1},r^{-1})}  \Id_{\bbC_{rsgs^{-1}r^{-1}}}.
\]
For arbitrary objects, these maps are defined componentwise with respect to the grading decomposition. Explicitly, if $x\in a_g$ and $y\in b_h$, then
\[
(\gamma_t^2)_{a,b}(x\otimes y) = \frac{\nu(g,h,t^{-1})\nu(t^{-1},tgt^{-1},tht^{-1})} {\nu(g,t^{-1},tht^{-1})} (x\otimes y),
\]
viewed as a vector in ${}^t a\otimes {}^t b$, and if $x\in a_g$, then
\[
(\gamma_{r,s})_a(x) = \frac{\nu(s^{-1},sgs^{-1},r^{-1})} {\nu(s^{-1},r^{-1},rsgs^{-1}r^{-1})\nu(g,s^{-1},r^{-1})} x,
\]
viewed as a vector in ${}^r({}^s a)$.

\item ($G$-crossed braiding) With respect to the $G$-action $\gamma_g$ and the above choice of
$G$-action tensorators and compositors, the canonical $G$-crossed braiding on $\Vect(G,\nu)$ is given as follows:
For homogeneous $a\in \Vect(G,\nu)_g$ and $b=\oplus_{h} b_h$, the unitary $c_{a,b}$ is defined componentwise by the ordinary Hilbert-space flip into the regraded target:
\[
c_{a,b_h}\colon a\otimes b_h \longrightarrow {}^g b_h\otimes a,
\qquad
x\otimes y \longmapsto y\otimes x,
\]
where ${}^g b_h$ denotes the same underlying Hilbert space as $b_h$, now placed in component $ghg^{-1}$. 
\end{itemize}

In summary, these structures make $\Vect(G,\nu)$ into a $G$-crossed braided $\rmW^*$-tensor category.
Different choices of the cocycle $\nu$ can yield inequivalent $\rmW^*$-tensor categories. 
However, once the monoidal structure is specified using the cocycle $\nu$, the remaining parts of the $G$-crossed braided monoidal structure are uniquely specified up to $G$-crossed braided monoidal equivalence (see for instance \cite[Cor.~7.7]{jones2022extension}).

\section{Useful results for the Levin-Gu SPT}

We recall the automorphism $\alpha \colon \cstar \to \cstar$ from Definition \ref{def:LevinGuEntangler}.

\begin{lem}
\label{lem:alpha_theta is a bounded spread automorphism}
    The automorphism $\alpha$ is a finite depth quantum circuit. 
    In particular, $\alpha$ is a quasi-factorizable QCA with spread $s = 1$.
\end{lem}
\begin{proof}

We group the triangles in $\Gamma$ into elementary hexagons that tile the entire plane.  
Note that this tiling of elementary hexagons can be colored using three colors, which we take to be red, blue, and green. An example is shown in Figure \ref{fig:example tiling of lattice into Hexagons}. We let $\scrR, \scrB, \scrG$ denote the collections of red, blue, and green hexagons, respectively.  
We now define $\cU^1$, $\cU^2$, and $\cU^3$ to be the following collections of unitaries: 
\[
\cU^1
\coloneqq
\left\{\prod_{\triangle \subseteq H} U_{\triangle} : H \in \scrR \right\},
\quad
\cU^2
\coloneqq
\left\{\prod_{\triangle \subseteq H} U_{\triangle} : H \in \scrB \right\},
\quad
\cU^3
\coloneqq
\left\{\prod_{\triangle \subseteq H} U_{\triangle} : H \in \scrG \right\}.
\]
Note that for $d = 1, 2, 3$, if $U_1, U_2 \in \cU^d$ with $U_1 \neq U_2$, then $\supp(U_1) \cap \supp(U_2) = \emptyset$.  
Furthermore, each $U \in \bigcup_{d = 1}^3 \cU^d$ acts only on a collection of seven vertices in an elementary hexagon.
Therefore, the collection $\{\cU^1, \cU^2, \cU^3\}$ defines a depth 3 quantum circuit.  
If we define $\widehat{\alpha}_d \colon \cstar\to \cstar$ by 
\[
\widehat{\alpha}_d(A)
=
\Ad\!\left(\prod_{U \in \cU^d}U \right)\!(A)
\]
for $A \in \cstar^{\loc}$, then we have that for $A \in \cstar^{\loc}$
\[
\alpha^{-1}(A)
=
\Ad\!\left(\prod_{\triangle \subseteq \Gamma} U_{\triangle}\right)\!(A)
=
\widehat{\alpha}_3 \circ \widehat{\alpha}_2 \circ \widehat{\alpha}_1(A).
\]
Thus, $\alpha^{-1}$ is a finite depth quantum circuit, and hence $\alpha$ is as well.

\begin{figure}[!ht]
    \centering
    \includegraphics[width=0.2\linewidth]{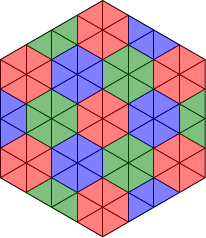}
    \caption{An example tiling of the triangular lattice into red, blue, green Hexagons. $\cU^1, \cU^2, \cU^3$ are supported on the red, blue, green colored triangles respectively. Each unitary $U$ on the hexagon $H$ centered at vertex $v$ is supported on $v^{+1}$.}
    \label{fig:example tiling of lattice into Hexagons}
\end{figure}

By Lemmas \ref{lem:FDQCBoundedSpread} and \ref{lem:FDQCQuasi-Factorizable}, $\alpha$ is a quasi-factorizable QCA.
To see that $\alpha$ has spread $1$, note that for $A \in \fA^{\loc}$, we have that $\alpha(A)$ is the conjugation of $A$ by commuting unitaries that act on triangles.
Since every vertex on a face is distance 1 away from every other vertex, any vertex in the support $\alpha(A)$ is distance at most 1 from a vertex in $\supp(A)$.  
Thus $\supp(\alpha(A)) \subseteq \supp(A)^{+1}_\Gamma$.
\end{proof}

\begin{lem}
    \label{lem:automorphism connecting trivial para GS and Levin Gu GS}
    For all $v \in \Gamma$, we have that $$\alpha(\sigma^x_v) = B_v,$$
    In particular, this implies that $$\alpha(H^0) = H.$$ This is the result of \cite[Appendix A]{PhysRevB.86.115109} in the infinite volume setting.
\end{lem}
\begin{proof}
Let $\triangle_v$ be the set of triangles containing vertex $v \in \Gamma$. We denote $<vpq>$ to explicitly refer to a face $<vpq>\in \triangle_v$ with vertices $v,p,q \in \Gamma$.
Since the notation becomes quite involved, we identify each face $\triangle$ with the set of vertices in the face $\triangle$.
    \begin{align*}
        \alpha(\sigma^x_v) &= \prod_{\mathclap{\triangle \in \triangle_v}} e^{-\frac{i\pi}{24}(3 \prod_{v' \in \triangle} \sigma^z_{v'} - \sum_{v' \in \triangle} \sigma^z_{v'})} \sigma_v^x \prod_{\mathclap{\triangle \in \triangle_v}}e^{\frac{i\pi}{24}(3 \prod_{v \in \triangle} \sigma^z_v - \sum_{v \in \triangle} \sigma^z_v)}\\
        &= \sigma_v^x \prod_{\mathclap{<vpq> \in \triangle_v}} e^{\frac{i\pi}{8} \sigma^z_v \sigma^z_p \sigma^z_q + \frac{i\pi}{24}( -\sigma^z_v + \sigma^z_p + \sigma^z_q) + \frac{i\pi}{8} \sigma^z_v \sigma^z_p \sigma^z_q + \frac{i\pi}{24}( -\sigma^z_v - \sigma^z_p - \sigma^z_q)}\\
        &= \sigma_v^x \prod_{\mathclap{\triangle \in \triangle_v}} e^{\frac{i\pi}{4} \prod_{v' \in \triangle} \sigma^z_{v'} - \frac{i\pi}{12}\sigma^z_v} = \sigma_v^x  i^{(- \sigma^z_v)} \prod_{\mathclap{<vpq> \in \triangle_v}} i^{\frac{1}{2}(\sigma^z_v \sigma^z_p \sigma^z_q)}\\
        \intertext{In the previous equality we used the identity $e^{i\pi/2} = i$ and that there are 6 faces having $v$ as a vertex, so $\sum_{{\triangle \in \triangle_v}} \frac{i\pi}{12}\sigma^z_v = \frac{i\pi}{2}\sigma^z_v $. }
        &= \sigma_v^x  i^{3 \sigma^z_v} \prod_{\mathclap{<vpq> \in \triangle_v}} i^{\frac{1}{2}(\sigma^z_v \sigma^z_p \sigma^z_q)} = \sigma_v^x   \prod_{\mathclap{<vpq> \in \triangle_v}} i^{\frac{1}{2}(\sigma^z_v + \sigma^z_v \sigma^z_p \sigma^z_q)} = \sigma_v^x   \prod_{\mathclap{<vpq> \in \triangle_v}} i^{ \frac{1}{2}\sigma^z_v(1+ \sigma^z_p \sigma^z_q)} = \sigma^x_v M,
        \intertext{where $M =  \prod_{{<vpq> \in \triangle_v}} i^{ \frac{1}{2}\sigma^z_v(1+ \sigma^z_p \sigma^z_q)}$.
        We claim that $M = \prod_{{<vpq> \in \triangle_v}} i^{ \frac{1}{2}(1+ \sigma^z_p \sigma^z_q)}$.
        Indeed, since the eigenvalues of $\sigma^z_v, \sigma^z_p, \sigma^z_q$ are $\pm 1$, the eigenvalues of $i^{ \frac{1}{2}\sigma^z_v(1+ \sigma^z_p \sigma^z_q)}$ are exactly the same as the eigenvalues of $i^{ \frac{1}{2}(1+ \sigma^z_p \sigma^z_q)}$ (with exactly the same eigenvectors).
        Therefore, $i^{ \frac{1}{2}\sigma^z_v(1+ \sigma^z_p \sigma^z_q)} = i^{ \frac{1}{2}(1+ \sigma^z_p \sigma^z_q)}$, so $M = \prod_{{<vpq> \in \triangle_v}} i^{ \frac{1}{2}(1+ \sigma^z_p \sigma^z_q)}$. 
        Additionally, $M = M^{-1}$, since $i^{ \frac{1}{2}(1+ \sigma^z_p \sigma^z_q)}$ has eigenvalues $\pm 1$ and thus $M$ has eigenvalues $\pm 1$.
        This can be thought of as a gauge redundancy. 
        Using this fact we get}
        &= \sigma^x_v M^{-1} =  \sigma_v^x   \prod_{\mathclap{<vpq> \in \triangle_v}} i^{- \frac{1}{2}(1+ \sigma^z_p \sigma^z_q)} = \sigma^x_v i^{-3} \prod_{\mathclap{<vpq> \in \triangle_v}} i^{-\frac{1}{2}\sigma^z_p \sigma^z_q}\\
        &= -\sigma^x_v i^{3} \prod_{\mathclap{<vpq> \in \triangle_v}} i^{-\frac{1}{2}\sigma^z_p \sigma^z_r} = -\sigma^x_v \prod_{\mathclap{<vpq> \in \triangle_v}} i^{\frac{1}{2}(1-\sigma^z_p \sigma^z_r)} = B_v
    \end{align*}
    The statement of the lemma trivially follows from this result, since the Hamiltonians are a summation of these individual terms.

    The statement $\alpha(H_S^0) = H_S$ now trivially follows for all $S \in \Gamma_f$.
\end{proof}

We recall the automorphism $\alpha^\gamma$ for $\gamma \in \bar P_R(\Gamma)$ defined in Section \ref{sec:Defect auts Hamiltonian Levin-Gu}, as well as the representations $\tilde \pi$ and $\tilde \pi_\gamma$ defined in Section \ref{sec:Defect sectors in the Levin-Gu SPT}.
We further recall the notation $V(\gamma)$ and $\widehat{V}(\gamma)$ from Notation \ref{nota:VerticesInDualPath}.

\begin{lem}
\label{lem:LGGSDefectStateInequivalent}
The representations $\tilde \pi$ and $\tilde \pi_\gamma$ are not unitarily equivalent
(equivalently, $\tilde \omega \not \simeq \tilde \omega_\gamma$).
\end{lem}
\begin{proof}
Note that since $\tilde \omega$ and $\tilde \omega_{\gamma}$ are pure states, $\tilde \omega$ and $\tilde \omega_{\gamma}$ are unitarily equivalent if and only if they are quasi-equivalent \cite[Prop.~10.3.7]{MR1468230}.
We can therefore apply \cite[Cor.~2.6.11]{MR887100}. 
Let $S \in \Gamma_f$.  
Then since $\gamma$ is a half-infinite dual path, there exists $v \in \widehat{V}(\gamma)$ such that every $\triangle \in \triangle_v$ satisfies that $\triangle \subseteq V^c$.  
Note that the last condition implies that $\supp(\widehat{B}_v) \subseteq V^c$.  
We let 
\[
\triangle_v^\gamma 
\coloneqq
\{<vqq'> \in \triangle_v : \text{$\gamma$ intersects the edge between $q$ and $q'$} \}.
\]
We therefore have that 
\begin{align*}
\widehat{B}_v
&=
-\sigma^x_v 
\prod_{<v q q'> \in \triangle_v^\gamma} i^{\frac{1 + \sigma^z_q \sigma^z_{q'}}{2}}
\prod_{<v q q'> \in \triangle_v \setminus \triangle_v^\gamma} i^{\frac{1 - \sigma^z_q \sigma^z_{q'}}{2}}
\\&=
-\sigma^x_v 
\prod_{<v q q'> \in \triangle_v^\gamma} i^{\sigma^z_q \sigma^z_{q'}} i^{\frac{1 - \sigma^z_q \sigma^z_{q'}}{2}}
\prod_{<v q q'> \in \triangle_v \setminus \triangle_v^\gamma} i^{\frac{1 - \sigma^z_q \sigma^z_{q'}}{2}}
\\&=
B_v \prod_{<v q q'> \in \triangle_v^\gamma}i^{\sigma^z_q \sigma^z_{q'}}.
\end{align*}
Therefore, since $\tilde \omega(B_v) = 1$ and $B_v \leq \mathds1$, we have by Lemma \ref{lem:can freely insert and remove P from the ground state.} that 
\[
\tilde\omega(\widehat{B}_v)
=
\tilde\omega(B_v\widehat{B}_v)
=
\tilde\omega\!\left(B_v^2 \prod_{<v q q'> \in \triangle_v^\gamma}i^{\sigma^z_q \sigma^z_{q'}}\right)
=
\tilde\omega\!\left(\prod_{<v q q'> \in \triangle_v^\gamma}i^{\sigma^z_q \sigma^z_{q'}}\right),
\]
where in the last step we used that $B_v^2 = 1$.  

Now, since $\gamma \in \bar{P}_R(\Gamma)$, we may assume that there exists $p \in \Gamma$ such that $p$ is contained in exactly one face $\triangle \coloneqq <vpp'> \in \triangle_v^\gamma$ (if such a $p \in \Gamma$ does not exist, then there is a different choice for $v$ for which such a $p$ does exist).
Therefore, we have that 
\begin{align*}
\tilde\omega(\widehat{B}_v)
&=
\tilde\omega\!\left(\prod_{<v q q'> \in \triangle_v^\gamma}i^{\sigma^z_q \sigma^z_{q'}}\right)
=
\tilde\omega\!\left(B_p \prod_{<v q q'> \in \triangle_v^\gamma}i^{\sigma^z_q \sigma^z_{q'}}B_p\right)
\\&=
\tilde\omega\!\left(
\sigma_p^x 
\prod_{<p r r'> \in \triangle_p} i^{\frac{1 - \sigma^z_r \sigma^z_{r'}}{2}} i^{\sigma^z_p \sigma^z_{p'}}
\prod_{<v q q'> \in \triangle_v^\gamma \setminus \{\triangle\}}i^{\sigma^z_q \sigma^z_{q'}} 
\sigma_p^x 
\prod_{<p r r'> \in \triangle_p} i^{\frac{1 - \sigma^z_r \sigma^z_{r'}}{2}} 
\right)
\\&=
\tilde\omega\!\left(
i^{-\sigma^z_p \sigma^z_{p'}}
\prod_{<v q q'> \in \triangle_v^\gamma \setminus \{\triangle\}}i^{\sigma^z_q \sigma^z_{q'}} 
\sigma_p^x 
\prod_{<p r r'> \in \triangle_p} i^{\frac{1 - \sigma^z_r \sigma^z_{r'}}{2}}
\sigma_p^x 
\prod_{<p r r'> \in \triangle_p} i^{\frac{1 - \sigma^z_r \sigma^z_{r'}}{2}} 
\right)
\\&=
\tilde\omega\!\left(
i^{-\sigma^z_p \sigma^z_{p'}}
\prod_{<v q q'> \in \triangle_v^\gamma \setminus \{\triangle\}}i^{\sigma^z_q \sigma^z_{q'}} 
B_p^2
\right)
=
\tilde\omega\!\left(
i^{-\sigma^z_p \sigma^z_{p'}}
\prod_{<v q q'> \in \triangle_v^\gamma \setminus \{\triangle\}}i^{\sigma^z_q \sigma^z_{q'}} 
\right).
\end{align*}
Now, since the eigenvalues of $\sigma^z_p$ are $1$ and $-1$, we have that $i^{-\sigma^z_p \sigma^z_{p'}} = (i^{\sigma^z_p \sigma^z_{p'}})^{-1} = -i^{\sigma^z_p \sigma^z_{p'}}$.
Therefore, we have that 
\begin{align*}
\tilde\omega(\widehat{B}_v)
&=
\tilde\omega\!\left(
i^{-\sigma^z_p \sigma^z_{p'}}
\prod_{<v q q'> \in \triangle_v^\gamma \setminus \{\triangle\}}i^{\sigma^z_q \sigma^z_{q'}} 
\right)
=
\tilde\omega\!\left(
-i^{\sigma^z_p \sigma^z_{p'}}
\prod_{<v q q'> \in \triangle_v^\gamma \setminus \{\triangle\}}i^{\sigma^z_q \sigma^z_{q'}} 
\right)
\\&=
-\tilde\omega\!\left(\prod_{<v q q'> \in \triangle_v^\gamma}i^{\sigma^z_q \sigma^z_{q'}}\right)
=
-\tilde\omega(\widehat{B}_v).
\end{align*}
Thus, $\tilde\omega(\widehat{B}_v) = 0$.  
However, $\tilde\omega_\gamma(\widehat{B}_v) = 1$.  
Therefore, we have that 
\[
|\tilde\omega(\widehat{B}_v) - \tilde\omega_\gamma(\widehat{B}_v)|
=
1
=
\|\widehat{B}_v\|,
\]
so by \cite[Cor.~2.6.11]{MR887100}, $\tilde\omega \not \simeq \tilde \omega_\gamma$.
\end{proof}

\begin{rem}
A similar argument to the proof of Lemma \ref{lem:LGGSDefectStateInequivalent} can be used to show the following result. 
Suppose $\gamma^1$ and $\gamma^2$ are self-avoiding half-infinite dual paths such that there exists $N \in \bbN$ such that $\gamma^1_i \neq \gamma^2_j$ for every $i, j \geq N$. 
Then $\tilde \omega_{\gamma^1} \not \simeq \tilde\omega_{\gamma^2}$, where the definition of $\tilde \omega_{\gamma^i}$ is analogous to the definition of $\omega_\gamma$ for $\gamma \in \bar P_R(\Gamma)$.
\end{rem}

\section{\texorpdfstring{$F$}{F}-symbols for Levin-Gu SPT using \texorpdfstring{$G$}{G}-defect automorphisms}
\label{sec:LGDefectSectorComputationAppendix}
Our aim in this section is to compute the $F$-symbols for the Levin-Gu SPT explicitly.
To do this, we will explicitly construct for the Levin-Gu SPT the $g$-defect automorphism $\alpha_\gamma^g$ from Definition \ref{def:defect automorphisms} with $g \in \bbZ_2$ being the non-trivial element.

Let $\gamma \in \bar P_R(\Gamma)$ and let $L_\gamma$ be a dividing completion of $\gamma$.
Then $L_\gamma$ divides $\Gamma$ into two halves, denoted by $r(L_\gamma), \ell(L_\gamma)$ following the convention in Section \ref{sec:paths and dual paths}. 
In what follows, we assume $(L_\gamma)_{dual} \subset \bbR^2$ is a straight line such that $R \subset (L_\gamma)_{dual}$ (see Figure \ref{fig:ProductOfBvTermsLevinGu}).
We first write down the automorphism $\tilde \beta_g^{r(L_\gamma)} = \alpha \circ \beta_g^{r(L_\gamma)} \circ \alpha^{-1}$. We can compute for all $A \in \cstar[loc]$ the following expression $\tilde \beta_g^{r(L_\gamma)}(A)$. Using the explicit form of $\alpha, \beta_g^{r(L_\gamma)}$ for the Levin-Gu SPT we get
\[
\tilde \beta_g^{r(L_\gamma)}(A) 
= 
\left(\prod_{v \in r(L_\gamma)} B_v\right)A\left(\prod_{v \in r(L_\gamma)} B_v\right)^*.
\]
Observe that $\tilde \beta_g^{r(L_\gamma)}$ extends to a well-defined automorphism of $\cstar$. 

 Let $A \in \cstar[loc]$. We consider a hexagon $S\subseteq r(L_\gamma)$ such that one of the sides of the hexagon lies along $(L_\gamma)_{dual}$ (see Figure \ref{fig:ProductOfBvTermsLevinGu}) and take $S$ to be large enough so that 
\[\tilde \beta_g^{r(L_\gamma)}(A)
=
\left(\prod_{v \in S} B_v\right)A\left(\prod_{v \in S} B_v\right)^*.
\]
We now compute the above expression. 
To do so, we notice that the lattice $\Gamma$ is tripartite.  
See Figure \ref{fig:ProductOfBvTermsLevinGu} to see the tripartite structure as well as the hexagon $S$ considered.
We let $a, b, c$ denote the labels of the vertices in $\Gamma$ according to the tripartite structure, and for $j = a, b, c$, we let $S_j \subseteq S$ be the collection of vertices labeled by $j$.  
Additionally, for $j = a, b, c$, we let $Q_j \coloneqq \prod_{v \in S_j} B_v$.  
Therefore, we obtain that 
\[
\tilde \beta_g^{r(L_\gamma)}(A)
=
(Q_aQ_bQ_c)A(Q_aQ_bQ_c)^*.
\]

\begin{figure}[!ht]
\centering
\begin{tikzpicture}[scale=0.7, rotate=90]
    \foreach \x in {0,...,4}{
    \foreach \y in {0,...,5}{
    \filldraw[draw=red,thick,fill=red!100] (\x*3,2*\y) circle(.1cm);
    \filldraw[draw=cyan,thick,fill=cyan!100] (\x*3+1,2*\y) circle(.1cm);
     \filldraw[draw=orange,thick,fill=orange!100] (\x*3+2,2*\y) circle(.1cm);
     \filldraw[draw=red,thick,fill=red!100] (\x*3+1.5,2*\y+1) circle(.1cm);
     \filldraw[draw=cyan,thick,fill=cyan!100] (\x*3+2.5,2*\y+1) circle(.1cm);
     \filldraw[draw=orange,thick,fill=orange!100] (\x*3+.5,2*\y+1) circle(.1cm);
}}
\draw[thick,red](3.5,9)--(4,10)--(5,10)--(5.5,9)--(6.5,9)--(7,10)--(8,10)--(8.5,9)--(9.5,9)--(10,10)--(11,10)--(11.5,9)--(11,8)--(11.5,7)--(12.5,7)--(13,6)--(12.5,5)--(11.5,5)--(11,4)--(11.5,3)--(11,2)--(10,2)--(9.5,3)--(8.5,3)--(8,2)--(7,2)--(6.5,3)--(5.5,3)--(5,2)--(4,2)--(3.5,3)--(4,4)--(3.5,5)--(2.5,5)--(2,6)--(2.5,7)--(3.5,7)--(4,8)--(3.5,9);
\draw[thick,cyan](4.5,9)--(5,10)--(6,10)--(6.5,9)--(7.5,9)--(8,10)--(9,10)--(9.5,9)--(10.5,9)--(11,8)--(12,8)--(12.5,7)--(12,6)--(12.5,5)--(12,4)--(11,4)--(10.5,3)--(9.5,3)--(9,2)--(8,2)--(7.5,3)--(6.5,3)--(6,2)--(5,2)--(4.5,3)--(3.5,3)--(3,4)--(3.5,5)--(3,6)--(3.5,7)--(3,8)--(3.5,9)--(4.5,9);
\draw[thick,orange](4.5,9)--(5.5,9)--(6,10)--(7,10)--(7.5,9)--(8.5,9)--(9,10)--(10,10)--(10.5,9)--(11.5,9)--(12,8)--(11.5,7)--(12,6)--(11.5,5)--(12,4)--(11.5,3)--(10.5,3)--(10,2)--(9,2)--(8.5,3)--(7.5,3)--(7,2)--(6,2)--(5.5,3)--(4.5,3)--(4,4)--(3,4)--(2.5,5)--(3,6)--(2.5,7)--(3,8)--(4,8)--(4.5,9);
\draw[thick,black,dashed](2.5,6)--(4.25,9.5)--(10.75,9.5)--(12.5,6)--(10.75,2.5)--(4.25,2.5)--(2.5,6);
\draw[thick, violet, dashed](0, 9.5) -- (14.5, 9.5);
\node[violet] at (15, 9.5) {$(L_\gamma)_{dual}$};
\draw[thick, dashed, brown] (6.75, 10.5) -- (6,8.5) -- (6.5, 8.5) -- (6.5, 7.5) -- (6, 7.5) -- (6, 6.5) -- (10.25, 6.5) -- (8.25, 10.5) -- (6.75, 10.5);
\node[brown] at (7.5, 8) {\footnotesize $ \supp(A)$};
\end{tikzpicture}
\caption{A hexagon $V$ to the right of the line $(L_\gamma)_{dual}$ over which we take the product of $B_v$ terms.
The hexagon is taken to be large enough that $\supp(A)$ does not intersect the corners of the hexagon.
The colors for the vertices and the edges correspond to the labels $a, b, c$.}
\label{fig:ProductOfBvTermsLevinGu}
\end{figure}

We now compute $Q_j$ for each $j = a, b, c$.  
We observe that
\[
Q_j
=
\prod_{v \in S_j}B_v
=
\prod_{v \in S_j}\prod_{<vqq'> \in \triangle_v}-\sigma^x_v i^{\frac{1 - \sigma^z_q\sigma^z_{q'}}{2}}.
\]
Now, if $v \in S_j$ and $<vqq'> \in \triangle_v$, then $q, q'$ correspond to the other labels (not $j$).
Therefore, we have that 
\[
Q_j
=
\prod_{v \in S_j}\prod_{<vqq'> \in \triangle_v}-\sigma^x_v i^{\frac{1 - \sigma^z_q\sigma^z_{q'}}{2}}
=
Z_j X_j,
\]
where $X_j \coloneqq \prod_{v \in S_j} \sigma^x_v$ and $Z_j$ is some function of $\sigma^z_v$ for $v \in \Gamma$ corresponding to the non-$j$ labels.  
We compute $Z_j$ by considering every edge $qq'$ such that $<vqq'> \in \triangle_v$ for some $v \in S_j$.
There are two cases to consider. 
First, suppose $qq'$ is an edge labeled by the color corresponding to $j$ in Figure \ref{fig:ProductOfBvTermsLevinGu}.  
Then $i^{\frac{1 - \sigma^z_q\sigma^z_{q'}}{2}}$ shows up as a factor in $B_v$ for exactly one $v \in S_j$.
Now, suppose $qq'$ is an edge between two non-$j$ vertices in $S$ that does not lie along the boundary of $S$, so that $qq'$ is not a colored edge in Figure \ref{fig:ProductOfBvTermsLevinGu}.
Then $i^{\frac{1 - \sigma^z_q\sigma^z_{q'}}{2}}$ shows up in $B_v$ for two different $v \in S_j$.
We observe that 
\[
\left(i^{\frac{1 - \sigma^z_q\sigma^z_{q'}}{2}}\right)^2
=
i^{1 - \sigma^z_q\sigma^z_{q'}}
=
i(i^{-1})^{\sigma^z_q\sigma^z_{q'}}.
\]
Since the eigenvalues of $\sigma^z_q\sigma^z_{q'}$ are $\pm 1$, we have that 
\[
\left(i^{\frac{1 - \sigma^z_q\sigma^z_{q'}}{2}}\right)^2
=
i(i^{-1})^{\sigma^z_q\sigma^z_{q'}}
=
i(i^{-1})\sigma^z_q\sigma^z_{q'}
=
\sigma^z_q\sigma^z_{q'}.
\]
We therefore have that 
\[
Z_j 
=
\prod_{v \in S_{\widehat{j}}} \sigma^z_v \prod_{qq' \in {L_\gamma}_{\widehat{j}}} i^{\frac{1 - \sigma^z_q\sigma^z_{q'}}{2}},
\]
where ${L_\gamma}_{\widehat{j}}$ is the path corresponding to $j$ illustrated in Figure \ref{fig:ProductOfBvTermsLevinGu} and $V_{\widehat{j}}$ is the collection of non-$j$ vertices in $V$ that have an edge $e$ to another non-$j$ vertex in $V$ where $e \notin {L_\gamma}_{\widehat{j}}$.

Using this computation, we have that 
\[
\tilde \beta_g^{r(L_\gamma)}(A)
=
(Q_aQ_bQ_c)A(Q_aQ_bQ_c)^*
=
(Z_aX_aZ_bX_bZ_cX_c)A(Z_aX_aZ_bX_bZ_cX_c)^*.
\]
We compute the quantity on the right.  
We first observe that 
\[
X_a Z_b
=
\prod_{v \in S_a} \sigma^x_v \prod_{v \in S_{\widehat{b}}} \sigma^z_v \prod_{qq' \in {L_\gamma}_{\widehat{b}}} i^{\frac{1 - \sigma^z_q\sigma^z_{q'}}{2}}
=
(-1)^k \prod_{v \in S_{\widehat{b}}} \sigma^z_v \prod_{qq' \in {L_\gamma}_{\widehat{b}}} i^{\frac{1 - (-1)^{\varepsilon_a}\sigma^z_q\sigma^z_{q'}}{2}} \prod_{v \in S_a} \sigma^x_v,
\]
where $k \in \bbN$ and $\varepsilon_a = 1$ if one of $q, q' \in V_a$ and $\varepsilon_a = 0$ otherwise.  
Note that when we conjugate by the above operator, the factor of $(-1)^k$ cancels.  
Similarly, we have that
\[
X_aX_bZ_c
=
\prod_{v \in S_a \cup S_c} \sigma^x_v \prod_{v \in S_{\widehat{c}}} \sigma^z_v \prod_{qq' \in {L_\gamma}_{\widehat{c}}} i^{\frac{1 - \sigma^z_q\sigma^z_{q'}}{2}}
=
(-1)^k \prod_{v \in S_{\widehat{c}}} \sigma^z_v \prod_{qq' \in {L_\gamma}_{\widehat{c}}} i^{\frac{1 - (-1)^{\varepsilon_s}\sigma^z_q\sigma^z_{q'}}{2}}
\prod_{v \in S_a \cup S_c} \sigma^x_v,
\]
where again $k \in \bbN$ and $\varepsilon_s = 1$ if exactly one of $q, q' \in V$ and $\varepsilon_s = 0$ otherwise.
We then have that 
\begin{align*}
\tilde \beta_g^{r(L_\gamma)}(A)
&=
(Z_aX_aZ_bX_bZ_cX_c)A(Z_aX_aZ_bX_bZ_cX_c)^*
\\&=
\Ad\!\left(
\prod_{v \in S_{\widehat{a}}} \sigma^z_v \prod_{v \in S_{\widehat{b}}} \sigma^z_v \prod_{v \in S_{\widehat{c}}} \sigma^z_v 
\prod_{qq' \in {L_\gamma}_{\widehat{a}}} i^{\frac{1 - \sigma^z_q\sigma^z_{q'}}{2}} 
\prod_{qq' \in {L_\gamma}_{\widehat{b}}} i^{\frac{1 - (-1)^{\varepsilon_a}\sigma^z_q\sigma^z_{q'}}{2}} 
\prod_{qq' \in {L_\gamma}_{\widehat{c}}} i^{\frac{1 - (-1)^{\varepsilon_s}\sigma^z_q\sigma^z_{q'}}{2}}
\right)\! \circ \beta_g^{r({L_\gamma})}(A).
\end{align*}

Now, because $S$ is large relative to the support of $A$, we can ignore effects that occur at the corners of the hexagon $S$ and simplify the above expression.  
In particular, we have that 
\[
\Ad\!\left(\prod_{v \in S_{\widehat{a}}} \sigma^z_v \prod_{v \in S_{\widehat{b}}} \sigma^z_v \prod_{v \in S_{\widehat{c}}} \sigma^z_v \right)\!(A')
=
\Ad\!\left(\prod_{v \in \partial S} \sigma^z_v\right)\!(A'),
\]
where $\partial S$ is the collection of vertices along the boundary of $S$.  
Here $A'$ is defined by 
\[
A'
\coloneqq
\Ad\!\left(
\prod_{qq' \in {L_\gamma}_{\widehat{a}}} i^{\frac{1 - \sigma^z_q\sigma^z_{q'}}{2}} 
\prod_{qq' \in {L_\gamma}_{\widehat{b}}} i^{\frac{1 - (-1)^{\varepsilon_a}\sigma^z_q\sigma^z_{q'}}{2}} 
\prod_{qq' \in {L_\gamma}_{\widehat{c}}} i^{\frac{1 - (-1)^{\varepsilon_s}\sigma^z_q\sigma^z_{q'}}{2}}
\right)\! \circ \beta_g^{r({L_\gamma})}(A)
\]
Therefore, we have that 
\[
\tilde \beta_g^{r(L_\gamma)}(A)
=
\Ad\!\left(
\prod_{v \in \partial S} \sigma^z_v
\prod_{qq' \in {L_\gamma}_{\widehat{a}}} i^{\frac{1 - \sigma^z_q\sigma^z_{q'}}{2}} 
\prod_{qq' \in {L_\gamma}_{\widehat{b}}} i^{\frac{1 - (-1)^{\varepsilon_a}\sigma^z_q\sigma^z_{q'}}{2}} 
\prod_{qq' \in {L_\gamma}_{\widehat{c}}} i^{\frac{1 - (-1)^{\varepsilon_s}\sigma^z_q\sigma^z_{q'}}{2}}
\right)\! \circ \beta_g^{r(L_\gamma)}(A).
\]
Now, we can replace $S$ with $r({L_\gamma})$ in the above equation, and we can also redefine ${L_\gamma}_{\widehat{a}}, {L_\gamma}_{\widehat{b}}, {L_\gamma}_{\widehat{c}}$ to refer to their continuations along the path ${L_\gamma}$.
We then have that for any $A \in \cstar^{\loc}$, 
\begin{align*}
    \tilde \beta_g^{r(L_\gamma)}(A)
&=
\Ad\!\left(
\prod_{v \in \partial r({L_\gamma})} \sigma^z_v
\prod_{qq' \in {L_\gamma}_{\widehat{a}}} i^{\frac{1 - \sigma^z_q\sigma^z_{q'}}{2}} 
\prod_{qq' \in {L_\gamma}_{\widehat{b}}} i^{\frac{1 - (-1)^{\varepsilon_a}\sigma^z_q\sigma^z_{q'}}{2}} 
\prod_{qq' \in {L_\gamma}_{\widehat{c}}} i^{\frac{1 - (-1)^{\varepsilon_s}\sigma^z_q\sigma^z_{q'}}{2}}
\right)\! \circ \beta_g^{r({L_\gamma})}(A)
\end{align*}

Let $\xi$ be the half-infinite dual path defined by $\xi_k \coloneqq (L_\gamma)_{1 - k}$.
Note that $\xi \in \bar P(\Gamma)$ since $L_\gamma$ is a completion of $\gamma$. 
For $j = a, b, c$, we define $\xi_{\widehat{j}}, \gamma_{\widehat{j}}$ analogously to ${L_\gamma}_{\widehat{j}}$ above (see Figure \ref{fig:DecorationForLGDefectSectorTripartiteColor}). Now we note that $$\tilde \beta_g^{r(L_\gamma)} \circ (\beta_g^{r(L_\gamma)})^{-1}(A) = \eta^\gamma \otimes \eta^\xi(A)$$ where

\begin{align*}
    \eta^\xi(A) &\coloneqq  \Ad\!\left(
\prod_{v \in \partial r(\xi)} \sigma^z_v
\prod_{qq' \in \xi_{\widehat{a}}} i^{\frac{1 - \sigma^z_q\sigma^z_{q'}}{2}} 
\prod_{qq' \in \xi_{\widehat{b}}} i^{\frac{1 - (-1)^{\varepsilon_a}\sigma^z_q\sigma^z_{q'}}{2}} 
\prod_{qq' \in \xi_{\widehat{c}}} i^{\frac{1 - (-1)^{\varepsilon_s}\sigma^z_q\sigma^z_{q'}}{2}}
\right)\!(A)\\
\eta^\gamma(A) &\coloneqq  \Ad\!\left(
\prod_{v \in \partial r(\gamma)} \sigma^z_v
\prod_{qq' \in \gamma_{\widehat{a}}} i^{\frac{1 - \sigma^z_q\sigma^z_{q'}}{2}} 
\prod_{qq' \in \gamma_{\widehat{b}}} i^{\frac{1 - (-1)^{\varepsilon_a}\sigma^z_q\sigma^z_{q'}}{2}} 
\prod_{qq' \in \gamma_{\widehat{c}}} i^{\frac{1 - (-1)^{\varepsilon_s}\sigma^z_q\sigma^z_{q'}}{2}}
\right)\!(A)
\end{align*}
Now we get the defect automorphism $\alpha^\gamma \coloneqq  \alpha^g_\gamma = \eta^\xi \circ \beta_g^{r(L_\gamma)}$ (Definition \ref{def:defect automorphisms}) as
\[
\alpha^{\gamma}(A)=
\Ad\!\left(
\prod_{v \in \partial r(\xi)} \sigma^z_v
\prod_{qq' \in \xi_{\widehat{a}}} i^{\frac{1 - \sigma^z_q\sigma^z_{q'}}{2}} 
\prod_{qq' \in \xi_{\widehat{b}}} i^{\frac{1 - (-1)^{\varepsilon_a}\sigma^z_q\sigma^z_{q'}}{2}} 
\prod_{qq' \in \xi_{\widehat{c}}} i^{\frac{1 - (-1)^{\varepsilon_s}\sigma^z_q\sigma^z_{q'}}{2}}
\right)\! \circ \beta_g^{r(L_{\gamma})}(A)
\]
for $A \in \cstar^{\loc}$.
Here $\partial r(\xi)$ is the portion of $\partial r(L_{{\gamma}})$ along $\xi$.

\begin{figure}[!ht]
    \centering
    \begin{tikzpicture}[scale=0.6]
    \foreach \x in {0,...,4}{
    \foreach \y in {0,...,2}{
    \filldraw[draw=red,thick,fill=red!100] (2*\x,\y*3) circle(.1cm);
    \filldraw[draw=cyan,thick,fill=cyan!100] (2*\x,\y*3+1) circle(.1cm);
    \filldraw[draw=orange,thick,fill=orange!100] (2*\x,\y*3+2) circle(.1cm);
    \filldraw[draw=orange,thick,fill=orange!100] (2*\x+1,\y*3+.5) circle(.1cm);
    \filldraw[draw=red,thick,fill=red!100] (2*\x+1,\y*3+1.5) circle(.1cm);
    \filldraw[draw=cyan,thick,fill=cyan!100] (2*\x+1,\y*3+2.5) circle(.1cm);
}}
\draw[thick,red](5,5.5)--(4,5)--(4,4)--(5,3.5)--(5,2.5)--(4,2)--(4,1)--(5,.5)--(5,0);
\draw[thick,orange](5,5.5)--(5,4.5)--(4,4)--(4,3)--(5,2.5)--(5,1.5)--(4,1)--(4,0);
\draw[thick,cyan](4,5)--(5,4.5)--(5,3.5)--(4,3)--(4,2)--(5,1.5)--(5,.5)--(4,0);
\draw[thick,dashed](4.5,8.5)--(4.5,5);
\draw[thick, violet, dotted] (4.5, 5) -- (4.5, 0);
\end{tikzpicture}
    \caption{An illustration of the notation used to define $\alpha^\gamma$.
    The dashed black path is $\gamma$, and the dotted purple path is $\xi$.
    The red, blue, and orange colors for the vertices correspond to the labels $a$, $b$, and $c$, and the red, blue, and orange edges denote the paths $\xi_{\widehat{a}}, \xi_{\widehat{b}}, \xi_{\widehat{c}}$ respectively. 
    The region $\partial r(\xi)$ consists of those edges to the right of the dashed/dotted line with an adjacent colored edge.}
    \label{fig:DecorationForLGDefectSectorTripartiteColor}
\end{figure}

We now write down a simplified form of the formula for $\alpha^\gamma$.  
We define $N(\xi)$ to be the subgraph of $\Gamma$ consisting of all vertices in $\xi$ and edges between them. 
Note the vertices in $N(\xi)$ form two half-infinite paths on the lattice (not the dual lattice!).
We have that 
\[
\alpha^{{\gamma}}(A)
=
\Ad\!\left(
\prod_{v \in \partial r(\xi)} \sigma^z_v
\prod_{qq' \in N(\xi)} i^{\frac{1 - (-1)^{\varepsilon_{qq'}}\sigma^z_q\sigma^z_{q'}}{2}} 
\right)\! \circ \beta_g^{r(L_{{\gamma}})}(A)
\]
for $A \in \cstar^{\loc}$.
Here the precise value of $\varepsilon_{qq'} \in \{0, 1\}$ is determined by the table \eqref{eq:SignChartForLGDefectAutomorphism}: 
Letting $r(L_\gamma)_a$ denote the vertices in $r(L_\gamma)$ labeled by $a$, we have that 
\begin{equation}
\label{eq:SignChartForLGDefectAutomorphism}
\varepsilon_{q q'}
=
\begin{cases}
0 & \text{if } qq' \in \xi_{\widehat{a}}
\\
1 & \text{if } qq' \in \xi_{\widehat{b}} \text{ and one of } q, q' \in r(L_\gamma)_a,
\\
0 & \text{if } qq' \in \xi_{\widehat{b}} \text{ and } q, q' \notin r(L_\gamma)_a,
\\
1 & \text{if } qq' \in \xi_{\widehat{c}} \text{ and exactly one of } q, q' \in r(L_\gamma),
\\
0 & \text{otherwise.}
\end{cases}
\end{equation}

\begin{lem}
\label{lem:LevinGuCocycleComputation}
We have that $\alpha^{\gamma} \circ \alpha^{\gamma} = \Ad(\sigma^z_{\partial \xi_{\mathrm{in}}} \sigma^z_{\partial \xi_{\mathrm{out}}})$, where $\partial \xi_{\mathrm{in}}$ and $\partial \xi_{\mathrm{out}}$ are the endpoints of $\xi_{\mathrm{in}}$ and $\xi_{\mathrm{out}}$ respectively.
\end{lem}

\begin{proof}
It can be easily verified that 
\begin{align*}
\alpha^{{\gamma}} \circ \alpha^{{\gamma}}
&=
\Ad\!\left(
\prod_{v \in \partial r(\xi)} \sigma^z_v \beta_g^{r(L_{{\gamma}})}(\sigma^z_v)
\prod_{qq' \in N(\xi)} i^{\frac{1 - (-1)^{\varepsilon_{qq'}}\sigma^z_q\sigma^z_{q'}}{2}} 
\beta_g^{r(L_{{\gamma}})}\!\left(i^{\frac{1 - (-1)^{\varepsilon_{qq'}}\sigma^z_q\sigma^z_{q'}}{2}}\right)
\right)
\\&=
\Ad\!\left(
\prod_{v \in \partial r(\xi)} \sigma^z_v (-\sigma^z_v)
\prod_{qq' \in N(\xi)} i^{\frac{1 - (-1)^{\varepsilon_{qq'}}\sigma^z_q\sigma^z_{q'}}{2}} 
\cdot i^{\frac{1 - (-1)^{\varepsilon_{qq'}}(-1)^{\epsilon_s}\sigma^z_q\sigma^z_{q'}}{2}}
\right)
\\&=
\Ad\!\left(
\prod_{qq' \in N(\xi)} i^{\frac{1 - (-1)^{\varepsilon_{qq'}}\sigma^z_q\sigma^z_{q'}}{2}} 
\cdot i^{\frac{1 - (-1)^{\varepsilon_{qq'}}(-1)^{\epsilon_s}\sigma^z_q\sigma^z_{q'}}{2}}
\right).
\end{align*}
Here, $\epsilon_s = 1$ if exactly one of $q, q' \in r(L_\gamma)$ and $\epsilon_s = 0$ otherwise.
In particular, $\epsilon_s = 1$ if and only if $qq' \notin \xi_{\mathrm{in}}$ and $qq' \notin \xi_{\mathrm{out}}$.

Note that if $\varepsilon_s = 1$, then 
\[
i^{\frac{1 - (-1)^{\varepsilon_{qq'}}\sigma^z_q\sigma^z_{q'}}{2}} 
\cdot i^{\frac{1 - (-1)^{\varepsilon_{qq'}}(-1)^{\epsilon_s}\sigma^z_q\sigma^z_{q'}}{2}}
=
i^{\frac{1 - (-1)^{\varepsilon_{qq'}}\sigma^z_q\sigma^z_{q'}}{2}} 
\cdot i^{\frac{1 + (-1)^{\varepsilon_{qq'}} \sigma^z_q\sigma^z_{q'}}{2}}
=
i^{\frac{1 - {(\sigma^z_q)}^2 {(\sigma^z_{q'})}^2}{4}}
=
1.
\]
On the other hand, if $\varepsilon_s = 0$, then 
\[
i^{\frac{1 - (-1)^{\varepsilon_{qq'}}\sigma^z_q\sigma^z_{q'}}{2}} 
\cdot i^{\frac{1 - (-1)^{\varepsilon_{qq'}}(-1)^{\epsilon_s}\sigma^z_q\sigma^z_{q'}}{2}}
=
i^{\frac{1 - (-1)^{\varepsilon_{qq'}}\sigma^z_q\sigma^z_{q'}}{2}} 
\cdot i^{\frac{1 - (-1)^{\varepsilon_{qq'}}\sigma^z_q\sigma^z_{q'}}{2}}
=
i^{1 - (-1)^{\varepsilon_{qq'}}\sigma^z_q\sigma^z_{q'}}. 
\]
We therefore have that
\begin{align*}
\alpha^{{\gamma}} \circ \alpha^{{\gamma}}
&=
\Ad\!\left(
\prod_{qq' \in N(\xi)} i^{\frac{1 - (-1)^{\varepsilon_{qq'}}\sigma^z_q\sigma^z_{q'}}{2}} 
\cdot i^{\frac{1 - (-1)^{\varepsilon_{qq'}}(-1)^{\epsilon_s}\sigma^z_q\sigma^z_{q'}}{2}}
\right)
\\&=
\Ad\!\left(
\prod_{qq' \in \xi_{\mathrm{in}} \cup \xi_{\mathrm{out}}} i^{1 - (-1)^{\varepsilon_{qq'}}\sigma^z_q\sigma^z_{q'}} 
\right).
\end{align*}
Finally, we simplify $i^{1 - (-1)^{\varepsilon_{qq'}}\sigma^z_q\sigma^z_{q'}}$.
Observe that 
\[
i^{1 - (-1)^{\varepsilon_{qq'}}\sigma^z_q\sigma^z_{q'}} 
=
i\left( i^{(-1)^{\varepsilon_{qq'}+ 1}}\right)\sigma^z_q\sigma^z_{q'}
=
i\cdot i^{(-1)^{\varepsilon_{qq'}+ 1}} \sigma^z_q\sigma^z_{q'},
\]
where the last equality follows since the two operators have exactly the same eigenvalues and eigenvectors.  
Therefore, we have that 
\begin{align*}
\alpha^{{\gamma}} \circ \alpha^{{\gamma}}
&=
\Ad\!\left(
\prod_{qq' \in \xi_{\mathrm{in}} \cup \xi_{\mathrm{out}}} i^{1 - (-1)^{\varepsilon_{qq'}}\sigma^z_q\sigma^z_{q'}} 
\right)
=
\Ad\!\left(
\prod_{qq' \in \xi_{\mathrm{in}} \cup \xi_{\mathrm{out}}} i\cdot i^{(-1)^{\varepsilon_{qq'}+ 1}} \sigma^z_q\sigma^z_{q'}
\right)
\\&=
\Ad\!\left(
\prod_{qq' \in \xi_{\mathrm{in}} \cup \xi_{\mathrm{out}}} \sigma^z_q\sigma^z_{q'}
\right)
=
\Ad(\sigma^z_{\partial \xi_{\mathrm{in}}} \sigma^z_{\partial \xi_{\mathrm{out}}}).
\qedhere
\end{align*}
\end{proof}

\begin{lem}
\label{lem:F-symbols are cocycles Levin Gu}
    Using the notation of Section \ref{sec:OtherCoherenceData}, we have that the $F$-symbols of the Levin-Gu SPT are given by: \begin{align*}
        F(g,h,k) = \begin{cases}
            -1 & g=h=k \text{ and non-trivial}\\
            1 & \emph{otherwise}
        \end{cases}
    \end{align*}
\end{lem}
\begin{proof}
To compute the $F$-symbols, we must compute $\Omega_{g, h}$ (in the notation of Section \ref{sec:OtherCoherenceData}). Since $\tilde \pi$ is a strict tensor unit, it follows that $\Omega_{1,1} = \Omega_{1, g} = \Omega_{g, 1} = \mathds1$. This straightaway gives that all $F$-symbols except $F(g, g, g)$ are guaranteed to be $1$. We now must find $F(g,g,g)$, given by $$F(g, g, g) = \Omega_{1, g} \Omega_{g, g}(\Omega_{g, 1}\alpha^{{\gamma}}(\Omega_{g, g}))^{-1}.$$ 

From Lemma \ref{lem:LevinGuCocycleComputation} we see that $\Omega_{g, g} = \sigma^z_{\partial \xi_{\mathrm{in}}} \sigma^z_{\partial \xi_{\mathrm{out}}}$.
We now compute $\alpha^{{\gamma}}(\Omega_{g, g})$.  
The key observation is that $\partial \xi_{\mathrm{in}} \in r(L_{{\gamma}})$ but $\partial \xi_{\mathrm{out}} \notin r(L_{{\gamma}})$, so $\beta_g^{r(L_{{\gamma}})}(\sigma^z_{\partial \xi_{\mathrm{in}}} \sigma^z_{\partial \xi_{\mathrm{out}}}) = -\sigma^z_{\partial \xi_{\mathrm{in}}} \sigma^z_{\partial \xi_{\mathrm{out}}}$.
Therefore, we have that 
\begin{align*}
\alpha^{{\gamma}}(\Omega_{g, g})
&=
\Ad\!\left(
\prod_{v \in \partial r(\xi)} \sigma^z_v
\prod_{qq' \in N(\xi)} i^{\frac{1 - (-1)^{\varepsilon_{qq'}}\sigma^z_q\sigma^z_{q'}}{2}} 
\right)\! \circ \beta_g^{r(L_{{\gamma}})}
(\sigma^z_{\partial \xi_{\mathrm{in}}} \sigma^z_{\partial \xi_{\mathrm{out}}})
\\&=
\Ad\!\left(
\prod_{v \in \partial r(\xi)} \sigma^z_v
\prod_{qq' \in N(\xi)} i^{\frac{1 - (-1)^{\varepsilon_{qq'}}\sigma^z_q\sigma^z_{q'}}{2}} 
\right)\!
(-\sigma^z_{\partial \xi_{\mathrm{in}}} \sigma^z_{\partial \xi_{\mathrm{out}}})
=
-\sigma^z_{\partial \xi_{\mathrm{in}}} \sigma^z_{\partial \xi_{\mathrm{out}}}
=
-\Omega_{g, g}
\end{align*}
Hence $F(g, g, g) = -1$, giving us the result.
\end{proof}

\section{Toric Code with ancillary vertex spins}
\label{app:TC with ancillary vertex spins}

We now recall the Hamiltonian $H^0$ for this system to be given by $$H_S^0 \coloneqq  H_S^{TC} + \sum_{v \in V(S)} \dfrac{\mathds1 - \tau_v^x}{2}, \qquad\qquad S \in \Gamma_f$$
where $H_S^{TC} \in \cstar^E$ is the Toric Code Hamiltonian on $S$. Let $\delta^0$ be the corresponding derivation. It is easy to see that $H^0$ is still a commuting projector Hamiltonian. Let $\omega_0$ be a state on $\cstar$ defined by $$\omega_0 \coloneqq  \omega^E_{TC} \otimes \omega^V_0$$ where $\omega^E_{TC}$ (defined on $\cstar^E$) is the Toric Code frustration-free ground state and $\omega^V_0$ is defined on $\cstar^V$ as a product state given by $\omega^V_0(A) \coloneqq  \bigotimes_{v \in \Gamma} \inner{\psi_v}{A \psi_v}$ and $\ket{\psi_v} \in \hilb_v$ satisfies $\ket{\psi_v} = \tau_v^x \ket{\psi_v}$.

Then it is easy to see that $\omega_0 $ is a frustration-free ground state of $H^0$. In fact, we have the following Lemma.

\begin{lem}
\label{lem:TC extension FF GS is unique}
    The state $\omega_0$ is the unique state satisfying for all $v,f$ $$\omega_0(A_v) = \omega_0(B_f) = \omega_0(\tau_v^x) = 1$$
    In particular, this means that $\omega_0$ is pure.
\end{lem}
\begin{proof}
We first observe that $\omega_0$ satisfies the above equation.  
Indeed, $\omega_0 = \omega_{TC}^E \otimes \omega_0^V$, and for every $v, f$, we have that $\omega_{TC}^E(A_v) = \omega_{TC}^E(B_f) = 1$ and $\omega_0^V(\tau^x_v) = 1$.
Now, suppose $\omega \colon \cstar \to \bbC$ is a state satisfying that $\omega(A_v) = \omega(B_f) = \omega(\tau^x_v) = 1$ for all $v, f$.  
We first claim that $\omega = \omega^E \otimes \omega_0^V$ for some $\omega^E \colon \cstar^E \to \bbC$.  
Indeed, let $A \in \cstar^{\loc}$ be a simple tensor.  
Then $A = A^E \otimes A^V$ for some $A^E \in \cstar^E$ and $A^V \in \cstar^V$.  
Furthermore, since $A$ is a simple tensor, we have that $A^V = \bigotimes_{v \in \supp(A^V)} A^V_v$ for some $A^V_v \in \cstar[v]$.
For each vertex $v$, we define $P_v \coloneqq (\mathds1 + \tau^x_v)/2 \in \cstar[v]$.
Note that $P_v$ is a rank-1 projection, and $\omega(P_v) = 1$.
Using Lemma \ref{lem:can freely insert and remove P from the ground state.}, we then have that 
\[
\omega(A)
=
\omega(A^E \otimes A^V)
=
\omega\!\left(A^E \otimes \bigotimes_{v \in \supp(A^V)} P_v A^V_v P_v \right).
\]
Now, $P_vA^V_vP_v \in \bbC P_v$ for all $v \in \supp(A^V)$ since $P_v \in \cstar[v]$ is a rank-1 projection, so 
\[
\bigotimes_{v \in \supp(A^V)}P_vA^V_vP_v = \lambda \bigotimes_{v \in \supp(A^V)}P_v
\]
for some $\lambda \in \bbC$.
We therefore have that 
\[
\omega(A)
=
\omega\!\left(A^E \otimes \bigotimes_{v \in \supp(A^V)} P_v A^V_v P_v \right)
=
\lambda \omega\!\left(A^E \otimes \bigotimes_{v \in \supp(A^V)} P_v \right)
=
\lambda\omega(A^E)
=
\omega(A^E)\omega_0^V(A^V).
\]
Since the simple tensors span a dense subspace of $\cstar$, we get that $\omega = \omega^E \otimes \omega_0^V$ for some $\omega^E \in \cS(\cstar^E)$.  

It remains to show that $\omega^E = \omega^E_{TC}$.
However, this follows from Lemma \ref{lem:UniqueFrustrationFreeGSToricCode}.
\end{proof}

Define $\pi_0$ to be the GNS representation of $\omega_0$ and let $\pi_0^{TC}$ be the GNS representation of $\omega^E_{TC}$. Let also $\pi^V_0$ be the GNS representation of $\omega_0^V$.

Since $\omega^E_{TC}, \omega_0^V$ are both pure, it follows that $\omega_0$ is also pure. The corresponding GNS representations are all irreducible.

\begin{lem}
\label{lem:TC extension Haag duality}
    The representation $\pi_0$ satisfies Haag duality.
\end{lem}
\begin{proof}
    First, note that $\omega_0 = \omega^E_{TC} \otimes \omega_0^V$ and observe that $\pi_0 \simeq \pi_0^{TC} \otimes \pi_0^V$ by uniqueness of the GNS representation.
    In fact, without loss of generality, we may assume $\pi_0 = \pi_0^{TC} \otimes \pi_0^V$.
    Let $\Lambda$ be a cone.  
    Then since $\pi_0^{TC}$ and $\pi_0^V$ both satisfy Haag duality for cones, we have that
    \[
    \pi_0(\cstar[\Lambda])'
    =
    (\pi_0^{TC}(\cstar[\Lambda]^E) \otimes \pi_0^{V}(\cstar[\Lambda]^V))'
    =
    \pi_0^{TC}(\cstar[\Lambda]^E)' \otimes \pi_0^{V}(\cstar[\Lambda]^V)'
    =
    \pi_0^{TC}(\cstar[\Lambda^c]^E)'' \otimes \pi_0^{V}(\cstar[\Lambda^c]^V)''
    =
    \pi_0(\cstar[\Lambda^c])''.
    \qedhere
    \]
\end{proof}

\begin{lem}
\label{lem:TC extension anyon sectors bound}
There are at most four irreducible anyon sectors with respect to $\pi_0$.
\end{lem}
\begin{proof}
By Lemma \ref{lem:TC extension Haag duality}, $\pi_0$ satisfies Haag duality for cones.  
It is also easy to verify that $\pi_0$ satisfies the approximate split property for cones \cite[Def.~5.1]{MR2804555}.
This follows quickly from the fact that $\pi_0 \simeq \pi_0^{TC} \otimes \pi_0^V$, $\pi_0^{TC}$ satisfies the approximate split property, and $\pi_0^V$ satisfies the split property.
Therefore, by \cite[Thm.~3.6]{MR3135456}, the number of distinct irreducible anyon sectors can be bounded by computing the index of the following subfactor. 
Let $\Upsilon = \Lambda_1 \cup \Lambda_2$, where $\Lambda_1$ and $\Lambda_2$ are disjoint cones that are sufficiently far apart.  
Then the number of distinct irreducible anyon sectors is at most $[\pi_0(\cstar[\Upsilon^c])' : \pi_0(\cstar[\Upsilon])'']$.
Now, we have that 
\begin{gather*}
\pi_0(\cstar[\Upsilon])''
=
\pi_0^{TC}(\cstar[\Upsilon])'' \otimes \pi_0^{V}(\cstar[\Upsilon])'',
\\
\pi_0(\cstar[\Upsilon^c])'
=
\pi_0^{TC}(\cstar[\Upsilon^c])' \otimes \pi_0^{V}(\cstar[\Upsilon^c])'
=
\pi_0^{TC}(\cstar[\Upsilon^c])' \otimes \pi_0^{V}(\cstar[\Upsilon])'',
\end{gather*}
where the last step follows since $\pi_0^V$ is the GNS representation of a product state.
Therefore, applying \cite[Thm.~4.9]{MR3135456}, we have that 
\begin{align*}
[\pi_0(\cstar[\Upsilon^c])' : \pi_0(\cstar[\Upsilon])'']
&=
[\pi_0^{TC}(\cstar[\Upsilon^c])' \otimes \pi_0^{V}(\cstar[\Upsilon])'' : \pi_0^{TC}(\cstar[\Upsilon])'' \otimes \pi_0^{V}(\cstar[\Upsilon])'']
\\&= 
[\pi_0^{TC}(\cstar[\Upsilon^c])': \pi_0^{TC}(\cstar[\Upsilon])'']
=
4.
\qedhere
\end{align*}
\end{proof}

We now show that there are at least 4 irreducible anyon sectors with respect to $\pi_0$. 
To do this, we inherit the previously defined automorphisms of $\cstar^E$ given by $\alpha_\gamma^\epsilon, \alpha_{\bar \gamma}^m, \alpha^\psi_{\gamma, \bar \gamma}$ (Definition \ref{def:ToricCodeAnyonAutomorphisms}).

\begin{lem}
    \label{lem:TC extension anyon sectors constructed}
    Let $\zeta \in \{\Id, \alpha_\gamma^\epsilon, \alpha_{\bar \gamma}^m, \alpha^\psi_{\gamma, \bar \gamma}\}$ be an automorphism of $\cstar^E$. Then the representations $\pi^\zeta \coloneqq  (\pi_0^{TC} \circ \zeta) \otimes \pi_0^V$ are mutually disjoint anyon representations with respect to $\pi_0$.
\end{lem}
\begin{proof}
    The representation $\pi^\zeta$ is obviously irreducible, since $\pi_0^{TC}, \pi_0^V$ are both irreducible and $\zeta$ is an automorphism. We now check if it satisfies the superselection criterion. 
    To do so, we show that it is localized and transportable.
    
    We first check that it is localized in some cone $\Lambda$.
    Let $\Lambda$ be a cone containing $\gamma, \bar{\gamma}$. 
    We show that $\pi^\zeta$ is localized in $\Lambda$.
    It suffices to check that $\pi^\zeta(A) = \pi_0(A)$ for all simple tensors $A \in \cstar[\Lambda^c]$
    Let $A \in \cstar[\Lambda^c]$ be a simple tensor. We then have that $A = A^V \otimes A^E$ where $A^V \in \cstar[\Lambda^c]^V$ and $A^E \in \cstar[\Lambda^c]^E$, so we have that
    \begin{align*}
        \pi^\zeta(A) = (\pi_0^{TC} \circ \zeta) (A^E) \otimes \pi_0^V(A^V) = \pi_0^{TC} (A^E) \otimes \pi_0^V(A^V) = \pi_0 (A).
    \end{align*}
    Thus $\pi^\zeta$ is localized in $\Lambda$.

    We now check transportability.
    Let $\Lambda'$ be another cone. Since $\pi_0^{TC} \circ \zeta$ is transportable, there exists some automorphism $\zeta' \colon \cstar^E \to \cstar^E$ such that $\pi_0^{TC} \circ \zeta'$ is localized in $\Lambda'$ and $\pi_0^{TC} \circ \zeta' \simeq \pi_0^{TC} \circ \zeta$. 
    By the above argument, $(\pi_0^{TC} \circ \zeta') \otimes \pi_0^V$ is localized in $\Lambda'$, and 
    \[
    (\pi_0^{TC} \circ \zeta') \otimes \pi_0^V \simeq (\pi_0^{TC} \circ \zeta)  \otimes \pi_0^V = \pi^\zeta.
    \]
    Thus $\pi^\zeta$ is transportable.
    
    Finally, we show that the representations $\{\pi^\zeta\}_\zeta$ are mutually unitarily inequivalent. We consider the case $\zeta = \alpha_\gamma^\epsilon$ and $\zeta' = \alpha^\psi_{\gamma, \bar \gamma}$ as an example. The other cases proceed similarly. Let $\omega^\zeta \coloneqq  \omega_0 \circ \zeta$ and $\omega^{\zeta'} \coloneqq  \omega_0 \circ {\zeta'}$. We use corollary \cite[2.6.11]{MR887100} along with \cite[Prop.~10.3.7]{MR1468230}. Choose a finite simply connected region $S$. Then consider $F_C \in \cstar[S^c]$ for a big enough loop $C \in \Gamma_f$ such that $\partial \bar \gamma$ is in the interior of $C, \bar C$. We then have that $\omega^{\zeta'}(F_C) = 1$ while $\omega^\zeta(F_C) = -1$. Since for every chosen $S$, there exists a big enough loop $C$ such that $F_C \in \cstar[S^c]$, we have that $\pi^\zeta \not \simeq \pi^{\zeta'}$. 
    This shows the full result.
\end{proof}

\begin{cor}
    \label{cor:TC extension anyon sector classification}
    The representations $\{\pi^\zeta\}_\zeta$ defined in Lemma \ref{lem:TC extension anyon sectors constructed} comprise the distinct irreducible anyon sectors with respect to $\pi_0$.
    
\end{cor}

\begin{proof}
    This follows straightforwardly from Lemmas \ref{lem:TC extension anyon sectors bound} and \ref{lem:TC extension anyon sectors constructed}.
\end{proof}

In fact, we have the following stronger result.  

\begin{prop}
\label{prop:TC extension braided monoidal}
Given a cone $\Lambda$, the category $\DHR_{\pi_0}$ of anyon representations with respect to $\pi_0$ localized in $\Lambda$ is braided monoidally equivalent to the category $\DHR_{\pi_0^{TC}}$ of anyon representations with respect to $\pi_0^{TC}$ localized in $\Lambda$.
\end{prop}

\begin{proof}
By Lemma \ref{lem:TC extension Haag duality}, $\pi_0$ satisfies Haag duality, so anyon representations localized in $\Lambda$ have a canonical extension to the auxiliary algebra $\cstar^a$ \cite{MR660538, MR2804555}.
Now, the automorphisms we use to construct the anyon representations in Lemma \ref{lem:TC extension anyon sectors constructed} are exactly those used in \cite{MR2804555}.
Therefore, the category $\DHR_{\pi_0}$ of anyon representations with respect to $\pi_0$ localized in $\Lambda$ is the same as the category of anyon representations with respect to $\pi_0^{TC}$ constructed in \cite{MR2804555}.
\end{proof}

\section{Relating SET Toric Code \texorpdfstring{$G$}{G}-defect representations to Hamiltonian terms}
\label{sec:SETToricCodeDefectHamiltonian}

Recall the SET Toric Code model discussed in Section \ref{sec:SET_model}.
The goal of this section is to prove some results concerning the symmetry action on the terms of the SET Toric Code Hamiltonian. 
This will relate the analysis for the SET Toric Code to our analysis of SPTs in sections \ref{sec:Defect auts Hamiltonian Levin-Gu} and \ref{sec:defects using auts}.

\begin{lem}
\label{lem:action of half plane sym on SET terms}
    Choose an infinite self-avoiding dual path $\bar L$. We have, $$\beta_g^{r (\bar L)}(A_v) = A_v \qquad \beta_g^{r(\bar L)}(\tilde Q_v) = \tilde Q_v \qquad \beta_g^{r(\bar L)}(\tilde B_f) = \left(\prod_{v \in r(\bar L)\cap f}\prod_{e \ni v} i g(e,v) \sigma^x_e\right)\tilde B_f$$ where $g(e,v) = +1$ if $\partial_1 e = v$ and $g(e,v) = -1$ if $\partial_0 e = v$.
\end{lem}
\begin{proof}
    The first identity is obvious. For the second identity, we observe that,
    \begin{align*}
        \tilde Q_v &= \frac{\mathds1+A_v}{2}Q_v = \frac{\mathds1+A_v}{2} Q_v \frac{\mathds1+A_v}{2} = \frac{\mathds1+A_v}{2} \left(\tau_v^x i^{-\tau^z_v \sum_{e \ni v} f(e,v) \sigma^x_e/2}\right)\frac{\mathds1+A_v}{2}
        \intertext{Now, on the $+1$ eigenspace of $A_v$, we have that $\sum_{e \ni v} f(e,v) \sigma^x_e/2 $ has eigenvalues $\pm 2,0$. Therefore, we can drop $-\tau^z_v$ from the exponent. We thus get,}
        &= \frac{\mathds1+A_v}{2}\left(\tau_v^x i^{ \sum_{e \ni v} f(e,v) \sigma^x_e/2}\right)\frac{\mathds1+A_v}{2}.
    \end{align*}
    Now it is obvious from the above explicit form of $\tilde Q_v$ that $\beta_g^{r(\bar L)}(\tilde Q_v) = \tilde Q_v$.

    For the third identity, we rewrite $\tilde B_f$ as follows:
    \[
    \tilde B_f = \prod_{e \in f}i^{-\sigma^x_e (\tau^z_{\partial_1 e} - \tau^z_{\partial_0 e})/2} B_f = \prod_{v \in f} i^{-\tau^z_v(\sum_{e \ni v} g(e,v)\sigma^x_e)/2} B_f.
    \]
    We now observe that
    \begin{align*}
        \beta_g^{r(\bar L)}(\tilde B_f)&=  \prod_{v \in f; v \in r(\bar L)} i^{\tau^z_v(\sum_{e \ni v}g(e,v) \sigma^x_e)/2}\prod_{v \in f; v \notin r(\bar L)} i^{-\tau^z_v(\sum_{e \ni v} g(e,v)\sigma^x_e)/2} B_f\\
        &= \prod_{v \in f; v \in r(\bar L)} i^{\tau^z_v(\sum_{e \ni v}g(e,v) \sigma^x_e)} \tilde B_f
        \intertext{Now we note that $\sum_{e \ni v}g(e,v) \sigma^x_e$ has eigenvalues in $\{\pm 2, 0\}$ and thus we can drop $\tau^z_v$. Now,}
        &= \prod_{v \in f; v \in r(\bar L)} i^{\sum_{e \ni v}g(e,v) \sigma^x_e} \tilde B_f = \prod_{v \in r(\bar L) \cap f} \prod_{e \ni v} i g(e,v) \sigma^x_e \tilde B_f,
    \end{align*}
    where in the last equality we've used that if $A^2= 1$, then $i^A = iA$. This shows the result.
\end{proof}

Now recall the automorphisms $\alpha^\sigma_{\bar \gamma}$ defined in Section \ref{sec:SET_defects}.
Let $\bar L$ be a dividing completion of $\bar \gamma$.
The following lemma shows that the action of the symmetry along $r(\bar L)$ on $A_v, \tilde B_f, \tilde Q_v$ can be erased using the automorphism $\alpha^\sigma_{\bar \gamma}$ acting along a part of $\bar L$. 

\begin{lem}
\label{lem:erasure string erases SET}
    Let $\bar \gamma \in \bar P_R(\Gamma)$ be a half-infinite path and $\bar L$ a completion of $\bar \gamma$. 
    Let $\bar \eta \in \bar P(\Gamma)$ be the path defined by $\bar \eta_k = \bar L_{1 - k}$. Choose a cone $\Lambda$ such that $\bar \gamma$ is contained in $\Lambda$. Then for all sites $s$ and $C_s \in \{A_v, \tilde B_f, \tilde Q_v\}$ such that $\supp(C_s) \subset \Gamma \cap \Lambda^c$, we have that $$\alpha_{\bar \eta}^\sigma \circ \beta_g^{\bar L}(C_s) = C_s.$$
\end{lem}
\begin{proof}
    For all $s$ sufficiently far away from $\bar L$ this Lemma follows immediately from \ref{lem:SET Hamiltonian terms are symmetric}. 
    
    Now we note that for $A_v, \tilde Q_v$ having supports overlapping with $\bar L$, the result immediately follows from \ref{lem:action of half plane sym on SET terms} and the fact that $\alpha^\sigma_{\bar \eta}$ only consists of $\sigma^x_e$ terms. 
    
    All that remains is to check for the $\tilde B_f$ terms whose support overlaps with $\bar L$. We have from Lemma \ref{lem:action of half plane sym on SET terms} that $$\beta_g^{r(\bar L)}(\tilde B_f) = \left(\prod_{v \in r(\bar L)\cap f}\prod_{e \ni v,e\in f} i g(e,v) \sigma^x_e\right)\tilde B_f$$ where $g(e,v) = +1$ if $\partial_1 e = v$ and $g(e,v) = -1$ if $\partial_0 e = v$. Therefore,

    \begin{align*}
        \alpha_{\bar \eta}^\sigma \circ \beta_g^{\bar L}(\tilde B_f) &= \alpha^\sigma_{\bar \eta}\left(\left(\prod_{v \in r(\bar L)\cap f}\prod_{e \ni v,e\in f} i g(e,v) \sigma^x_e\right)\tilde B_f\right) = \left(\prod_{v \in r(\bar L)\cap f}\prod_{e \ni v,e\in f} i g(e,v) \sigma^x_e\right) \alpha^\sigma_{\bar \eta}(\tilde B_f)\\
        &= \left(\prod_{v \in r(\bar L)\cap f}\prod_{e \ni v,e\in f} i g(e,v) \sigma^x_e\right) \left(\prod_{e \in \bar L \cap f} e^{-i \pi p(e) \sigma^x_e/2}\right) \tilde B_f\\
        &= \left(\prod_{v \in r(\bar L)\cap f}\prod_{e \ni v,e\in f} i g(e,v) \sigma^x_e\right) \left(\prod_{e \in \bar L \cap f} (-i){ p(e) \sigma^x_e}\right) \tilde B_f = B_f,        
    \end{align*}
    where in the last equality comes from the fact that 
    $$\left(\prod_{v \in r(\bar L)\cap f}\prod_{e \ni v} i g(e,v) \sigma^x_e\right) = \left(\prod_{e \in \bar L \cap f} i{ p(e) \sigma^x_e}\right)$$ This identity may be verified by checking all cases of how the line $\bar L$ can intersect $f$. We have now shown the required result.
\end{proof}

Finally, we recall the defect state $\tilde \omega^\sigma_{\bar \gamma} = \tilde \omega \circ \tilde \alpha_{\bar \gamma}^\sigma$ for a dual path $\bar \gamma$, where $\tilde \alpha_{\bar \gamma}^\sigma$ is given in Definition \ref{def:SETToricCodeDefectAutomorphism}.

\begin{lem}
    Pick a dual path $\bar \gamma$. For all sites $s$ outside $\bar \gamma$, the state $\tilde \omega^\sigma_{\bar \gamma}$ looks like the ground state. Specifically, pick $C_s \in \{A_v, B_f, Q_v\}$. Then for all sites $s$ outside $\bar \gamma$, we have that $$\tilde \omega^\sigma_{\bar \gamma}(C_s) = 1$$
\end{lem}
\begin{proof}
    Follows immediately using the definition of $\tilde \alpha_{\bar \gamma}^\sigma$ and Lemmas \ref{lem:erasure string erases SET}, \ref{lem:SET toric code unique FF GS}.
\end{proof}

\end{appendix}

\addtocontents{toc}{\protect\setcounter{tocdepth}{0}}  

\section*{Conflict of interest}

The authors have no conflicts of interest to disclose.

\section*{Data availability}

This manuscript has no associated data. 

\addtocontents{toc}{\protect\setcounter{tocdepth}{1}}  

\bibliographystyle{alpha}
\bibliography{bibliography}
    
\end{document}